\documentclass{aa}

\usepackage[varg]{txfonts}

\usepackage{graphicx}
\usepackage{commath}
\usepackage{mathabx}
\usepackage{rotating}
\usepackage{multirow}
\usepackage{lscape}
\usepackage{booktabs}
\usepackage{natbib}
\bibpunct{(}{)}{;}{a}{}{,}

\usepackage[version=4]{mhchem}

\begin{document}

\title{
Measuring the ratio of the gas and dust emission radii of protoplanetary disks in the Lupus star-forming region
}

\author{E. Sanchis\inst{1,2}
  \and L. Testi\inst{1, 3, 4}
  \and A. Natta\inst{4, 5}
  \and S. Facchini\inst{1} 
  \and C. F. Manara\inst{1} 
  \and A. Miotello\inst{1} 
  \and B. Ercolano\inst{2, 3}
  \and Th. Henning\inst{6}
  \and T. Preibisch\inst{2}
  \and J. M. Carpenter\inst{7}
  \and I. de Gregorio-Monsalvo\inst{8, 9} 
  \and R. Jayawardhana\inst{10}
  \and C. Lopez\inst{11}
  \and K. Mu\v{z}i\'c\inst{12}
  \and I. Pascucci\inst{13, 14}
  \and A. Santamar\'ia-Miranda\inst{9}
  \and S. van Terwisga\inst{6}
  \and J. P. Williams\inst{15}
  }

\offprints{E. Sanchis, \email{esanchis@eso.org}}

\institute{European Southern Observatory, Karl-Schwarzschild-Strasse 2, D-85748 Garching bei M{\"u}nchen, Germany
  \and Universit{\"a}ts-Sternwarte, Ludwig-Maximilians-Universit{\"a}t M{\"u}nchen, Scheinerstrasse 1, D-81679 M{\"u}nchen, Germany
  \and Excellence Cluster Origins, Boltzmannstrasse 2, D-85748 Garching bei M{\"u}nchen, Germany
  \and INAF/Osservatorio Astrofisico di Arcetri, Largo E. Fermi 5, I-50125 Firenze, Italy
  \and School of Cosmic Physics, Dublin Institute for Advanced Studies, 31 Fitzwilliams Place, Dublin 2, Ireland
  \and Max Planck Institute for Astronomy, K{\"o}nigstuhl 17, D-69117, Heidelberg, Germany
  \and Joint ALMA Observatory, Av. Alonso de C\'ordova 3107, Vitacura, Santiago, Chile
  \and N\'ucleo Milenio Formaci\'on Planetaria - NPF, Universidad de Valpara\'iso, Av. Gran Breta\~na 1111,  Valpara\'iso, Chile
  \and European Southern Observatory, 3107, Alonso de C\'ordova, Santiago de Chile
  \and Department of Astronomy, Cornell University, Ithaca, NY 14853, USA
  \and Atacama Large Millimeter/Submillimeter Array, Joint ALMA Observatory, Alonso de C\'ordova 3107, Vitacura 763-0355, Santiago, Chile
  \and CENTRA, Faculdade de Ci\^{e}ncias, Universidade de Lisboa, Ed. C8, Campo Grande, 1749-016 Lisboa, Portugal
  \and Lunar and Planetary Laboratory, The University of Arizona, Tucson, AZ $85721$, USA
  \and Earths in Other Solar Systems Team, NASA Nexus for Exoplanet System Science, USA
  \and Institute for Astronomy, University of Hawaii, Honolulu, HI $96822$, USA
  }

\date{Received ... / Accepted ...}

\abstract{
We perform a comprehensive demographic study of the CO extent relative to dust of the disk population in the Lupus clouds, in order to find indications of dust evolution and possible correlations with other disk properties. 
We increase up to 42 the number of disks of the region with measured $R_{\mathrm{CO}}$ and $R_{\mathrm{dust}}$ from observations with the Atacama Large Millimeter/submillimeter Array (ALMA), based on the gas emission in the $\ce{^{12}CO}$ $J = 2-1$ rotational transition and large dust grains emission at $\sim 0.89$ mm. 
The CO integrated emission map is modeled with an elliptical Gaussian or Nuker function, depending on the quantified residuals; the continuum is fitted to a Nuker profile from interferometric modeling. 
The CO and dust sizes, namely, the radii enclosing a certain fraction of the respective total flux (e.g., $R_{68\%}$) are inferred from the modeling. 
The CO emission is more extended than the dust continuum, with a $R_{68\%}^{\mathrm{CO}}$/$R_{68\%}^{\mathrm{dust}}$ median value of 2.5, for the entire population and for a sub-sample with high completeness. 
6 disks, around $15\%$ of the Lupus disk population have a size ratio above 4. 
Based on thermo-chemical modeling, this value can only be explained if the disk has undergone grain growth and radial drift. These disks do not have unusual properties, spreading across the disk population's ranges of stellar mass ($M_{\star}$), disk mass ($M_{\mathrm{disk}}$), CO and dust sizes  ($R_{\mathrm{CO}}$, $R_{\mathrm{dust}}$), and mass accretion of the entire population. 
We search for correlations between the size ratio and $M_{\star}$, $M_{\mathrm{disk}}$, $R_{\mathrm{CO}}$ and $R_{\mathrm{dust}}$: only a weak monotonic anti-correlation with the $R_{\mathrm{dust}}$ is found, this would imply that dust evolution is more prominent in more compact dusty disks. 
The lack of strong correlations is remarkable: the sample covers a wide range of stellar and disk properties, and the majority of the disks have very similar size ratios. 
This result suggests that the bulk of the disk population may have a similar behavior and evolutionary stage, independent of the stellar and disk properties. 
These results should be further investigated, since the optical depth difference between CO and dust continuum might play a major role in the observed size ratios of the population. 
Lastly, we find a monotonic correlation between the CO flux and the CO size. The results for the majority of the disks are consistent with optically thick emission and an average CO temperature of around 30 K, however, the exact value of the temperature is difficult to constrain.
}

\maketitle


\section{Introduction}\label{sec:intro}
Planets form around stars during their pre-main sequence phase, when still surrounded by a circumstellar disk of gas and dust. 
Setting observational constraints on the gas and dust properties of these disks is crucial in order to understand what are the ongoing physical processes in the disk. These processes shape the planet formation mechanisms, and ultimately tell us about the disk’s ability to form planets \citep[see, e.g.,][]{mordasini+2012}.

The advent of the Atacama Large Millimeter/submillimeter Array (ALMA) allowed for a characterization of the dust properties in large populations of disks \citep[e.g.,][]{tazzari+2017,andrews2018A,hendler+2020}, based on surveys targeting nearby star-forming regions \citep{ansdell+2016,ansdell+2017,barenfeld+2016,pascucci+2016,cox+2017,cieza+2018,cazzoletti+2019,williams+2019}. 
However, demographic studies of the gas disk properties in these regions are scarcer \citep[e.g.,][]{long+2017,ansdell+2018,najita+2018,boyden+2020}, due to the fewer detections, the difficulty of finding reliable gas tracers \citep[e.g.,][]{miotello+2016,miotello+2017}, and frequently, cloud contamination.

A key diagnostic of the evolutionary stage of a disk is its size. Dust and gas evolve differently, thus, we can learn about the undergone physical processes in the disk by studying the relative extent between gas and dust \citep[e.g.,][]{sellek+2020a,sellek+2020b}. 
Initially, the dust grains have sizes below 1 $\mu m$ and are kinematically coupled to the gas \citep[e.g.,][]{fouchet+2007,birnstiel+2010}. 
The pressure gradient of a disk, which generally points outward, exerts an additional force that causes gas to orbit in a slightly sub-keplerian speed. Dust grains grow by coagulation, and when large enough --of the order of mm sizes-- orbiting grains are no longer supported by the outward pressure force. 
A frictional force is induced on the large grains, and by angular momentum conservation, a drift inwards that results in piled-up large grains in a compact configuration \citep[e.g.,][]{weidenschilling1977,pinilla+2012b,canovas+2016}. 
On the other hand, gas, in viscous disks, spreads out to conserve angular momentum and enable gas close in to accrete onto the star \citep[e.g.,][]{Lynden-BellPringe1974,nakamoto&nakagawa1994,HuesoGuillot2005}. 
In wind-driven accretion models \citep[for a review, see][]{turner+2014}, the gas extent will also be larger than the dust extent: dust still drifts inwards, while the gas extent does not vary significantly. 
Observations at (sub-)mm wavelengths typically trace the large dust grains (sizes up to cm sizes) decoupled from the gas \citep{testi+2014,andrews2015}, hence, disks with undergone dust evolution will appear more extended in gas than in dust continuum from ALMA observations.

A difference in size between the gas and dust content has been confirmed from observations of individual Young Stellar Objects \citep[YSOs; e.g.,][]{isella+2007,andrews+2012}, and thanks to ALMA, also from larger samples \citep[e.g.,][]{ansdell+2018,boyden+2020}. Besides the effect of dust evolution, the optical depth difference between dust continuum and gas rotational lines may also contribute to the disparity in the observed gas and dust sizes \citep[e.g.,][]{trapman+2019}. 
While dust thermal emission in the outer disk is optically thin or only partially thick \citep[e.g.,][]{huang+2018}, the gas emission is, in general, optically thicker \citep[e.g.,][]{GuilloteauDutrey1998}. A difference in optical depth implies the dust emission to be fainter than the gas rotational line, thus the emission of the dust outer disk would fall below the sensitivity limit of the instrument at a smaller radius compared to the gas outer emission. 

Consequently, identifying the effect that dominates the size ratio is not easy. \cite{trapman+2019} showed that disks with gas/dust size ratios above 4 can only be explained if grain growth and subsequent radial drift has occurred. 
Such high size ratios between gas and dust have already been observed  \citep{facchini+2019}. 
The existence of pressure bumps can also limit the study of dust evolution based on the gas/dust size ratio. In such scenario, dust grains from the outer disk would only drift inwards down to the bump location. This might result in a larger observed dust size, thus a lower size ratio.

In this work, we expand over the previous study of the gas and dust content in the protoplanetary disk population of Lupus \citep{ansdell+2018}. 
The gas extent is measured based on emission of $\ce{^{12}CO}$ rotational lines at (sub-)mm  wavelengths, while the dust extent is obtained from the continuum emission of large grains. 
The CO emission from these lines is appropriate for the study of the gas extent due to its abundance. These lines are optically thick at low CO column densities \citep{vanDishoeckBlack1988}, allowing CO to self-shield and avoid photodissociation from UV photons. 
Extremely low CO temperatures (of $\sim20$ K) limits the study of gas based on these lines, since CO may freeze out onto the dust grains' surface, no longer emitting at these rotational lines.

The integrated CO emission is modeled to empirical functions, this allows us to increase the number of disks with characterized CO compared to previous studies. 
In addition, disks surrounding brown dwarfs (BDs) from more recent observations \citep{sanchis+2020a} are added to the studied sample. 
Dust disk sizes are estimated from fitting empirical models in the visibility plane. 
This manuscript is organized as follows: in Section~\ref{sec:sampleselection} we describe the Lupus disk sample and the observations used; the modeling of the CO and dust continuum emission is presented in Section~\ref{sec:modeling}; the resulting sizes are summarized in Section~\ref{sec:sizeresults}; in Section~\ref{sec:discussion}, we perform the demographic analysis of the CO and dust sizes, and discuss what the results entail; finally, in Section~\ref{sec:conclusions}, we summarize the main findings of this study.


\section{Sample selection}\label{sec:sampleselection}
The objects studied in this work belong to the Lupus clouds (I-IV), a low-mass Star-Forming Region (SFR) that is part of the Scorpius-Centaurus OB association \citep{comeron2008}. 
Lupus is one of the closest SFRs, at a median distance of $158.5$ pc \citep[from individual Gaia parallaxes of the Lupus members,][]{gaiacollaboration2018}. The age of the region is approximately $1$-$3$ Myr \citep{comeron2008,alcala+2017}.

The sample includes young stellar objects with confirmed protoplanetary disks, down to the BD regime (we define as BD systems of spectral type equal or later than M6, and whose central object mass is $<0.1$ $M_{\odot}$). 
The sources were selected from the catalogs of the clouds \citep{hughes+1994,mortier+2011,merin+2008,comeron2008,dunham+2015,bustamante+2015,muzic+2014,muzic+2015}, their infrared (IR) excess estimated from \textit{Spitzer} \citep['Cores to Discs' legacy project,][]{evans+2009} and 2MASS \citep[][]{cutri+2003} data. Details on the sample selection for the ALMA surveys are to be found in \citet{ansdell+2016,ansdell+2018} for the stellar objects, and \citet{sanchis+2020a} for the BDs. 
All objects are confirmed members of the Lupus clouds from radial velocity analysis \citep{frasca+2017}. The stellar properties were taken from \citet{alcala+2014,alcala+2017} and \citet{muzic+2014}, while stellar luminosities ($L_{\star}$) and masses ($M_{\star}$) have been recalculated taking into account the distance from the precise Gaia DR$2$ parallaxes \citep{gaiacollaboration2018,manara+2018a,alcala+2019}. The stellar mass is obtained from the position in the Hertzsprung–Russell (HR) diagram set by the effective temperature and the updated $L_{\star}$. 
The stellar mass is primarily interpolated from the pre-main sequence models of \citet{baraffe+2015}, which provide accurate estimates of $M_{\star}$ for BDs, M dwarfs and low mass stars up to $1.4$ $M_{\odot}$. These models are ideal for our sample, since the great majority of Lupus objects are within this mass range. 
For the very few objects above $1.4$ $M_{\odot}$ (only 3 in the entire Lupus sample) the \citet{siess+2000} models are used instead. 
The stellar mass uncertainty is obtained from a Monte Carlo procedure as in \citet{alcala+2017}.

Following these criteria, the selected ALMA dataset is composed by $100$ protoplanetary disks around YSOs in the Lupus clouds, 9 of which are BDs. However, our analysis  concentrates on the 42 disks whose CO and dust radii could be measured.

\subsection{Observations}\label{sec:obsresults}
The CO radial extent of the disks is measured from archival ALMA observations covering the $\ce{^{12}CO}$ $J = 2-1$ rotational line in Band~6 (at $230.538$ GHz). For $3$ sources, the $J = 3-2$ rotational transition in Band~7 (at $345.796$ GHz) is used instead. 
The dust sizes are obtained based on modeling of archival observations of dust continuum in ALMA Band~$7$ (centered at $\sim 0.89$ mm). 
The $\ce{^{12}CO}$ channel maps are built after subtracting the continuum and cleaning with a Briggs weighting and robustness = +0.5. 
In Table~\ref{tab:observations}, the details of the line and continuum ALMA observations used in this study are summarized; including information of the ALMA project IDs, the number of Lupus sources targeted at each ALMA project, angular resolution, and the corresponding references that describe the observations and the instrument configuration. 
\begin{table*}
\footnotesize
\caption{
Summary of the archival ALMA projects used in this work for the gas and dust modeling.
}
\label{tab:observations}      
\centering              
\begin{tabular}{lcclll}   
\hline\hline            
Observation & Frequency & Ang. resolution       & Sources / Survey name & ALMA Project ID & References  \\
            & [GHz] & [${}^{\prime\prime} \times {}^{\prime\prime}$] &  &                 &             \\    
\hline 
$\ce{^{12}CO}$ (2-1) & $230.538$ & $\sim0.24 \times 0.23$ & 86, Band~6 Lupus Disk & 2015.1.00222.S, PI: Williams & \citet{ansdell+2018} \\
$\ce{^{12}CO}$ (2-1) & $230.538$ & $\sim0.26 \times 0.22$ & 7, Lupus Completion Disk & 2016.1.01239.S, PI: van Terwisga & \citet{vanterwisga+2018} \\
$\ce{^{12}CO}$ (2-1) & $230.538$ & $\sim0.53 \times 0.39$ & Sz~82 & 2013.1.00226.S, PI: Oberg & \citet{cleeves+2016} \\
$\ce{^{12}CO}$ (2-1) & $230.538$ & $\sim0.25 \times 0.22$ & Sz~91 & 2013.1.01020.S, PI: Tsukagoshi & \citet{canovas+2016} \\
$\ce{^{12}CO}$ (3-2) & $345.796$ & $\sim0.36 \times 0.33$ & 5, Lupus BD Disks & 2017.1.01243.S, PI: Testi & \citet{sanchis+2020a} \\

Cont. Band 7 & $\sim 335$ & $\sim0.34 \times 0.30$ & 86, Band~7 Lupus Disk & 2013.1.00220.S, PI: Williams & \citet{ansdell+2016} \\
Cont. Band 7 & $\sim 335$ & $\sim0.19 \times 0.17$ & 7, Lupus Completion Disk & 2016.1.01239.S, PI: van Terwisga & \citet{vanterwisga+2018} \\
Cont. Band 7 & $\sim 335$ & $\sim0.37 \times 0.29$ & Sz~82 & 2013.1.00694.S, PI: Cleeves & \citet{cleeves+2016} \\
Cont. Band 7 & $\sim 335$ & $\sim0.21 \times 0.15$ & Sz~91 & 2013.1.00663.S, PI: Canovas & \citet{canovas+2016} \\
Cont. Band 7 & $\sim 335$ & $\sim0.36 \times 0.34$ & 5, Lupus BD Disks & 2017.1.01243.S, PI: Testi & \citet{sanchis+2020a} \\

\hline
\multicolumn{6}{c}{Complementary data} \\
\hline
$\ce{^{12}CO}$ (2-1) & $230.538$ & $\sim0.10 \times 0.08$ & 7, DSHARP project & 2016.1.00484.L, PI: Andrews & \citet{andrews+2018dsharp}  \\

Cont. Band 6 & $\sim 225.4$ & $\sim0.27 \times 0.26$ & 86, Band~6 Lupus Disk & 2015.1.00222.S, PI: Williams & \citet{ansdell+2018} \\
\hline
\end{tabular}
\end{table*}

In order to test our method for determining the CO disk radial extent for a few disks in the dataset described above, we have analyzed available ALMA data at higher resolution and better sensitivity. 
These additional data are part of the DSHARP large program \citep[][for a general description of the project; see also the other DSHARP publications, II-X]{andrews+2018dsharp}, that also covered the $\ce{^{12}CO}$ J = $2$-$1$ rotational transition for all targets. 
The Lupus disks targeted in DSHARP are: Sz~68 (HT~Lup), Sz~71 (GW~Lup), Sz~82 (IM~Lup), Sz~83 (RU~Lup), Sz~114, Sz~129, and MY~Lup. 
Lastly, the continuum dataset of the Band 6 Lupus disk Survey was used to test 
the dust sizes results between this and previous work \citep[][]{ansdell+2018}.


\section{Modeling}\label{sec:modeling}
The methodology employed to measure the gas and dust sizes of the Lupus disk population is described in this Section. 

\subsection{CO modeling}\label{sec:gasmodeling}
The CO emission of each disk is primarily modeled by fitting the integrated line map to an elliptical Gaussian function in the image plane. For disks in which a Gaussian model does not conveniently describe the observed CO emission, the so-called Nuker profile model \citep[e.g.,][]{lauer+1995,tripathi+2017} is used instead. 
We assess the quality of the Gaussian fit by comparing its radii results to those from high angular resolution and sensitivity observations, and by quantifying the residuals between observation and model. This is explained in detail in Section~\ref{sec:dsharpradii}.

This modeling is appropriate for the CO disks characterization due to the low S/N for the bulk of the sample. 
The integrated map is obtained by summing up all the channels showing emission above noise level around the known position of the object; the range of channels are selected based on visual examination of channel map and spectrum. 
For the elliptical Gaussian modeling, the \texttt{imfit} task from CASA software \citep{mcmullin+2007} was used. 
The task provides the parameter values with uncertainties of the Gaussian fit to the observed emission. 
The Nuker profile modeling is performed by fitting\footnote{using \texttt{scipy.optimize} Python module, \texttt{https://www.tutorialspoint.com/scipy/scipy}$\_$\texttt{optimize.htm}} the azimuthally averaged CO emission to this function, centered at the optimal position from the \texttt{imfit} results. The outer edge of the Nuker model is set as the radius in which the azimuthally average profile first reaches zero.

\subsubsection{Size definition}\label{sec:rgasmodeling}
The size definition used in this work is the radius enclosing a certain fraction of the total modeled flux, for the CO ($R_{\mathrm{CO}}$) and for the dust ($R_{\mathrm{dust}}$) components separately. This definition has been recently used to characterize large samples of disks from ALMA observations \citep[e.g.,][]{tazzari+2017,andrews2018A,hendler+2020}, and for theoretical modeling of disks \citep[e.g.,][]{rosotti+2019,trapman+2019,trapman+2020}. 
The fractions considered are the 68, 90 and $95\%$ for easy comparison with previous works. 
To estimate the CO radii, we first obtain the deprojected model emission profile: from the deconvolved major-axis full-width-half-maximum (FWHM) of the elliptical Gaussian model, or built from the optimal values of the parameters in the Nuker fitting. We then produce the cumulative distribution functions \citep[$f_{\mathrm{cumul}}$, following e.g., Eq.~A.1 in][]{sanchis+2020a}. 
The radius (e.g., $R_{68\%}$) is inferred from the expression $f_{\mathrm{cumul}} (R_{68\%}) = 0.68 \cdot F_{\mathrm{tot}}$, where $F_{\mathrm{tot}}$ is the total integrated line emission of the model. For the elliptical Gaussian models, the $R_{68\%}$ can be obtained from the standard deviation ($\sigma$) of the Gaussian function: 
\begin{equation}\label{eq:r68gaussian}
R_{68\%} = \sigma \cdot \sqrt{- 2 \cdot \ln ( 1-0.68) } \simeq 1.51 \cdot \sigma  \mathrm{,}
\end{equation}
The uncertainty of the CO sizes are obtained from the major-axis FWHM error on the Gaussian fits. For the Nuker fitting of CO, the size uncertainties are acquired from a Monte Carlo procedure: 1000 realizations of the free parameters are drawn from a random normal distribution defined by the parameters' optimal values and their standard deviation; from these set of values we build 1000 Nuker models and measure their $R_{68\%}$, $R_{90\%}$, and $R_{95\%}$. Their associated standard deviation are taken as the size uncertainty of the Nuker models.

The method to infer CO sizes of the Lupus disk population differs from the approach in \citet{ansdell+2018}. In that work, the CO size was estimated from the curve of growth of keplerian masked moment zero maps. 
The keplerian masking assumes a physical model in which gas kinematics are described by keplerian rotation. Their moment zero map is built from selected emission on each channel that is expected to come from the disk. 
We avoid this approach in order to keep our analysis as general as possible, without any assumptions on the disk physics. 
Additionally, sizes of fainter sources are difficult to measure using the curve of growth, since there is no clear end of the disk emission in the curve of growth. 
In Section~\ref{sec:rgasresults} we compare our sizes to the results of \citet{ansdell+2018}.

Lastly, we note that the sizes for 3 BD disks (SSTc2d J154518.5-342125, 2MASS J16085953-3856275 and Lup706) are obtained from the emission of a different line ($\ce{^{12}CO}$ $J = 3-2$). Differences in the measured radii between this line and the $\ce{^{12}CO}$ $J = 2-1$ line are expected to be negligible, since the two lines are being emitted from essentially the same layer in the disk atmosphere, therefore with almost identical temperatures. 

\subsubsection{CO size uncertainties, from comparison to the DSHARP survey}\label{sec:dsharpradii}
The purpose of this Section is to assess the systematic errors of the CO modeling used, and to find a reliable criterion to determine in which cases the CO emission can be modeled to an elliptical Gaussian or to a Nuker profile model instead. 
To accomplish these goals, we compare the radii of six disks from our sample to the radii from additional $\ce{^{12}CO}$ ($J = 2-1$) observations of the same objects at higher angular resolution and sensitivity \citep[DSHARP project, details in][]{andrews+2018dsharp}.

In order to perform this comparison, we need a reliable measurement of the CO disk sizes from the DSHARP data, which are treated as the fiducial sizes of these disks. 
This is accomplished by interferometric modeling of the $\ce{^{12}CO}$ line visibilities: channels with line emission are continuum-subtracted and then spectrally integrated; the resulting visibilities are then modeled by a Nuker profile model. A comprehensive description of this modeling can be found in Appendix~\ref{sec:appendix_visfit}. 
The resulting sizes are tabulated in Table~\ref{tab:dsharpradii} of Appendix~\ref{sec:appendix_comparingcosizes}, where we summarize the CO sizes using different methodology on the two datasets. 

We compare the $R_{68\%}$ from the elliptical Gaussian modeling of the Lupus disk survey with the fiducial sizes from the interferometric modeling of the DSHARP data (Figure~\ref{fig:r68comparison}, in  Appendix~\ref{sec:appendix_visfit}). For all the disks except one, the elliptical Gaussian modeling yields smaller sizes than the fiducial values. One disk (MY~Lup) has nearly identical size results between the two datasets, a second disk (Sz~114) has a size deviation below $20\%$, two other objects (Sz~71 and Sz~129) have $\sim 30\%$ difference between the inferred sizes, and the last two sources (Sz~82 and Sz~83) have a discrepancy above $40\%$. When inspecting the $R_{90\%}$ radii, the discrepancies are slightly increased, with only three disks with a size deviation below $30\%$, and discrepancies beyond $40\%$ for the remaining disks. 

Several factors might contribute to the difference in the measured sizes. 
Firstly, the difference in sensitivity between observations can affect the detection of emission in the outermost regions of the disk. In addition, the different angular resolution may also have an impact: in general, the better resolved the disk, the better the size measurement. 
Another possible cause is the fact of modeling the Lupus disk population in the image plane, while the fiducial sizes are obtained from modeling in the \textit{uv}-plane. 
Lastly, the size difference could be due to the elliptical Gaussian model not being able to reproduce the true CO emission. 
To understand the impact of these effects, we study them separately. 

The sensitivity difference is tested by using the exact same method to model the two datasets, that is, fitting elliptical Gaussian models to the disk survey and to the DSHARP sets. The results are included in Table~\ref{tab:dsharpradii} of Appendix~\ref{sec:appendix_comparingcosizes}. The measured sizes between the two datasets are very similar, with only $\sim5\%$ difference. Therefore, sensitivity has a minor effect on the inferred CO sizes of the Lupus disk dataset. 
The effect of the angular resolution can also be inspected from this comparison. The angular resolution has a stronger effect on smaller disks (i.e., of the order of the beam size). The two smallest disks (Sz~129 and MY~Lup) show a slightly larger size difference of $\sim15\%$ compared to the aforementioned difference of the sample. Although the sample considered is very limited, our results show that resolution effect might be relevant, especially in disks of size of the order of the beam size.

The effect of measuring the CO radial extent from modeling the emission in the image or in the \textit{uv}-plane is investigated by modeling in the two planes the same dataset with the same empirical function (i.e., elliptical Gaussian). For each disk of the DSHARP dataset, we reconstruct the moment zero maps from the line visibilities; the \texttt{imfit} task is then used for the Gaussian modeling in the image plane. The interferometric modeling is analogous to the methodology described in Appendix~\ref{sec:appendix_visfit}, but using a Gaussian function instead of the Nuker function. The size results are included in Table~\ref{tab:dsharpradii}. The difference in size is negligible for every disk, $2\%$ on average, thus modeling the emission in the image plane has a negligible effect on the inferred size.

Lastly, we test the accuracy of the Gaussian modeling with respect to the Nuker profile modeling. We compare the interferometric modeling results when fitting the DSHARP data to a Gaussian or a Nuker profile. The results (Table~\ref{tab:dsharpradii}) show a size difference of $\sim 20\%$ on average. Two disks (MY~Lup and Sz~129) have size differences below 5\%; one disk (Sz~114) has a difference of $\lesssim 15\%$; another disk (GW~Lup) has a difference around $30\%$, and the remaining two disks have differences beyond $40\%$.

Hence, the Gaussian modeling not reproducing the observed emission of certain objects is the most limiting effect on the CO size determination. It yields accurate CO sizes in several disks, but in other disks (typically those with extended emission) the inferred sizes can differ significantly with respect to the true CO extent. 
For those disks, the Nuker model is able to describe the extended emission of the disk accurately. 
In order to determine which CO disks can be described by an elliptical Gaussian model, we developed a criterion that evaluates the quality of the model, based on the amount of residuals (difference between observed and modeled emission). This criterion is described in detail in Appendix~\ref{sec:appendix_residualscriteria}. 

Based on this criterion, the CO emission is fitted to an elliptical Gaussian for those disk models with negligible residuals (i.e., when the quantified residuals are outside the $\mu \pm \sigma$ range of the entire population), otherwise the emission is fitted to a Nuker function.

In summary, our modeling in the image plane typically allows to measure the CO sizes for the Lupus disk sample with an uncertainty $\lesssim30\%$, based on the comparison with available observations at higher resolution and sensitivity. 
Due to its simplicity and its ability to reproduce the observed CO emission, we use the elliptical Gaussian modeling for the cases in which the measured $R_{\mathrm{CO}}$ is reliable. 
For CO disks with Gaussian model residuals outside the valid range, the Nuker modeling in the image plane is used instead.

\subsection{Dust modeling}\label{sec:rdustmodeling}
The dust disks are modeled in the \textit{uv}-plane to an empirical function, the Nuker profile. 
We refer to \citet{sanchis+2020a} for the detailed description of the interferometric modeling, in which the \texttt{Galario} package \citep{tazzari+2018} was used in combination with a Monte Carlo Markov Chain procedure to model the continuum emission of the BD disks and sources from the Lupus disk completion survey. 
In the present work, we take the $R_{\mathrm{dust}}$ results of \citet{sanchis+2020a} for the 10 disks with detected $\ce{^{12}CO}$, and model the remaining disks of the Lupus population using identical methodology. 
The dust sizes considered are the radii enclosing the $68$, $90$, and $95\%$ of the total disk emission, analogous to the size definition of the CO disk.

Performing the modeling in the \textit{uv}-plane may reduce possible uncertainties associated to the image reconstruction process. Nevertheless, we tested the resulting dust sizes when modeling to a Nuker function in the image plane for a number of resolved disks. The results are in very good agreement with the dust sizes obtained from fitting the visibilities (deviation of $\sim$$5\%$). 
Thus, modeling the continuum emission in the image or in the \textit{uv}-plane does not have a significant impact in the size results. Only for very compact sources, the sizes obtained from the image plane modeling may be affected by the beam.

\section{Disk size results}\label{sec:sizeresults}

\subsection{CO size results}\label{sec:rgasresults}
The CO-disk size results of the Lupus disk population are presented in this Section. We exclude the results from disks with model peak below 3 times the rms of the observed moment zero map, those with maps partially covered by clouds, and disks that belong to binary systems with angular separation below $2^{\prime\prime}$. 
The resulting CO disk sizes ($R_{68\%}$) are summarized in Table~\ref{tab:rgasresults}. The uncertainties in the table are associated to the fitting method employed. Nevertheless, we warn that the inferred CO sizes may have a discrepancy of $0\sim30\%$ with respect to the true CO extent, based on our tests described in Section~\ref{sec:dsharpradii}. 

By definition of the Gaussian function, there is a constant relation between the $R_{68\%}$ and the two other radii ($R_{90\%}$, $R_{95\%}$):
\begin{equation}\label{eq:r90gaussian}
R_{90\%} \simeq 1.42 \cdot R_{68\%}  \mathrm{,}
\end{equation}
and
\begin{equation}\label{eq:r95gaussian}
R_{95\%} \simeq 1.62 \cdot R_{68\%}  \mathrm{,}
\end{equation}
The above relations can be used to obtain the $R_{90\%}$ and $R_{95\%}$ radii for the CO Gaussian models. For the disks modeled with the Nuker function, we provide the optimal parameters of the fit in Table~\ref{tab:COnukerparams} of the Appendix~\ref{sec:appendix_COnukerparams}.

Out of $51$ disks detected in CO, three are partially covered by clouds (J15450634-3417378, J15450887-3417333, J16011549-4152351), other three yield models whose S/N is too low (Sz~98, J16085324-3914401, J16095628-3859518), and three belong to close binary systems (Sz~68, Sz~74, Sz~123A). Thus  
our methodology allowed us to model the emission and size of $42$. 
Three of these CO sizes are provided as upper limits (with tabulated value being the $95\%$ confidence level), since the deconvolved FWHM of their elliptical Gaussian models exhibit a point-like nature. Additionally, two of these objects with CO size upper limits are disks around BDs. 
Table~\ref{tab:rgasresults} includes a column stating the CO model used to infer the CO sizes (elliptical Gaussian model referred as 'G', Nuker model as 'N'). 
In Appendix~\ref{sec:appendix_obsmodres} we include the observed, modeled, and residual CO maps, together with the line spectrum and the modeled intensity profile of every disk with measured CO size. Cloud absorption is seen on the line spectrum for a considerable number of sources. This reduces the integrated flux of the line. However, it should not have significant incidence in the measured CO radii \citep{ansdell+2018}. 

Lastly, molecular outflows from $\ce{^{12}CO}$ observations have been reported in at least 3 of the tabulated sources based on the dynamical analysis of the CO emission \citep[EX~Lup, V1192~Sco, Sz~83;][]{hales+2018,santamaria-miranda+2020,huang+2020}. 
The outflows of the first two objects are within the reported CO sizes in Table~\ref{tab:rgasresults}. Our sizes are obtained by modeling the total integrated emission detected, thus a fraction of the modeled emission does not belong to the disk but to the molecular outflows. Therefore, we consider the inferred CO sizes of EX~Lup and V1192~Sco as upper limits. 
On the other hand, Sz~83 shows a very intricate structure with spirals, jets, and clumps of emission \citep{herczeg+2005,ansdell+2018,andrews+2018dsharp,huang+2020}. We discuss this disk in greater detail in  Appendix~\ref{sec:appendix_singularobjects}, together with other singular systems of the sample. 
Our CO size reported in Table~\ref{tab:rgasresults} is larger than the keplerian disk size, the surrounding non-keplerian emission, and might contain a fraction of the emission from the spiral arms \citep{huang+2020}. For consistency, we use the CO size measured by our methodology, although we warn that the true value of the CO disk size might differ. 
\begin{table*}
\scriptsize
\caption{Results of the CO and dust sizes for the Lupus disks sample. The table includes all objects which gas extent could be estimated following the methodology described in Sections~\ref{sec:gasmodeling} and ~\ref{sec:dsharpradii}. Distances of the sources are estimated as the inverse of the parallax \citep[from Gaia DR2][]{gaiacollaboration2018}; the tabulated distance of objects with uncertain parallax is the mean distance of the Lupus clouds (158.5 pc). Dust mass obtained assuming optically thin emission of the ALMA Band~7 continuum observations, with dust optical depth of $\kappa_{890\mu m}$ = 2 $\mathrm{cm}^2 \mathrm{g}^{-1}$, and average dust temperature of $T_{\mathrm{dust}}$ = 20 K.}
\label{tab:rgasresults}      
\centering              
\begin{tabular}{llccccccccccr}   
\hline\hline            
$\#$ & Object & Dist. & SpT & $T_{\mathrm{eff}}$ & $M_{\star}$ & 
Model & 
$M_{\mathrm{dust}}$ & $R_{68\%}^{\mathrm{CO}}$  & $R_{68\%}^{\mathrm{dust}}$  & $R_{90\%}^{\mathrm{dust}}$ & $R_{95\%}^{\mathrm{dust}}$ & $\frac{R_{68\%}^{\mathrm{CO}}}{R_{68\%}^{\mathrm{dust}}}$ \\
 &  & [pc] &  & [K] & [$M_{\odot}$] & 
CO & 
[$M_{\oplus}$] & [${}^{\prime\prime}$] & [${}^{\prime\prime}$] & [${}^{\prime\prime}$] & [${}^{\prime\prime}$] &  \\    
\hline 

1  &  EXLup   & $ 157.7 $ &  M$0$ & $ 3850 $ & $ 0.53 $ & N &$  19.1 $ 
& $ < 0.70 $
& $ 0.19 \pm 0.01 $  & $ 0.23 \pm 0.01 $  & $ 0.25 \pm 0.01 $ 
& $ < 3.7 $ \\ 
2  &  Lup706   & $ 158.5 $ &  M$7.5$ & $ 2795 $ & $ 0.06 $ & G &$  0.4 $ 
& < $ 0.39 $ 
& < $ 0.36 $  & < $ 0.51 $  & < $ 0.58 $ 
& - \\
3  &  MYLup   & $ 156.6 $ &  K$0$ & $ 5100 $ & $ 1.09 $ & G &$  76.5 $ 
& $ 0.81 \pm 0.10 $ 
& $ 0.38 \pm 0.01 $  & $ 0.51 \pm 0.01 $  & $ 0.58 \pm 0.02 $ 
& $ 2.1 \pm 0.3 $ \\
4  &  RXJ1556.1-3655   & $ 158.0 $ &  M$1$ & $ 3705 $ & $ 0.49 $ & G &$  24.8 $ 
& $ 0.53 \pm 0.05 $ 
& $ 0.19 \pm 0.01 $  & $ 0.25 \pm 0.01 $  & $ 0.28 \pm 0.01 $ 
& $ 2.8 \pm 0.3 $ \\
5  &  RYLup   & $ 159.1 $ &  K$2$ & $ 4900 $ & $ 1.53 $ & N &$  123.0 $ 
& $ 1.15 \pm 0.10 $ 
& $ 0.61 \pm 0.01 $  & $ 0.80 \pm 0.01 $  & $ 0.93 \pm 0.03 $ 
& $ 1.9 \pm 0.2 $ \\
6  &  Sz65   & $ 155.3 $ &  K$7$ & $ 4060 $ & $ 0.70 $ & G &$  27.4 $ 
& $ 0.76 \pm 0.09 $ 
& $ 0.16 \pm 0.01 $  & $ 0.24 \pm 0.01 $  & $ 0.28 \pm 0.02 $ 
& $ 4.8 \pm 0.6 $ \\
7  &  Sz66   & $ 157.3 $ &  M$3$ & $ 3415 $ & $ 0.29 $ & G &$  6.5 $ 
& < $ 0.61 $ 
& $ 0.11 \pm 0.02 $  & $ 0.16 \pm 0.05 $  & $ 0.19 \pm 0.08 $ 
& <$ 5.7 $ \\
8  &  Sz69   & $ 154.6 $ &  M$4.5$ & $ 3197 $ & $ 0.20 $ & G &$  7.1 $ 
& $ 0.56 \pm 0.08 $ 
& < $ 0.09 $  & < $ 0.20 $  & < $ 0.31 $ 
& >$ 5.9 $ \\
9  &  Sz71   & $ 155.9 $ &  M$1.5$ & $ 3632 $ & $ 0.41 $ & G &$  71.2 $ 
& $ 0.84 \pm 0.12 $ 
& $ 0.45 \pm 0.01 $  & $ 0.70 \pm 0.01 $  & $ 0.83 \pm 0.02 $ 
& $ 1.9 \pm 0.3 $ \\
10  &  Sz72   & $ 155.9 $ &  M$2$ & $ 3560 $ & $ 0.37 $ & G &$  6.0 $ 
& $ 0.15 \pm 0.05 $ 
& $ 0.08 \pm 0.02 $  & $ 0.12 \pm 0.05 $  & $ 0.14 \pm 0.07 $ 
& $ 1.8 \pm 0.7 $ \\
11  &  Sz73   & $ 156.8 $ &  K$7$ & $ 4060 $ & $ 0.78 $ & G &$  13.2 $ 
& $ 0.48 \pm 0.06 $ 
& $ 0.27 \pm 0.04 $  & $ 0.47 \pm 0.08 $  & $ 0.55 \pm 0.12 $ 
& $ 1.8 \pm 0.3 $ \\
12  &  Sz75   & $ 151.8 $ &  K$6$ & $ 4205 $ & $ 0.80 $ & N &$  31.9 $ 
& $ 0.97 \pm 0.11 $ 
& $ 0.12 \pm 0.01 $  & $ 0.17 \pm 0.01 $  & $ 0.19 \pm 0.01 $ 
& $ 8.0 \pm 0.9 $ \\
13  &  Sz76   & $ 159.5 $ &  M$4$ & $ 3270 $ & $ 0.23 $ & G &$  4.9 $ 
& $ 0.62 \pm 0.06 $ 
& $ 0.26 \pm 0.02 $  & $ 0.41 \pm 0.05 $  & $ 0.47 \pm 0.08 $ 
& $ 2.4 \pm 0.3 $ \\
14  &  Sz77   & $ 154.8 $ &  K$7$ & $ 4060 $ & $ 0.75 $ & G &$  2.1 $ 
& $ 0.17 \pm 0.08 $ 
& < $ 0.36 $  & < $ 0.75 $  & < $ 0.88 $ 
& >$ 0.5 $ \\
15  &  Sz82   & $ 158.4 $ &  K$5$ & $ 4350 $ & $ 0.95 $ & N &$  264.0 $ 
& $ 3.75 \pm 0.67 $ 
& $ 1.15 \pm 0.01 $  & $ 1.64 \pm 0.01 $  & $ 1.82 \pm 0.01 $ 
& $ 3.3 \pm 0.6 $ \\
16  &  Sz83*   & $ 159.6 $ &  K$7$ & $ 4060 $ & $ 0.67 $ & N &$  191.7 $ 
& $ 1.44 \pm 0.13 $ 
& $ 0.29 \pm 0.01 $  & $ 0.39 \pm 0.01 $  & $ 0.46 \pm 0.01 $ 
& $ 5.0 \pm 0.4 $ \\
17  &  Sz84   & $ 152.6 $ &  M$5$ & $ 3125 $ & $ 0.17 $ & G &$  13.4 $ 
& $ 0.89 \pm 0.12 $ 
& $ 0.24 \pm 0.01 $  & $ 0.34 \pm 0.03 $  & $ 0.40 \pm 0.04 $ 
& $ 3.7 \pm 0.5 $ \\
18  &  Sz90   & $ 160.4 $ &  K$7$ & $ 4060 $ & $ 0.78 $ & G &$  9.9 $ 
& $ 0.33 \pm 0.10 $ 
& $ 0.12 \pm 0.01 $  & $ 0.16 \pm 0.03 $  & $ 0.19 \pm 0.04 $ 
& $ 2.7 \pm 0.8 $ \\
19  &  Sz91   & $ 159.1 $ &  M$1$ & $ 3705 $ & $ 0.51 $ & N &$  27.7 $ 
& $ 1.36 \pm 0.10 $ 
& $ 0.61 \pm 0.01 $  & $ 0.00 \pm 0.01 $  & $ 0.00 \pm 0.01 $ 
& $ 2.2 \pm 0.2 $ \\
20  &  Sz96   & $ 156.6 $ &  M$1$ & $ 3705 $ & $ 0.45 $ & G &$  1.8 $ 
& $ 0.15 \pm 0.07 $ 
& < $ 0.14 $  & < $ 0.28 $  & < $ 0.37 $ 
& >$ 1.0 $ \\
21  &  Sz100   & $ 136.9 $ &  M$5.5$ & $ 3057 $ & $ 0.14 $ & G &$  18.1 $ 
& $ 0.66 \pm 0.12 $ 
& $ 0.26 \pm 0.01 $  & $ 0.32 \pm 0.01 $  & $ 0.35 \pm 0.02 $ 
& $ 2.5 \pm 0.5 $ \\
22  &  Sz102   & $ 158.5 $ & K$2$ & $ 4900 $ & $ - $ & G &$  6.1 $ 
& $ 0.33 \pm 0.02 $ 
& $ 0.13 \pm 0.01 $  & $ 0.22 \pm 0.02 $  & $ 0.27 \pm 0.03 $ 
& $ 2.6 \pm 0.2 $ \\
23  &  Sz111   & $ 158.3 $ &  M$1$ & $ 3705 $ & $ 0.51 $ & N &$  79.3 $ 
& $ 2.08 \pm 0.39 $ 
& $ 0.44 \pm 0.01 $  & $ 0.58 \pm 0.01 $  & $ 0.69 \pm 0.02 $ 
& $ 4.7 \pm 0.9 $ \\
24  &  Sz114   & $ 162.2 $ &  M$4.8$ & $ 3175 $ & $ 0.19 $ & G &$  44.8 $ 
& $ 0.74 \pm 0.15 $ 
& $ 0.25 \pm 0.01 $  & $ 0.35 \pm 0.01 $  & $ 0.39 \pm 0.01 $ 
& $ 2.9 \pm 0.6 $ \\
25  &  Sz118   & $ 163.9 $ &  K$5$ & $ 4350 $ & $ 1.04 $ & G &$  30.0 $ 
& $ 0.63 \pm 0.14 $ 
& $ 0.37 \pm 0.01 $  & $ 0.44 \pm 0.02 $  & $ 0.48 \pm 0.04 $ 
& $ 1.7 \pm 0.4 $ \\
26  &  Sz129   & $ 161.7 $ &  K$7$ & $ 4060 $ & $ 0.78 $ & G &$  83.5 $ 
& $ 0.76 \pm 0.16 $ 
& $ 0.30 \pm 0.01 $  & $ 0.41 \pm 0.01 $  & $ 0.46 \pm 0.01 $ 
& $ 2.5 \pm 0.5 $ \\
27  &  Sz130   & $ 160.3 $ &  M$2$ & $ 3560 $ & $ 0.39 $ & G &$  2.8 $ 
& $ 0.53 \pm 0.12 $ 
& < $ 0.32 $  & < $ 0.69 $  & < $ 0.83 $ 
& >$ 1.6 $ \\
28  &  Sz131   & $ 160.3 $ &  M$3$ & $ 3415 $ & $ 0.30 $ & G &$  3.9 $ 
& $ 0.56 \pm 0.15 $ 
& $ 0.08 \pm 0.02 $  & $ 0.13 \pm 0.05 $  & $ 0.16 \pm 0.08 $ 
& $ 7.2 \pm 2.8 $ \\
29  &  Sz133   & $ 153.1 $ & K$5$ & $ 4350 $ & $ - $ & G &$  28.5 $ 
& $ 0.95 \pm 0.11 $ 
& $ 0.46 \pm 0.01 $  & $ 0.68 \pm 0.04 $  & $ 0.80 \pm 0.07 $ 
& $ 2.1 \pm 0.2 $ \\
30  &  SSTc2d J154518.5-342125   & $ 151.8 $ &  M$6.5$ & $ 2935 $ & $ 0.08 $ & G &$  2.3 $ 
& $ 0.17 \pm 0.06 $ 
& $ 0.05 \pm 0.01 $  & $ 0.08 \pm 0.01 $  & $ 0.09 \pm 0.01 $ 
& $ 3.2 \pm 1.2 $ \\
31  &  SSTc2d J160002.4-422216   & $ 164.2 $ &  M$4$ & $ 3270 $ & $ 0.23 $ & G &$  57.0 $ 
& $ 1.12 \pm 0.13 $ 
& $ 0.51 \pm 0.01 $  & $ 0.71 \pm 0.02 $  & $ 0.83 \pm 0.05 $ 
& $ 2.2 \pm 0.3 $ \\
32  &  SSTc2d J160703.9-391112   & $ 158.5 $ &  M$4.5$ & $ 3200 $ & $ 0.16 $ & N &$  2.0 $ 
& $ 1.04 \pm 0.09 $ 
& $ 0.43 \pm 0.08 $  & $ 0.56 \pm 0.27 $  & $ 0.65 \pm 0.44 $ 
& $ 2.4 \pm 0.5 $ \\
33  &  SSTc2d J160830.7-382827   & $ 156.1 $ &  K$2$ & $ 4900 $ & $ 1.53 $ & N &$  58.2 $ 
& $ 1.39 \pm 0.13 $ 
& $ 0.56 \pm 0.01 $  & $ 0.67 \pm 0.01 $  & $ 0.75 \pm 0.03 $ 
& $ 2.5 \pm 0.2 $ \\
34  &  SSTc2d J160901.4-392512   & $ 164.3 $ &  M$4$ & $ 3270 $ & $ 0.23 $ & G &$  8.3 $ 
& $ 0.83 \pm 0.08 $ 
& $ 0.47 \pm 0.02 $  & $ 0.57 \pm 0.04 $  & $ 0.62 \pm 0.07 $ 
& $ 1.8 \pm 0.2 $ \\
35  &  SSTc2d J160927.0-383628   & $ 159.3 $ &  M$4.5$ & $ 3200 $ & $ 0.20 $ & G &$  1.7 $ 
& $ 0.50 \pm 0.09 $ 
& < $ 0.36 $  & < $ 0.67 $  & < $ 0.80 $ 
& >$ 1.4 $ \\
36  &  SSTc2d J161029.6-392215   & $ 163.2 $ &  M$4.5$ & $ 3200 $ & $ 0.20 $ & G &$  3.4 $ 
& $ 0.58 \pm 0.11 $ 
& $ 0.23 \pm 0.04 $  & $ 0.31 \pm 0.12 $  & $ 0.36 \pm 0.19 $ 
& $ 2.5 \pm 0.6 $ \\
37  &  SSTc2d J161243.8-381503   & $ 159.8 $ &  M$1$ & $ 3705 $ & $ 0.45 $ & G &$  13.5 $ 
& $ 0.30 \pm 0.11 $ 
& $ 0.10 \pm 0.01 $  & $ 0.16 \pm 0.03 $  & $ 0.19 \pm 0.05 $ 
& $ 3.0 \pm 1.1 $ \\
38  &  V1094Sco   & $ 153.6 $ &  K$6$ & $ 4205 $ & $ 0.83 $ & G &$  230.3 $ 
& $ 1.93 \pm 0.15 $ 
& $ 1.31 \pm 0.01 $  & $ 1.83 \pm 0.01 $  & $ 1.96 \pm 0.01 $ 
& $ 1.5 \pm 0.1 $ \\
39  &  V1192Sco   & $ 150.8 $ & M$4.5$ & $ 3197 $ & $ - $ & N &$  0.4 $ 
& $< 0.98 $ 
& < $ 0.83 $  & < $ 0.95 $  & < $ 0.97 $ 
& - \\ 
40  &  2MASS J16070854-3914075  & $ 175.8 $ & M$1.8$ & $ 4000 $ & $ - $ & G &$  50.2 $ 
& $ 1.36 \pm 0.19 $ 
& $ 0.62 \pm 0.01 $  & $ 0.90 \pm 0.04 $  & $ 1.06 \pm 0.07 $ 
& $ 2.2 \pm 0.3 $ \\
41  &  2MASS J16081497-3857145   & $ 158.5 $ &  M$5.5$ & $ 3060 $ & $ 0.10 $ & G &$  3.7 $ 
& $ 0.42 \pm 0.14 $ 
& $ 0.13 \pm 0.02 $  & $ 0.18 \pm 0.05 $  & $ 0.21 \pm 0.06 $ 
& $ 3.2 \pm 1.2 $ \\
42  &  2MASS J16085953-3856275   & $ 150.2 $ &  M$8.5$ & $ 2600 $ & $ 0.02 $ & G &$  0.2 $ 
& < $ 0.17 $ 
& < $ 0.12 $  & < $ 0.17 $  & < $ 0.19 $ 
& - \\

\hline
\end{tabular}
\tablefoot{
* The modeled emission of Sz~83 based on our methodology includes the keplerian emission, non-keplerian extended emission, and part of the spiral structure as reported in \citet{huang+2020}. 
Therefore, the true CO size of the Sz~83 disk might differ from the tabulated value.
}

\end{table*}

\begin{figure*}
  \begin{center}
  \includegraphics[width=.475\textwidth]{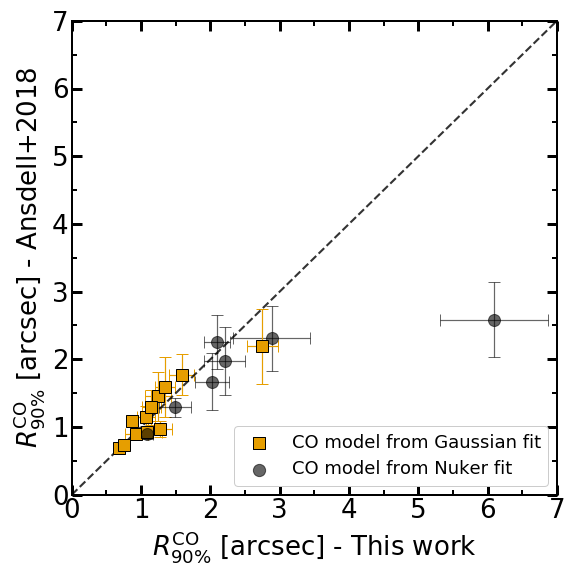}
  \includegraphics[width=.475\textwidth]{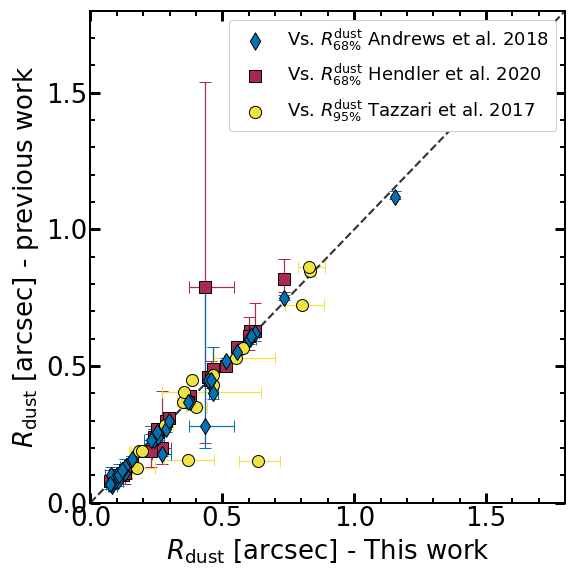}
  \end{center}
  \caption[]{
  \textit{Left panel}:
  Comparison between the $R_{90\%}^{\mathrm{CO}}$ sizes from \citet{ansdell+2018} and the sizes inferred in this work by fitting an elliptical Gaussian model (\text{orange}) or a Nuker function in the azimuthally averaged integrated emission (\text{black}).
  \textit{Right panel}:
  Comparison between the $R_{\mathrm{dust}}$ sizes from this work and the literature \citep[i.e.,][]{andrews2018A,hendler+2020,tazzari+2017}.
  }
  \label{fig:rgas_comparison_a18}
\end{figure*}

In the left panel of Figure~\ref{fig:rgas_comparison_a18}, we compare our results to the $22$ $R_{90\%}^{\mathrm{CO}}$ sizes from \citet{ansdell+2018}, derived using the curve of growth method on keplerian masked CO maps. 
The $R_{90\%}^{\mathrm{CO}}$ is used for this comparison, since is the only reported size in \citet{ansdell+2018}. 
Due to the different methodology between the two studies, the comparison between the two studies using $R_{68\%}^{\mathrm{CO}}$ and $R_{95\%}^{\mathrm{CO}}$ might differ from  Figure~\ref{fig:rgas_comparison_a18} due to the difference in methodology. However, the $R_{68\%}^{\mathrm{CO}}$, which is the radius used in the discussion section of this paper (Section~\ref{sec:discussion}), will typically show lower discrepancies, since it is less affected by the low sensitivity on the outermost regions of the disks.

The CO sizes from the two methods are in good agreement for the majority of disks, only one object (Sz~82) has a difference in radius above $30\%$. This object is the largest CO disk of the Lupus population, this size divergence is likely due to the contrasting approach of the methods. The radius from \citet{ansdell+2018} is inferred from a moment zero map built from selected emission at each channel expected by keplerian rotation of the gas, while in this work there is no assumption on the velocity structure of the observed CO. 
The Sz~82 disk has an extremely large tail of emission (as seen in the integrated maps of the object, Figure~\ref{fig:comodelresults_all_4} of Appendix~\ref{sec:appendix_obsmodres}) that was not captured in the modeling from \citet{ansdell+2018}, and explains the large size difference between the two studies. This extended emission was already observed previously \citep{cleeves+2016,pinte+2018a}. In Appendix~\ref{sec:appendix_singularobjects} we discuss in detail the Sz~82 disk, together with other singular objects of Lupus population.

\subsection{Dust size results}\label{sec:rdustresults}
The resulting radii from the dust modeling are summarized in Table~\ref{tab:rgasresults}, together with uncertainties. 
For disks in which the dust emission is not appropriately modeled, we provide upper limits of the sizes, estimated as the 95th percentile of the corresponding size. 

The protoplanetary disk sample of the Lupus region have been modeled in various studies  \citep[e.g.,][]{tazzari+2017,andrews2018A,hendler+2020} based on the same ALMA Band~7 surveys. 
Our dust disk results can be directly compared to those from the literature (see right panel in Figure~\ref{fig:rgas_comparison_a18}). In general, the $R_{68\%}^{\mathrm{dust}}$ results are in very good agreement with the results of \citet{andrews2018A} and \citet{hendler+2020}, the studies that characterized the dust sizes for a larger sample of Lupus disks. 
Only five disks have differences above $20\%$ with respect to the $R_{68\%}^{\mathrm{dust}}$ from \citet{andrews2018A}. Three of those disks (Sz~66, Sz~72, Sz~131) are marginally resolved in continuum, sizes between the three studies vary between $0.08$ to $0.11^{\prime\prime}$. The remaining two are: SSTc2d J$160703.9$-$391112$, which has large uncertainties in the three studies, nevertheless, our results are compatible within error bars;
and Sz~73, which $R_{68\%}^{\mathrm{dust}}$ from our modeling is in good agreement with \citet{hendler+2020}. Lastly, when comparing our $R_{95\%}^{\mathrm{dust}}$ with the outer radii results for the sub-sample of disks studied in \citet{tazzari+2017}, our results are in very good agreement, with only four objects with differences higher than $20\%$. In this case, the differences in radii might arise due to the modeling approach: instead of an empirical function, \citet{tazzari+2017} fitted the emission to a physical model, which can result in a different model emission profile. Besides, the $R_{95\%}^{\mathrm{dust}}$ used for the comparison is expected to have larger uncertainties than $R_{68\%}^{\mathrm{dust}}$, since it is more affected by the low signal of the outermost disk. 

The dust sizes presented in this work are based on (sub-)mm  continuum emission, which typically probes the population of large dust grains at the disk's mid-plane. These sizes are appropriate to constrain dust evolution of the disks. Observations in other wavelengths can be used as well to infer the size of the disks. 
For instance, scattered-light imaging in Near Infra-Red (NIR) wavelengths probes micron-sizes grains --dynamically more coupled to gas-- at the upper atmospheric layers of the disk. 
Five disks in our sample have been recently observed with VLT/SPHERE \citep{avenhaus+2018,garufi+2020}. 
We can compare the extent of the disks in NIR observations to our size results, by taking the outermost radius at which the signal in NIR is detected and our $R_{95\%}$. 
The sizes from NIR observations are on average $\sim40\%$ larger than our $R_{95\%}^{\mathrm{dust}}$, expected since the smaller grains are more dynamically bound to gas. The NIR sizes are, on the other hand, $\sim50\%$ smaller than our $R_{95\%}^{\mathrm{CO}}$. This comparison is limited due to the very different nature of the observations, the differing definition of the size, and the narrow sample of disks imaged in NIR.

\subsection{Gas/dust size ratio results}\label{sec:sizeratioresults}
In this and following sections we focus our analysis and discussion on the radii enclosing $68\%$ of the CO and dust fluxes ($R_{68\%}$) instead of $R_{90\%}$ or $R_{95\%}$. 
This is due to the moderate sensitivity of the observations, which could affect the detection of weak emission, typically in the outermost regions of the disk. This might have an impact in the outer slope of model emission when fitting to a Nuker profile. 
The $R_{68\%}$ radius is less affected than $R_{90\%}$ and $R_{95\%}$ by the outer slope of the model. Since our dataset is assembled by combining various surveys with different resolution and sensitivity, we favor the use of the $R_{68\%}$ to reduce this possible effect. We also warn that in the following analysis and figures, the size uncertainties used are the ones derived from the respective method employed. However, CO sizes based on this dataset might have a discrepancy with respect to the true CO size between 0 and $\sim30\%$, as explained in Section~\ref{sec:dsharpradii}. 

\begin{figure}
  \resizebox{\hsize}{!}{\includegraphics{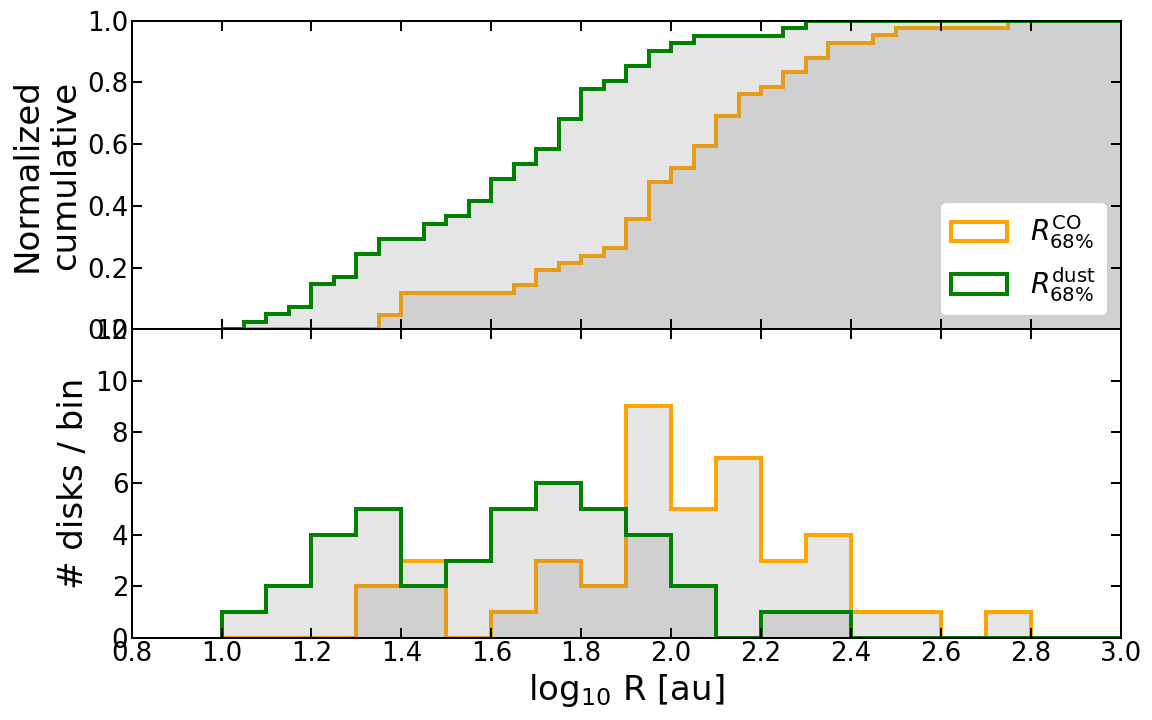}}
  \caption{Histograms and cumulative distributions of the radii enclosing $68\%$ of the total CO and dust continuum emission for the Lupus disks that have measurements of the two sizes. Upper limits of the $R_{\mathrm{CO}}$ and $R_{\mathrm{dust}}$ are included in the histograms, their value being the $95\%$ confidence level.}
  \label{fig:histogram_rgasrdust68}
\end{figure}

In Figure~\ref{fig:histogram_rgasrdust68}, we show the histograms and cumulative distributions of the radii ($R_{\mathrm{CO}}$ and $R_{\mathrm{dust}}$) of all the Lupus disks with measured CO and dust sizes. The radii are obtained for each disk following the methodology described in Section~\ref{sec:modeling}. A difference between the CO disk and the dust disk sizes becomes apparent from the figure. The Anderson--Darling test\footnote{using \texttt{scipy.stats} Python module, \texttt{https://docs.scipy.org/doc/scipy/reference/stats.html}} yields a <$0.001\%$ probability that the two radii histograms are drawn by the same parent distribution. 
There is a selection effect toward larger CO sizes, since it is more difficult to detect and measure CO sizes as small as the dust sizes. Nevertheless, the fraction of disks with measured $R_{\mathrm{dust}}$  and unknown $R_{\mathrm{CO}}$ is small (around $20\%$ of disks with known $R_{\mathrm{dust}}$), thus this effect would not change the observed size difference. This disparity in sizes was already reported in \citet{ansdell+2018} for a smaller sample of the Lupus disk population, and in other SFRs, such as Taurus \citep{najita+2018}, and Orion \citep{boyden+2020}. 

\begin{figure}
  \resizebox{\hsize}{!}{\includegraphics{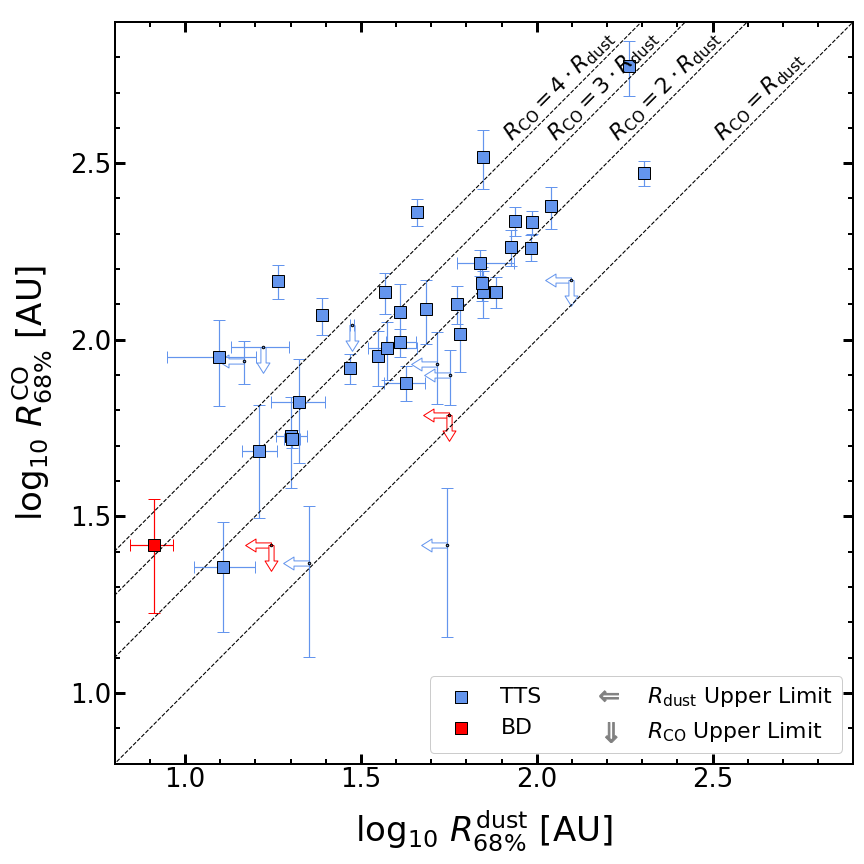}}
  \caption{Comparison between the $R_{68\%}$ CO and dust emission for the entire disk population.}
  \label{fig:rgasrdust68}
\end{figure}

In order to investigate the relative size of CO with respect to the dust continuum, we inspect the ratio between $R_{\mathrm{CO}}$ against $R_{\mathrm{dust}}$. In Figure~\ref{fig:rgasrdust68}, the radii enclosing $68\%$ of the respective total fluxes are shown, with dashed lines representing the 1, 2, 3, and 4 ratios between CO and dust radii. 
The median of the $R^{\mathrm{CO}}_{68\%}$ / $R^{\mathrm{dust}}_{68\%}$ ratio is $2.5$, excluding disks with an upper limit value in CO and/or dust size. 
The dispersion of this sample (considered as the standard deviation of the size ratio sample) is relatively high, of $1.5$, raised by the few disks with very high size ratios. The median and dispersion of the size ratio when using the $R_{90\%}$ CO and dust radii are slightly larger, 2.7 and 1.5 respectively. 
In Appendix~\ref{sec:appendix_singularobjects}, we describe in more detail disks from singular objects, namely, disks with very high size ratios (\ref{sec:sz75} - \ref{sec:sz65}), the brightest object of the sample (Sz~82, \ref{sec:sz82}), and the results of disks around BDs and very-low mass stars (\ref{sec:bdsvlms}). 
We also note that a few disks with measured sizes are orbiting a component of a binary or multiple system. We have only considered systems with relatively large angular separation between components (> 2 ${}^{\prime}{}^{\prime}$). The impact of binarity and effects such as tidal truncation cannot be constrained based on our limited sample of disks that are part of a multiple system. 

The measured size ratios might be even larger on compact objects: \cite{trapman+2019} showed that the measured size ratio is lower than the true value on disks with sizes similar to the beam size. 
On the other hand, the demographics analysis is affected by a lower completeness of fainter and non-detected CO disks. These disks would likely have small CO/dust size ratios. 
There is indeed a number of disks with measured $R_{\mathrm{dust}}$ but without $R_{\mathrm{CO}}$: these disks spread over the entire $M_{\star}$ range. Therefore, these disks with presumably low size ratio would appear along the full $M_{\star}$ range.

In the Lupus sample, most of the disks around more massive stars are detected in both CO and dust and the sizes could be characterized. The completeness level at the low $M_{\star}$ range of the sample is lower, since disks around less massive objects are generally fainter in continuum and line emission. 
Therefore, we focus on the solar mass range sub-sample in order to reduce the possible biases due to a lower completeness. In the stellar mass range between 0.7 and 1.1 $M_{\odot}$, the total number of protoplanetary disks in the Lupus sample considered (Section~\ref{sec:sampleselection}) is 10. All of them are detected in $\ce{^{12}CO}$ and dust continuum. One source (Sz~68) is excluded from the analysis since is a multiple system with angular separation below $2{}^{\prime}{}^{\prime}$. Another object (Sz~77) has an upper limit on the $R_{\mathrm{dust}}$. The remaining eight disks have measured radii in $\ce{^{12}CO}$ and dust continuum. The size ratio median in this mass range remains 2.5, with a dispersion of 2. If the $R_{90\%}$ sizes are used instead, the median for this sub-sample is 2.6, with a dispersion of 2.2. 

The median of the size ratios for the entire sample or the sub-sample considered are higher than the average value of 2 measured in \cite{ansdell+2018}. The CO sizes in this work are in good agreement with the 22 measured disk sizes in \cite{ansdell+2018}. 
A possible explanation of this difference is the larger sample of disks with measured sizes (42 disks in this work against 22). 
If we only consider the same sample of disks from  \cite{ansdell+2018} with measured sizes, the size ratio median is again $2.5$ (same median if the $R_{90\%}$ radii are used). Thus, the difference with respect the previous study is not due to a larger sample.

Another possible explanation is a difference in the measured dust size. Indeed, the $R_{90\%}^{\mathrm{dust}}$ in this work is $\sim22\%$ shorter than the tabulated sizes in \cite{ansdell+2018}. This, in combination with the uncertainty and scatter of the sample, accounts for the observed size ratio difference. 
This discrepancy in $R_{90\%}^{\mathrm{dust}}$ can be due to two main differences between the two studies. Firstly, this work makes use of continuum emission in ALMA Band~7 ($\sim 0.89$ mm), while \cite{ansdell+2018} used the continuum emission in ALMA Band~6 ($\sim 1.33$ mm). 
The datasets of the two bands differ also in spatial resolution and sensitivity (on average, Band 7 observations with beam size FWHM of $0.{}^{\prime}{}^{\prime}32$ and rms of $\sim$0.3 mJy, Band6 with $0.{}^{\prime}{}^{\prime}23$ and $\sim0.1$mJy). However, the size results in \cite{ansdell+2018} are mostly for bright and relatively large disks, thus sensitivity/resolution should not have a strong effect. 
The second difference is the method to infer the dust sizes; Nuker profile modeling by fitting the continuum visibilities (this work), and the curve of growth method in the image plane (previous work). 
In order to understand the origin of this dust size difference, we used the curve of growth method for the Band 6 and Band 7 continuum maps of the disks with measured $R_{90\%}^{\mathrm{dust}}$ in \citet{ansdell+2018}. 
In both cases, the sizes match the results from the previous study. The curve of growth sizes from Band 7 are marginally larger than the Band 6 sizes ($\sim6\%$), this is expected since disks observed in Band 7 are typically brighter, optically thicker, and probe slightly smaller grains (thus less affected by radial drift). 
These tests indicate that the cause of the size ratio difference between this work and \citet{ansdell+2018} is the method used to infer the dust size, rather than the different ALMA Band considered. The curve of growth method typically overestimates the dust extent.  Therefore, we favor the use of our method, which also provides dust sizes that concur with other recent works  \citep[e.g.,][]{tazzari+2017,andrews2018A,hendler+2020}.


\section{Discussion}\label{sec:discussion}
In this section we discuss the physical implications of the CO and dust continuum sizes that we found for the entire Lupus disk population. Additionally, thanks to the significant number of disks with measured CO and dust sizes, we search for possible correlations between the measured size ratio of the sample and various stellar and disk properties.

\subsection{Disk evolution: gas size relative to dust size}\label{sec:rgasrdust}
The relative size between gas and dust is a fundamental property of protoplanetary disks, since it can be linked to evolutionary processes of the disk  \citep[e.g.,][]{dutrey+1998,facchini+2017,trapman+2019}. If the disk has undergone dust evolution (understood as grain growth and subsequent radial drift), the dust emission at (sub-)mm  wavelength may appear much more compact than the gas emission. 
Gas evolution, on the other hand, cannot be constrained based only on the relative gas/dust size: the main two mechanisms of angular momentum transport (viscous evolution, wind-driven accretion) generally cause the gaseous disk to either increase in size or remain similar, thus, the two mechanisms contribute to a large gas/dust size if dust evolution has occurred. 

A second major effect that contributes to the observed size divergence between gas and dust is the optical depth. This effect is due to the larger optical depth of the $\ce{^{12}CO}$ rotational line with respect to the continuum emission at (sub-)mm  wavelengths. This causes the line emission to appear more extended than the optically thinner dust emission. 
Another way to understand the optical depth effect is by assuming a disk where gas and dust are equally distributed. If the dust emission is optically thin, the $R_{68\%}^{\mathrm{dust}}$ would trace the $68\%$ of the total disk mass. But the $R_{68\%}$ of an optically thick line would trace a larger fraction of the total disk mass, since the line emission from the innermost region is hidden due to the optical thickness. Thus, the measured $R_{68\%}^{\mathrm{CO}}$ of the optically thick line would necessarily be larger than the $R_{68\%}^{\mathrm{dust}}$. 

The presence of pressure bumps might also have an influence on the size ratio of a disk. If present, radial drift would stop at the location of the outermost bump, resulting in piled up dust, and likely a ring-like structure. Although very high sensitivity and resolution observations are needed in order to confirm the presence of bumps or rings, most of the Lupus disks targeted on the DSHARP project show rings or enhancements of dust emission \citep{huang+2018}. The existence of bumps might cause dust sizes to be larger, resulting in smaller size ratios. The existence of bumps does not necessarily produce small size ratios; it would ultimately depend on the location of the bump. 

As a result, disentangling between dust evolution and optical effect is very difficult: while the optical depth is almost certainly present, the dust evolution does not necessarily occur. \cite{trapman+2019} studied in detail the possible contributions of these and other effects to the gas/dust size ratio based on a large grid of thermo-chemical models \citep{facchini+2017}, and concluded that a size ratio higher than 4 is a clear sign of dust evolution. For disks below this threshold, dust evolution could still have occurred, but specific modeling of each disk is required in order to confirm it. 
In their study, the same radius definition as in the present work was used (a fraction of the total observable flux, not a physical radius), and their CO radii were obtained by measuring the flux extent of the same CO line ($\ce{^{12}CO}$ $J = 2-1$), with differences in the CO sizes below 10$\%$ when considering the $\ce{^{12}CO}$ $J = 3-2$ line. 
Therefore, their findings can be directly applied to our size ratio measurements. The population's mean value of 2.5 that we obtain is far below the ratio threshold of 4 suggested by \cite{trapman+2019}, thus radial drift cannot be confirmed as an ubiquitous process of the Lupus disk population. 
The threshold value of 4 might change slightly with a different setup of the thermo-chemical modeling. In \citet{trapman+2019}, the threshold was obtained for a standard disk with a number of assumptions (most significantly, the gas structure being set by a self-similar solution of a viscous accreting disk, and a local gas-to-dust ratio of 100). 

The fraction of disks with size ratios above the threshold value of 4 is $\sim$15$\%$ for the entire population ($\sim$13$\%$ if we only consider disks whose size ratio uncertainties are strictly above 4). 
If we examine the 0.7-1.1 $M_{\odot}$ sub-sample, the fraction is marginally higher, 2 out of 9 objects (1 out of 9 if only objects in this mass range with size ratio uncertainties above 4 are considered). 
These fractions of disks above the threshold remain the same if the $R_{90\%}$ radii are used instead of $R_{68\%}$, for the entire disk population, and for the 0.7-1.1 $M_{\odot}$ sub-sample. Following \citet{trapman+2019} results, these disks with size ratio above the threshold can only be explained if dust evolution took place.

The sources that we identified as having with size ratio above the threshold of 4 are (ordered from higher to lower size ratio): 
Sz~75, Sz~131, Sz~69, Sz~83, Sz~65, and Sz~111. 
Although this sub-set of sources is small, the main stellar and disk properties cover relatively wide ranges, for instance, the stellar masses are distributed throughout 0.2 and 0.8 $M_{\odot}$. 

In Appendix~\ref{sec:appendix_singularobjects}, we describe in detail each of these systems with high size ratios. 
Sz~83, one of the most active sources of the Lupus clouds, might have a lower size ratio when considering the dynamical size based on keplerian motion \citep{huang+2020}. On the other hand, for Sz~69 we only provide a lower bound the size of the dust emission cannot be accurately determined. 
We did not find any properties or features that these disks might share: specifically, their accretion signatures are ordinary \citep{alcala+2017}, and only one of them is a known transition disk \citep[Sz~111 disk,][]{vandermarel+2018}. 
Three of these disks belong to wide binary systems, at separations at which tidal truncation effects should not have any incidence. 

In summary, $\sim 15$-$20\%$ of the disk population in Lupus has a disk size ratio greater than 4. This result suggests that a considerable fraction of protoplanetary disks in Lupus have suffered radial drift and dust evolution, which is crucial to form the cores of planets.

\subsection{Possible correlations between the size ratio and other stellar and disk properties}\label{sec:possiblecorrelations}
\begin{figure*}
    \begin{center}
    \includegraphics[width=.495\textwidth]{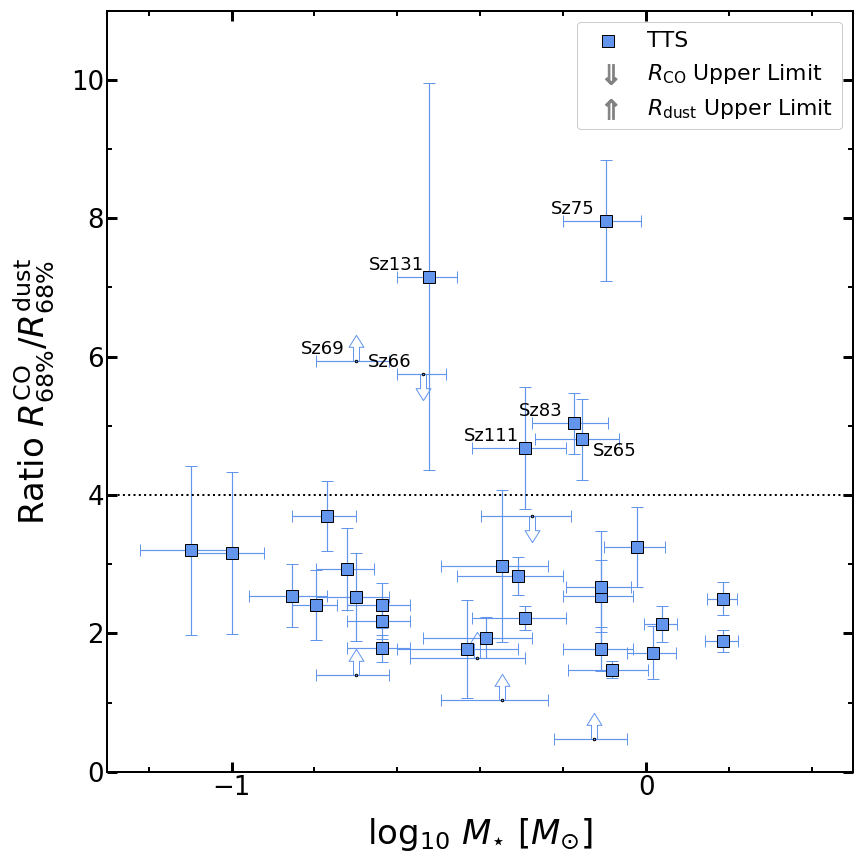}
    \includegraphics[width=.495\textwidth]{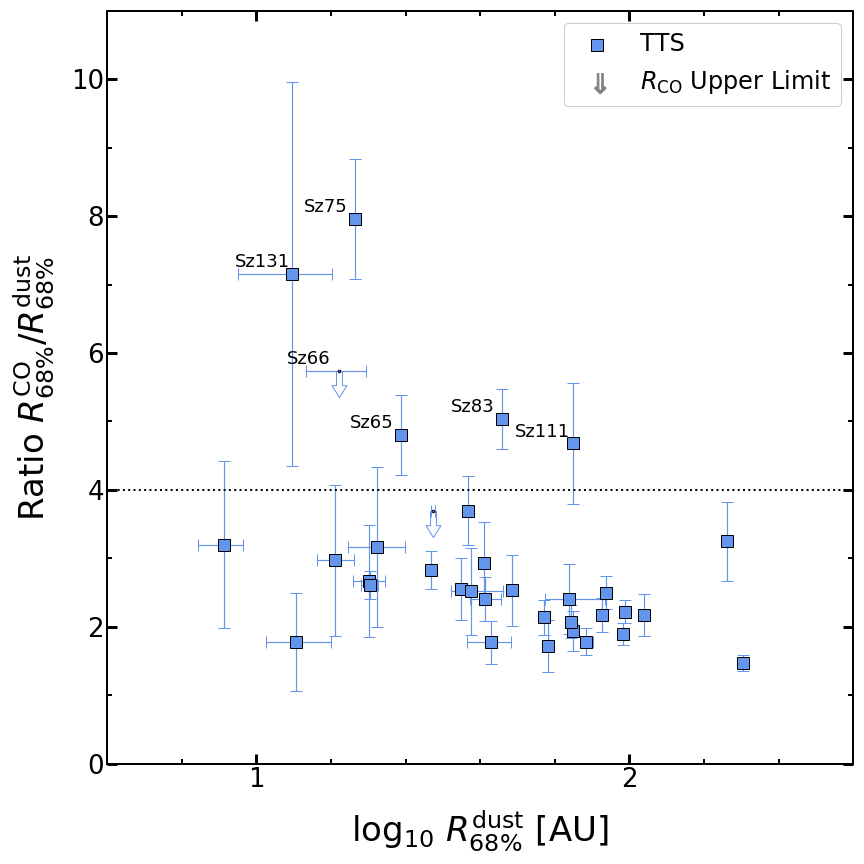}
    \end{center}
    \begin{center}
    \includegraphics[width=.495\textwidth]{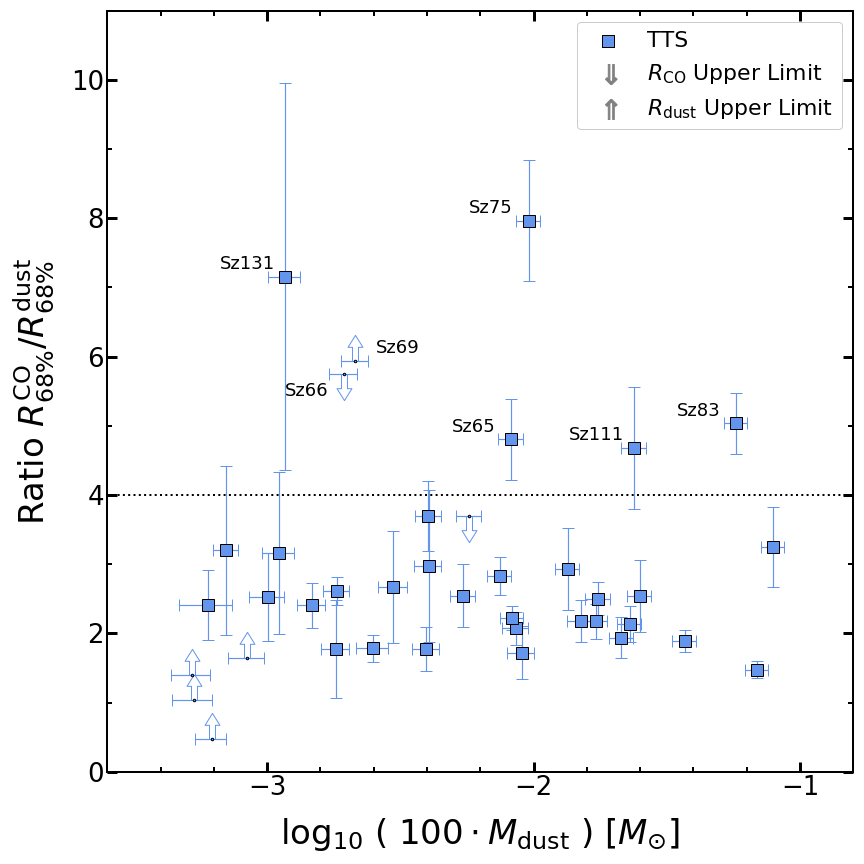}
    \includegraphics[width=.495\textwidth]{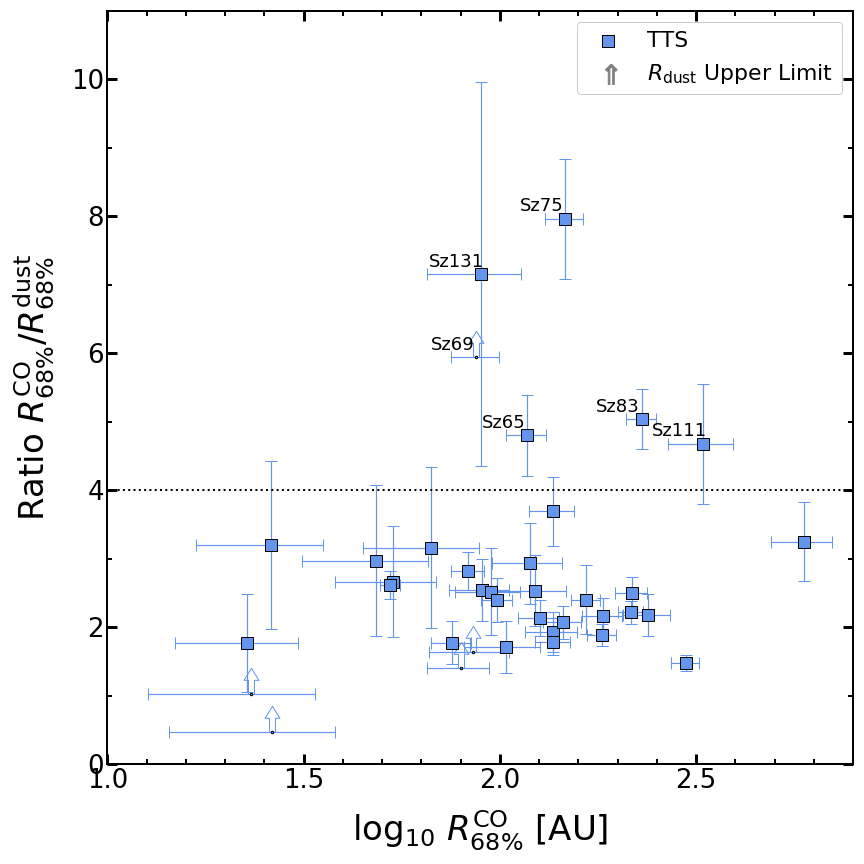}
    \end{center}
  \caption[]{
  Ratio between CO and dust sizes as a function of various stellar and disk properties: 
  (\textit{top left}) 
  as a function of the stellar mass of the central object ($M_{\star}$); 
  (\textit{top right}) 
  as a function of the dust size ($R_{68\%}^{\mathrm{dust}}$); 
  (\textit{bottom left}) 
  as a function of the disk mass, estimated from $M_{\mathrm{dust}}$ and assuming a gas-to-dust ratio of $100$; 
  (\textit{bottom right}) 
  as a function of the CO size ($R_{68\%}^{\mathrm{CO}}$). 
  Disks with a size ratio above the horizontal threshold cannot be explained without prominent dust evolution, based on disk evolution models of \citet{trapman+2019}.
  }
  \label{fig:correlations_rgasrdust}
\end{figure*}

The large population of disks with characterized CO and dust sizes allows us to search for possible correlations between the CO/dust size ratio and the main stellar and disk properties. We examined the relation between the size ratio and the stellar mass, the total disk mass, and the dust and CO sizes separately.  Figure~\ref{fig:correlations_rgasrdust} shows the size ratio as a function of each of these properties. The stellar mass and its uncertainty are obtained as explained in Section~\ref{sec:sampleselection}, while for the CO and dust sizes and uncertainties, we use the results from the modeling described in Section~\ref{sec:modeling} (sizes summarized in Table~\ref{tab:rgasresults}). 
The total disk mass is approximated from the dust disk mass, assuming a gas-to-dust ratio of 100. The dust disk mass is computed assuming that the emission is optically thin and in the Rayleigh-Jeans regime \citep{beckwith+1990}, with an average temperature on the dust mid-plane of 20 K \citep[as in][]{pascucci+2016,ansdell+2016,ansdell+2018,sanchis+2020a} and a dust optical depth of $\kappa_{890\mu m}$ = 2 $\mathrm{cm}^2 \mathrm{g}^{-1}$ \citep[as in][]{ricci+2014,testi+2016,sanchis+2020a}. The uncertainty considered for the $M_{\mathrm{dust}}$ is the $10\%$ associated to the flux calibrator uncertainty of the ALMA observations. The inferred $M_{\mathrm{dust}}$ values of each disk are included in Table~\ref{tab:rgasresults}. While this is a big approximation for the disk mass, it is useful in order to have an overall understanding of the available disk mass. 

For this examination we make use of the Spearman and Pearson correlation coefficients \citep[similar to the analysis of dust properties' correlations conducted in][]{hendler+2020}. The Spearman test measures the monotonicity of the relationship between two sets of variables (its null hypothesis is that the two sets are monotonically uncorrelated), while the Pearson test evaluates the linear relationship between the two sets (its null hypothesis being that the two sets are linearly uncorrelated). The Pearson test assumes that the two variables are normally distributed. Therefore, we also test the normality of each disk property by performing the Shapiro-Wilk test \citep{ShapiroWilk1965}, which null hypothesis is that the set of values is drawn from a normal distribution. The \texttt{scipy.stats} Python module\footnote{\texttt{https://docs.scipy.org/doc/scipy/reference/stats.html}} is used to perform the aforementioned tests. For each relationship, the tests are performed by excluding all objects with upper limits in the CO size or the dust size. If the p-value of a given test is below 0.05, the null hypothesis of the respective test is rejected.

\begin{table*}
\caption{Results of the statistical tests searching for possible correlations between the size ratio and various stellar and disk properties ($M_{\star}$, $M_{\mathrm{disk}}$, $R_{\mathrm{dust}}$, and $R_{\mathrm{CO}}$). The p-value of each test is included in parenthesis. 
The first four rows summarize the tests' results when considering the entire population of disks with characterized CO and dust sizes (excluding upper limits); the last four rows show the results when excluding disks with size ratios above 4.}
\label{tab:correlations}      
\centering              
\begin{tabular}{lccccr}   
\hline\hline            
X-axis & Y-axis & Spearman test & Shapiro test X-axis & Shapiro test Y-axis & Pearson test  \\
\hline 
$\log_{10} M_{\star}$ [$M_{\odot}$] & $R_{\mathrm{CO}} / R_{\mathrm{dust}}$ & Non-monotonic (0.33) & Normal (0.31) & Not normal (4e-5) & -  \\
$\log_{10} M_{\mathrm{disk}}$ [$M_{\odot}$] & $R_{\mathrm{CO}} / R_{\mathrm{dust}}$ & Non-monotonic (0.61)  & Normal (0.60) & Not normal (9e-6) & -  \\
$\log_{10} R_{\mathrm{dust}}$ [AU] & $R_{\mathrm{CO}} / R_{\mathrm{dust}}$ & Monotonic (0.002) & Normal (0.91) & Not normal (9e-6) & -  \\
$\log_{10} R_{\mathrm{CO}}$ [AU] & $R_{\mathrm{CO}} / R_{\mathrm{dust}}$ & Non-monotonic (0.75) & Normal (0.70) & Not normal (9e-6) & -  \\
\hline

$\log_{10} M_{\star}$ [$M_{\odot}$] & $R_{\mathrm{CO}} / R_{\mathrm{dust}}$ & Non-monotonic (0.06) & Normal (0.36) & Normal (0.39) & Non-linear (0.07)  \\
$\log_{10} M_{\mathrm{disk}}$ [$M_{\odot}$] & $R_{\mathrm{CO}} / R_{\mathrm{dust}}$ & Non-monotonic (0.22) & Normal (0.81) & Normal (0.38) & Non-linear (0.22)  \\
$\log_{10} R_{\mathrm{dust}}$ [AU] & $R_{\mathrm{CO}} / R_{\mathrm{dust}}$ & Monotonic (0.008) & Normal (0.93) & Normal (0.38) & Linear (0.025) \\
$\log_{10} R_{\mathrm{CO}}$ [AU] & $R_{\mathrm{CO}} / R_{\mathrm{dust}}$ & Non-monotonic (0.19) & Normal (0.71) & Normal (0.38) & Non-linear (0.53)  \\
\hline
\end{tabular}
\end{table*}
The results of the tests are summarized in Table~\ref{tab:correlations}. The size ratio of the sample is not normally distributed since the null hypothesis of the Shapiro-Wilk test is rejected. Therefore we cannot test for linearity between the size ratio and the other properties. However, the Spearman test can be performed independently of the normality of the properties. 
The relation between the size ratio and the $R_{\mathrm{dust}}$ is the only one that rejects the null hypothesis of the Spearman test, that is, it is unlikely that the size ratio and the $R_{\mathrm{dust}}$ are monotonically uncorrelated. Additionally, the measured size ratio of compact disks (those with sizes of the order of the beam size) may be lower than the true value due to the beam size \citep{trapman+2019}. In such case, the anti-correlation with $R_{\mathrm{dust}}$ might be steeper than what is seen in Figure~\ref{fig:correlations_rgasrdust}. 
However, this result should be taken with caution, since the Y-axis (the size ratio) is dependent on the X-axis ($R_{\mathrm{dust}}$ is the denominator in the size ratio), thus the anti-correlation found could be boosted by this dependence between the two axes. Besides, the test does not take into account uncertainties, which are large in the Y-axis. 
If the anti-correlation with $R_{\mathrm{dust}}$ is true, it would mean that compact dusty disks have higher size ratios than extended dusty disks. And, if we consider \citet{trapman+2019} findings, radial drift and dust evolution might be more prominent in these compact dusty disks. 
The Spearman test found no monotonicity between the size ratio and $R_{\mathrm{CO}}$, thus the size ratio is more tightly affected by the dust size than the CO size. For the remaining properties (i.e.,  stellar and disk masses), no correlations are found. 

The results plotted in Figure~\ref{fig:correlations_rgasrdust} show that disks with very large size ratios (e.g., above the threshold considered) appear along the full range of stellar masses, disk masses and CO sizes. These disks with exceptionally high size ratios may be in a different evolutionary stage compared to the bulk of the disk population. 
Therefore, we performed the correlation tests but excluding disks with size ratios above the considered threshold of 4. The results of the different tests are summarized in the bottom rows of Table~\ref{tab:correlations}. 
Based on the Shapiro test, the size ratio of this sub-sample is normally distributed, thus the Pearson test can be performed. 
In this sub-sample, the tests yield very low likelihood that the size ratio is uncorrelated with the dust size, analogous to the results for the entire sample. 
The tests do not confirm possible correlations with the remaining properties, although, the p-value of the Spearman and Pearson's test between the size ratio and the stellar mass are very low (0.06 and 0.07). This result might point towards a possible anti-correlation with $M_{\star}$; based on \citet{trapman+2019} results,
this would tentatively suggest that dust evolution could be more efficient in disks around less massive stars. 
This is in line with theoretical and observational work that suggested that radial drift is more effective in disks around low-mass stars \citep{pinilla+2013,pascucci+2016,mulders+2015}. 
In order to confirm or refute a tentative anti-correlation with $M_{\star}$, it is necessary to significantly increase the sample of disks with measured gas and dust sizes.

The results of these statistical tests show a remarkable lack of strong correlations between the size ratio and the investigated properties. The sample of disks with measured size ratios is considerable, and it covers a very wide range of stellar masses, disk masses, dust and CO sizes. And yet, the vast majority of the disks have similar ratios, between $\sim2$ and 4. This denotes that, aside from the small fraction of disks with exceptionally high size ratios, the bulk of the population behaves in a similar manner, independent of its stellar and disk properties. 
Extending the sample of disks with characterized gas and dust sizes is essential to confirm the results found. In particular, by expanding over other SFRs, the evolution of the size ratio over time can be investigated: this would help us to further constrain the ongoing and/or suffered physical processes and the evolutionary stage of the disks.

\subsection{Optically thick emission and CO temperature}\label{sec:coopticaldepth}
\begin{figure}
  \resizebox{\hsize}{!}{\includegraphics{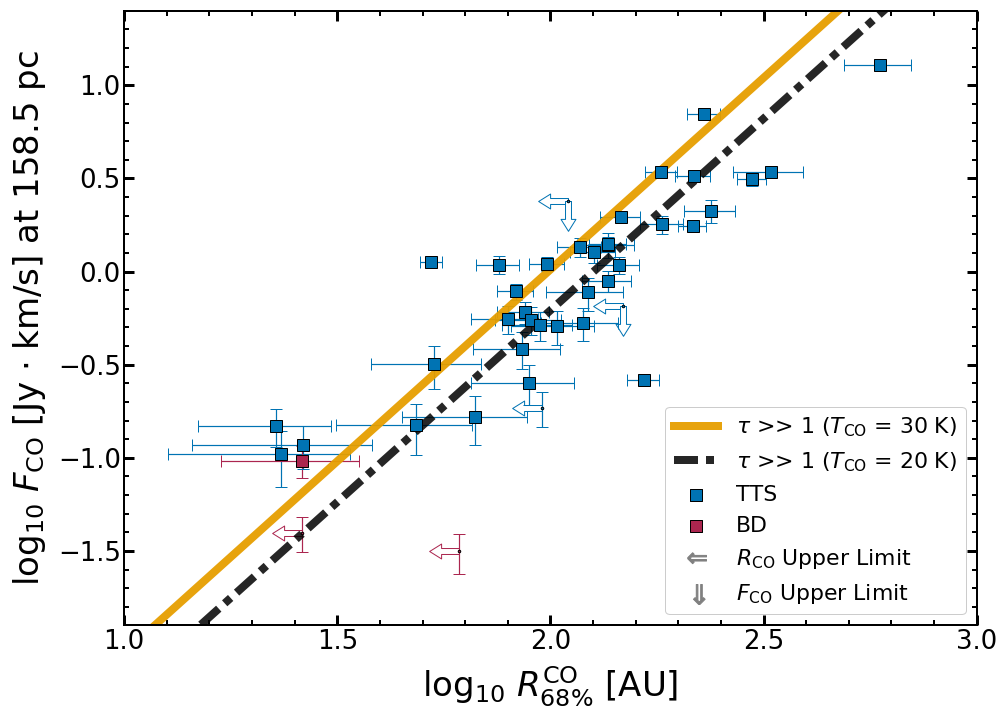}}
  \caption{Relation between the radius enclosing $68\%$ of the total CO flux (scaled at the median distance of the region, and deprojected with inclination) and the flux enclosed by that radius for the entire Lupus CO disk population. Lines represent optically thick emission of CO with an average temperature of 30 K (orange), and of 20 K (black dash-dotted). Objects with outflows within the measured radius are considered to be upper limits in $F_{\mathrm{CO}}$.}
  \label{fig:rgas68_fco}
\end{figure}

Lastly, we have investigated the CO emission as a function of the CO size, and tried to constrain the temperature of the CO emitting layer. Figure~\ref{fig:rgas68_fco} shows the modeled CO flux plotted against the CO size for the entire sample. 
First, we have performed statistical tests (as in Section~\ref{sec:possiblecorrelations}) searching for possible correlations between the two properties. The Spearman test provides a very low p-value (of 3e-5, obtained by excluding CO size upper limits). Thus its null hypothesis is rejected, and the two properties are monotonically correlated. 
On the other hand, linearity could not be tested since the CO flux sample is not normally distributed (its p-value from the Shapiro test is $<0.05$). 
The monotonic correlation found is expected due to the optically thick emission of the $\ce{^{12}CO}$ lines. 

Based on this result, it is also possible to examine the temperature of the CO emitting layer. In Figure~\ref{fig:rgas68_fco}, we have plotted an orange line representing optically thick emission with an average CO temperature ($T_{\mathrm{CO}}$) of 30 K. This line is composed by a grid of optically thick emission profiles with constant temperature. 
These profiles are constant with radius, thus  described as $I_{\mathrm{CO}}(R) = B_{\nu}(T_{CO})$, with $\nu$ being the frequency of the $\ce{^{12}CO}$ ($J = 2-1$) transition line, and $T_{\mathrm{CO}} = 30$ K. This $T_{\mathrm{CO}}$ is based on the results in \cite{pinte+2018a}, where the emission profile of the same CO line (among dust continuum and other transition lines) was studied in detail for the IM~Lup disk. 
The grid of profiles is assembled by taking increasing values of the outer disk edge, in order to cover the entire x-axis and populate the plot. For each profile, we computed the radius enclosing the 68$\%$ of the total intensity , and by plotting all the profiles we obtain the orange line. 

In the figure, a disk with optically thick emission and an average $T_{\mathrm{CO}} = 30$ K would be intersected by this line. Around one third of the sample is crossed by this line when considering their uncertainties in radius. 
Considering only systems whose errorbars do not cross the line, 
four disks lay on the left side of the optically thick line:
one disk (Sz~72) is among the faintest disks of the sample, and its size uncertainty is large. 
This source, together with other two objects on the left side (Sz~73, Sz~102) can be explained either by an underestimate of their CO size, by a higher CO temperature, or a combination of both. From the CO emission maps, these disks have a bright compact core of emission (thus, likely warmer than 30 K), and their outer disk emission is either faint or absent. This could happen if the outer emission is fainter than the sensitivity of the observations. 
For the last source on the left side of the line (EX~Lup), the presence of a blueshifted molecular outflow \citep{hales+2018} makes determination of the CO disk flux and size difficult, thus its exact position in the plot is uncertain. 

On the other hand, a considerable fraction of the population appears on the right side of the optically thick line at $30$ K (about half of the sample, only considering disks with errorbars not crossing the line). This can be explained by several factors. 
Firstly, cloud absorption, which can be seen in the line spectrum in a considerable number of disks (see  Appendix~\ref{sec:appendix_obsmodres}) can explain disks on the right side of the line. Absorption from clouds would decrease the total CO emission, while the measured radius would be mostly unaffected \citep{ansdell+2018}. 
Another possible explanation is that the average CO temperature of some of these disks could be below 30 K, this would be expected in very extended sources, since the regions further from the star are generally colder. 
Besides, the inclination of the disks might also have an effect. The optically thick line plotted assumes a face-on orientation: if inclined, the emission would appear fainter (the optically thick lines would shift downwards). Nevertheless, the CO fluxes of the Lupus disks shown in Figure~\ref{fig:rgas68_fco} are corrected accounting for the disk inclination, thus this effect should be minor. 
Lastly, partially optically thick CO emission in the outer disk can cause the emission to be fainter, thus appearing below the optically thick line. 

A second line representing optically thick emission at the typical freeze-out temperature of CO ($T_{\mathrm{CO}} = 20$ K) is included in the figure. 
Eight disks appear on the right side of the 20 K line, taking errorbars into account. These disks are likely explained by a combination of the aforementioned effects that shift the position of the disk to the right side of the line. 
However, it might be possible that the CO in some regions of these disks is indeed at temperatures lower than the freeze-out temperature, which could be explained by vertical mixing, as suggested in \cite{pietu+2007}. 

For the case of the three BDs in the sample, their radii are obtained from a different CO transition ($\ce{^{12}CO}$ 3--2). The optically thick lines of the transition used for the BDs would appear slightly above the drawn lines of Figure~\ref{fig:rgas68_fco}. 
The only BD with measured size lays within errorbars on the 30 K line. 

In conclusion, a monotonic correlation between the CO size and flux is found, as expected from optically thick emission. Our results for the temperature of the CO emitting layer are consistent with a temperature of around 30K as previous studies suggested, albeit a fraction of the sample might have slightly lower average temperatures. However, the exact determination of $T_{\mathrm{CO}}$ for the sample or for individual sources is difficult due to the size uncertainties, cloud absorption and other factors that limit this analysis.


\section{Conclusions}\label{sec:conclusions}
We have investigated the relative extent of gas and dust in a large sample of protoplanetary disks of the Lupus clouds, in order to constrain the evolutionary stage of the disk population. 
We have assembled the largest sample of protoplanetary disks of the region with characterized CO and dust sizes based on ALMA observations. To infer the gas disk sizes, we have modeled the integrated emission maps of the $\ce{^{12}CO}$ ($J = 2-1$) transition line from ALMA Band 6 observations using an elliptical Gaussian function, or a Nuker profile for models with considerable residuals. For the dust modeling, the continuum emission of large grains (at $\sim$0.89 mm wavelength) is modeled in the \textit{uv}-plane to a Nuker profile. 
The radii enclosing $68\%$, $90\%$, $95\%$ of the respective total flux, are estimated from the CO and dust models. The CO/dust size ratio ($R_{\mathrm{CO}}$/$R_{\mathrm{dust}}$) is then used to investigate the evolutionary stage of the disk population: prominent dust evolution (i.e., grain growth and radial drift) typically produces compact dust emission at these wavelengths, thus high size ratios; gas evolution, on the other hand, cannot be constrained based only on this size ratio. 

The median value of the size ratio is $2.5$ for the entire population and for a sub-sample with high completeness. 
$15\%$ of the population show a size ratio above 4 ($20\%$ when considering a sub-sample with high completeness), 
based on thermo-chemical modeling \citep{facchini+2017,trapman+2019}, 
such high values can only be explained if grain growth and subsequent radial drift has occurred. These disks with very high size ratios do not show unusual characteristics, and their stellar and disk properties cover wide ranges of the entire population. For the rest of the population, dust evolution cannot be ruled out, but individual thermo-chemical modeling is necessary. 

We have searched for possible correlations of the population' size ratio with other stellar and disk properties. Only a tentative monotonic anti-correlation with $R_{\mathrm{dust}}$ is suggested by the null hypothesis tests performed. 
The absence of strong correlations is very significant, the studied sample covers a wide range of stellar and disk properties, and the vast majority of the population has a very similar size ratio (between $\sim2$ and 4). This suggests that a large fraction of protoplanetary disks in Lupus behave similarly and may be in a similar evolutionary stage. These results are limited by the optical depth difference between continuum and $\ce{^{12}CO}$ ($J = 2-1$) line, which can affect each disk's measured size ratio differently, thus hiding their true behavior. 
Additionally, extending this analysis to the disk population in other SFRs is pivotal to learn about the temporal evolution and the evolutionary stages of protoplanetary disks. 
Finally, a monotonic correlation between the CO disk flux and size is found. The CO temperature for most of the disks, although difficult to determine accurately, is consistent with previous studies that suggest an average temperature of around 30 K.


\begin{acknowledgements}
We thank the anonymous referee for providing constructive comments that helped to improve the clarity and quality of the manuscript. 
This paper makes use of the following ALMA data: 
ADS/JAO.ALMA$\#$2017.1.01243.S, 
ADS/JAO.ALMA$\#$2013.1.00220.S, 
ADS/JAO.ALMA$\#$2013.1.00226.S, 
ADS/JAO.ALMA$\#$2013.1.00663.S, 
ADS/JAO.ALMA$\#$2013.1.00694.S, 
ADS/JAO.ALMA$\#$2013.1.00798.S, 
ADS/JAO.ALMA$\#$2013.1.01020.S, 
ADS/JAO.ALMA$\#$2015.1.00222.S, 
ADS/JAO.ALMA$\#$2016.1.00484.L, and
ADS/JAO.ALMA$\#$2016.1.01239.S. 
ALMA is a partnership of ESO (representing its member states), NSF (USA) and NINS (Japan), together with NRC (Canada) and NSC and ASIAA (Taiwan) and KASI (Republic of  Korea), in cooperation with the Republic of Chile. The Joint ALMA Observatory is operated by ESO, AUI/NRAO and NAOJ. 
This work was partly supported by the Italian Ministero dell Istruzione, Universit\`a e Ricerca through the grant Progetti Premiali 2012 – iALMA (CUP C$52$I$13000140001$), by the Deutsche Forschungs-gemeinschaft (DFG, German Research Foundation) - Ref no. FOR $2634$/$1$ TE $1024$/$1$-$1$, and by the DFG cluster of excellence Origins (www.origins-cluster.de). 
This project has received funding from the European Union's Horizon 2020 research and innovation programme under the Marie Sklodowska-Curie grant agreement No 823823 (DUSTBUSTERS) and from the European Research Council (ERC) via the ERC Synergy Grant {\em ECOGAL} (grant 855130). 
S. F. acknowledges an ESO Fellowship. 
T. H. acknowledges support from the European Research Council under the Horizon 2020 Framework Program via the ERC Advanced Grant Origins 83 24 28. 
IdG-M is partially supported by MCIU-AEI (Spain) grant AYA2017-84390-C2-R (co-funded by FEDER). 
K. M. acknowledges funding by the Science and Technology Foundation of Portugal (FCT), grants No. IF/$00194$/$2015$, PTDC/FIS-AST/$28731$/$2017$ and UIDB/$00099$/$2020$.
\end{acknowledgements}


\bibliographystyle{aa} 
\bibliography{gasdisks_lupus.bib} 


\begin{appendix}

\section{Interferometric modeling of DSHARP line emission}\label{sec:appendix_visfit}
While analysis in the image plane gives a first order insightful information of the emission, fitting the observations in the \textit{uv}-plane provides the most robust method to characterize the disk emission. By working in the \textit{uv}-plane, we avoid systematic errors from the image reconstruction process (e.g., dependency on the weighting, and masking applied during this process). 

In recent work, interferometric modeling allowed to characterize dust continuum in large samples of disks from ALMA observations \citep[e.g.,][]{tazzari+2017,tripathi+2017,andrews2018A,sanchis+2020a}. We explore this methodology to model the line emission.

The interferometric modeling of the gas can be accomplished by integrating all channels that show disk emission after the continuum is subtracted. This is analogous (but in the Fourier space) to the moment zero map in which all the channel maps are summed up. The resulting visibilities are then modeled to an empirical emission function, analogous to the interferometric modeling of dust continuum conducted in \citet{sanchis+2020a}. 

Gas line emission can be modeled using any preferred empirical function, in this work we used the Nuker function \citep[see][for details of this function]{tripathi+2017} fitted in the \textit{uv}-plane as the fiducial CO sizes of these objects. The advantage of using this profile resides in independently fitting the inner and outer slopes of the disk emission. In Section~\ref{sec:dsharpradii} we also test the interferometric modeling by fitting a Gaussian function \citep[as Equation~2 in][]{sanchis+2020a}. This methodology assumes a axi-symmetric emission of the disk, thus substructure or asymmetries are not modeled. Nevertheless, the size determination is not affected by the presence of substructure. More elaborated functions that account for these second order features can be used on the interferometric modeling described here. 

The \texttt{Galario} package \citep{tazzari+2018} is used to convert the empirical model into synthetic visibilities, and to compute the $\chi^{2}$ between observed and synthetic visibilities. In addition, the affine invariant Markov Chain Monte Carlo method from \citet{GoodmanWeare2010} is also used \citep[via the \texttt{emcee} package][]{foremanmackey+2013} to investigate the parameter space, optimizing for models with the lowest $\chi^{2}$. This is done for 200 independent walkers for thousands of steps. After a number of steps the values of the free parameters converge to those that provide the best fit between synthetic and observed visibilities. 
Due to expensive computation time required to model the large set of visibilities from DSHARP observations, baselines $> 1000$ $k\lambda$ are excluded from the fit. 

\begin{figure}
  \resizebox{\hsize}{!}{\includegraphics{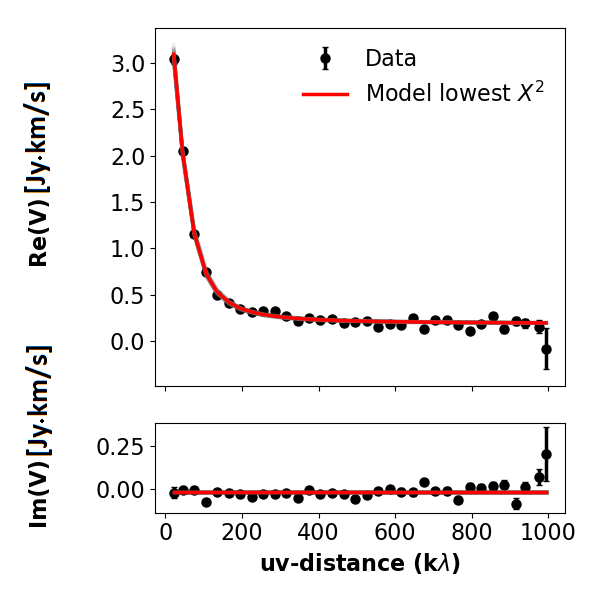}}
  \caption{Observed and model visibilities of Sz~71 CO emission, plotted as real and imaginary parts as a function of the baseline (in k$\lambda$). The data from the observations are plotted as black data points with error bars, the model with the lowest ${\chi}^{2}$ is shown as solid red curve, and a random set of converged models from the parameter space investigation are drawn as gray curves (mostly covered by the lowest $\chi^2$ model). This figure was made with the \texttt{uvplot} Python package \citep{uvplot_mtazzari}.}
  \label{fig:dsharpvisfit}
\end{figure}

\begin{figure}
  \resizebox{\hsize}{!}{\includegraphics{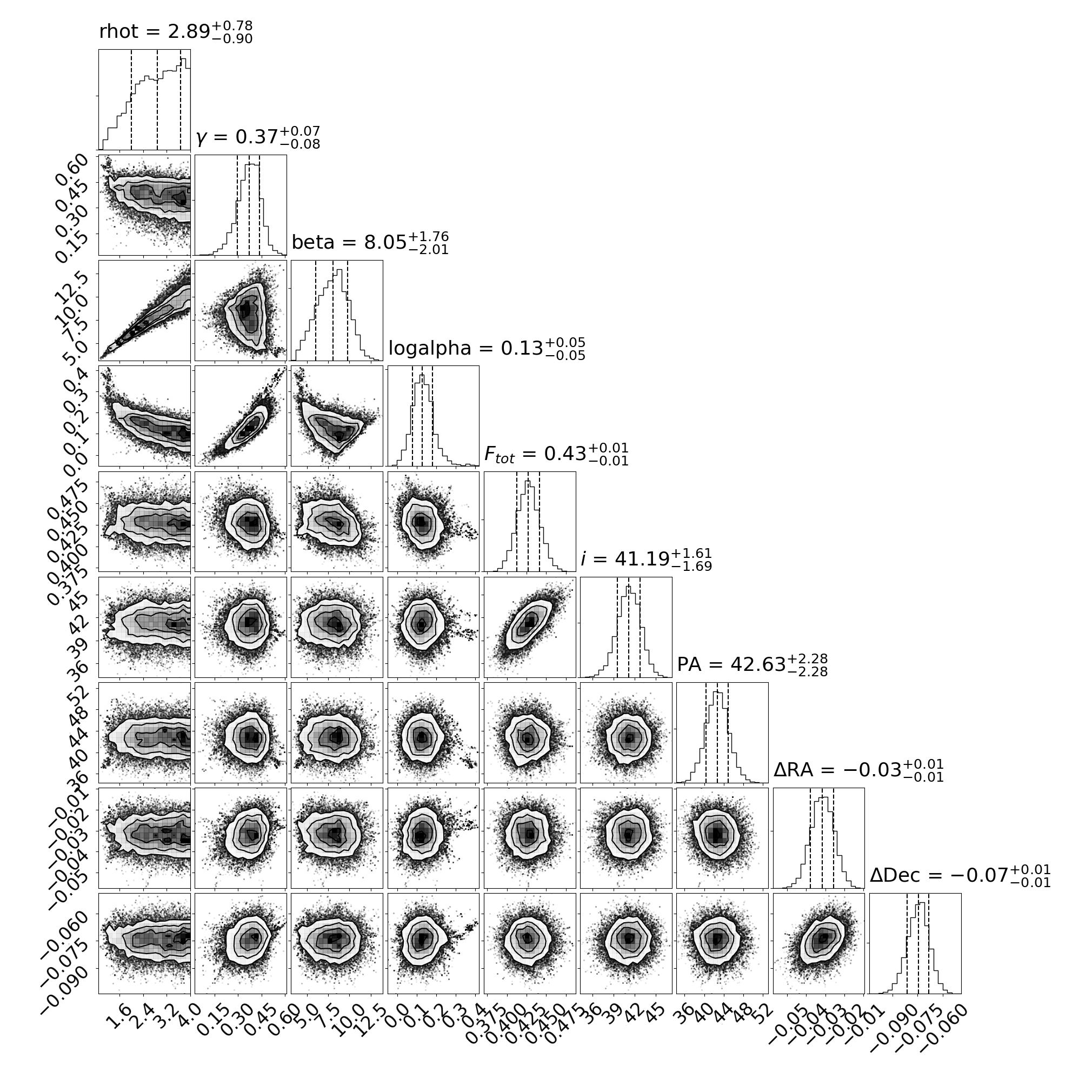}}
  \caption{Corner plot of the Nuker fitting of the observed CO visibilities. The top sub-panels show the histograms of the free parameters from the MCMC chains. The remaining sub-panels show the 2-dimensional histograms between pairs of parameters.}
  \label{fig:gwlupcorner}
\end{figure}

As an example, in the following figures of this Appendix we present the results of the interferometric modeling of the Sz~71 CO disk  fitted to a Nuker model. The fitting tool converges to the models with lowest $\chi^2$. In Figure~\ref{fig:dsharpvisfit}, we show the fit of the visibilities: real and imaginary part of observed and synthetic (for the model with lowest $\chi^2$) visibilities as a function of baseline. 
Once the fitting tool has converged, the subsequent Markov Chain Monte Carlo method (MCMC) chains store the values of the parameters that fit best to the observed visibilities. Histograms of the free parameters are build from these chains. The median of each parameter histogram are taken as the best value of the parameter, and the $16$th and $84$th percentiles are used as lower and upper uncertainties. The 1D and 2D histograms between model parameters are plotted in Figure~\ref{fig:gwlupcorner}.

\begin{figure*}
    \begin{center}
    \includegraphics[width=.8\textwidth]{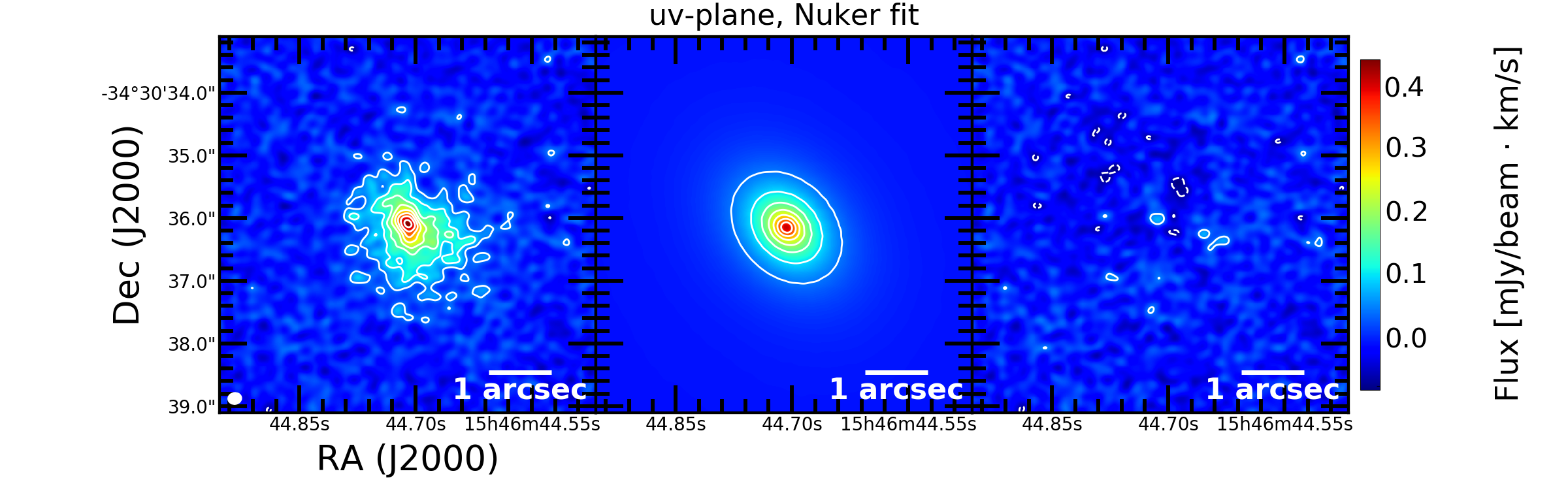}
    \end{center}
    \begin{center}
    \includegraphics[width=.8\textwidth]{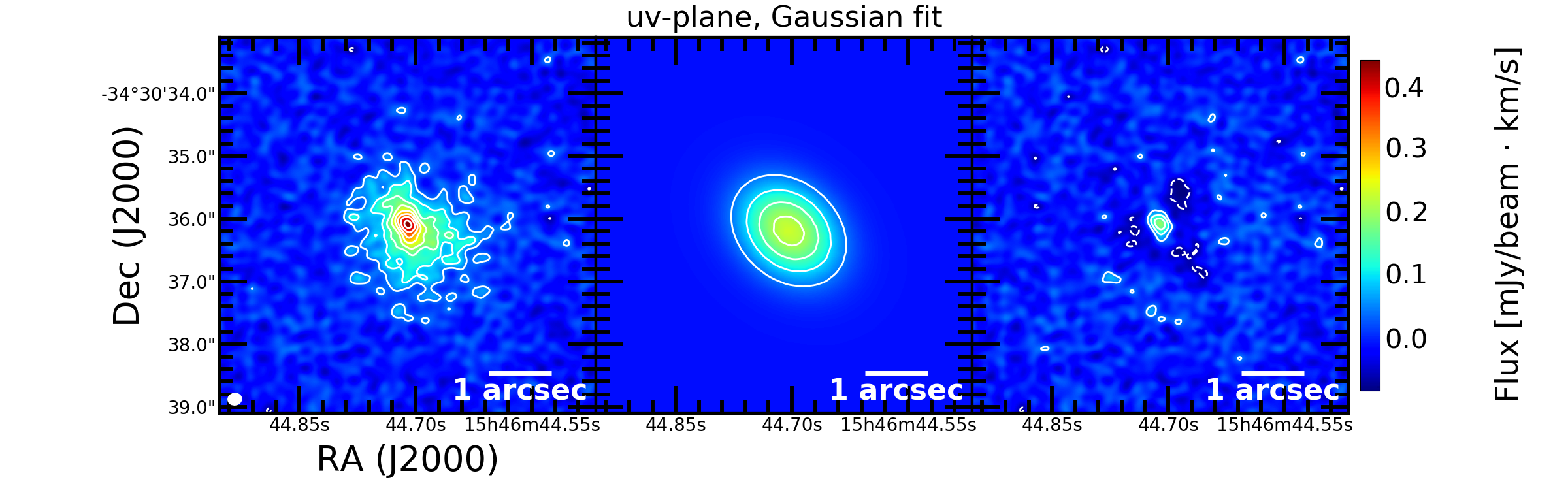}
    \end{center}
  \caption[]{
Reconstructed moment zero maps from fitting in the \textit{uv}-plane the observed CO emission around GW~Lup (Sz~71) using a Nuker profile model (\textit{top panels}) and an elliptical Gaussian function (\textit{bottom panels}). For each model, the sub-panels represent the observed (\textit{left}), modeled (\textit{center}) and residual (\textit{right}) reconstructed CO maps.}
  \label{fig:gwlupnukerobsmodres}
\end{figure*}

Additionally, in order to have an idea of the quality of the interferometric modeling, we show the observed, modeled and residual moment zero maps of CO reconstructed from the visibilities in Figure~\ref{fig:gwlupnukerobsmodres}. In order to visualize the differences between the Nuker modeling and the Gaussian modeling in the \textit{uv}-plane, we also include the reconstructed maps of the Gaussian fit in Figure~\ref{fig:gwlupnukerobsmodres}.

\begin{figure}
  \resizebox{\hsize}{!}{\includegraphics{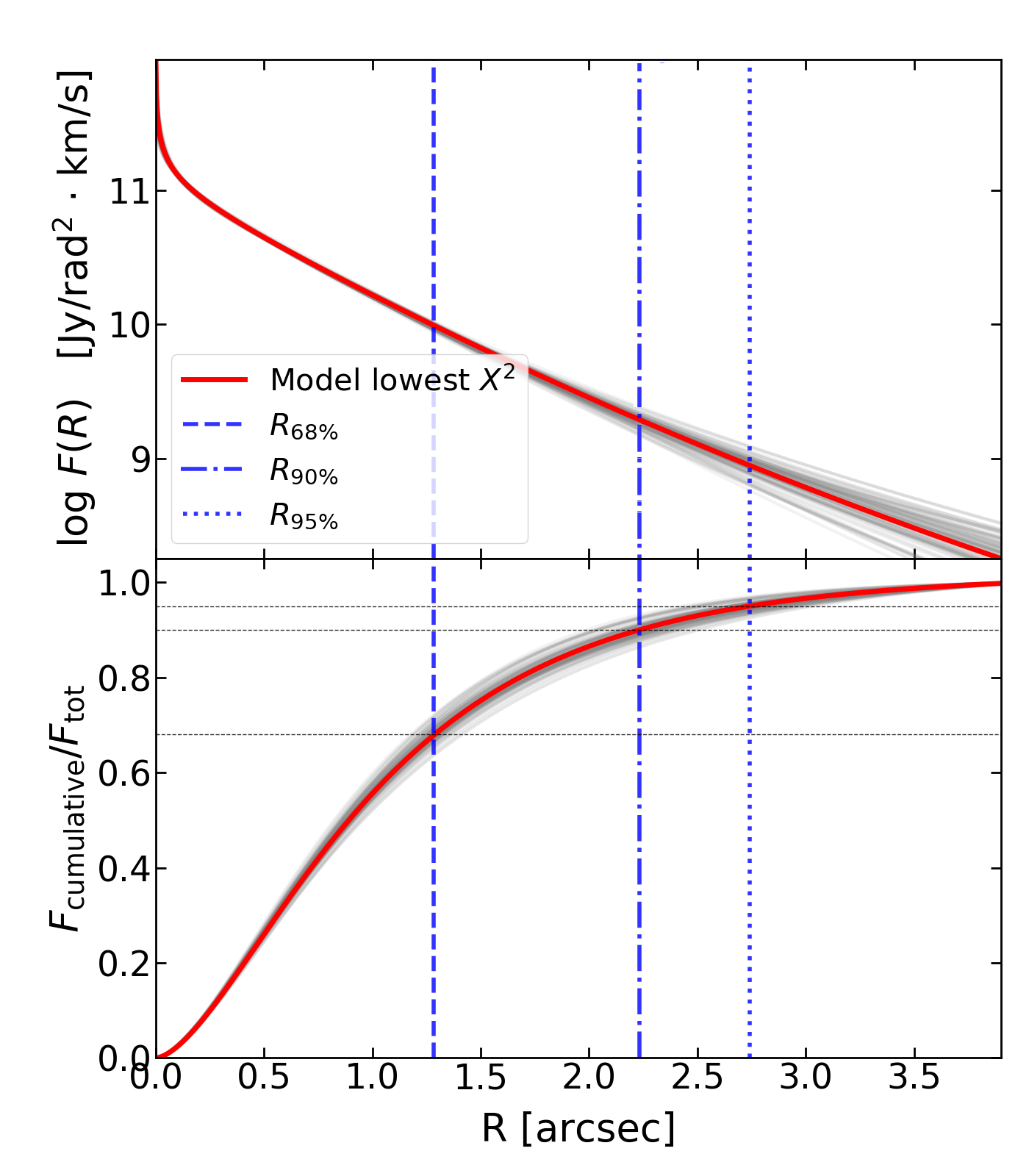}}
  \caption{Radial brightness profile (\textit{top panel}) and the associated cumulative flux (\textit{bottom panel}) modeled for the CO emission of Sz~71 fitted to a Nuker profile. The emission distribution of the model with lowest $\chi^2$ from the fit is drawn as red, while a subset of converged models are shown as thin gray curves.}
  \label{fig:dsharpfcumul}
\end{figure}

The cumulative distribution function ($f_{\mathrm{cumul}}$) 
for each model is built from the converged chains. From the cumulative distribution, we estimate the $R_{68\%}$, $R_{90\%}$, and $R_{95\%}$ CO radii of each model, and build histograms of each radius. The main value and lower/upper uncertainties of each radius are the median, 16th/84th percentiles of the respective histograms. The cumulative distribution and mean radii values for the modeled CO disk of Sz~71 are shown in Figure~\ref{fig:dsharpfcumul}. 

The CO radii results from the interferometric modeling (fitting a Nuker and a Gaussian function) of all the DSHARP disks can be found in Table~\ref{tab:dsharpradii} of Appendix~\ref{sec:appendix_comparingcosizes}. The Sz~68 CO disk was excluded from the interferometric modeling since it is part of a multiple system that is unresolved in the Lupus disk survey. Additionally, the CO integrated emission from the DSHARP survey is irregular; its shape does not resemble a smooth disk.

\section{Results of CO sizes using different methodology}\label{sec:appendix_comparingcosizes}
\begin{figure}
  \resizebox{\hsize}{!}{\includegraphics{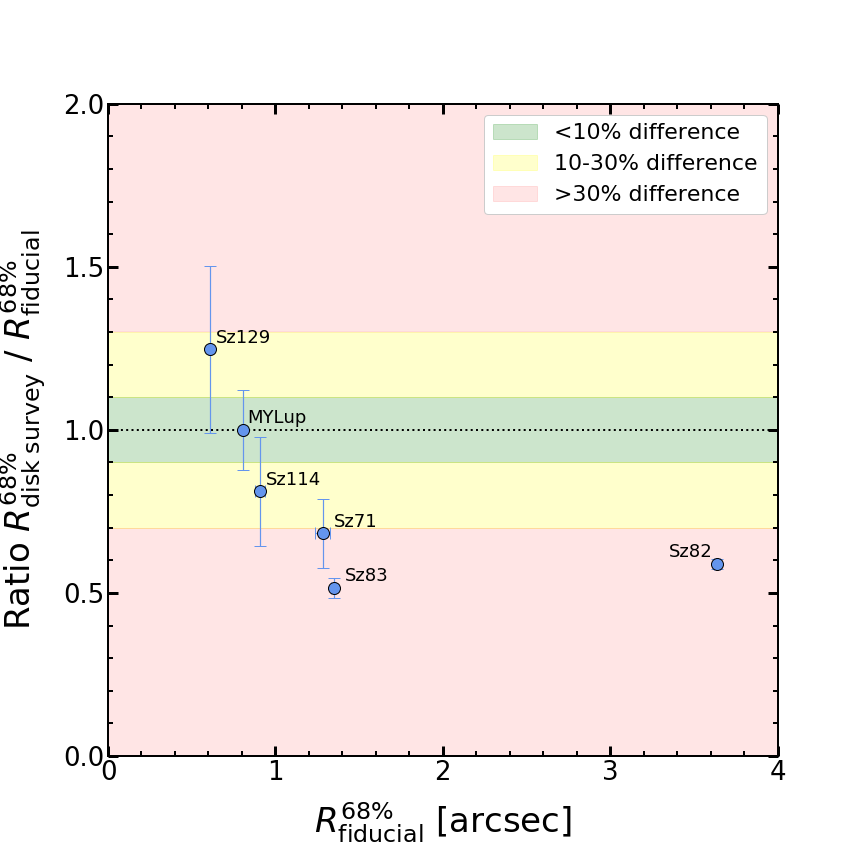}}
  \caption{Comparison between the CO radial extent of objects with two datasets (Lupus disk survey and DSHARP project). The Y-axis shows the ratio between the radius from the Lupus disk survey (obtained by fitting an elliptical Gaussian model in the moment zero map) divided by the fiducial radius of the CO disk. The fiducial CO size is considered as the size inferred from the DSHARP dataset by interferometric fitting of the line visibilities with a Nuker model (interferometric modeling described in Appendix~\ref{sec:appendix_visfit}). 
  The X-axis represents the fiducial size to ease the comparison. 
  The colored regions indicate disks with a divergence in sizes below $10\%$ (green region), between and $10$ and $30\%$ (yellow), and greater than $30\%$ (red).}
  \label{fig:r68comparison}
\end{figure}

In this Appendix we summarize the CO radii obtained by different modeling (Table~\ref{tab:dsharpradii}) for the two datasets (Lupus disk survey, and DSHARP project). The different models considered are: the interferometric modeling of the DSHARP visibilities fitting Nuker and Gaussian functions, and the elliptical Gaussian modeling in the image plane for the two datasets, and the Nuker fit in the image plane for the lower sensitivity and resolution dataset. These results are used to assess the systematic uncertainties of the modeling described in Section~\ref{sec:gasmodeling}. 
\begin{table*}
\caption{CO radii of protoplanetary disks in Lupus observed in the DSHARP survey, inferred from modeling in the image and in the \textit{uv}-plane, using two different datasets (Lupus disk survey, and DSHARP observations).}
\label{tab:dsharpradii}      
\centering 
\scriptsize
\begin{tabular}{lccccccccr}   
\hline\hline            
 & \multicolumn{4}{l}{\it Radii from Lupus disk survey} & \multicolumn{5}{l}{\it Radii from DSHARP survey} \\ 
\hline 
 & \multicolumn{1}{c}{\it image-plane, Gaussian fit} & 
 \multicolumn{3}{c}{\it image-plane, Nuker fit} & \multicolumn{1}{c}{\it image-plane, Gaussian fit} & \multicolumn{1}{c}{\it uv-plane, Gaussian fit} & \multicolumn{3}{c}{\it uv-plane, Nuker fit} \\ 
Object & $R_{68\%}$ [$ ^{\prime\prime}$] 
& $R_{68\%}$ [$ ^{\prime\prime}$]  & $R_{90\%}$ [$ ^{\prime\prime}$] & $R_{95\%}$ [$ ^{\prime\prime}$] 
& $R_{68\%}$ [$ ^{\prime\prime}$] 
& $R_{68\%}$ [$ ^{\prime\prime}$] 
& $R_{68\%}$ [$ ^{\prime\prime}$]  & $R_{90\%}$ [$ ^{\prime\prime}$] & $R_{95\%}$ [$ ^{\prime\prime}$]  \\ 
\hline 
MY~Lup & 
 $ 0.81 \pm  0.10 $ & 
 $ 0.85 \pm  0.07 $ & $ 1.22 \pm  0.10 $ & $ 1.35 \pm  0.11 $ & 
 $ 0.81 \pm  0.02 $ & 
 $ 0.81 ^{+ 0.01 }_{- 0.01 }$ & 
 $ 0.81 ^{+ 0.01 }_{- 0.01 }$ & $ 1.12 ^{+ 0.03 }_{- 0.03 }$ & $ 1.27 ^{+ 0.04 }_{- 0.04 }$  \\
Sz~71 & 
 $ 0.87 \pm  0.13 $ & 
 $ 0.97 \pm  0.10 $ & $ 1.39 \pm  0.19 $ & $ 1.51 \pm  0.22 $ & 
 $ 0.89 \pm  0.03 $ & 
 $ 0.88 ^{+ 0.02 }_{- 0.02 }$ & 
 $ 1.28 ^{+ 0.05 }_{- 0.04 }$ & $ 2.23 ^{+ 0.10 }_{- 0.10 }$ & $ 2.74 ^{+ 0.12 }_{- 0.13 }$  \\
Sz~82 & 
 $ 2.14 \pm  0.06 $ & 
 $ 3.75 \pm  0.67 $ & $ 6.09 \pm  0.78 $ & $ 6.98 \pm  0.76 $ & 
 $ 2.52 \pm  0.04 $ & 
 $ 2.64 ^{+ 0.01 }_{- 0.01 }$ & 
 $ 3.64 ^{+ 0.02 }_{- 0.02 }$ & $ 5.45 ^{+ 0.01 }_{- 0.02 }$ & $ 6.01 ^{+ 0.01 }_{- 0.01 }$  \\
Sz~83 & 
 $ 0.69 \pm  0.04 $ & 
 $ 1.44 \pm  0.13 $ & $ 2.18 \pm  0.21 $ & $ 2.38 \pm  0.23 $ & 
 $ 0.76 \pm  0.03 $ & 
 $ 0.71 ^{+ 0.01 }_{- 0.01 }$ & 
 $ 1.35 ^{+ 0.01 }_{- 0.01 }$ & $ 2.43 ^{+ 0.04 }_{- 0.04 }$ & $ 3.01 ^{+ 0.03 }_{- 0.04 }$  \\
Sz~$114$ & 
 $ 0.74 \pm  0.15 $ & 
 $ 0.86 \pm  0.08 $ & $ 1.09 \pm  0.10 $ & $ 1.15 \pm  0.11 $ & 
 $ 0.79 \pm  0.04 $ & 
 $ 0.79 ^{+ 0.02 }_{- 0.02 }$ & 
 $ 0.91 ^{+ 0.03 }_{- 0.03 }$ & $ 1.33 ^{+ 0.06 }_{- 0.05 }$ & $ 1.53 ^{+ 0.07 }_{- 0.07 }$  \\
Sz~$129$ & 
 $ 0.76 \pm  0.16 $ & 
 $ 0.75 \pm  0.06 $ & $ 0.99 \pm  0.09 $ & $ 1.05 \pm  0.10 $ & 
 $ 0.60 \pm  0.02 $ & 
 $ 0.58 ^{+ 0.01 }_{- 0.01 }$ & 
 $ 0.61 ^{+ 0.01 }_{- 0.01 }$ & $ 0.83 ^{+ 0.04 }_{- 0.03 }$ & $ 0.94 ^{+ 0.08 }_{- 0.04 }$  \\

\hline
\end{tabular}
\end{table*}
Lastly, in Figure~\ref{fig:r68comparison}, we compare CO radii obtained from the elliptical Gaussian modeling in the image plane of the lower sensitivity dataset (as described in Section~\ref{sec:gasmodeling}) with the CO radii from the interferometric fit of a Nuker modeling of the DSHARP data (explained in Section~\ref{sec:appendix_visfit}) for the same objects. 
The radii based on the interferometric modeling of the DSHARP data are considered as the fiducial CO sizes of the disks. This comparison is used in Section~\ref{sec:dsharpradii} to assess the quality of the elliptical Gaussian modeling.

\section{Criterion for the CO modeling in the image plane based on the residuals}\label{sec:appendix_residualscriteria}
\begin{figure}
  \resizebox{\hsize}{!}{\includegraphics{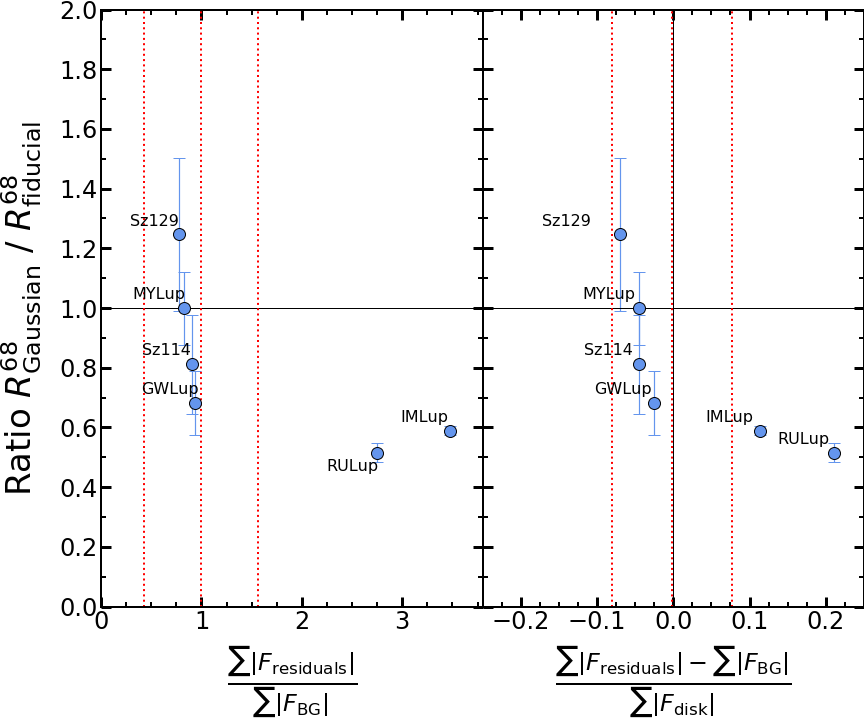}}
  \caption{
  CO size ratio compared to the residuals from the elliptical Gaussian modeling for the $6$ disks with two datasets available. The fiducial radii are the sizes obtained from the interferometric modeling of the visibilities from the DSHARP dataset, following the methodology described in Appendix~\ref{sec:appendix_visfit}.   
  (\textit{Left panel}): size ratio as a function of residuals over background fraction (criterion 1). 
  (\textit{Right panel}): size ratio vs. the difference between residuals and background over the observed disk emission (criterion 2). 
  Central red line represents the median value of the entire Lupus sample, left and right vertical lines are the $\mu \pm \sigma$ values.}
  \label{fig:res_imfit}
\end{figure}

The residuals are quantified as the sum of emission enclosed by the inferred $R_{68\%}$ and centered at the elliptical Gaussian centroid in the residual map. Absolute values are used to account for negative residuals. This quantity is also computed on the observed moment zero map centered on the object, and on an emission-free region of the sky (averaged over four random locations in the background). 
Two quantities are used as the criteria for the quality of the fit: the residuals over background fraction $\frac{\sum |F_{\mathrm{res}}|}{\sum |F_{\mathrm{BG}}|}$, and the difference between residuals and background over the observed disk emission $\frac{\sum |F_{\mathrm{res}}| - \sum |F_{\mathrm{BG}}|}{\sum |F_{\mathrm{disk}}|}$. 
A value of $1$ in the first quantity means that the residuals of the model are indistinguishable from the background emission. The second quantity gives an idea of what is the fraction of residuals over the observed disk emission; in this case, a 0 value represents a perfect model. 
These quantities are used together with the size ratio between the model and the fiducial size of each disk to evaluate the quality of each model. The results for elliptical Gaussian models are shown in the two sub-panels of Figure~\ref{fig:res_imfit}, for those disks in which we have the two (moderate and high resolution) datasets. The red lines in both panels represent the median ($\mu$), and $\mu \pm \sigma$ for the entire Lupus CO disks sample of the respective quantity. 
For both quantities, the four disks within the region delimited by the $\mu \pm \sigma$ region (Sz~71, Sz~114, Sz~129, and MY~Lup) have all a size divergence with the fiducial CO size between $0\%$ and $\sim30\%$. On the other hand, Sz~82 and Sz~83 fall clearly outside; these disks show a larger divergence in size, both $> 40\%$ difference respect to the fiducial size. 

\begin{figure}
  \resizebox{\hsize}{!}{\includegraphics{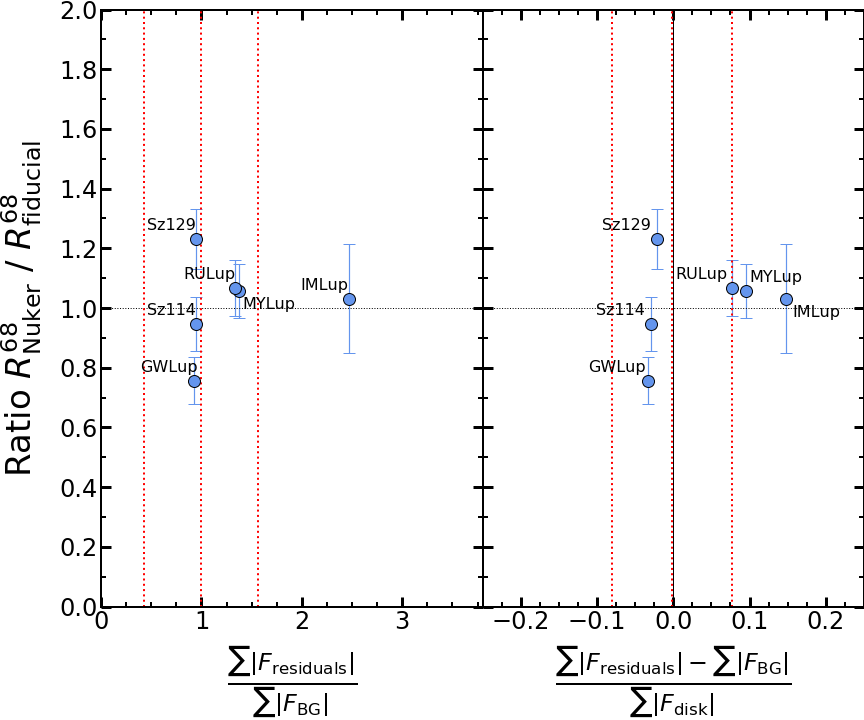}}
  \caption{
  Residuals criteria for the Nuker modeling of the $6$ disks with two datasets available. The fiducial radii are the sizes obtained from the interferometric modeling of the visibilities from the DSHARP dataset, following the methodology described in Appendix~\ref{sec:appendix_visfit}. 
  \textit{Left panel}: size ratio as a function of residuals over background fraction (criterion 1). 
  \textit{Right panel}: size ratio vs. the difference between residuals and background over the observed disk emission (criterion 2). 
  Red vertical lines identical to the ones in Figure~\ref{fig:res_imfit}.}
  \label{fig:res_imnuker}
\end{figure}

From this analysis, elliptical Gaussian modeling of disks with residuals within the $\mu \pm \sigma$ range of the sample (as in Figure~\ref{fig:res_imfit}) provide CO sizes with an accuracy between 0 to $\sim30\%$. Gaussian models of disks with residuals outside the valid range can differ more than $40\%$ with respect to the fiducial value. Therefore, for disks with residuals outside the $\mu \pm \sigma$, their CO emission should not be modeled with a Gaussian function, and instead is modeled by fitting their integrated maps with a Nuker profile (as described in Section~\ref{sec:gasmodeling}). 
We also tested the quality of the sizes inferred from the Nuker profile modeling in the image plane. The size ratio between the Nuker model in the image plane and the fiducial model is represented as a function of the Nuker model residuals in Figure~\ref{fig:res_imnuker}. The results show that the inferred radii are in better agreement with the fiducial sizes than the Gaussian modeling. The size difference is found to be $\sim 0-30\%$ for every disk, and this remains true for the three radii ($R_{68\%}$, $R_{90\%}$, and $R_{95\%}$). 

\section{Best fit parameters for CO disks modeled with a Nuker function.}\label{sec:appendix_COnukerparams}

\begin{table*}
\caption{Optimal parameters of the Nuker model used to fit the CO integrated emission of several sources from the full Lupus sample, based on the criteria described in Appendix~\ref{sec:appendix_residualscriteria}. The modeling of the integrated line emission is performed following the methodology from Section~\ref{sec:gasmodeling}. The parameters of the Nuker function are: the transition radius $\rho_t$, the inner and outer slopes $\gamma$ and $\beta$, the smoothing parameter $\alpha$, and the integrated total flux $F_{\mathrm{tot}}$. An additional parameter $\rho_{\mathrm{end}}$ was used as the outermost radius of the Nuker model, which is set to coincide with the radial distance at which the azimuthally averaged line emission reaches a zero value.}
\label{tab:COnukerparams}      
\centering              
\begin{tabular}{llcccccr}   
\hline\hline            
$\#$ & Object & $\rho_{t}$ & $\gamma$ & $\beta$ & $\alpha$ & $F_{\mathrm{tot}}$ & $\rho_{\mathrm{end}}$ \\
 &  & [$ ^{\prime\prime}$] & [-] & [-] & [-] & [mJy/beam] & [$ ^{\prime\prime}$]  \\    
\hline 

1  &  EXLup   & $ 1.20  \pm  0.15 $  & $ -0.09 \pm 0.01 $  & $ 6.86 \pm 0.63 $  & $ 1.23 \pm 0.04 $  & $ 296 \pm 1 $  & $ 1.4 $ \\
2  &  RYLup   & $ 0.37  \pm  0.01 $  & $ -0.01 \pm 0.01 $  & $ 2.38 \pm 0.05 $  & $ 2.25 \pm 0.08 $  & $ 952 \pm 9 $  & $ 2.7 $ \\
3  &  Sz75   & $ 0.58  \pm  0.02 $  & $ 0.10 \pm 0.01 $  & $ 2.83 \pm 0.10 $  & $ 2.61 \pm 0.09 $  & $ 448 \pm 3 $  & $ 1.9 $ \\
4  &  Sz82   & $ 6.93  \pm  1.42 $  & $ -0.10 \pm 0.01 $  & $ 6.21 \pm 0.74 $  & $ 0.93 \pm 0.03 $  & $ 2824 \pm 14 $  & $ 8.2 $ \\
5  &  Sz83   & $ 0.20  \pm  0.01 $  & $ 0.01 \pm 0.01 $  & $ 1.58 \pm 0.02 $  & $ 2.33 \pm 0.11 $  & $ 748 \pm 7 $  & $ 2.6 $ \\
6  &  Sz91   & $ 0.38  \pm  0.01 $  & $ 0.01 \pm 0.01 $  & $ 1.99 \pm 0.04 $  & $ 2.00 \pm 0.07 $  & $ 268 \pm 2 $  & $ 2.5 $ \\
7  &  Sz111   & $ 0.22  \pm  0.01 $  & $ 0.06 \pm 0.01 $  & $ 1.23 \pm 0.01 $  & $ 6.49 \pm 0.44 $  & $ 598 \pm 4 $  & $ 3.2 $ \\
8  &  V1192Sco   & $ 0.31  \pm  0.01 $  & $ -0.14 \pm 0.01 $  & $ 1.96 \pm 0.05 $  & $ 3.42 \pm 0.21 $  & $ 141 \pm 1 $  & $ 1.8 $ \\
9  &  SSTc2d J160703.9-391112   & $ 0.71  \pm  0.01 $  & $ 0.07 \pm 0.01 $  & $ 2.38 \pm 0.08 $  & $ 10.00 \pm 1.33 $  & $ 139 \pm 1 $  & $ 1.7 $ \\
10  &  SSTc2d J160830.7-382827   & $ 0.53  \pm  0.01 $  & $ -0.07 \pm 0.01 $  & $ 2.22 \pm 0.02 $  & $ 3.03 \pm 0.06 $  & $ 1187 \pm 4 $  & $ 2.7 $ \\

\hline
\end{tabular}
\end{table*}

\section{Observed, model and residual CO maps of the Lupus disk population}\label{sec:appendix_obsmodres}

\begin{figure*}
    \begin{center}
    \includegraphics[width=.490\textwidth]{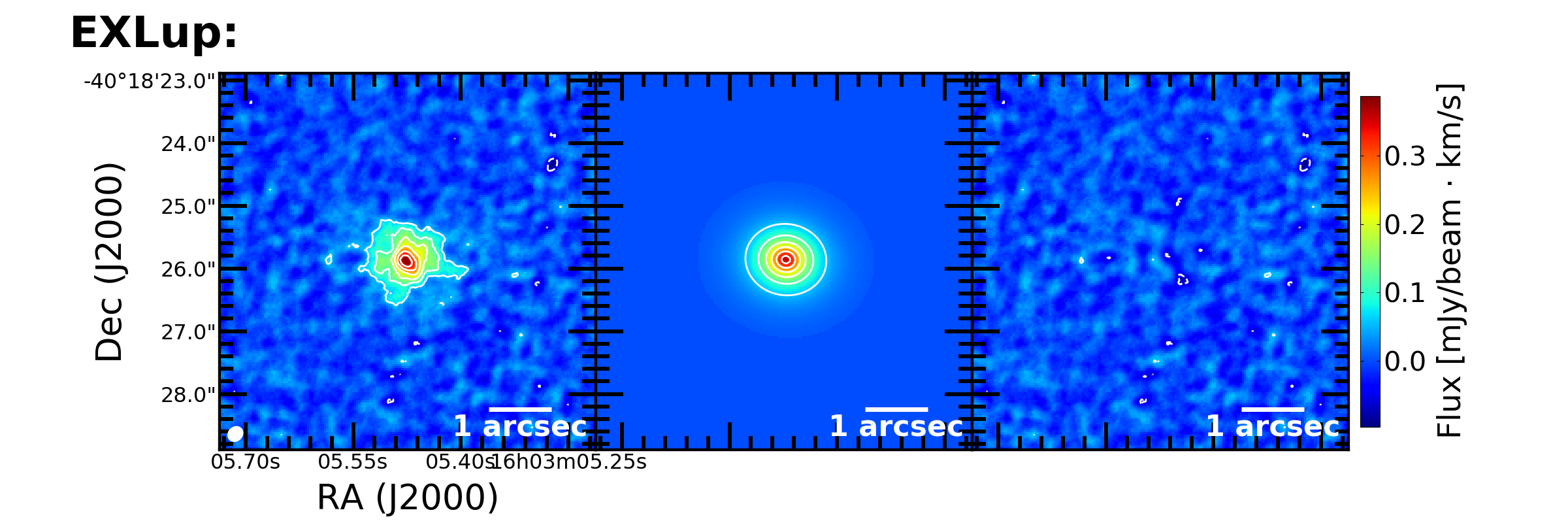}
    \includegraphics[width=.225\textwidth]{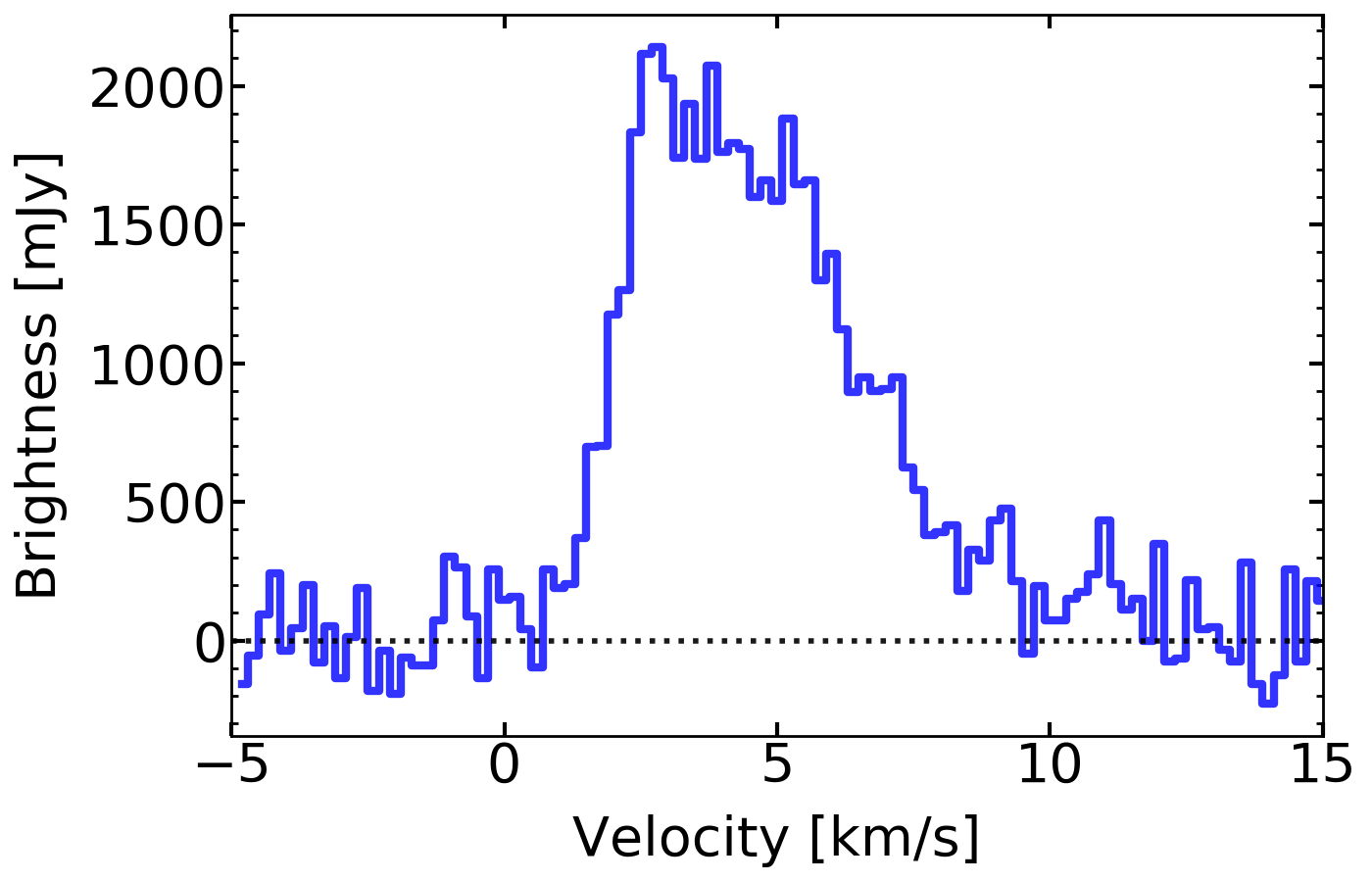}
    \includegraphics[width=.235\textwidth]{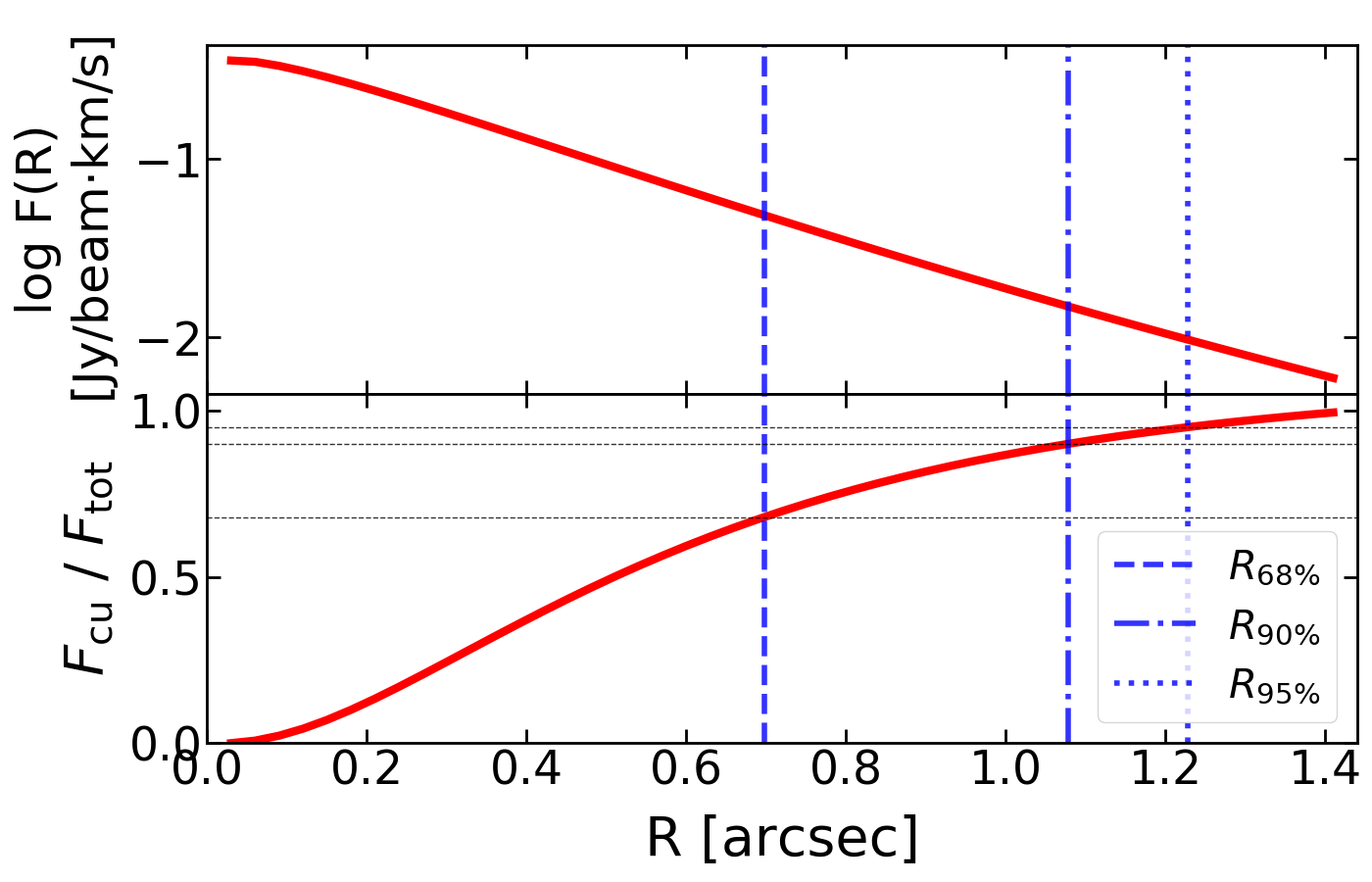}
    \end{center}
    \begin{center}
    \includegraphics[width=.490\textwidth]{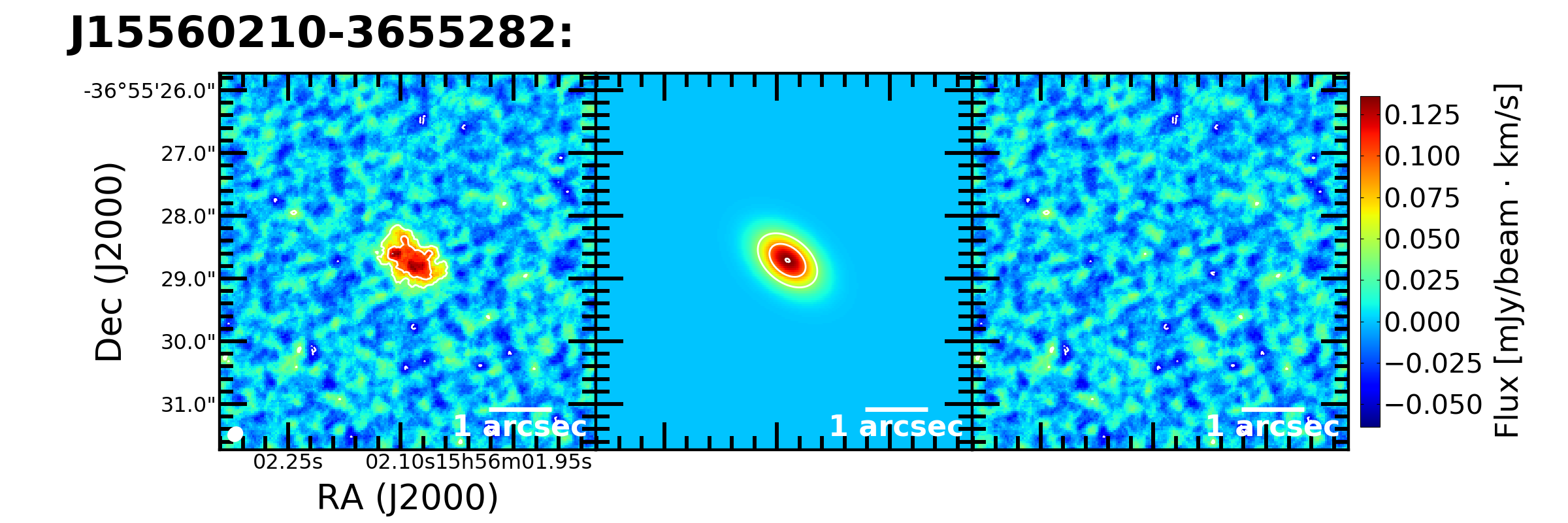}
    \includegraphics[width=.225\textwidth]{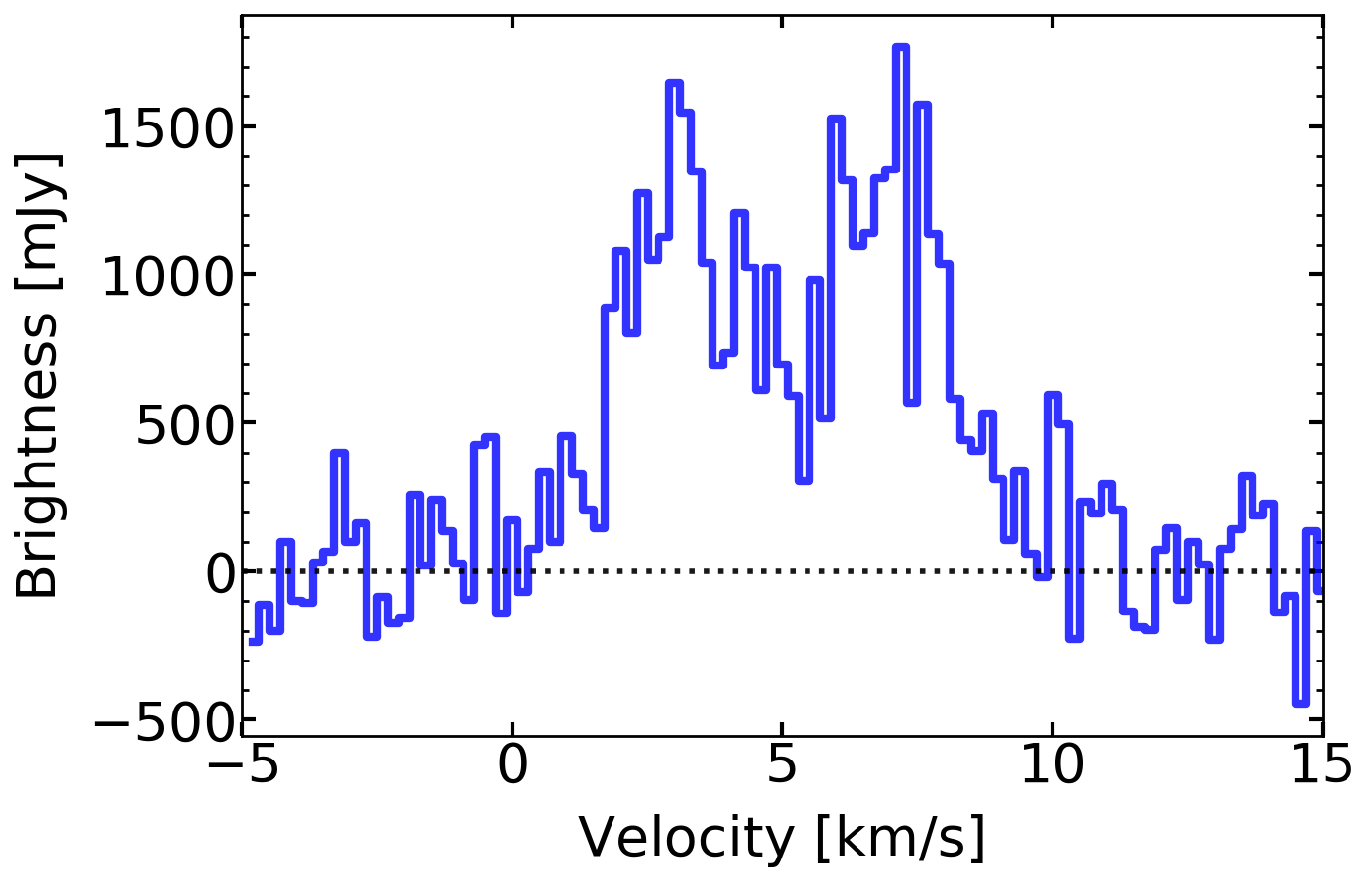}
    \includegraphics[width=.235\textwidth]{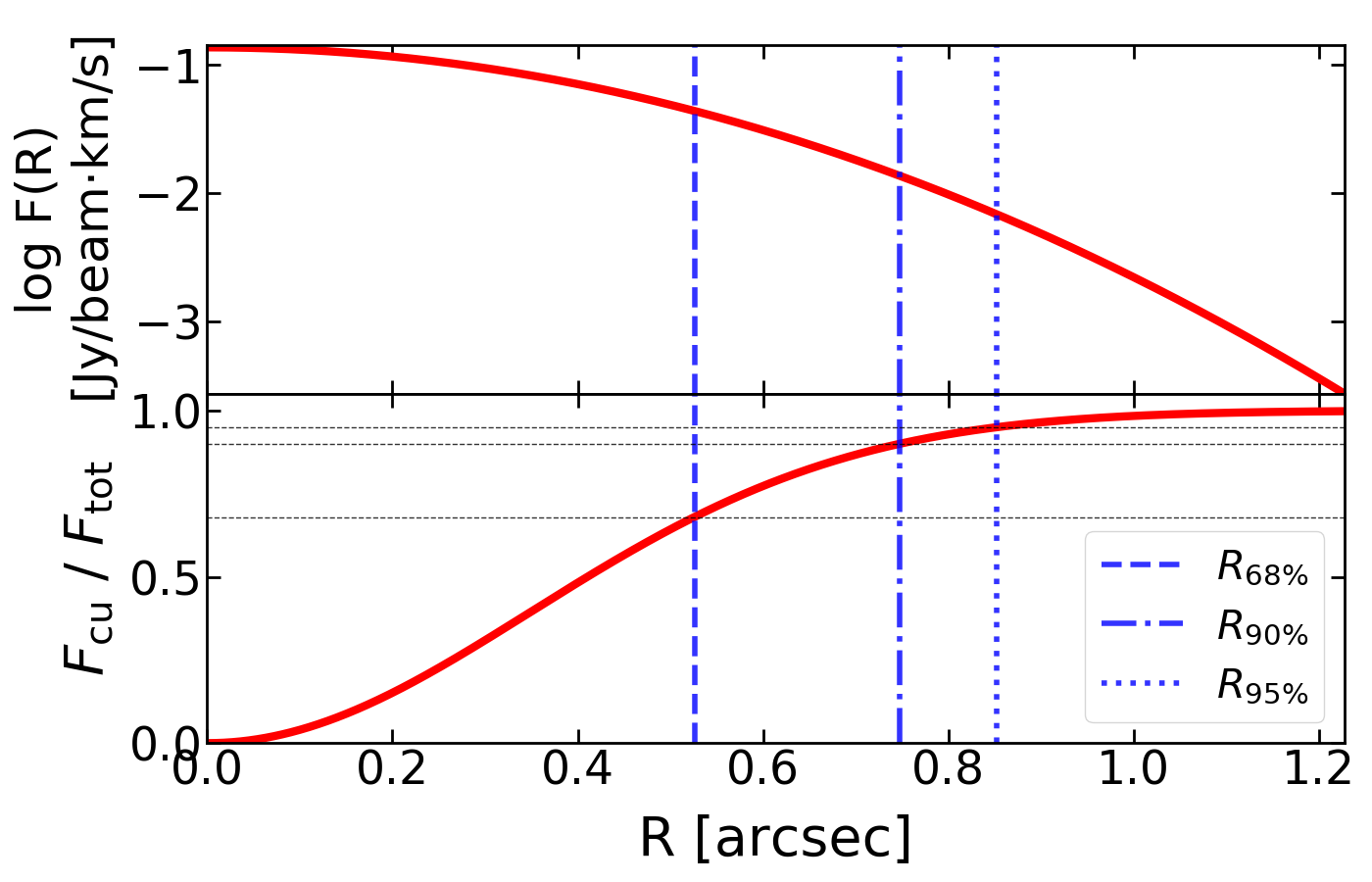}
    \end{center}
    \begin{center}
    \includegraphics[width=.490\textwidth]{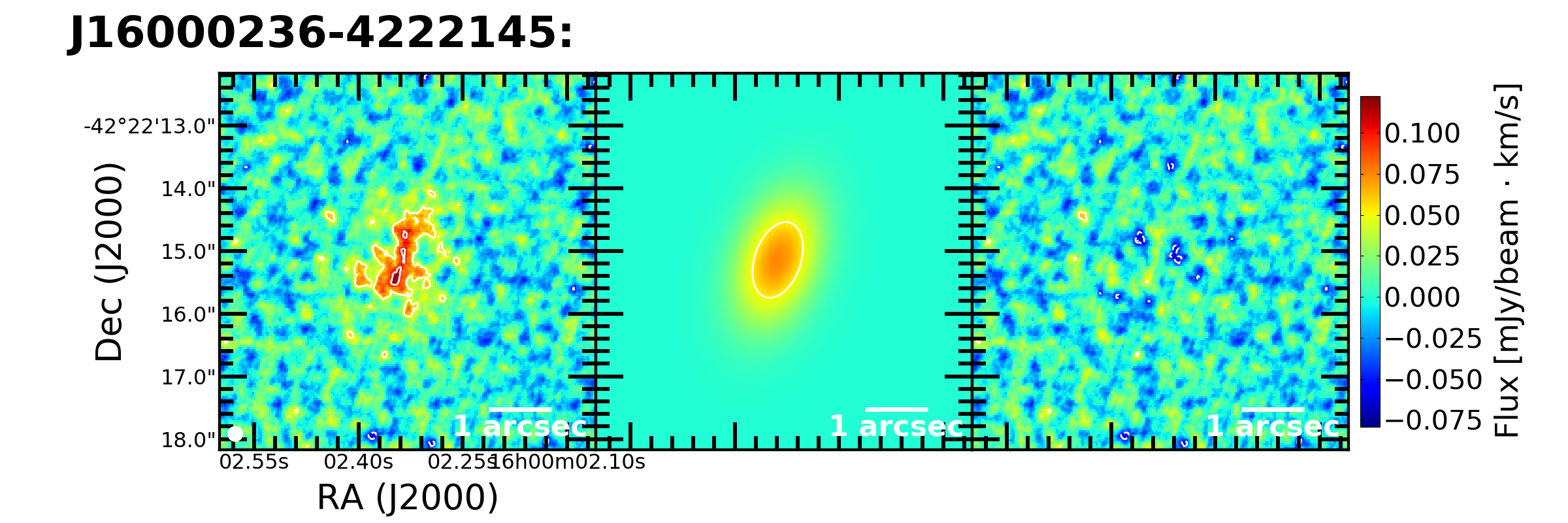}
    \includegraphics[width=.225\textwidth]{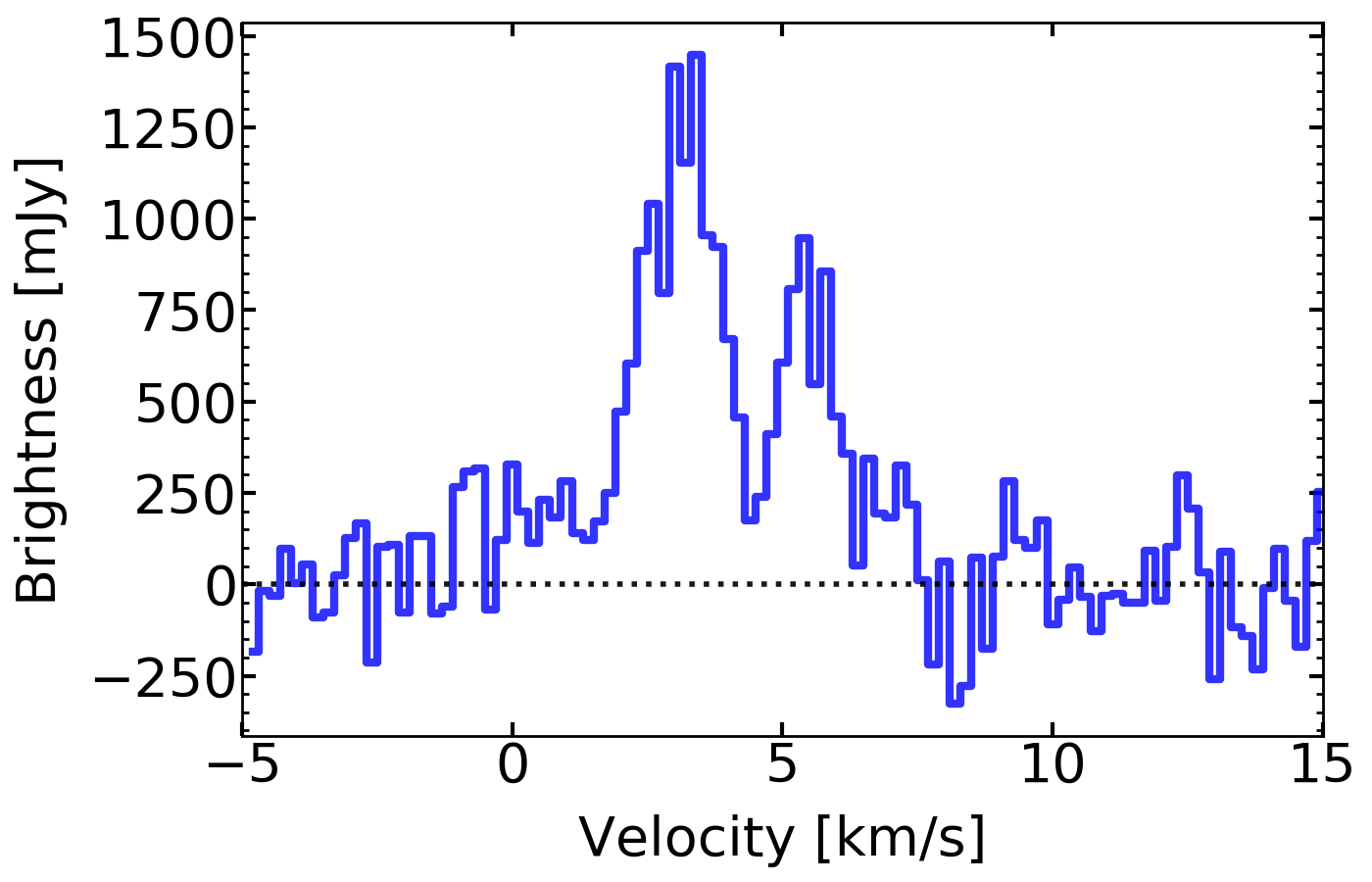}
    \includegraphics[width=.235\textwidth]{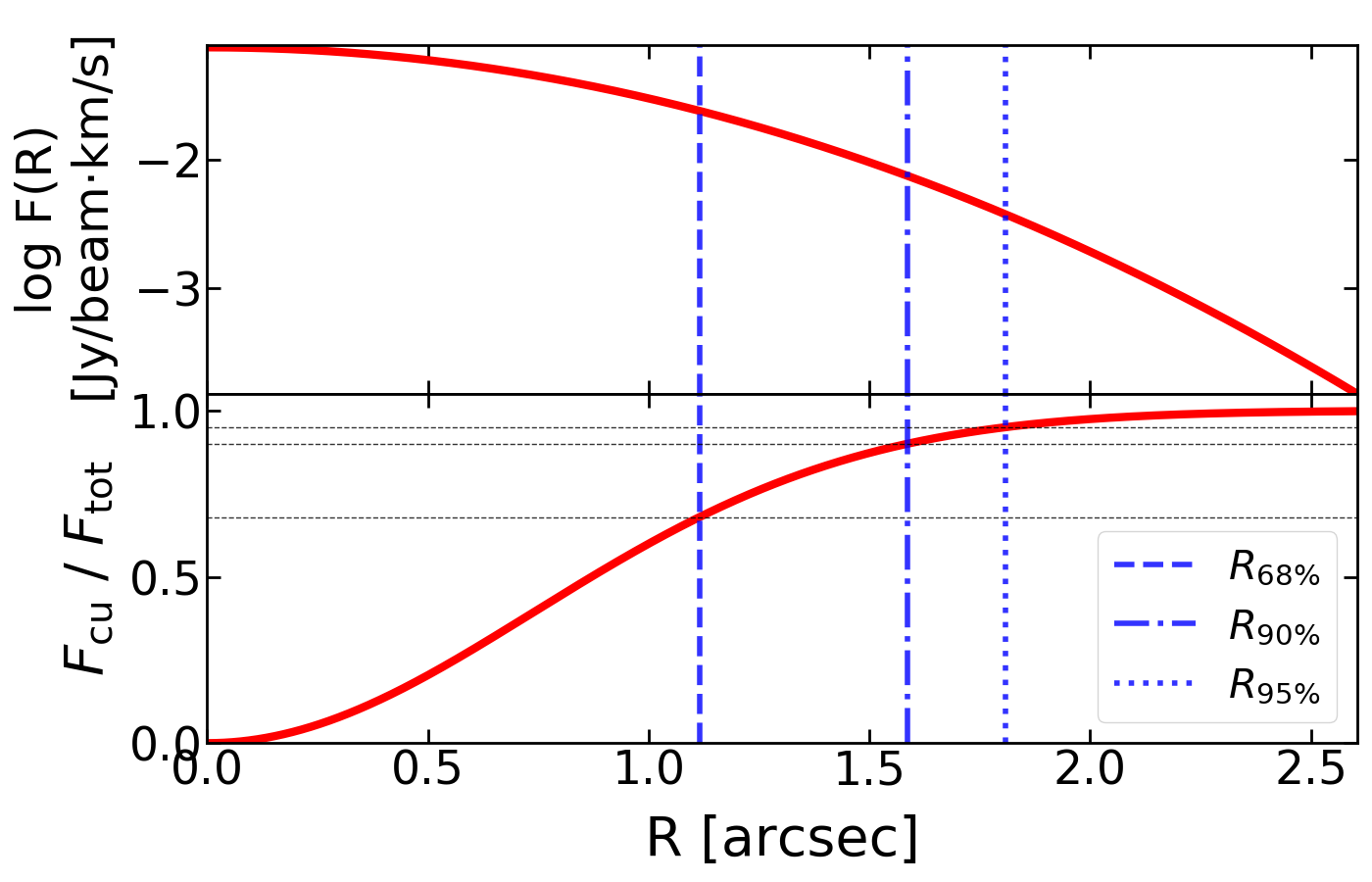}
    \end{center}
    \begin{center}
    \includegraphics[width=.490\textwidth]{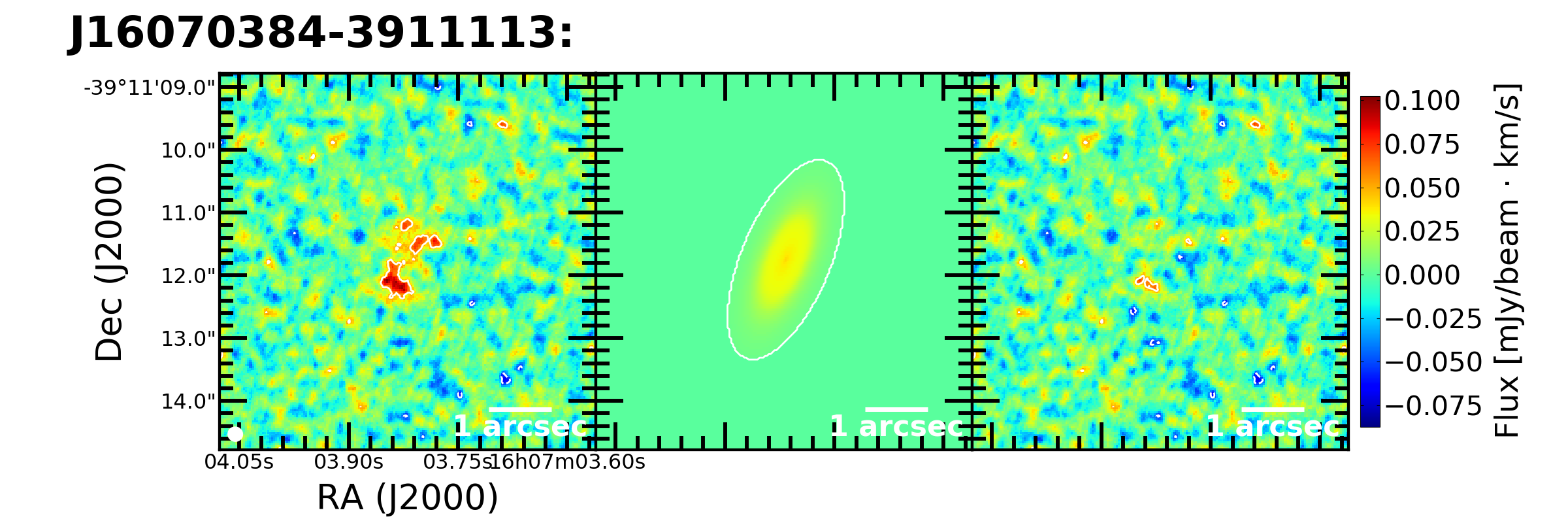}
    \includegraphics[width=.225\textwidth]{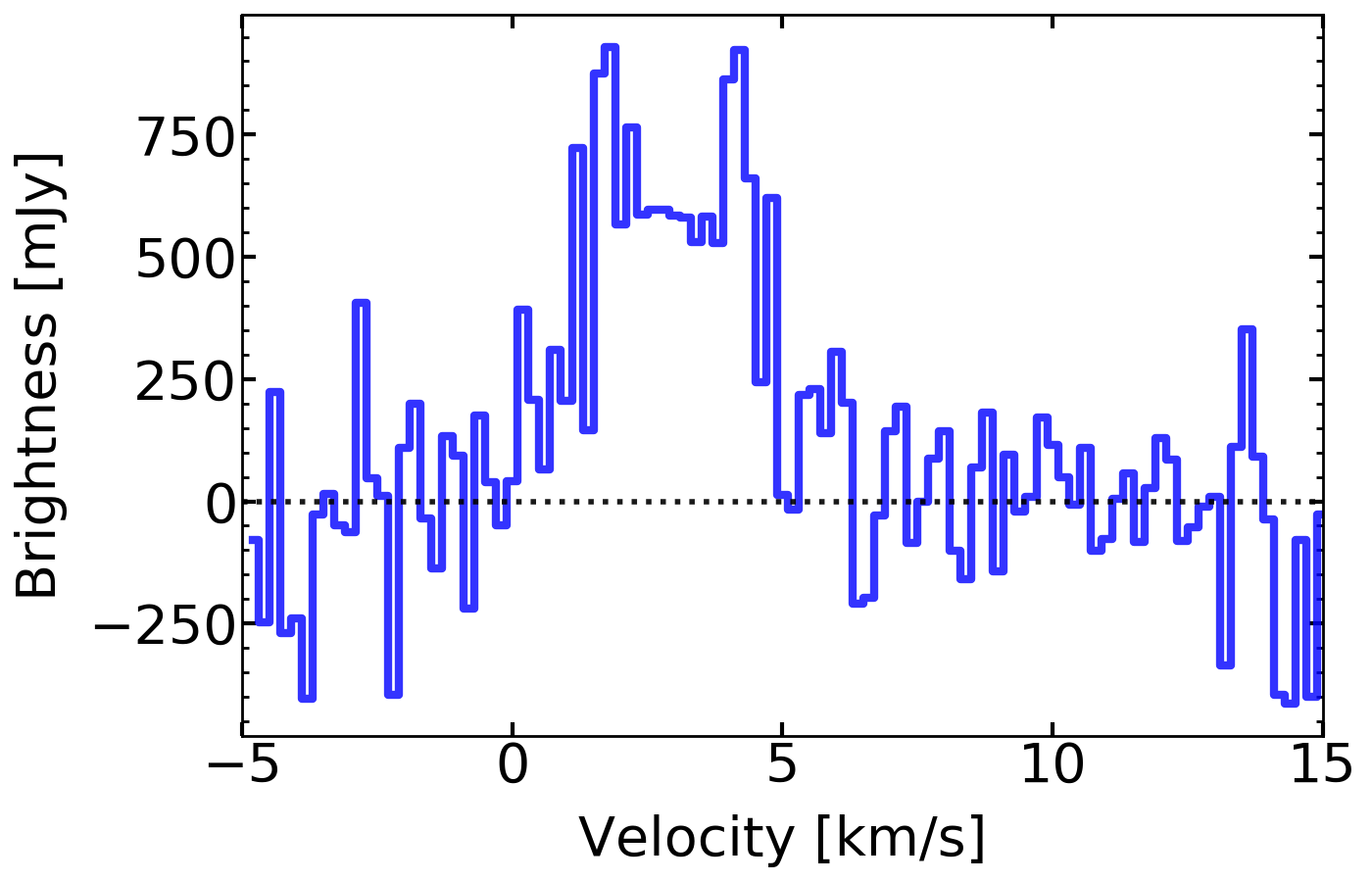}
    \includegraphics[width=.235\textwidth]{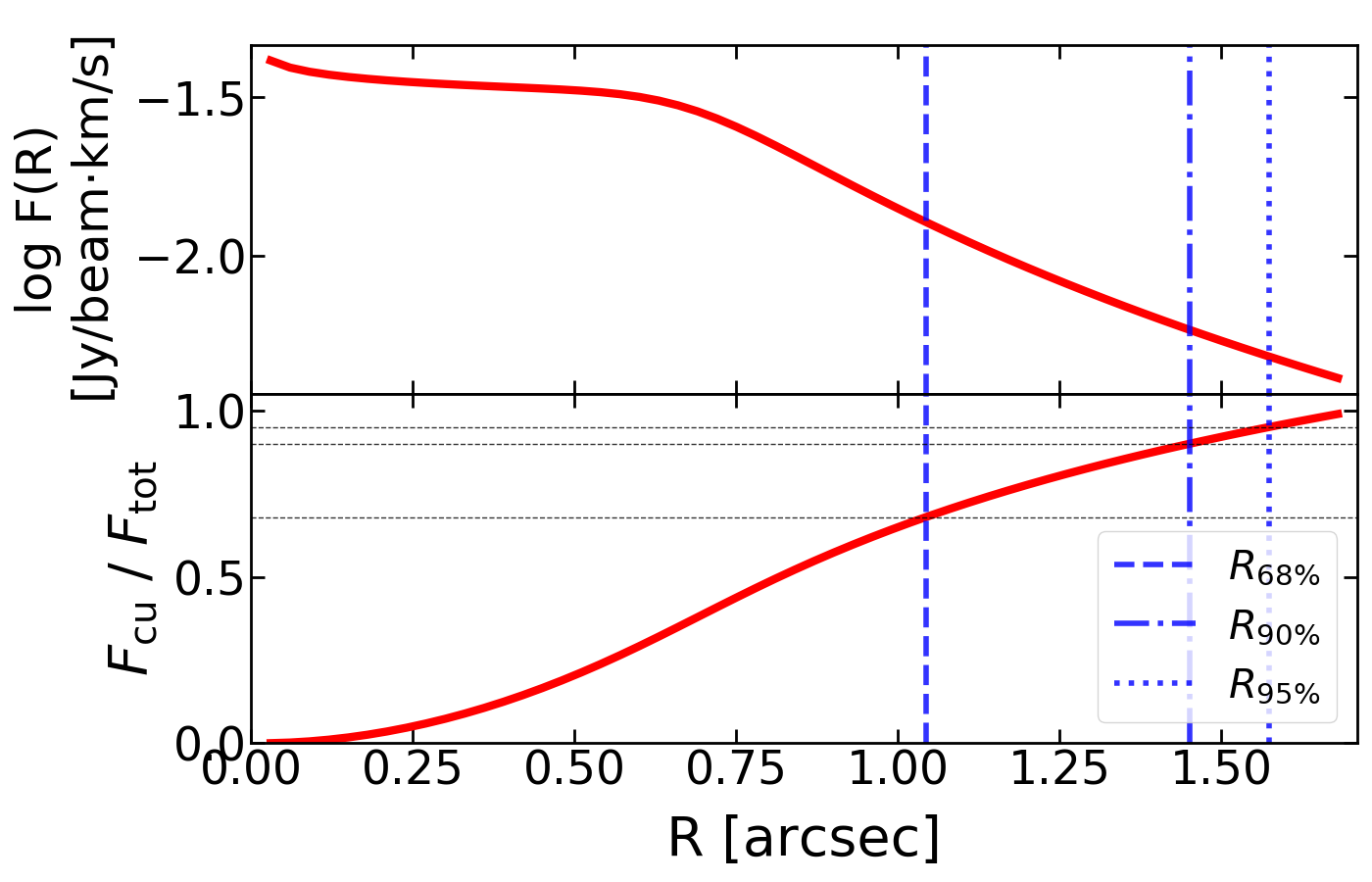}
    \end{center}
    \begin{center}
    \includegraphics[width=.490\textwidth]{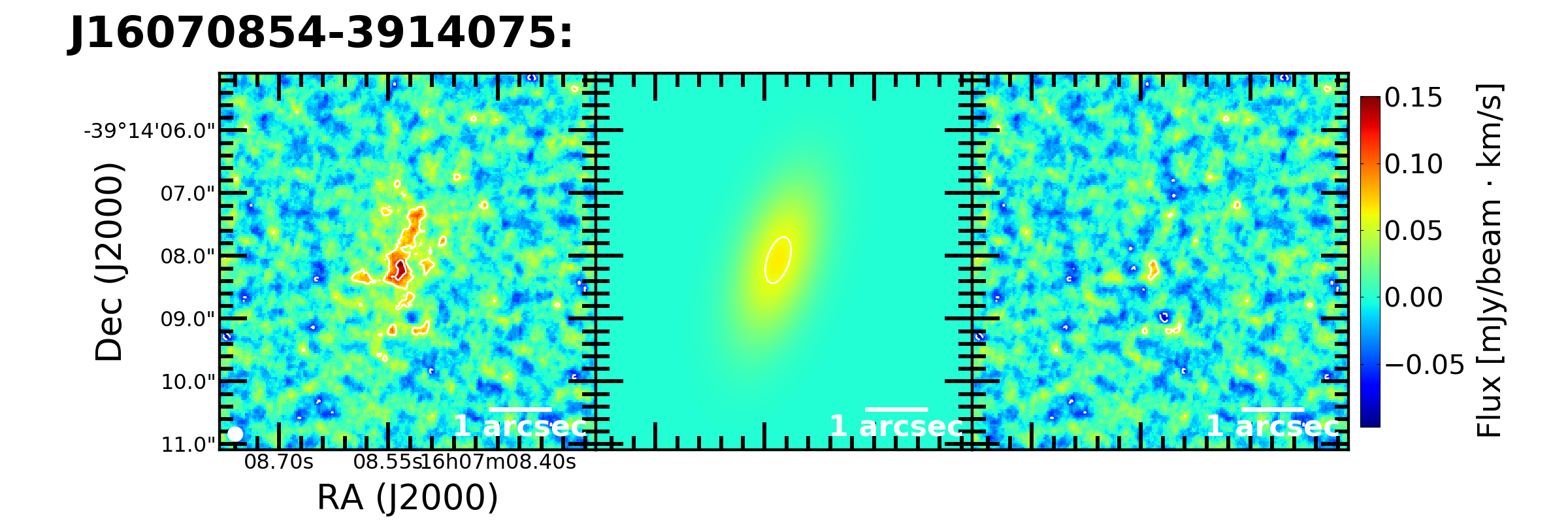}
    \includegraphics[width=.225\textwidth]{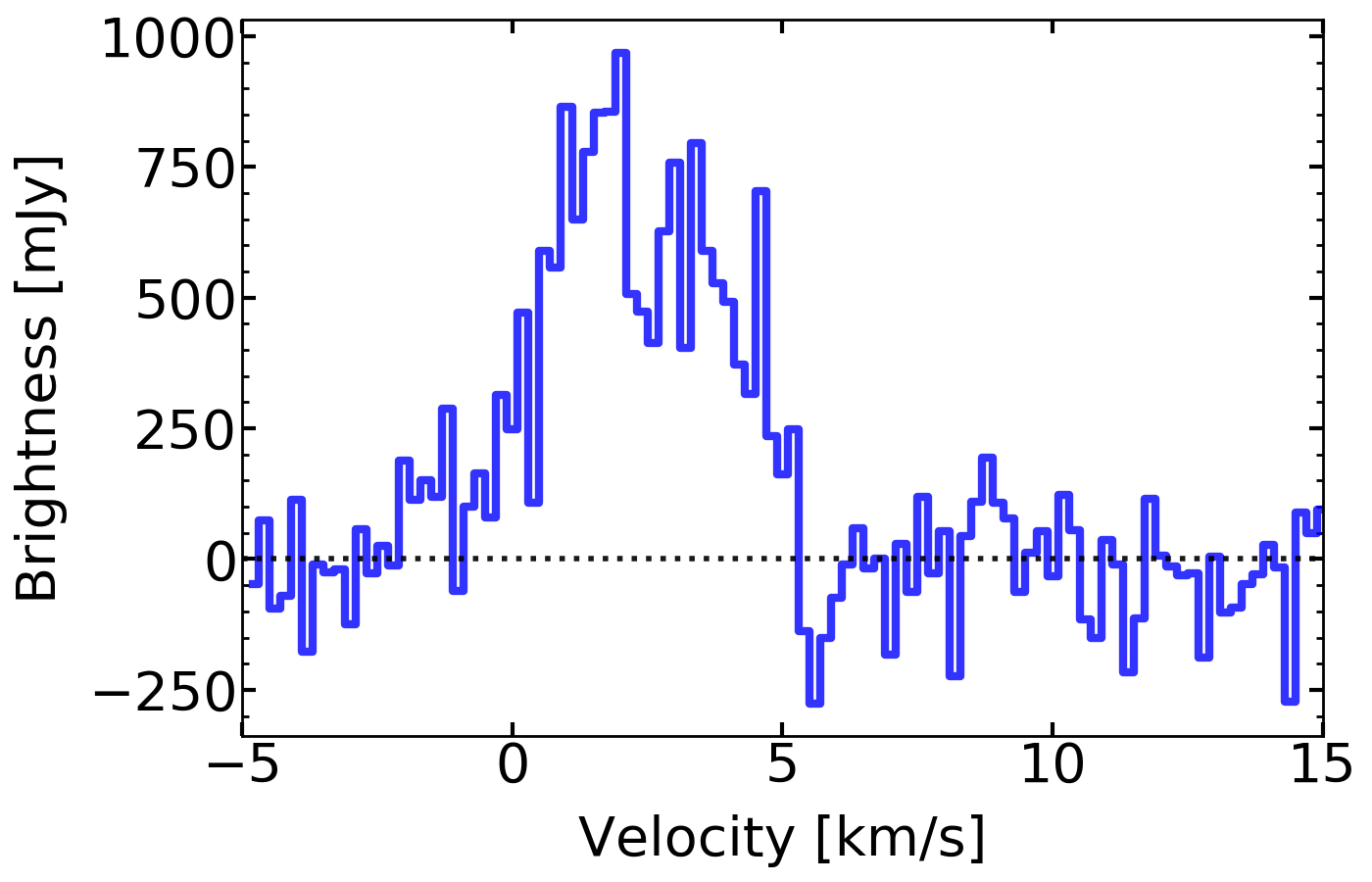}
    \includegraphics[width=.235\textwidth]{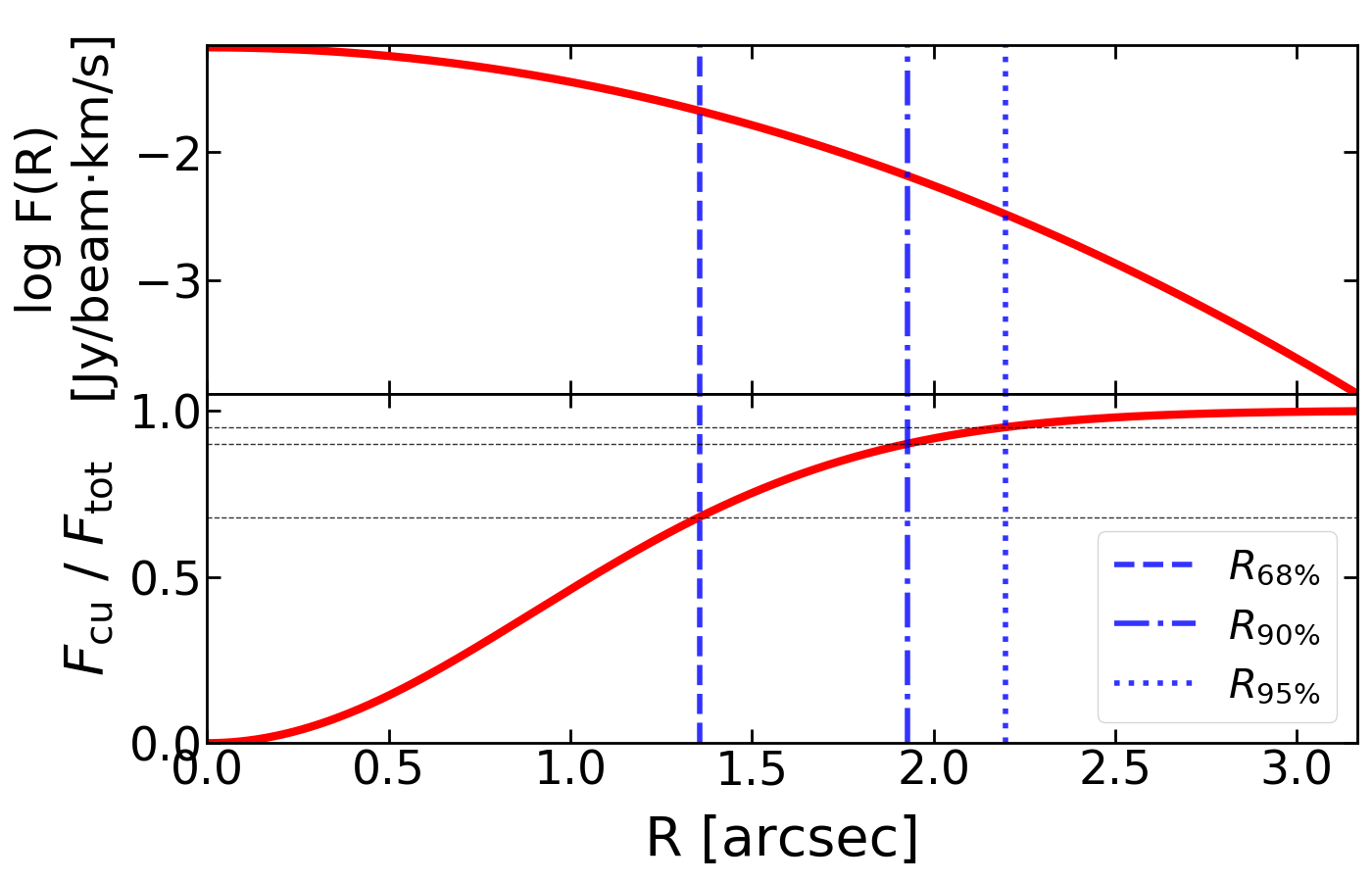}
    \end{center}
    \begin{center}
    \includegraphics[width=.490\textwidth]{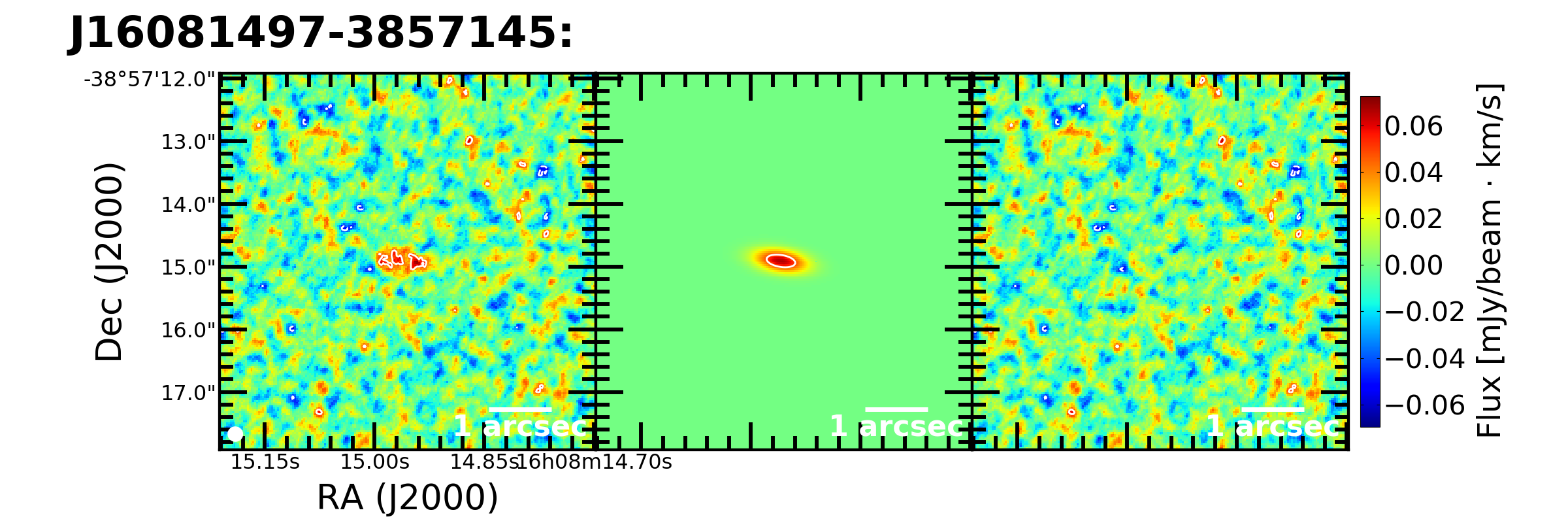}
    \includegraphics[width=.225\textwidth]{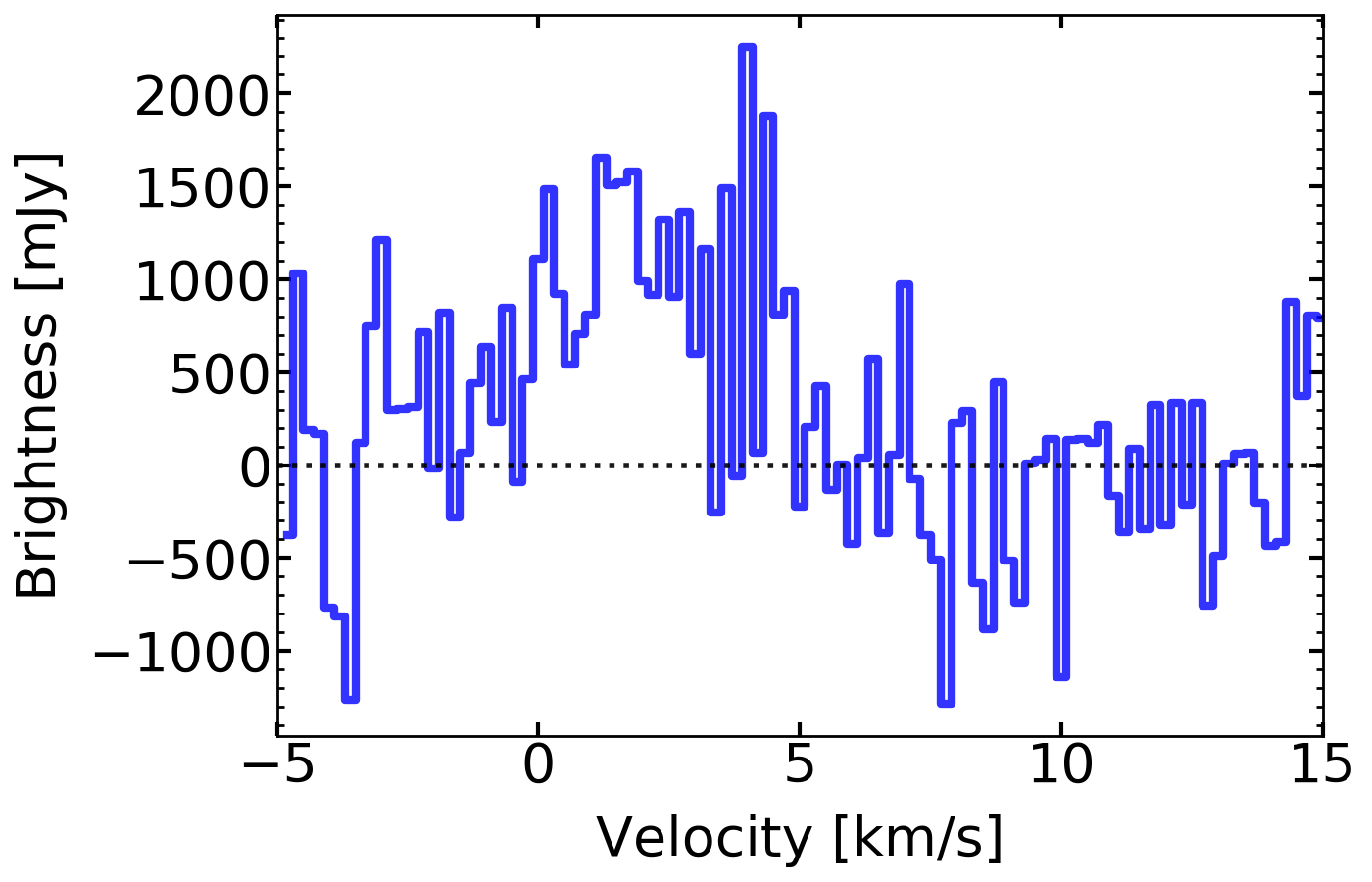}
    \includegraphics[width=.235\textwidth]{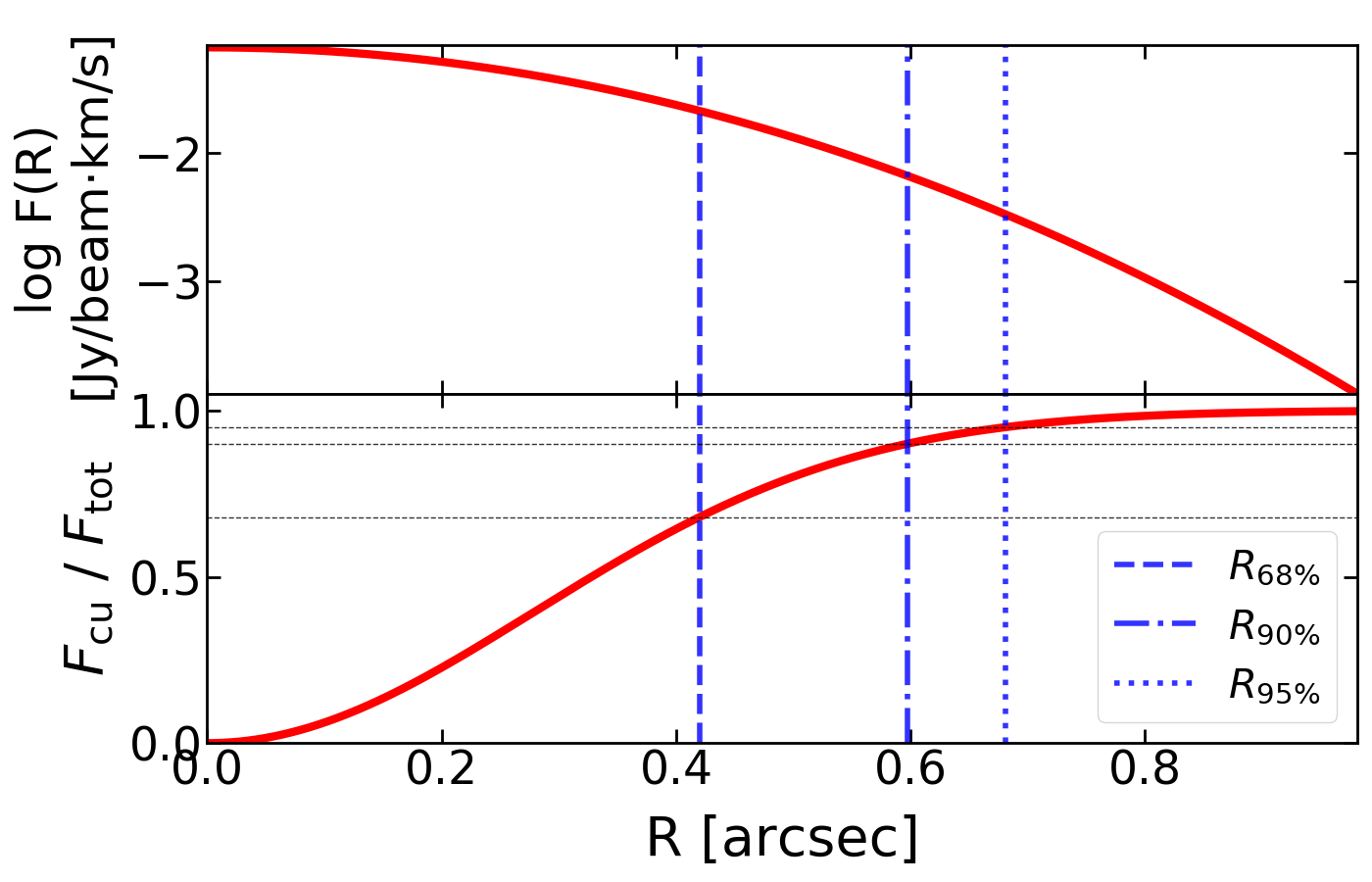}
    \end{center}
    \begin{center}
    \includegraphics[width=.490\textwidth]{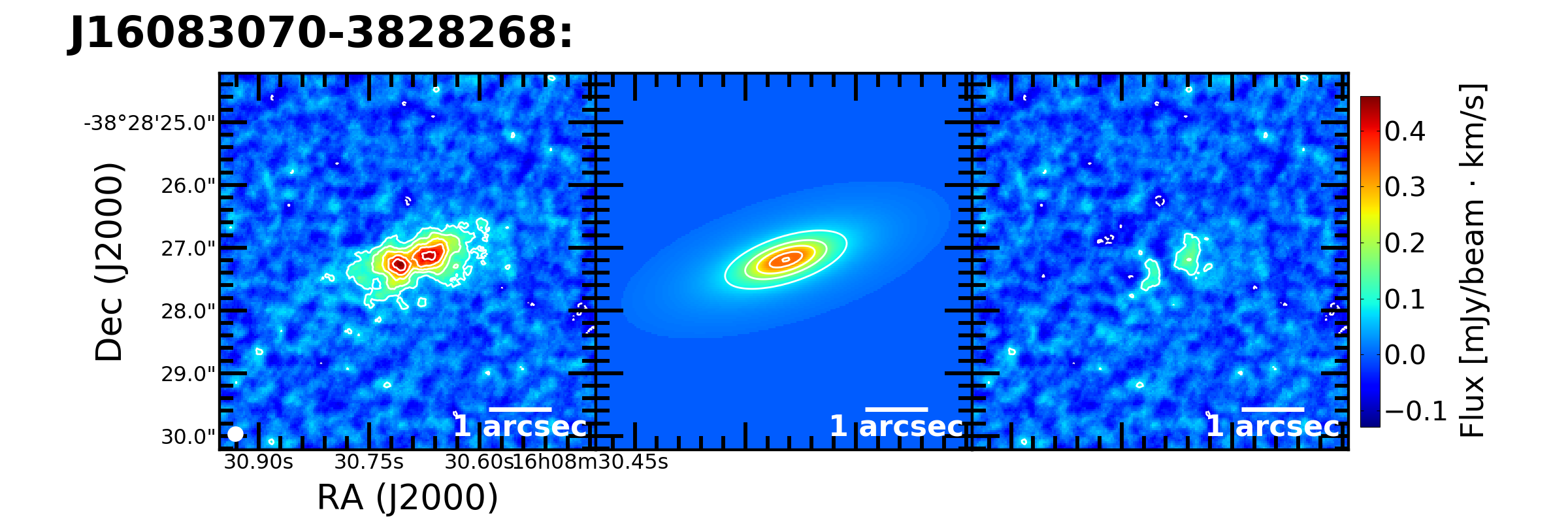}
    \includegraphics[width=.225\textwidth]{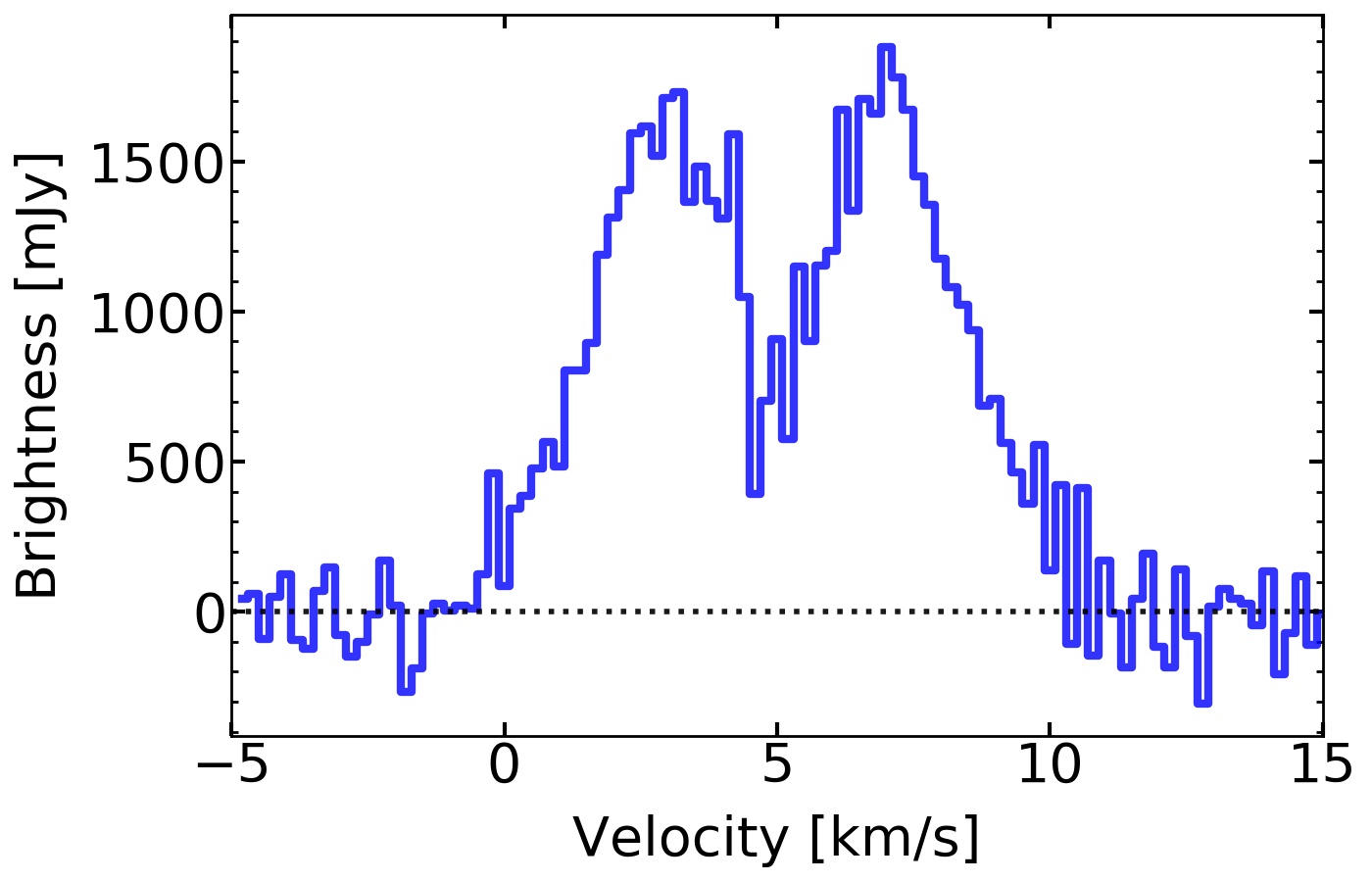}
    \includegraphics[width=.235\textwidth]{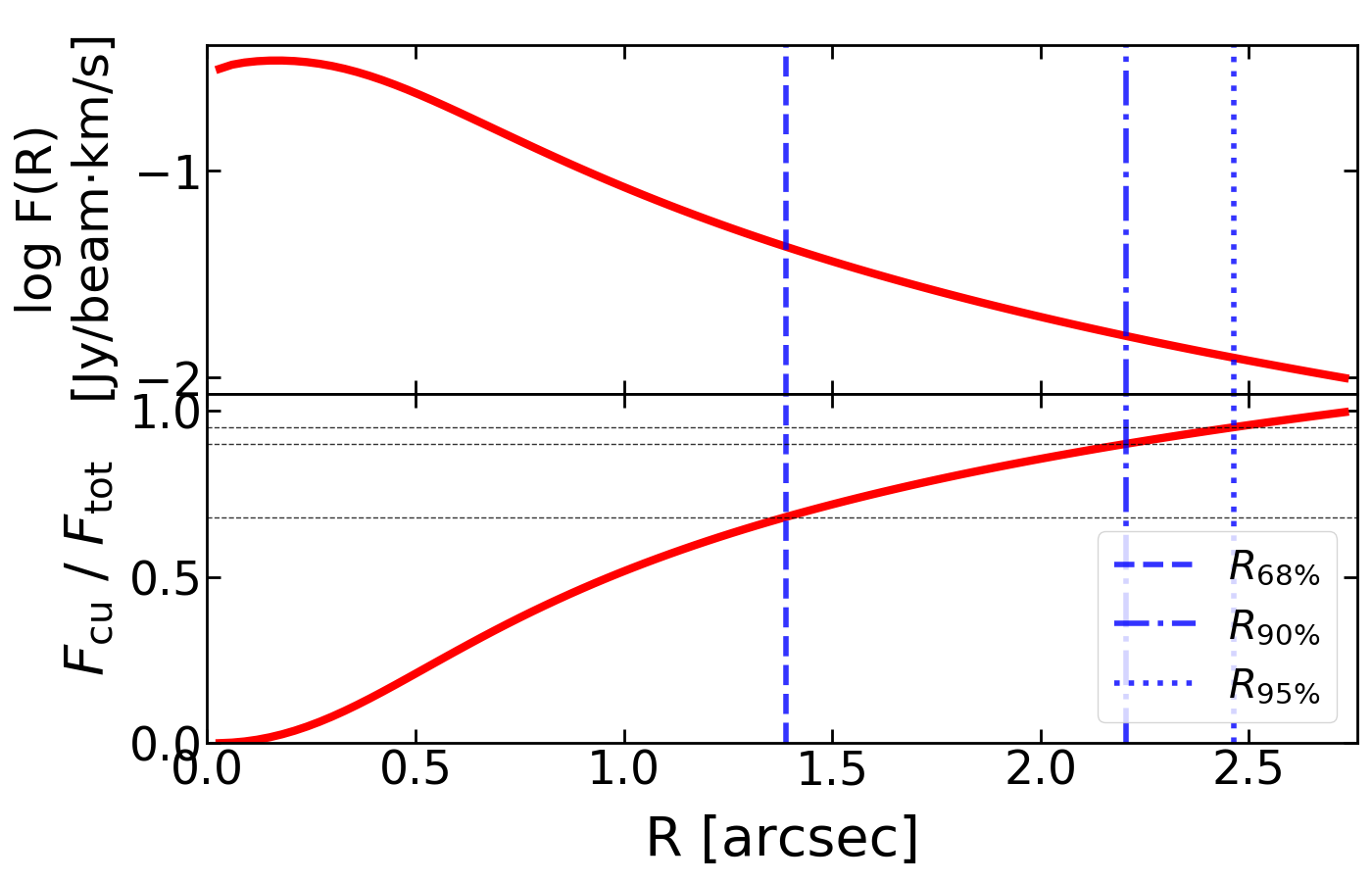}
    \end{center}
  \caption[]{
  Results of the CO modeling for every disk with measured CO size, following the methodology described in Section~\ref{sec:gasmodeling}. For each disk, the first three sub-panels show the observed, model and residual CO moment zero maps; solid (dashed) line contours are drawn at increasing (decreasing) $3\sigma$ intervals. The forth sub-panel represents the integrated spectrum enclosed by the $R_{68\%}^{\mathrm{CO}}$. Last sub-panel shows the radial brightness profile and the respective cumulative distribution of the CO model.
  }
  \label{fig:comodelresults_all_1}
\end{figure*}

\begin{figure*}
    \begin{center}
    \includegraphics[width=.490\textwidth]{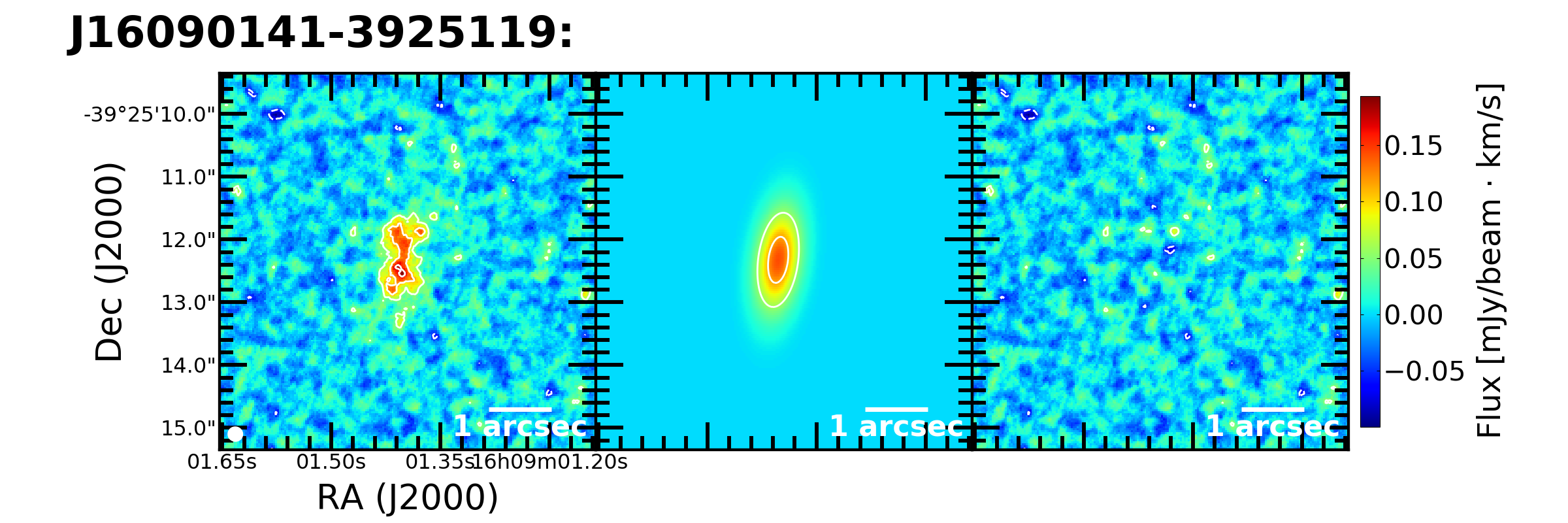}
    \includegraphics[width=.225\textwidth]{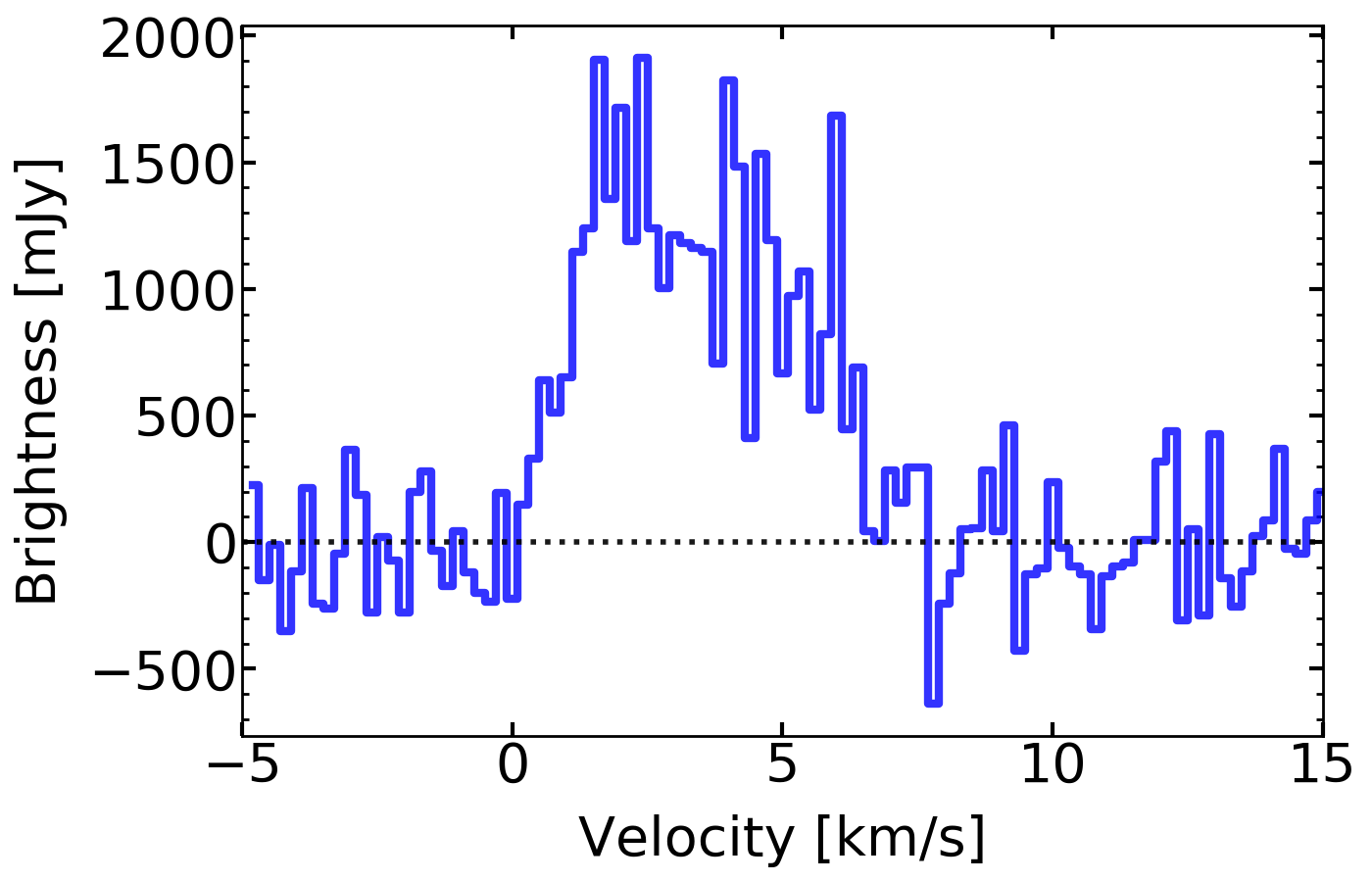}
    \includegraphics[width=.235\textwidth]{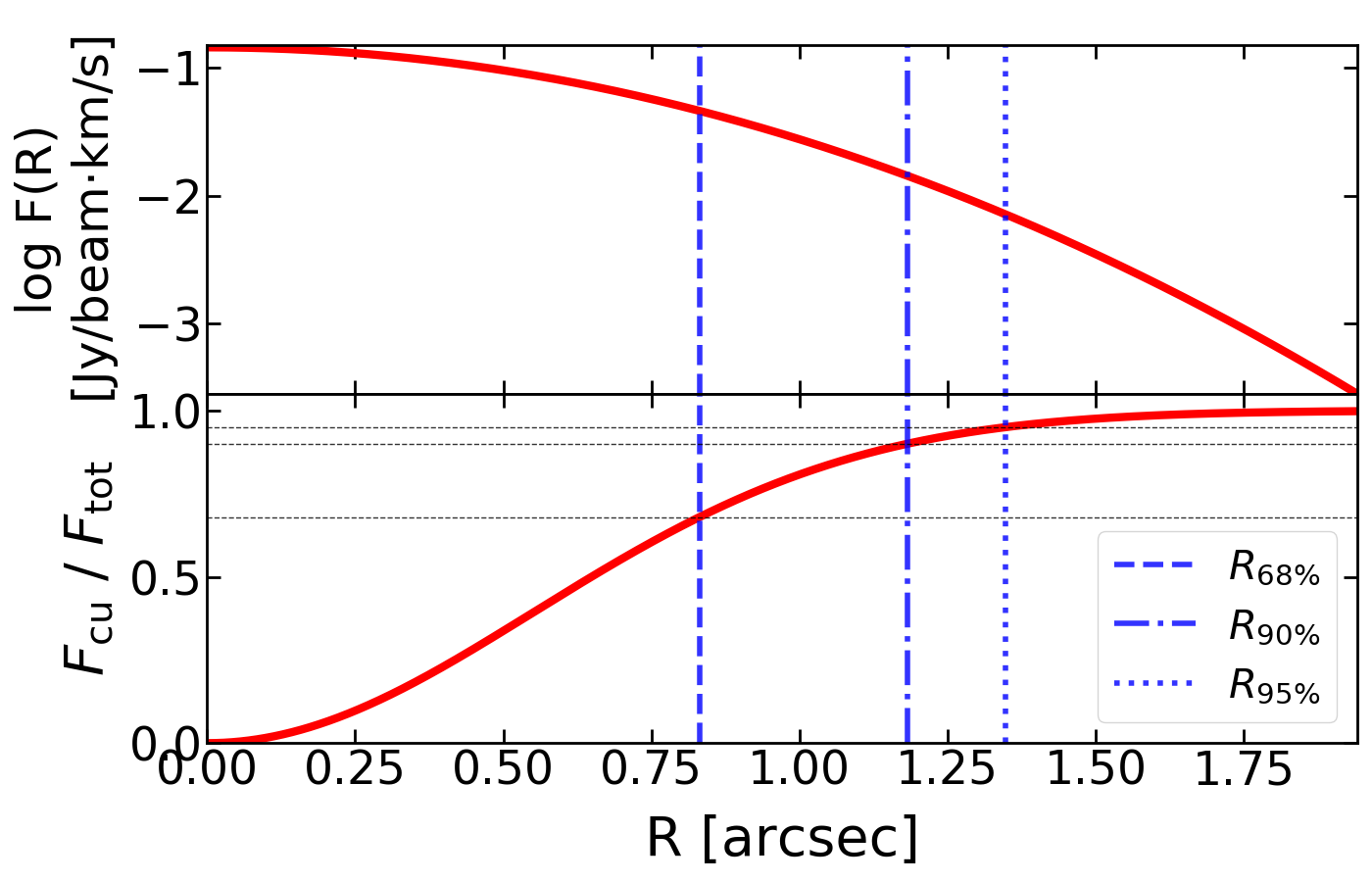}
    \end{center}
    \begin{center}
    \includegraphics[width=.490\textwidth]{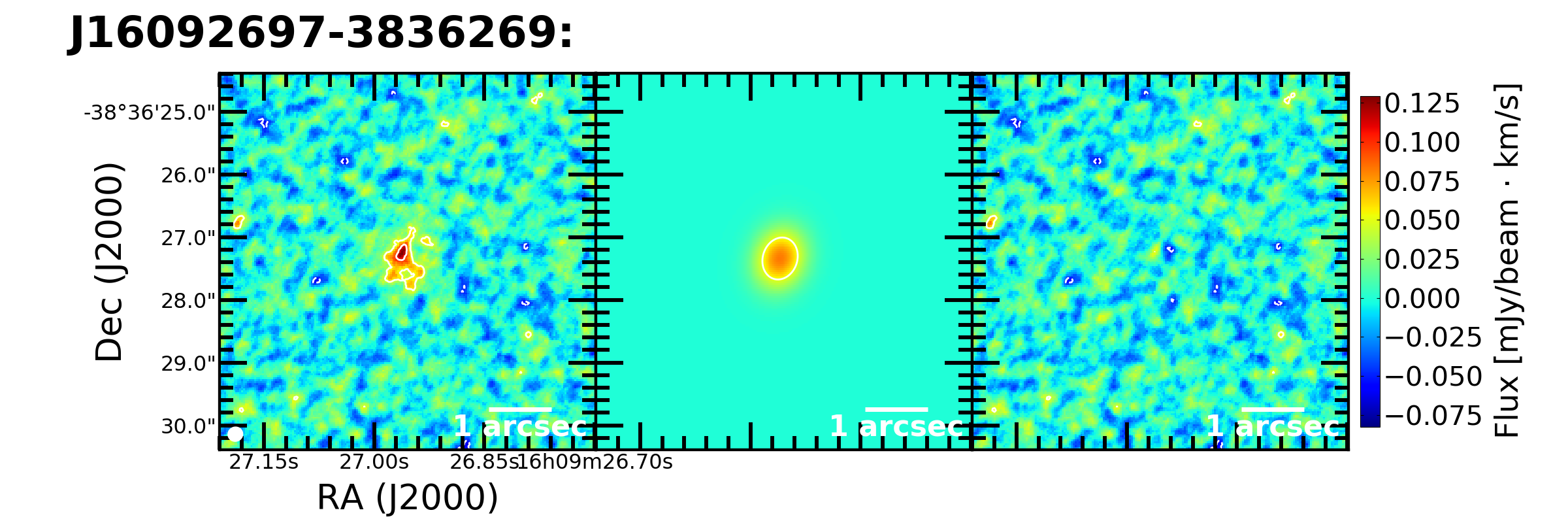}
    \includegraphics[width=.225\textwidth]{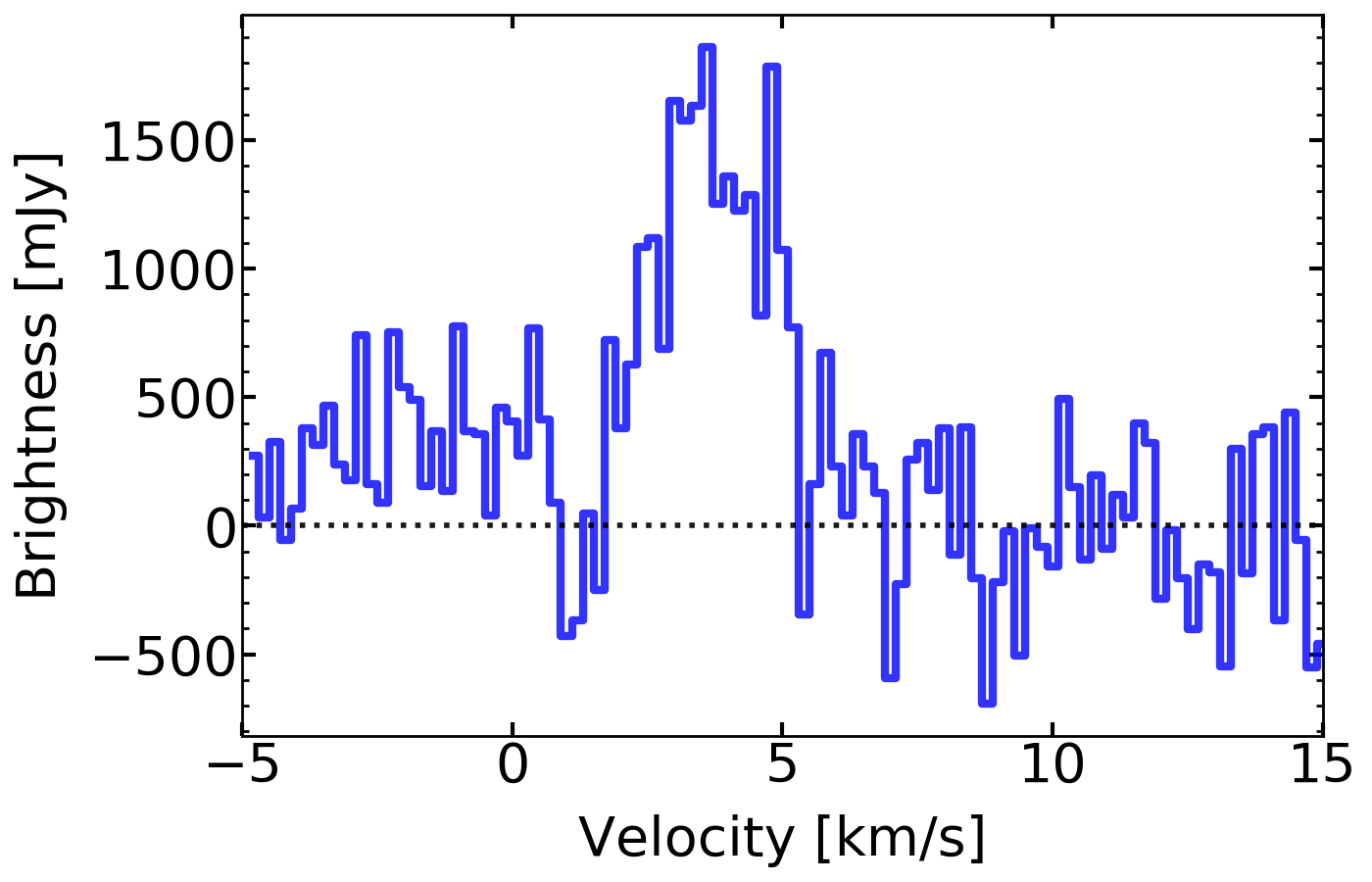}
    \includegraphics[width=.235\textwidth]{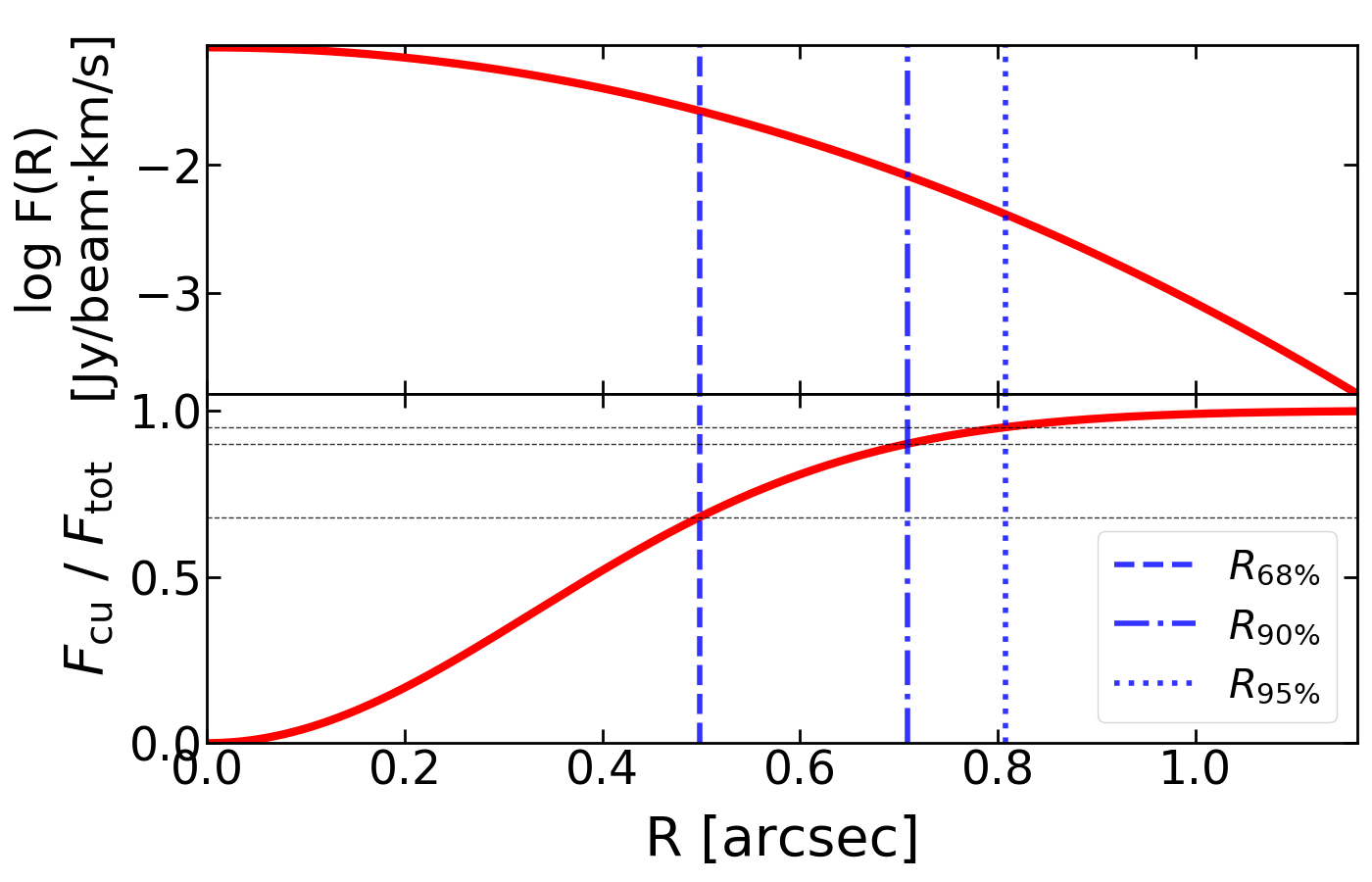}
    \end{center}
    \begin{center}
    \includegraphics[width=.490\textwidth]{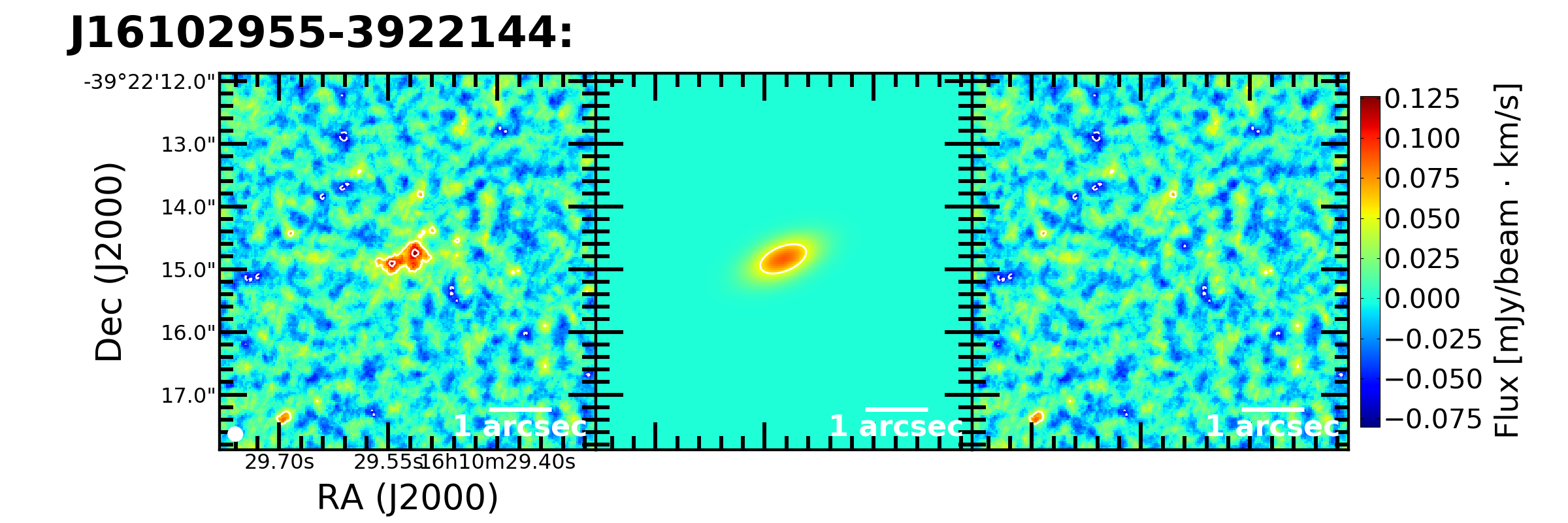}
    \includegraphics[width=.225\textwidth]{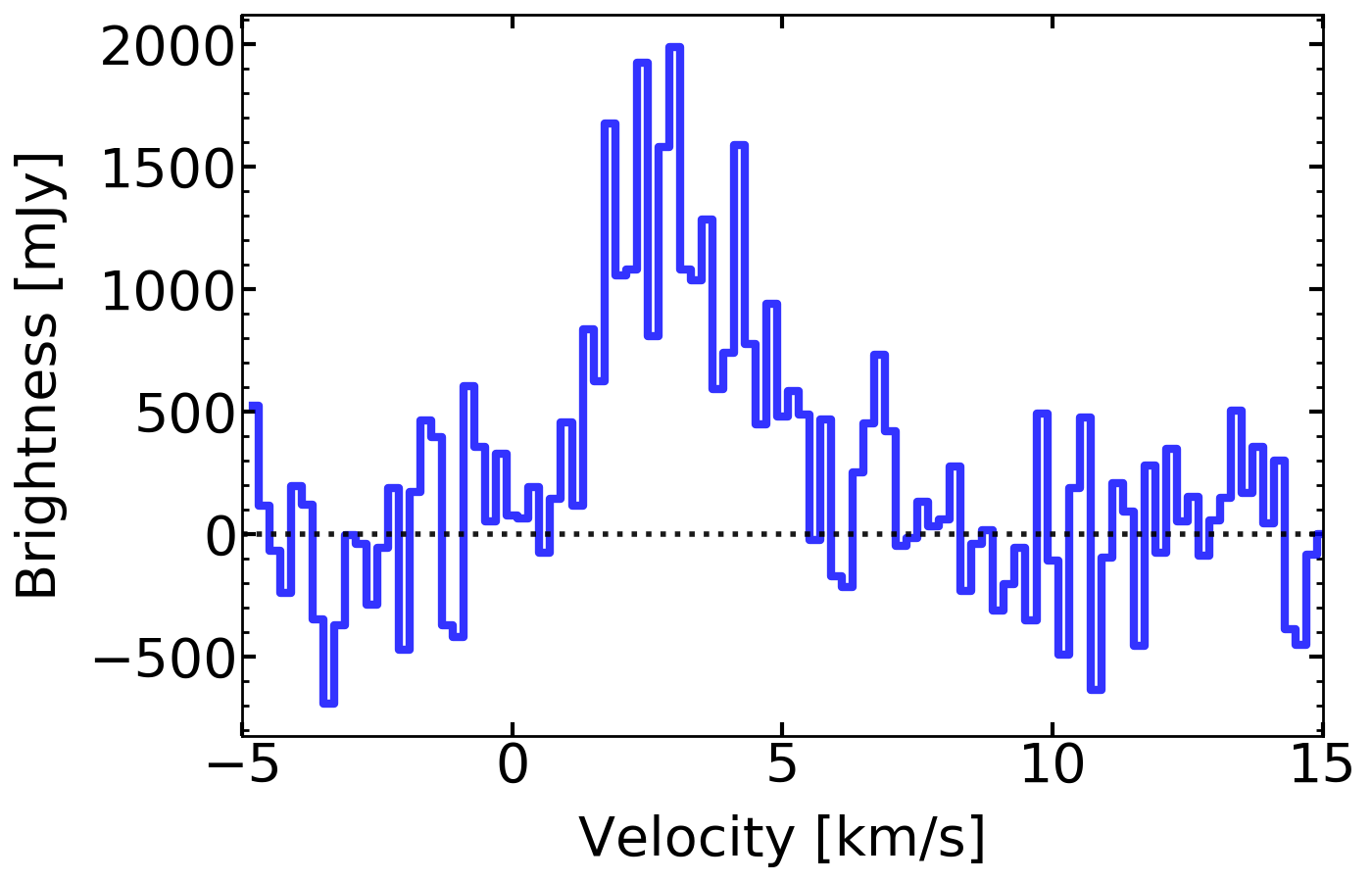}
    \includegraphics[width=.235\textwidth]{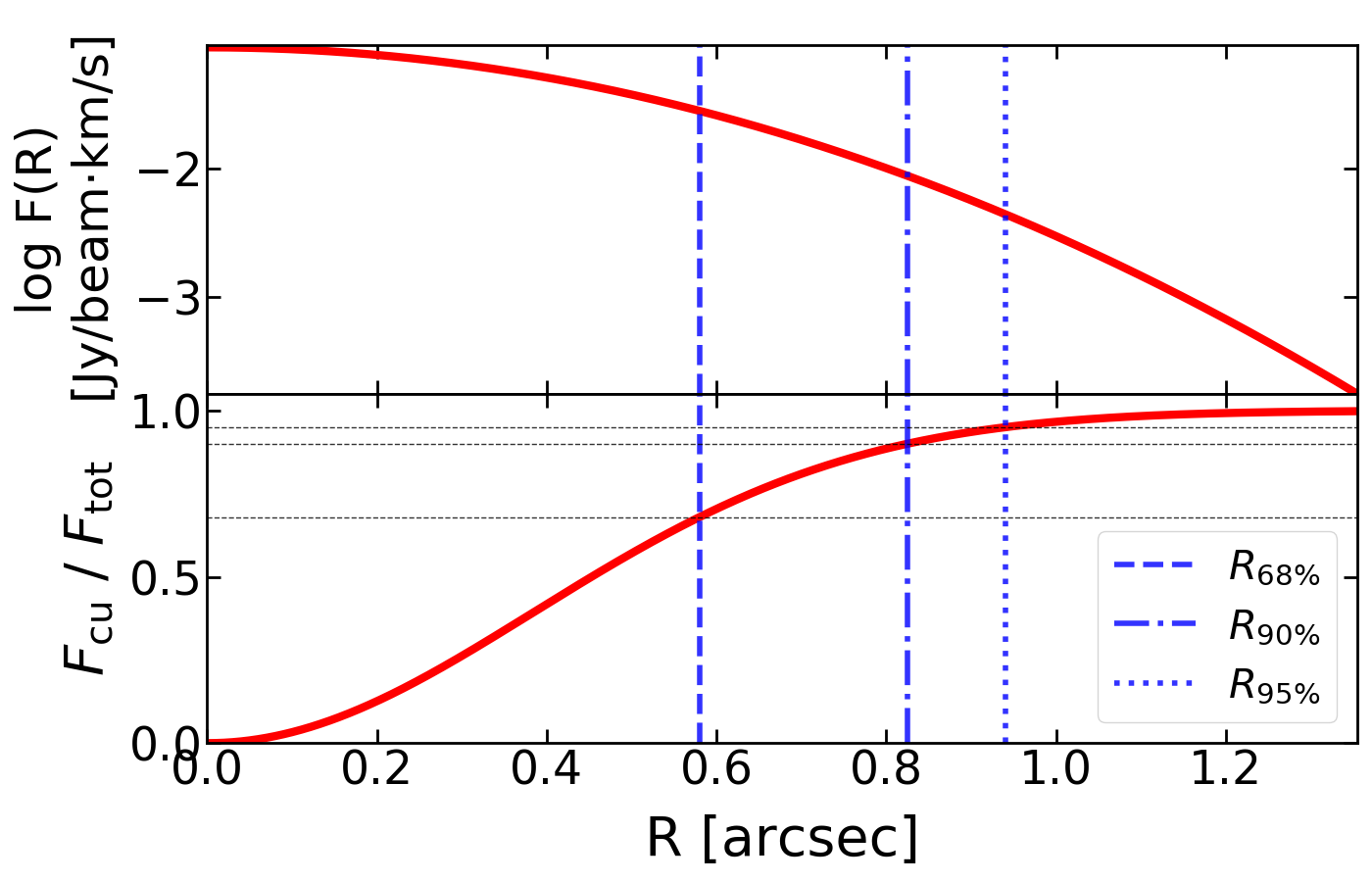}
    \end{center}
    \begin{center}
    \includegraphics[width=.490\textwidth]{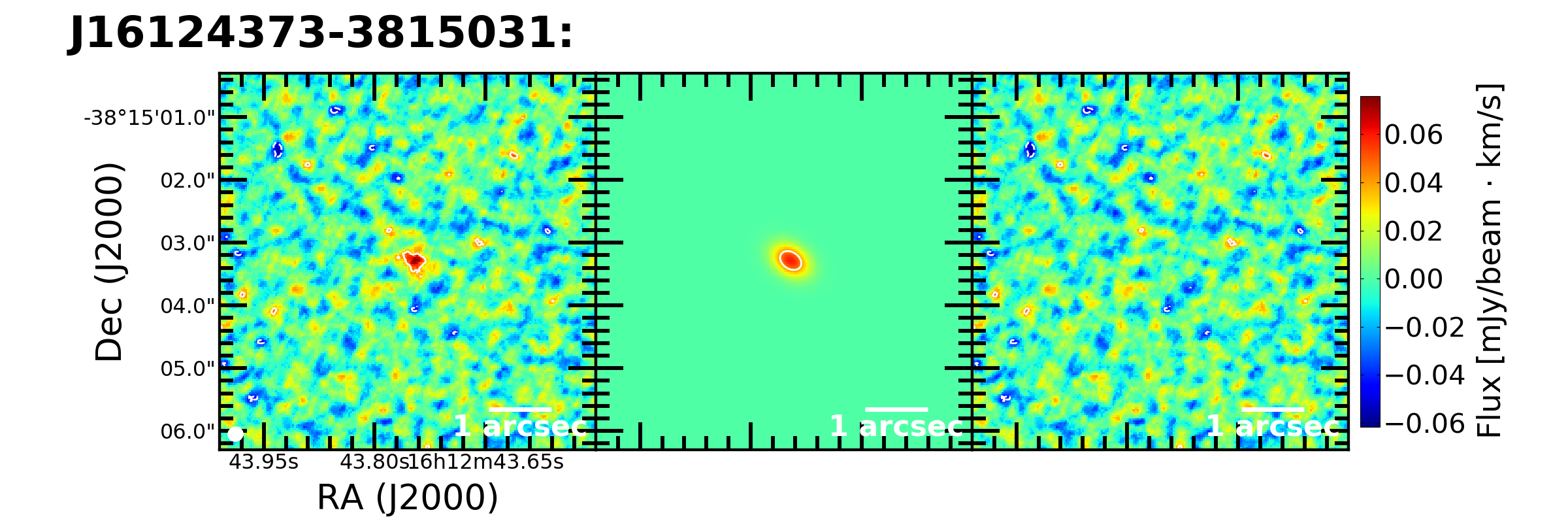}
    \includegraphics[width=.225\textwidth]{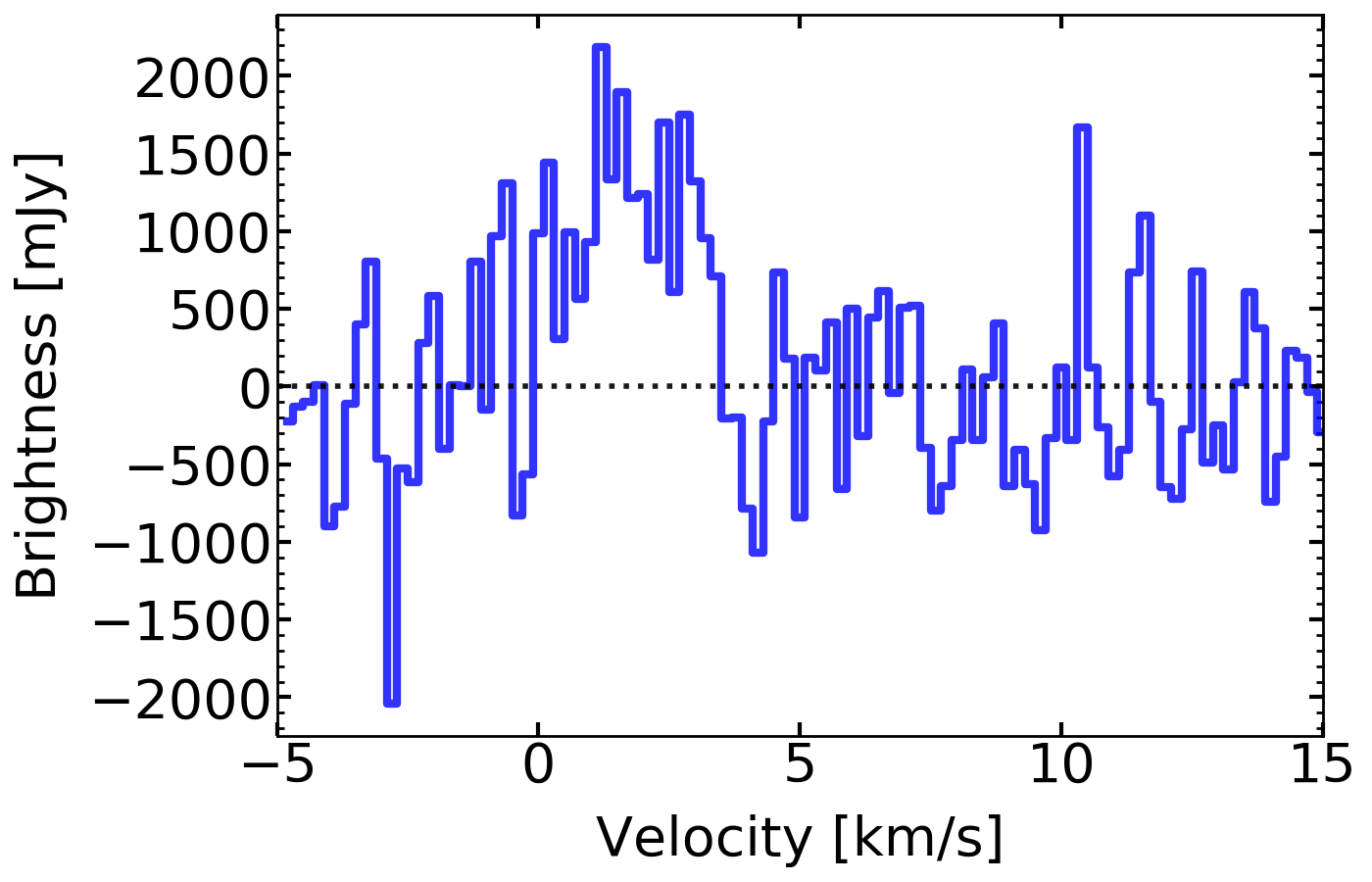}
    \includegraphics[width=.235\textwidth]{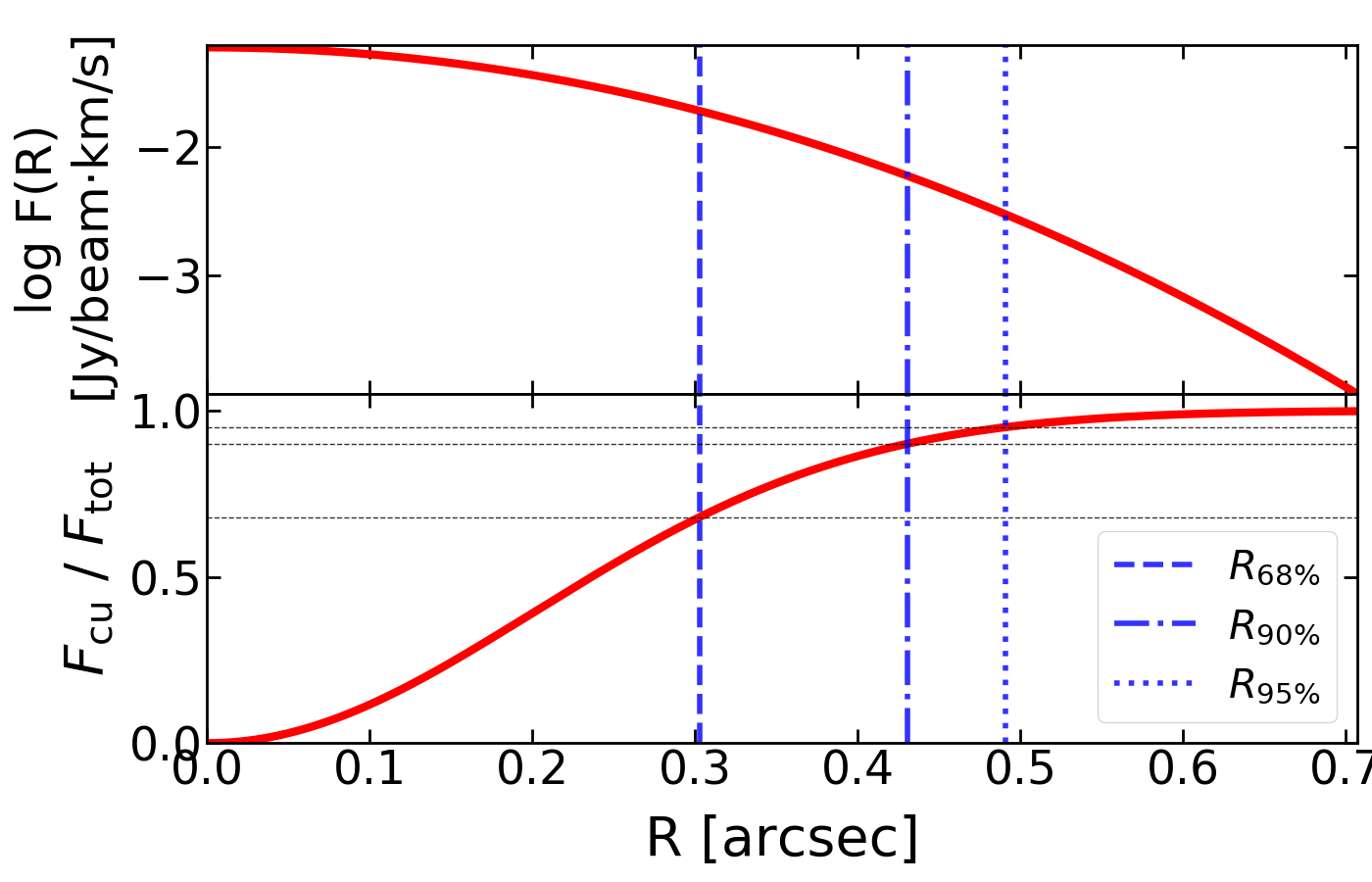}
    \end{center}
    \begin{center}
    \includegraphics[width=.490\textwidth]{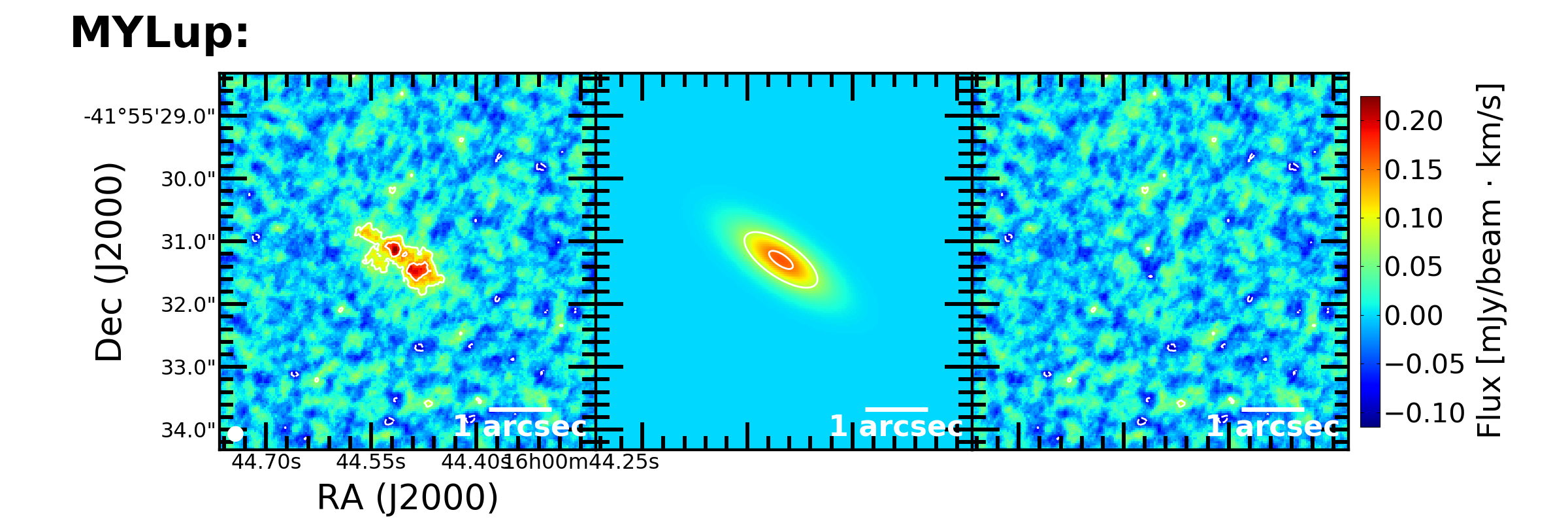}
    \includegraphics[width=.225\textwidth]{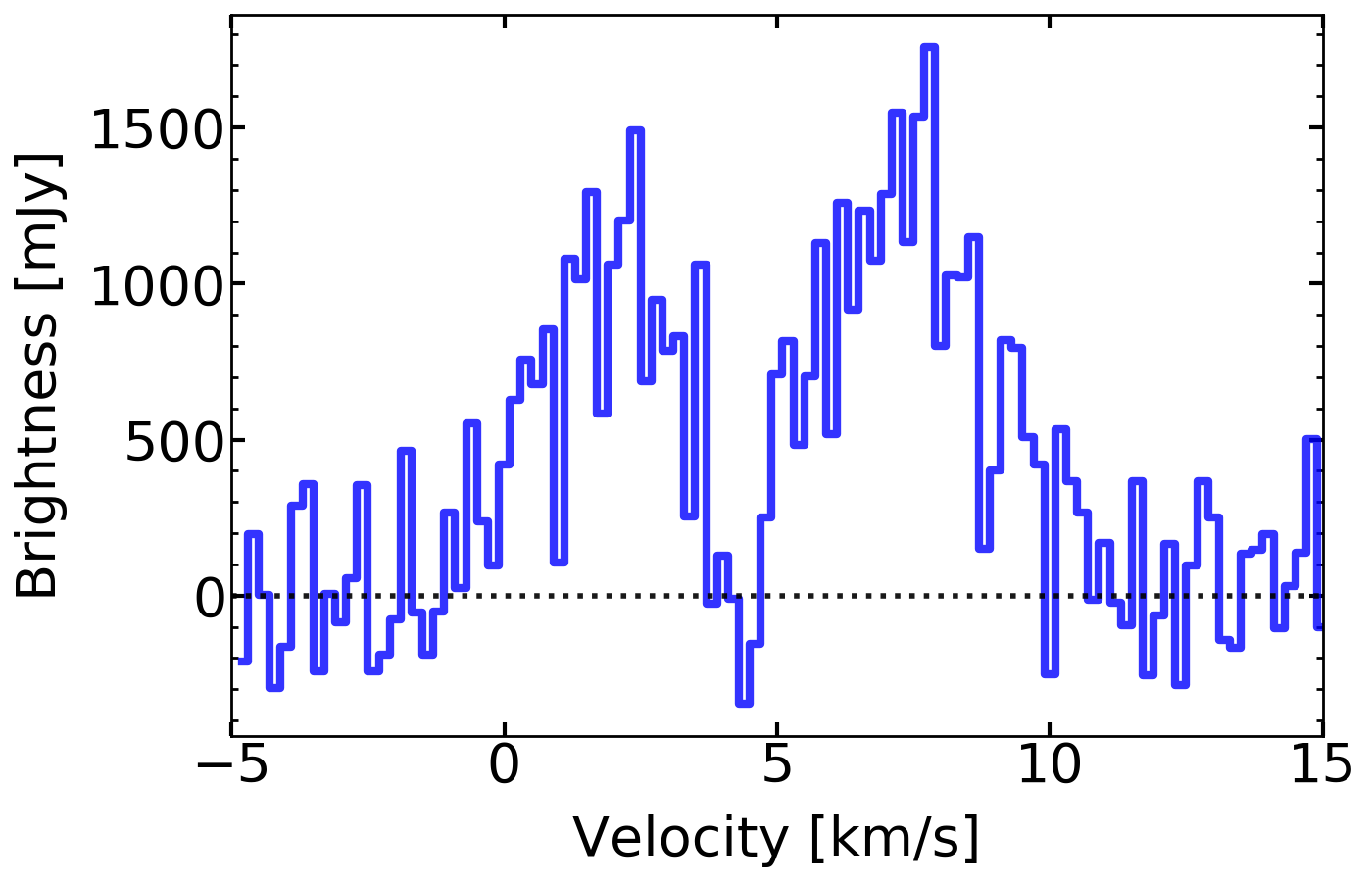}
    \includegraphics[width=.235\textwidth]{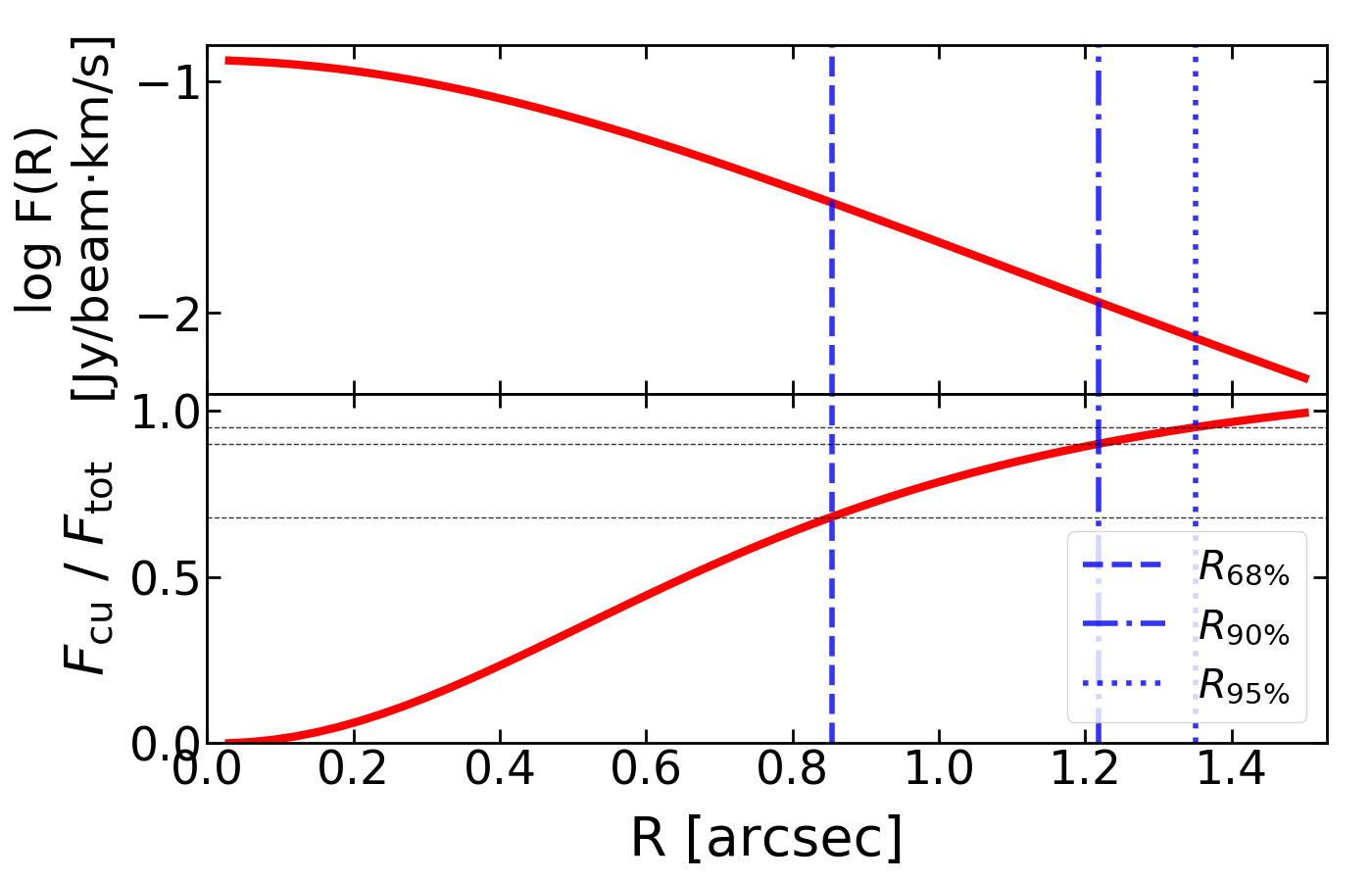}
    \end{center}
    \begin{center}
    \includegraphics[width=.490\textwidth]{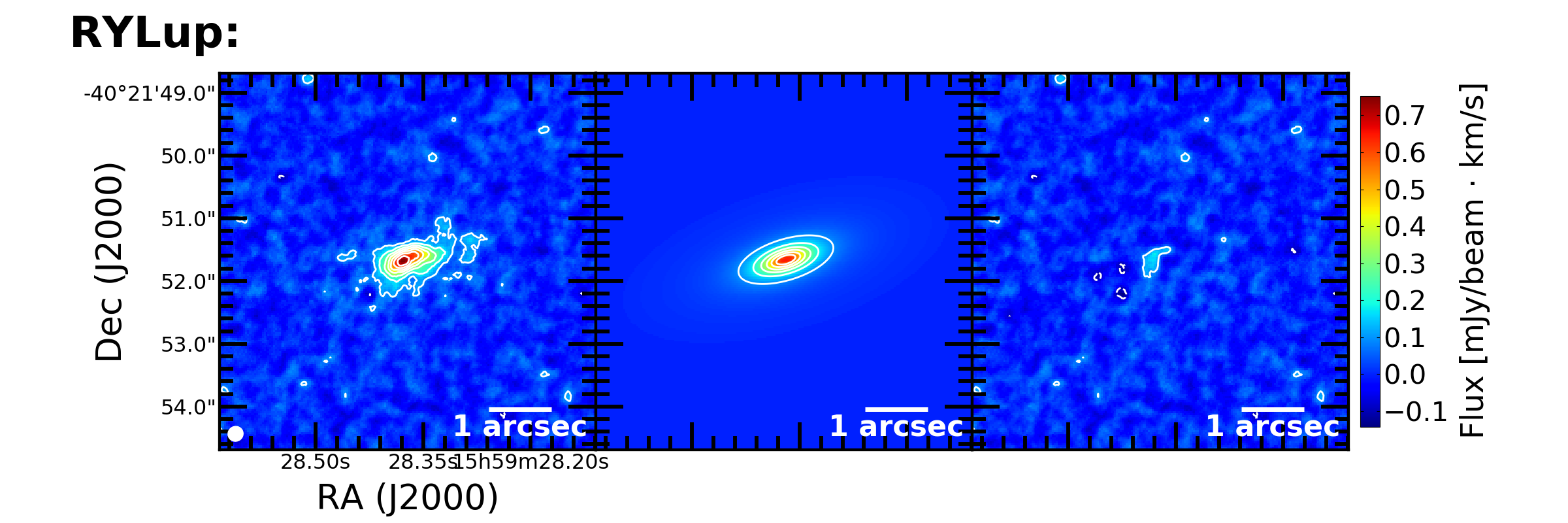}
    \includegraphics[width=.225\textwidth]{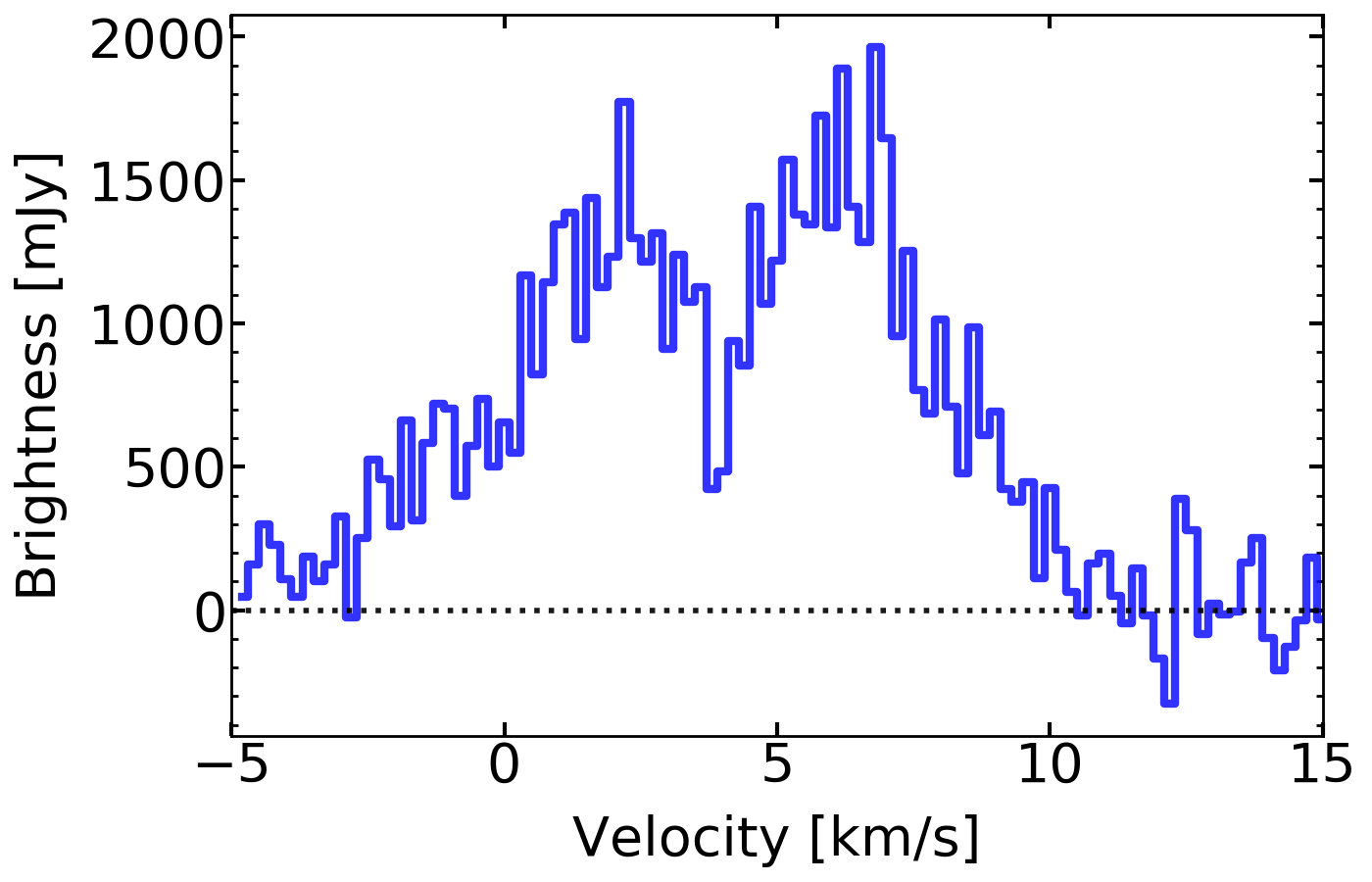}
    \includegraphics[width=.235\textwidth]{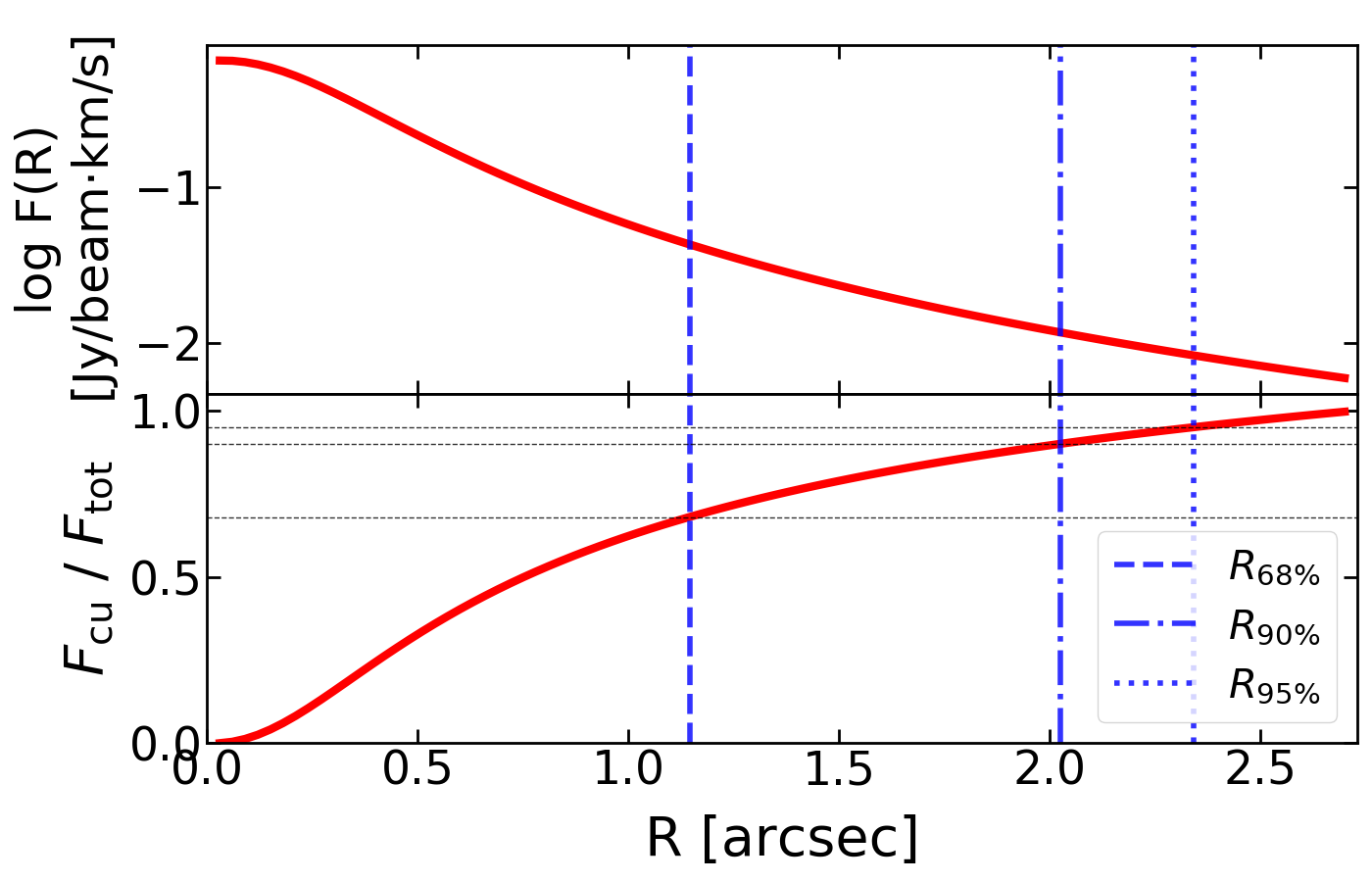}
    \end{center}
    \begin{center}
    \includegraphics[width=.490\textwidth]{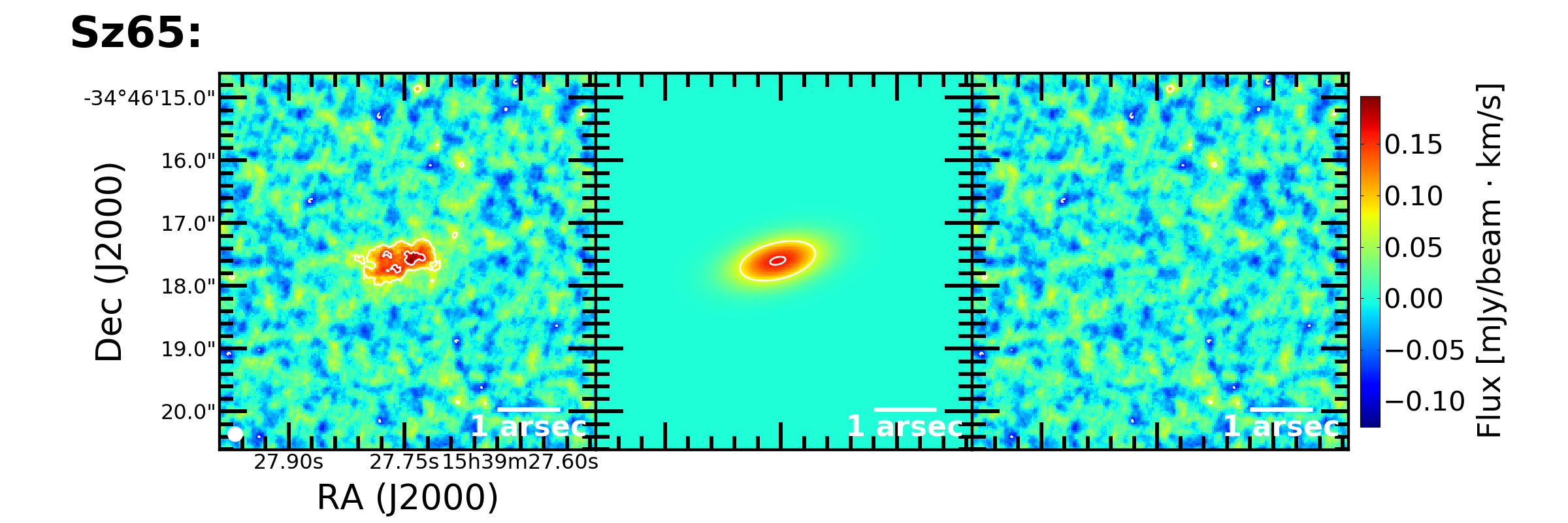}
    \includegraphics[width=.225\textwidth]{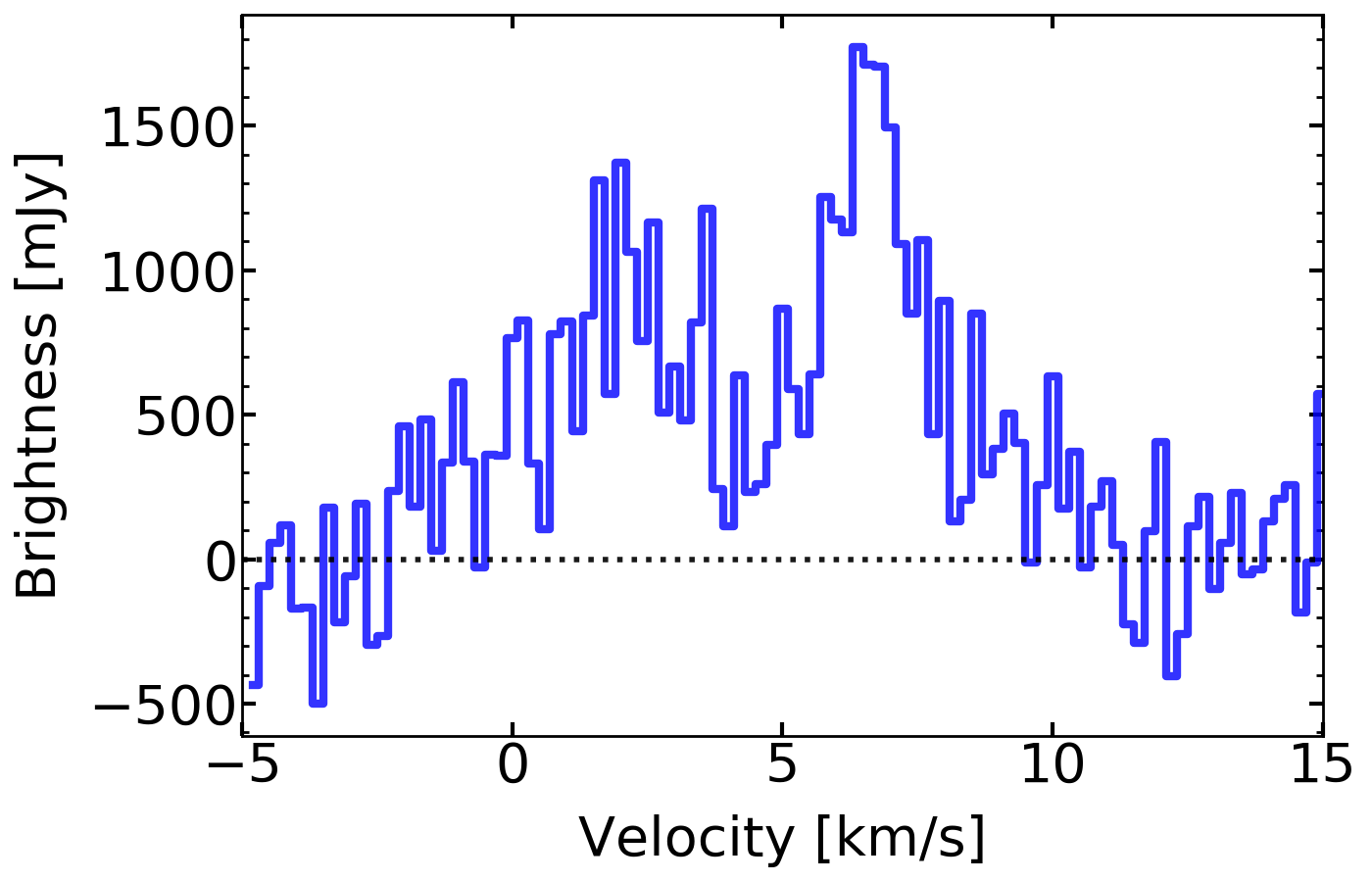}
    \includegraphics[width=.235\textwidth]{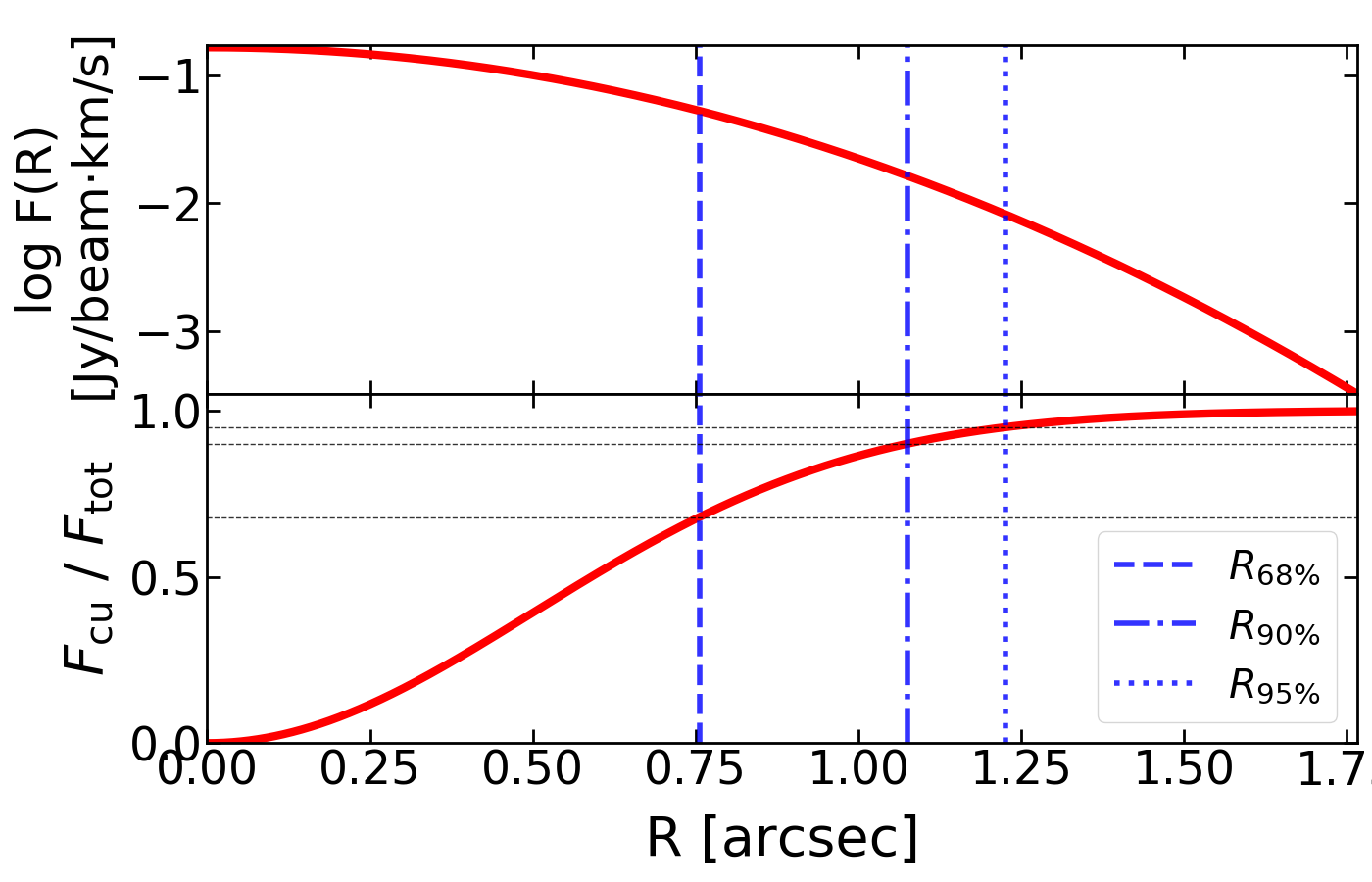}
    \end{center}
  \caption[]{
  Results of the CO modeling for every disk with measured CO size, following the methodology described in Section~\ref{sec:gasmodeling}. For each disk, the first three sub-panels show the observed, model and residual CO moment zero maps; solid (dashed) line contours are drawn at increasing (decreasing) $3\sigma$ intervals. The forth sub-panel represents the integrated spectrum enclosed by the $R_{68\%}^{\mathrm{CO}}$ of the source. Last sub-panel shows the radial brightness profile and the respective cumulative distribution of the CO model.
  }
  \label{fig:comodelresults_all_2}
\end{figure*}

\begin{figure*}
    \begin{center}
    \includegraphics[width=.490\textwidth]{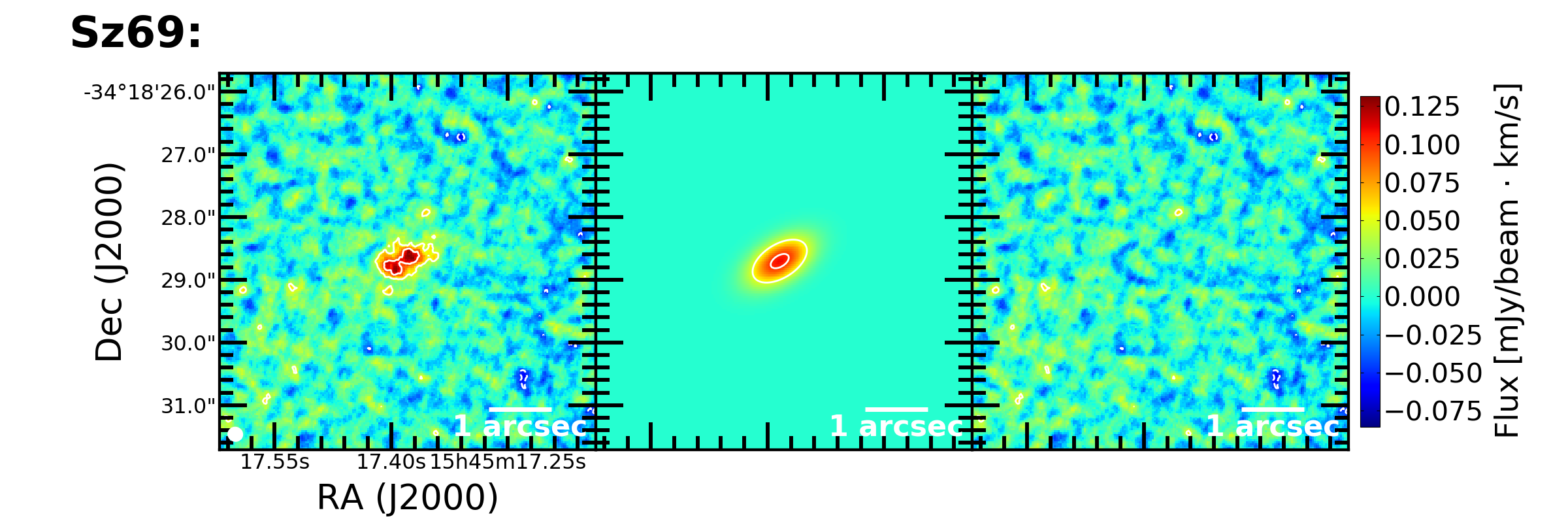}
    \includegraphics[width=.225\textwidth]{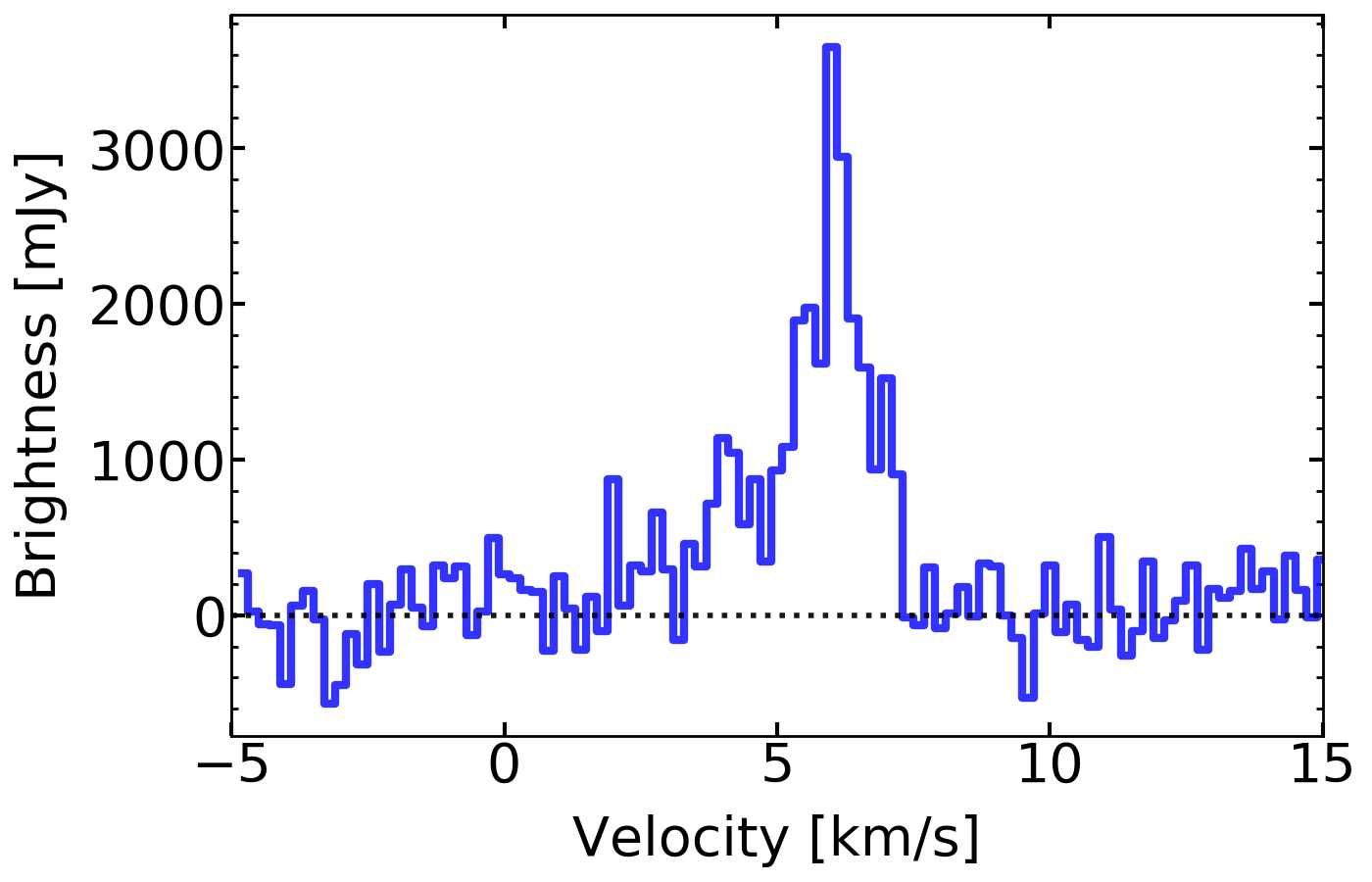}
    \includegraphics[width=.235\textwidth]{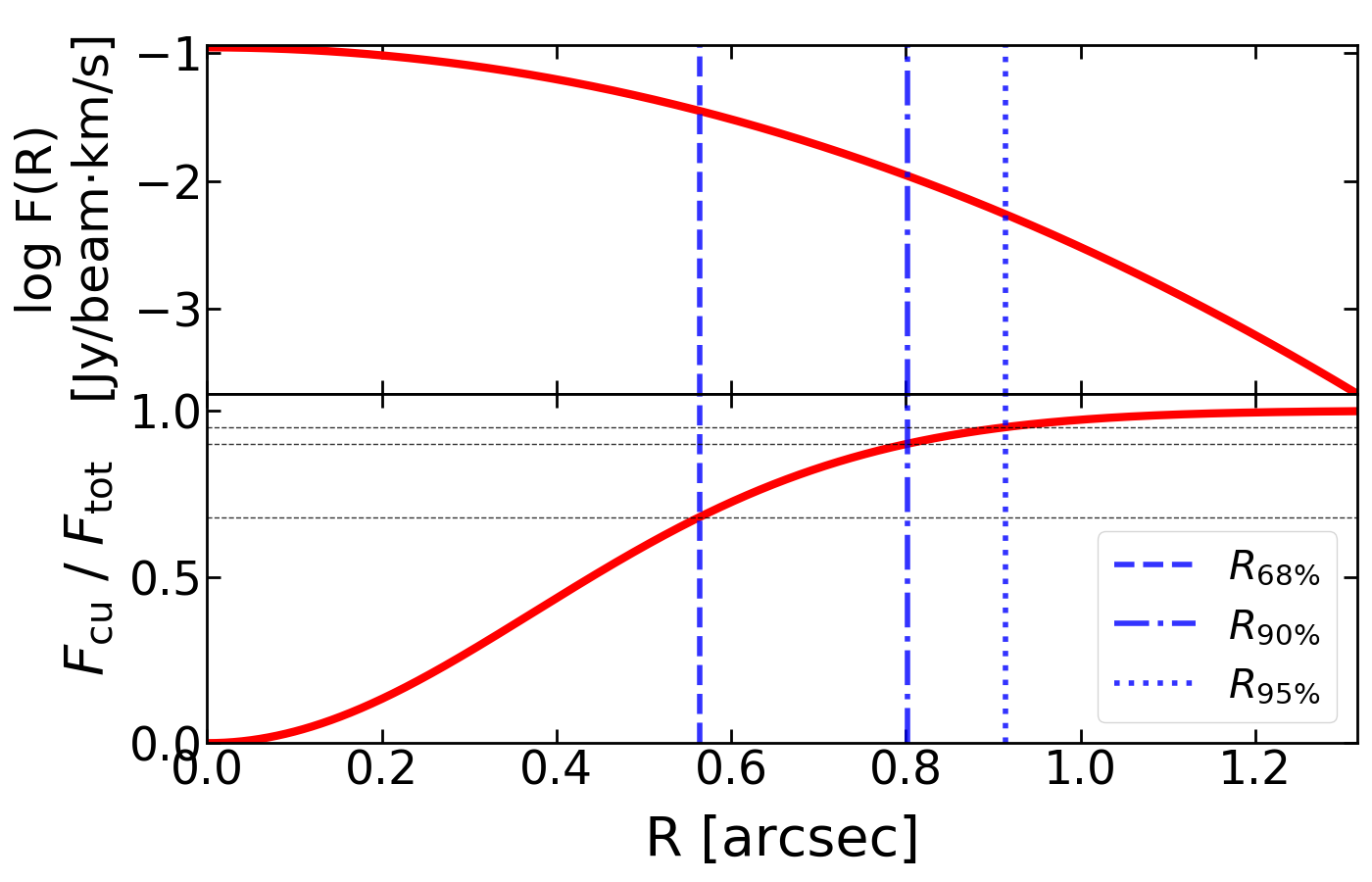}
    \end{center}
    \begin{center}
    \includegraphics[width=.490\textwidth]{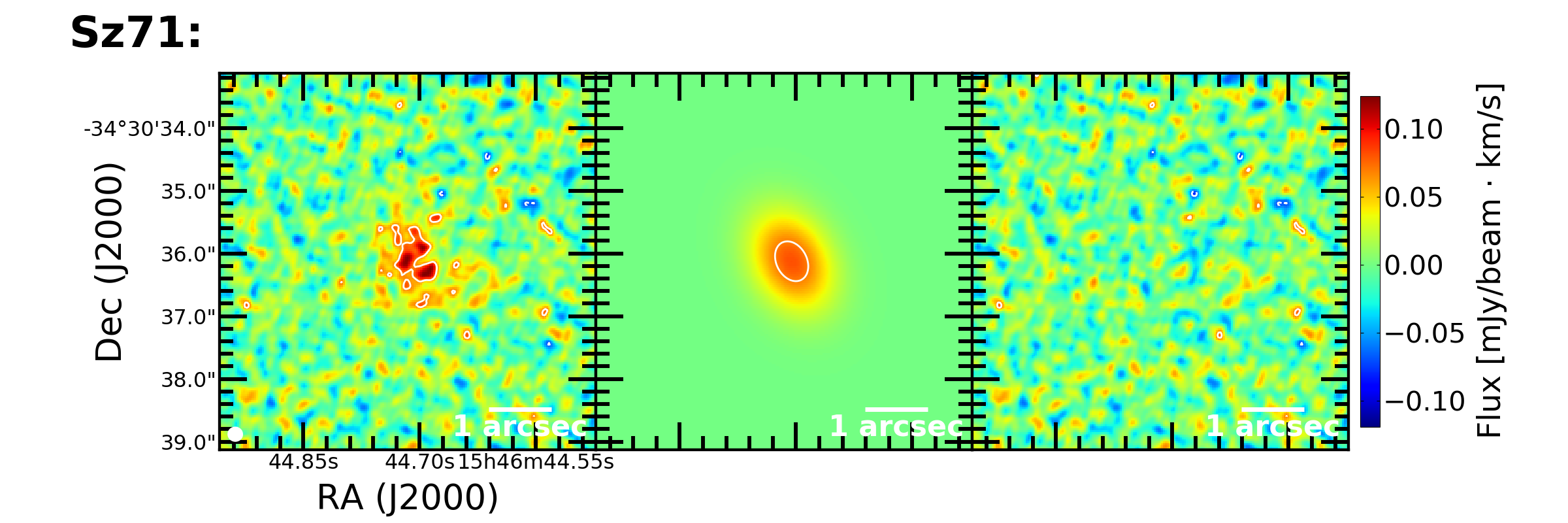}
    \includegraphics[width=.225\textwidth]{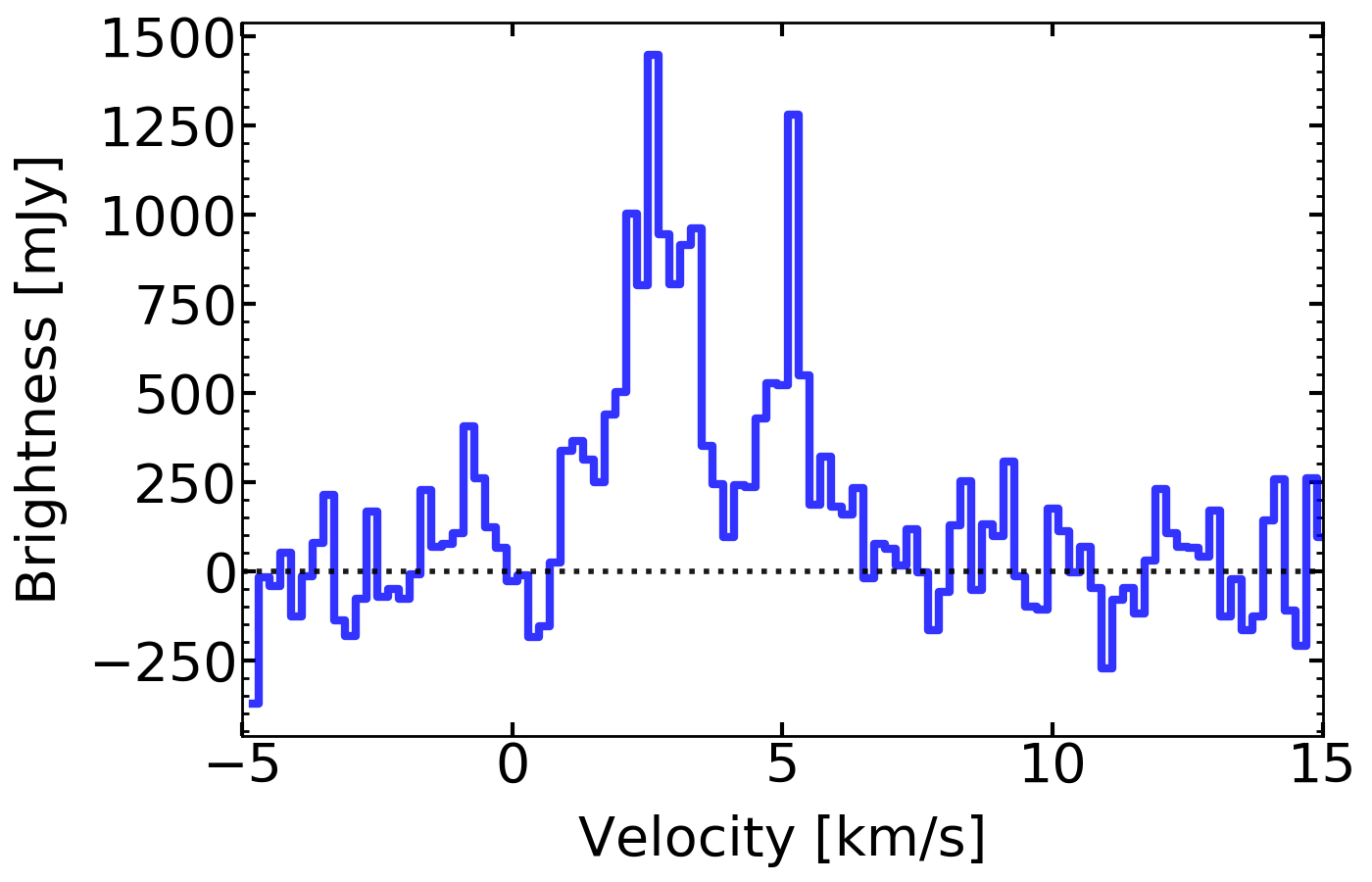}
    \includegraphics[width=.235\textwidth]{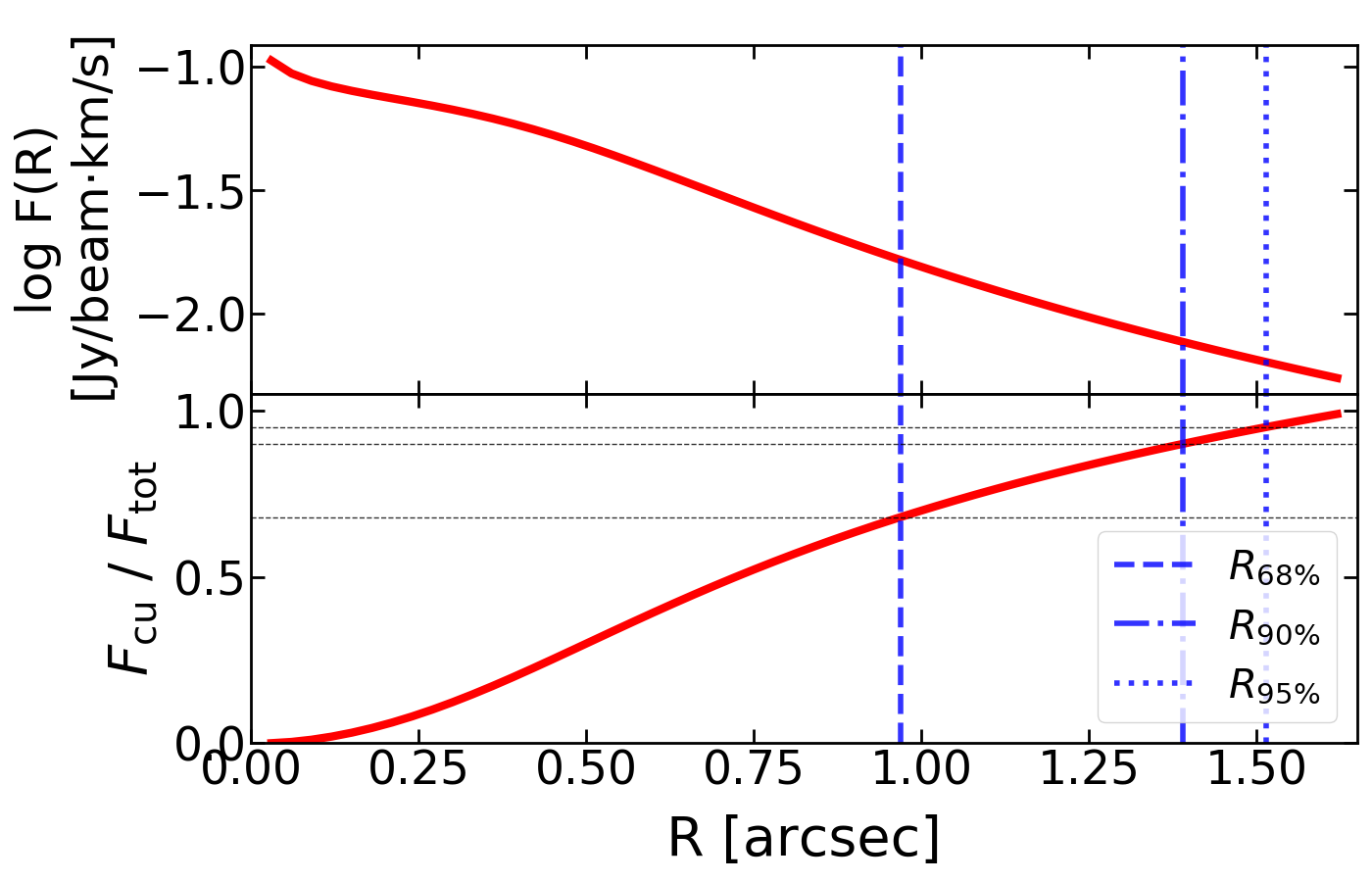}
    \end{center}
    \begin{center}
    \includegraphics[width=.490\textwidth]{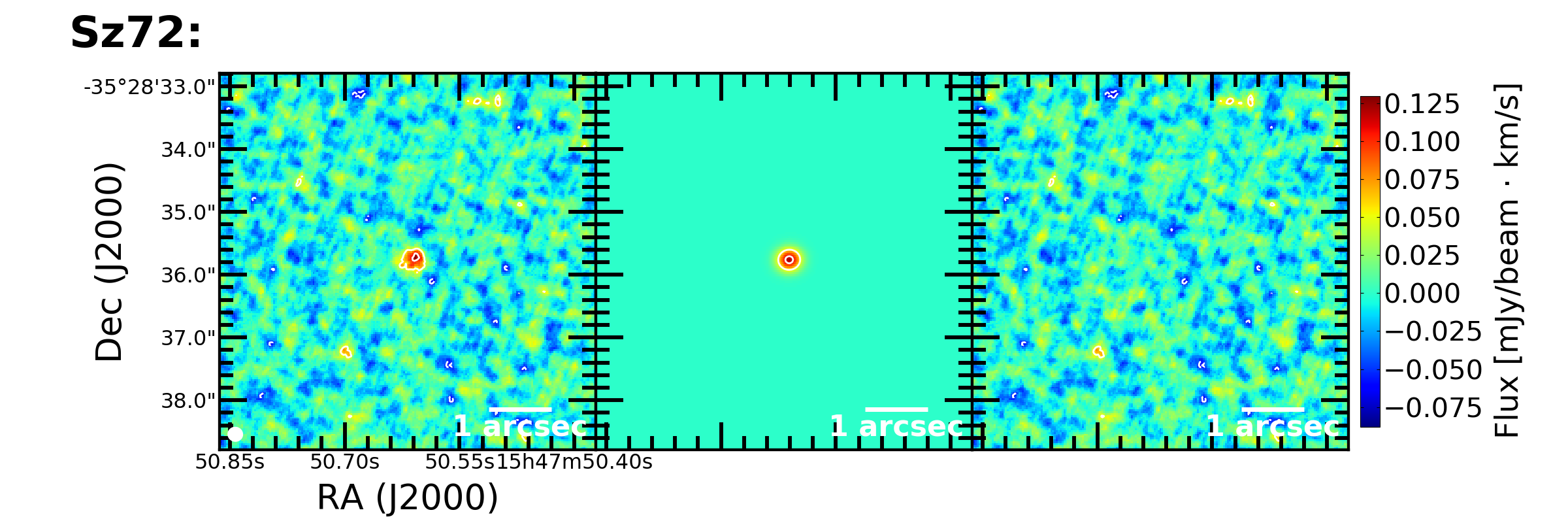}
    \includegraphics[width=.225\textwidth]{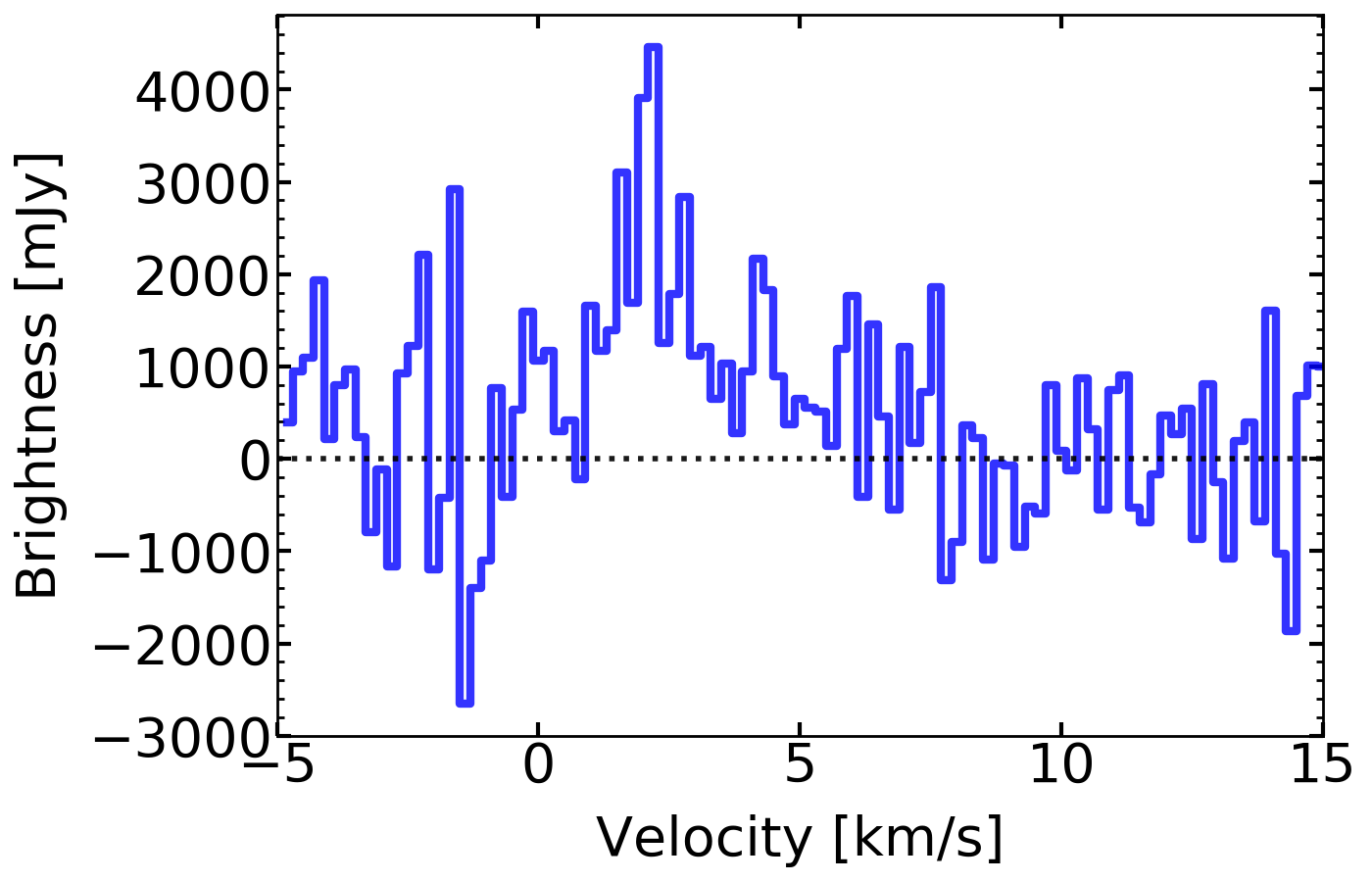}
    \includegraphics[width=.235\textwidth]{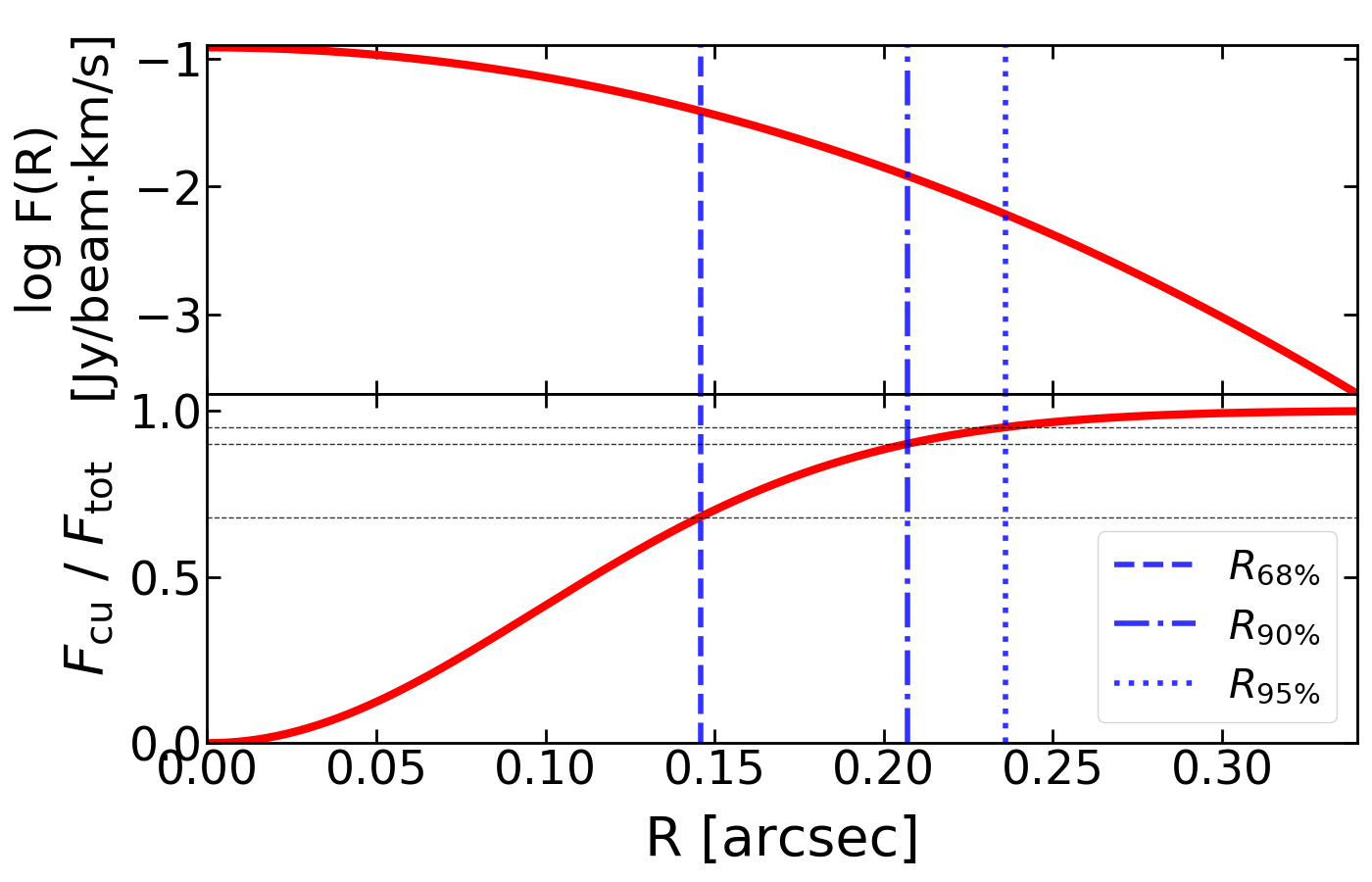}
    \end{center}
    \begin{center}
    \includegraphics[width=.490\textwidth]{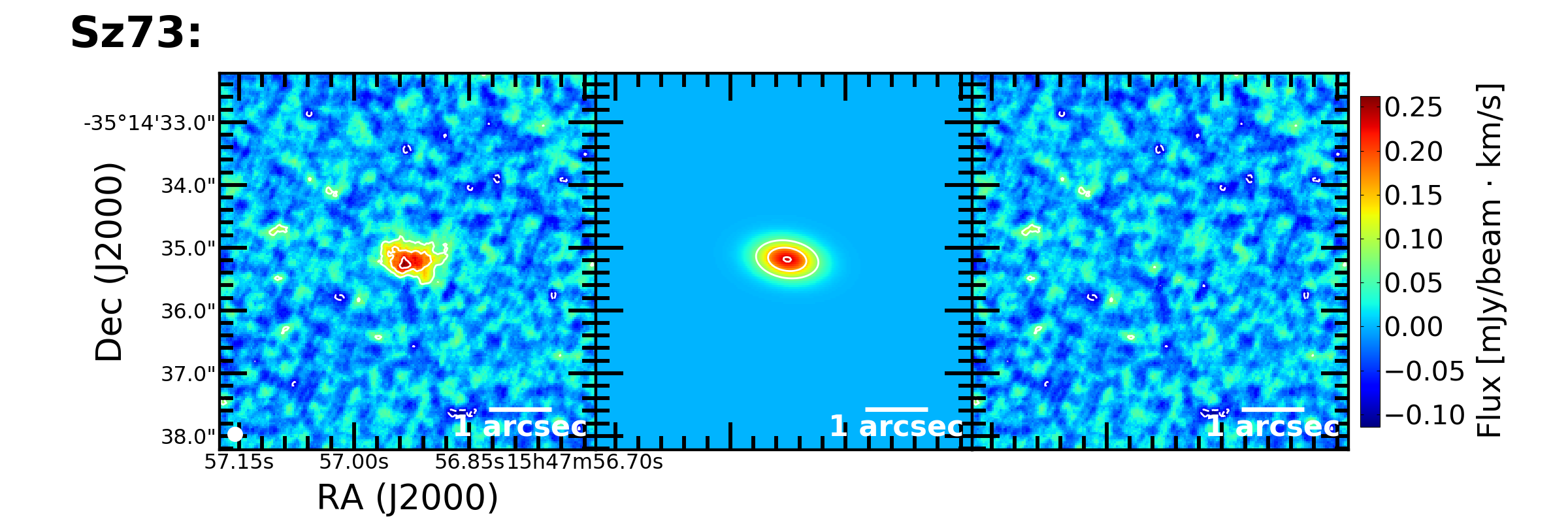}
    \includegraphics[width=.225\textwidth]{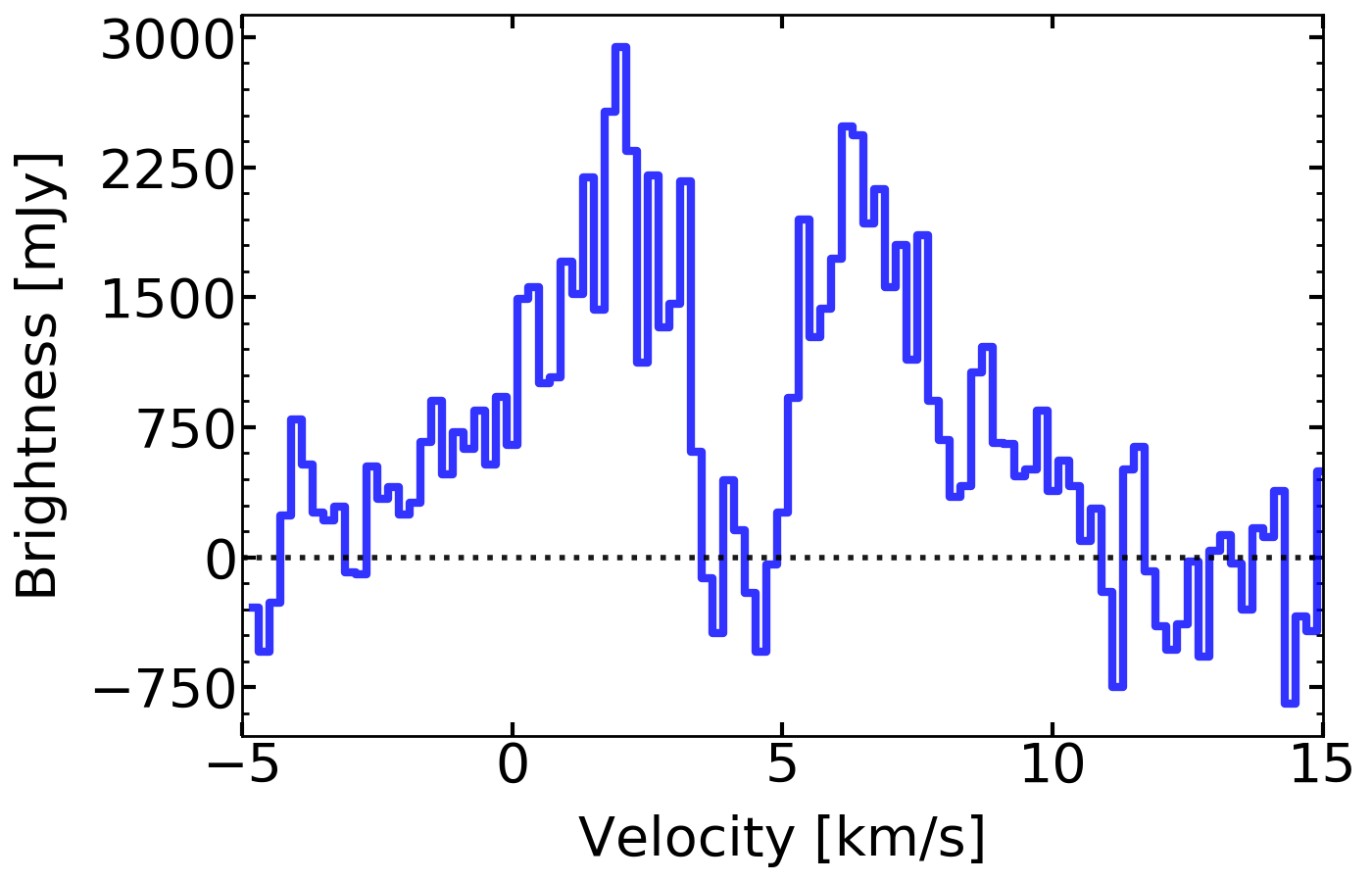}
    \includegraphics[width=.235\textwidth]{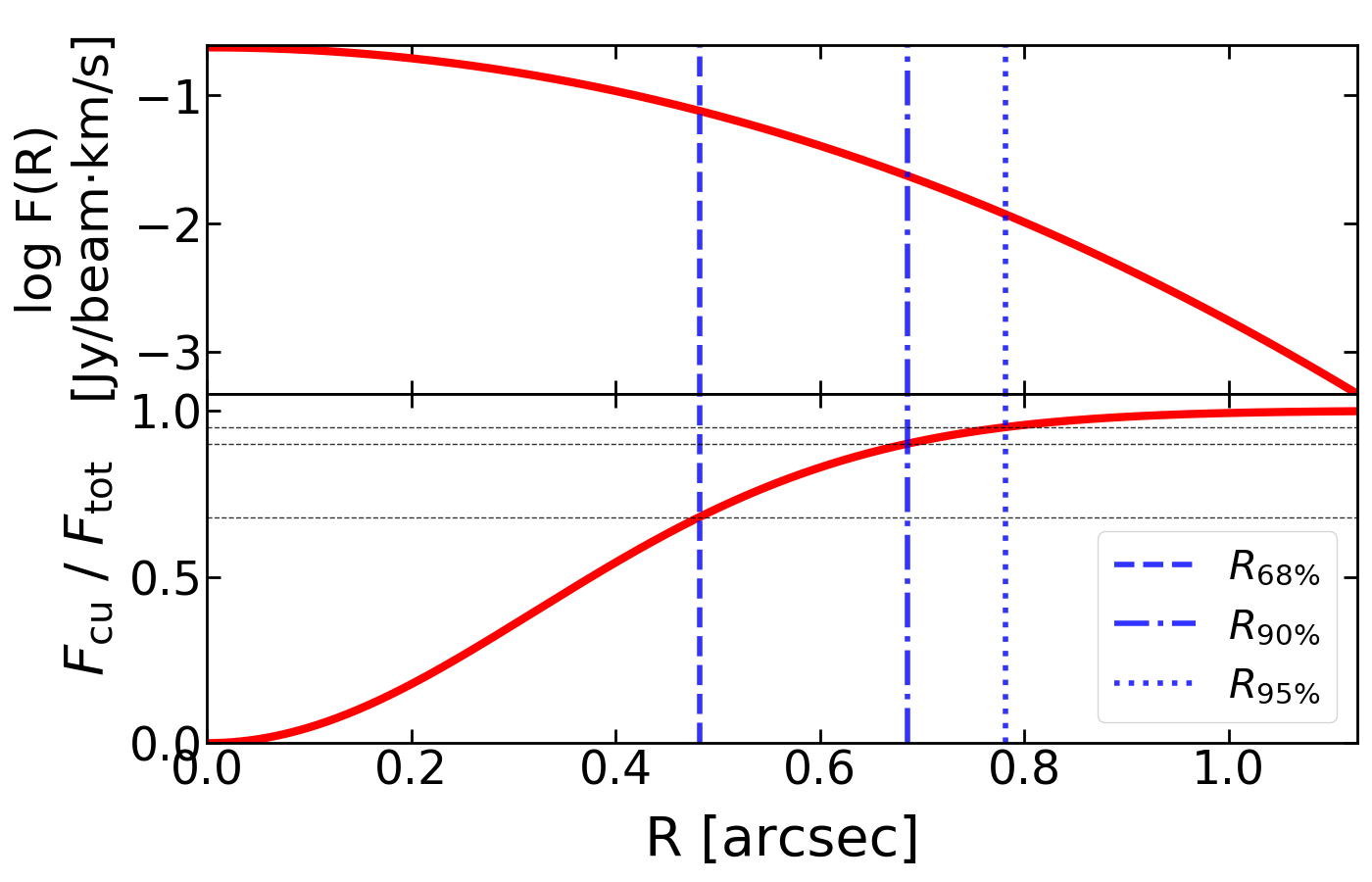}
    \end{center}
    \begin{center}
    \includegraphics[width=.490\textwidth]{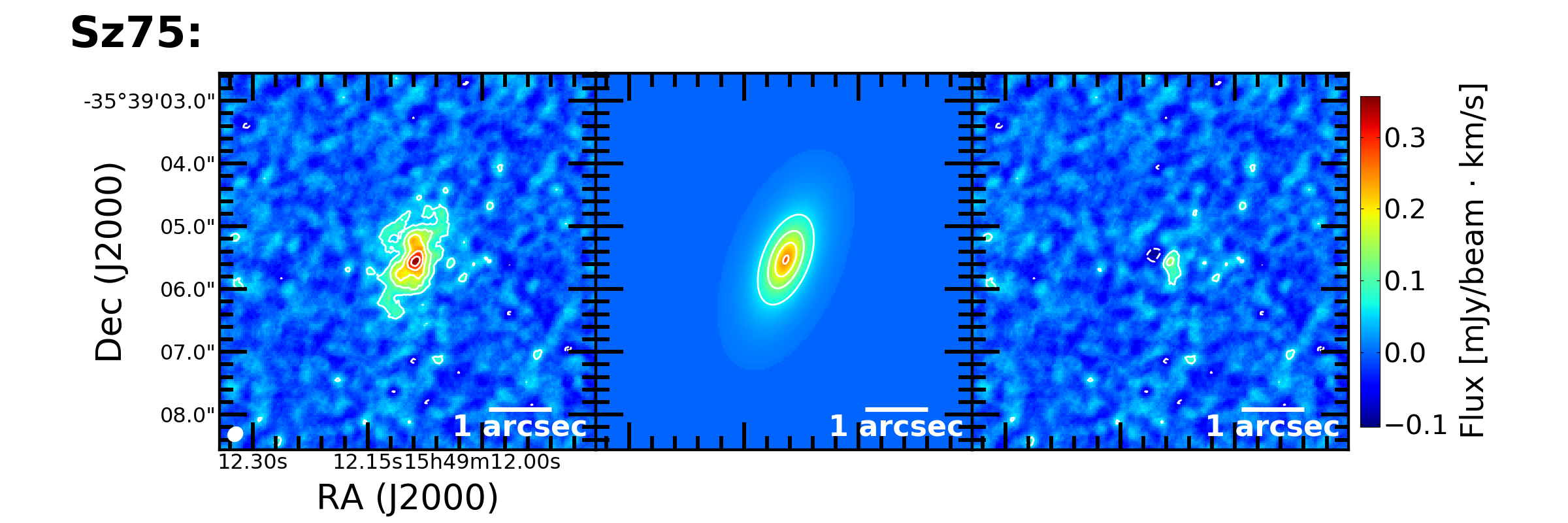}
    \includegraphics[width=.225\textwidth]{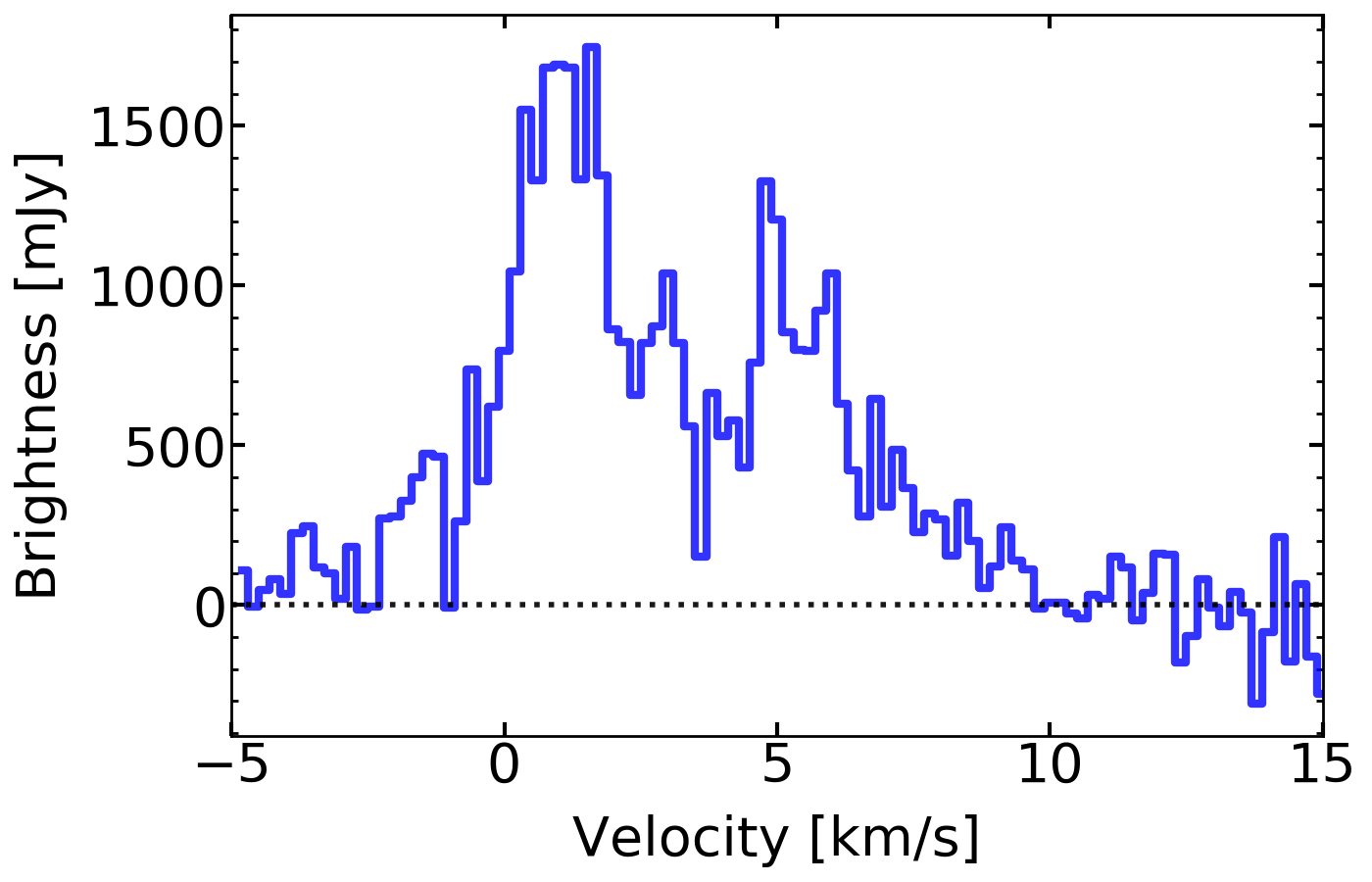}
    \includegraphics[width=.235\textwidth]{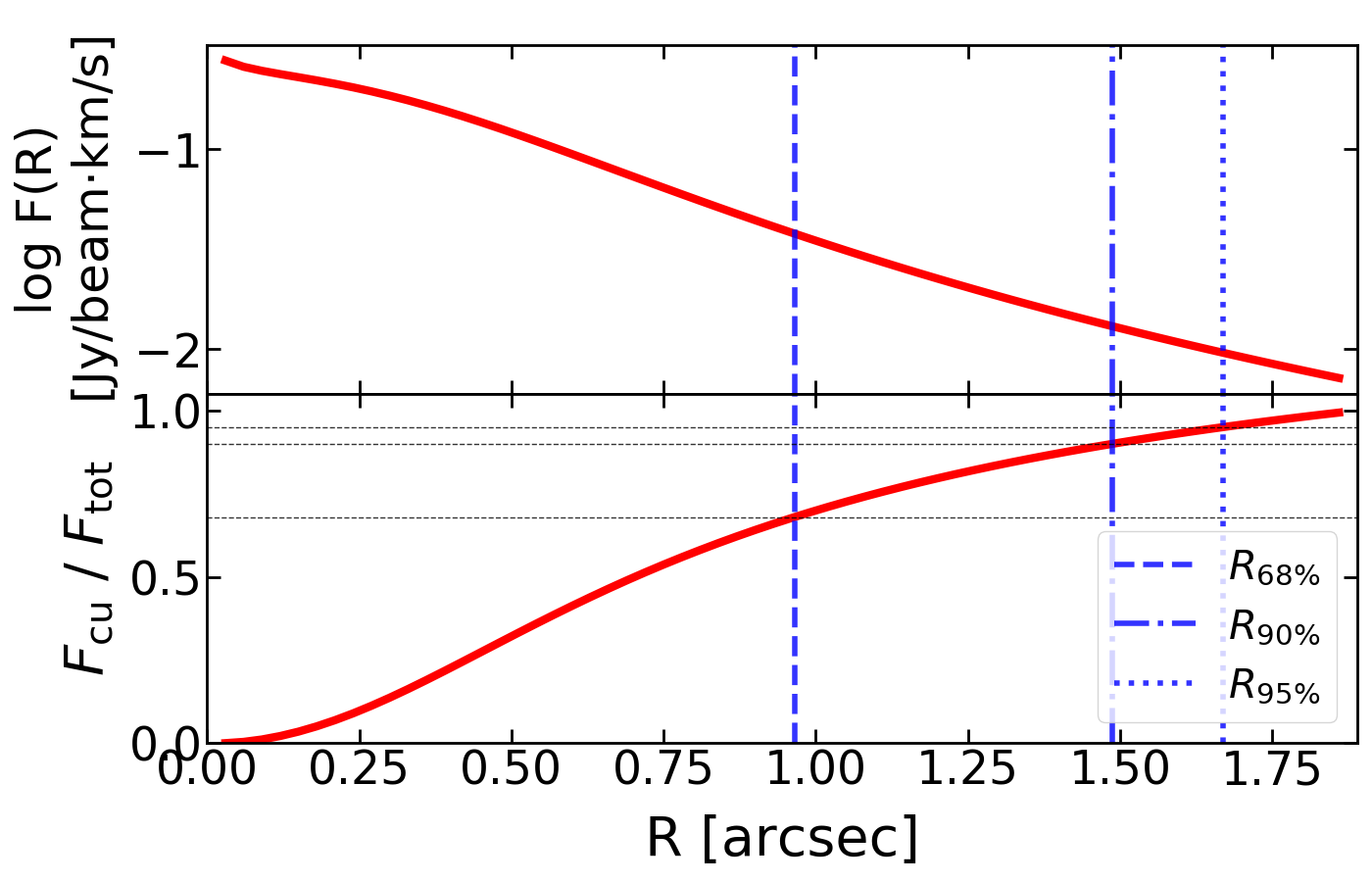}
    \end{center}
    \begin{center}
    \includegraphics[width=.490\textwidth]{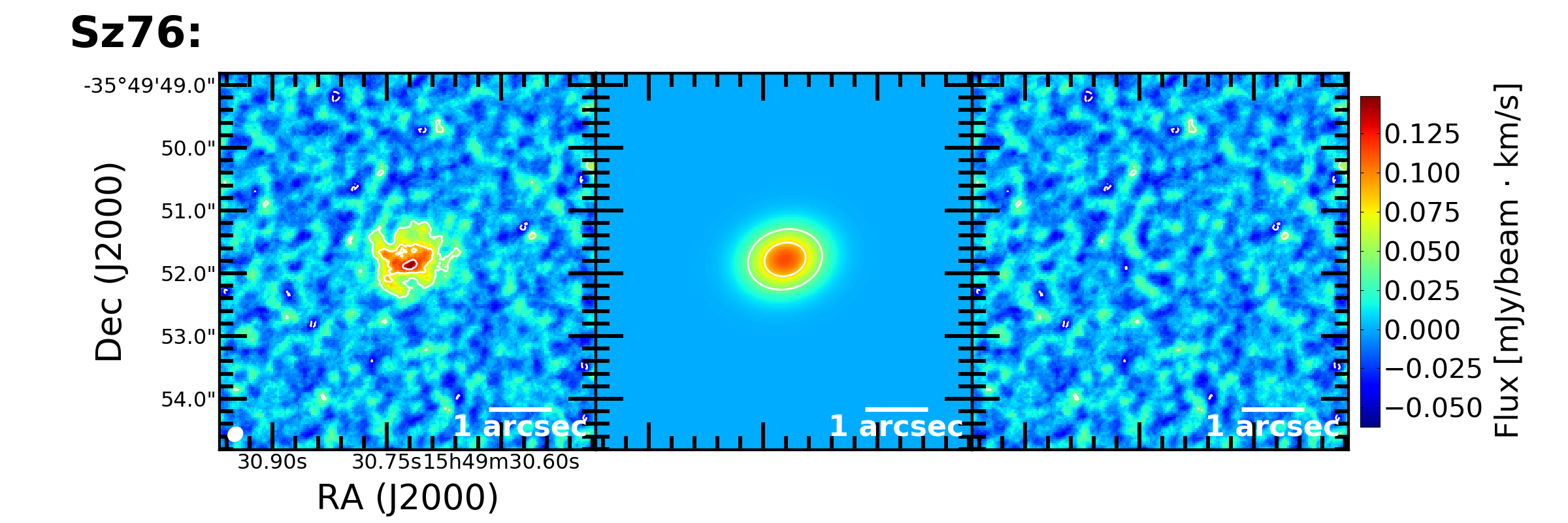}
    \includegraphics[width=.225\textwidth]{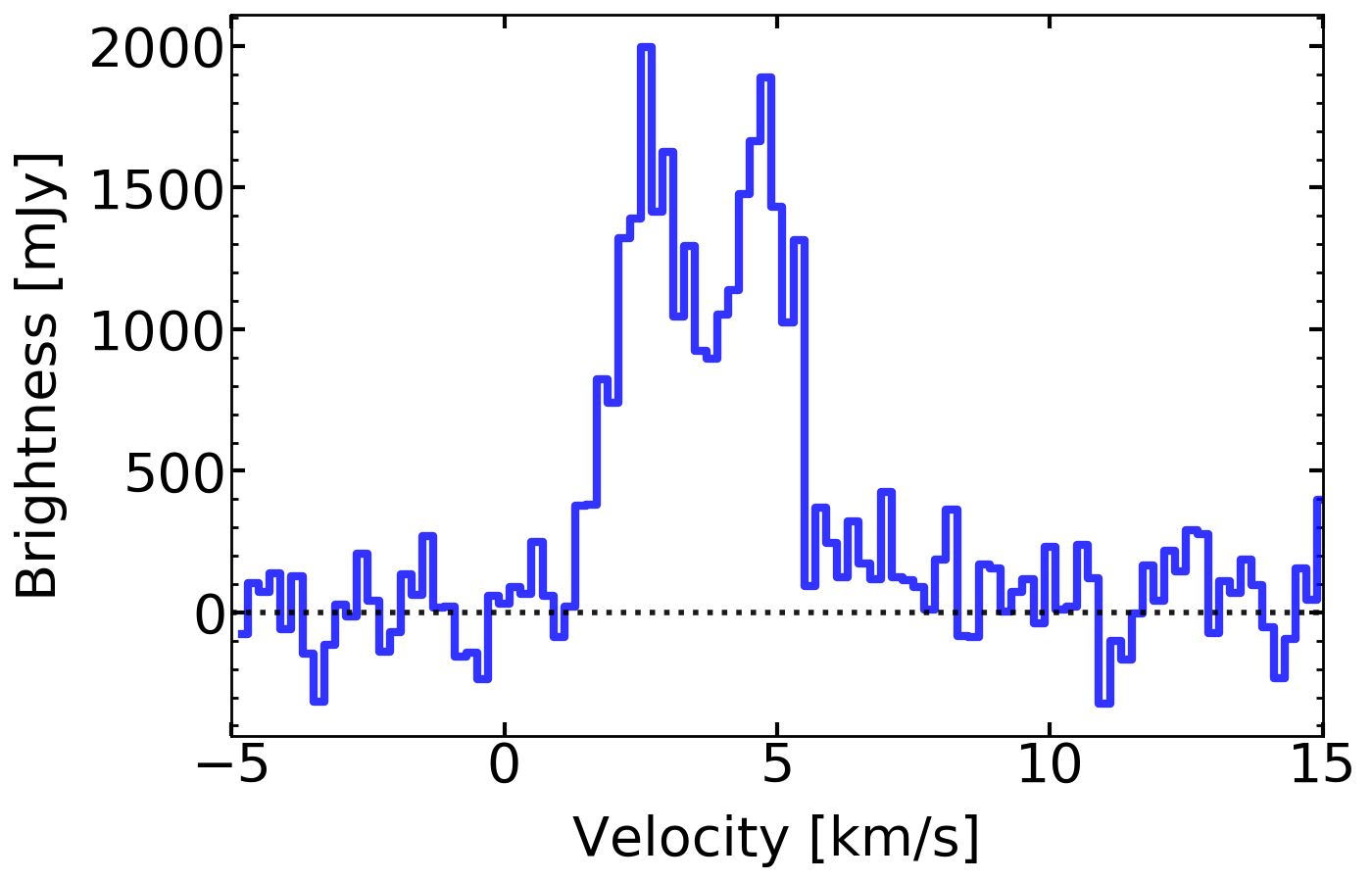}
    \includegraphics[width=.235\textwidth]{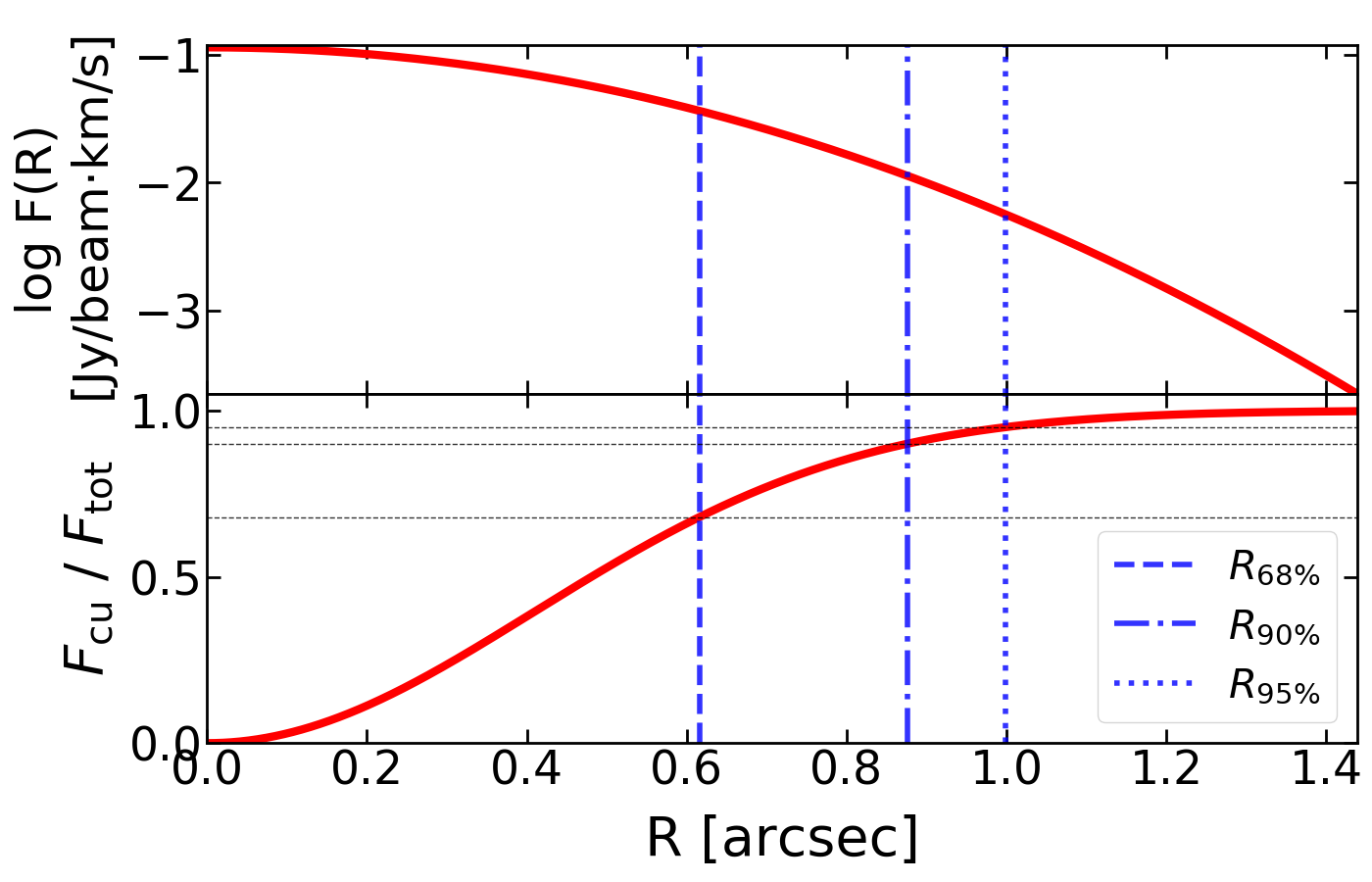}
    \end{center}
    \begin{center}
    \includegraphics[width=.490\textwidth]{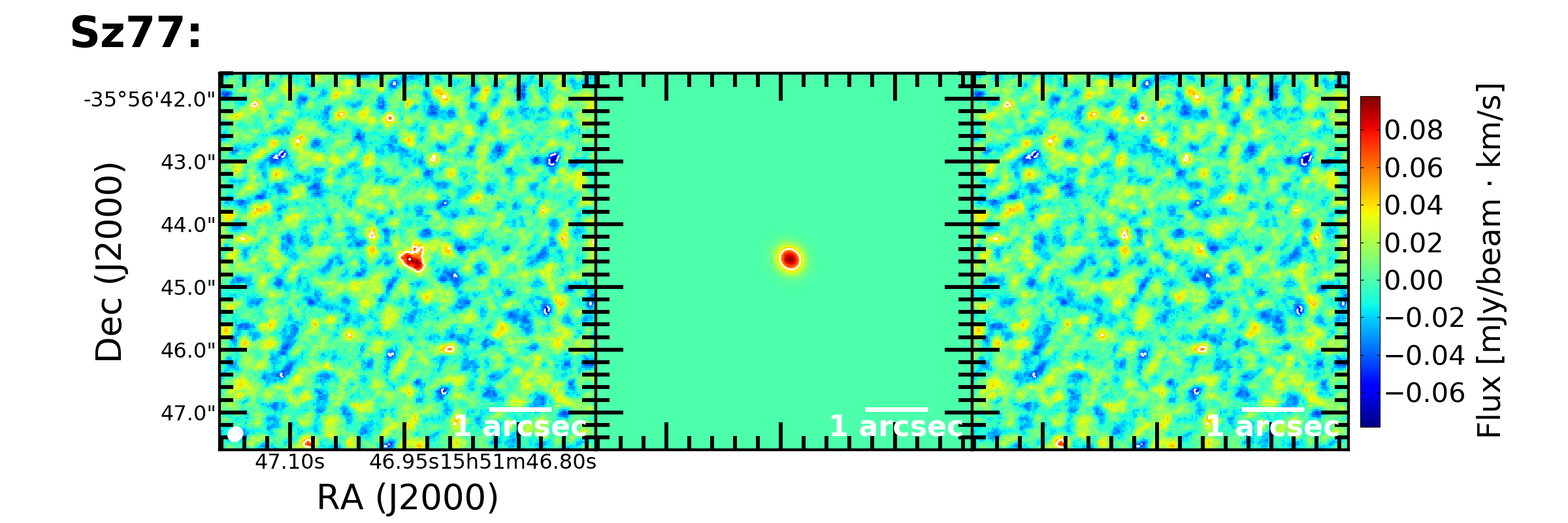}
    \includegraphics[width=.225\textwidth]{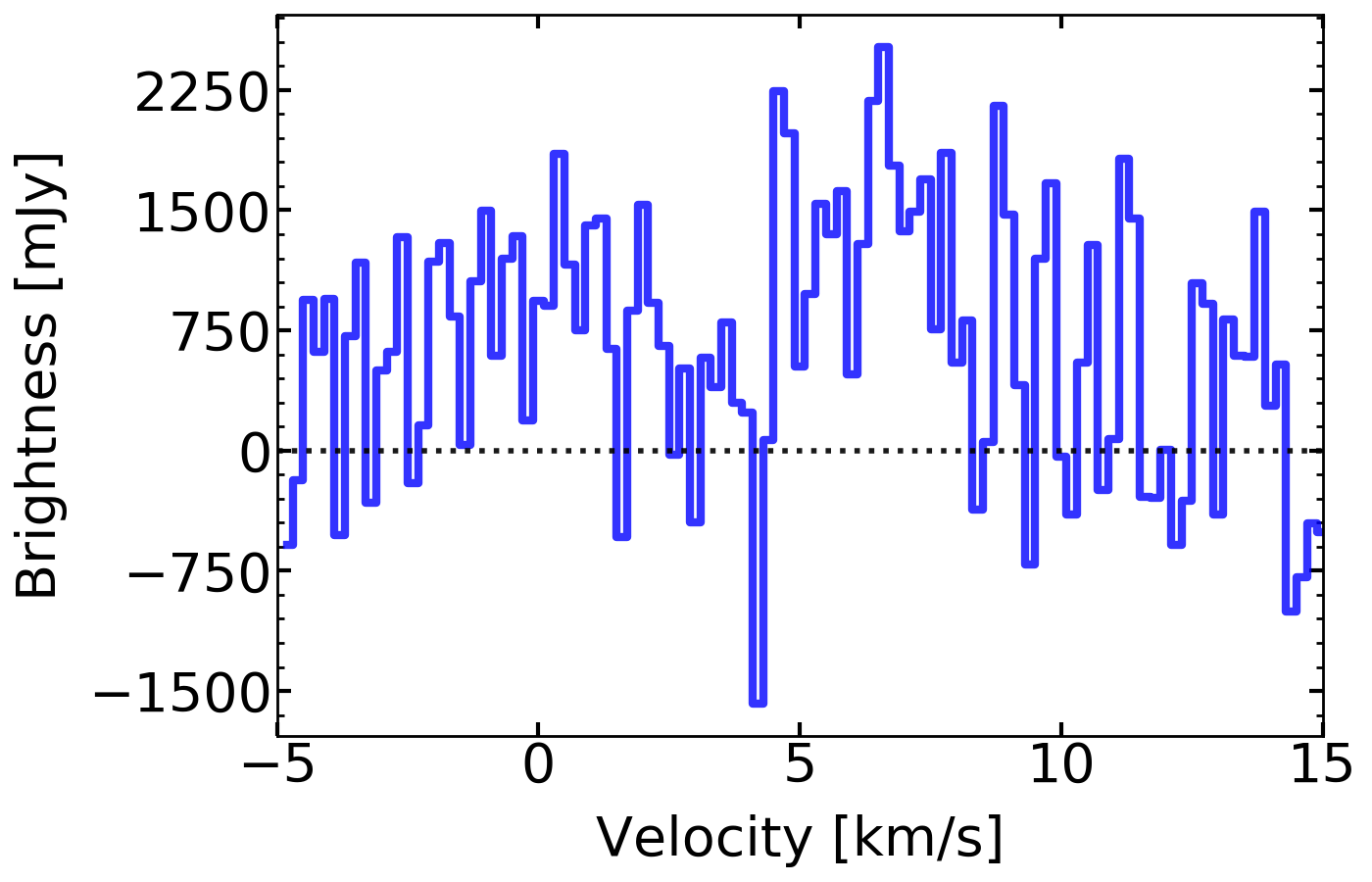}
    \includegraphics[width=.235\textwidth]{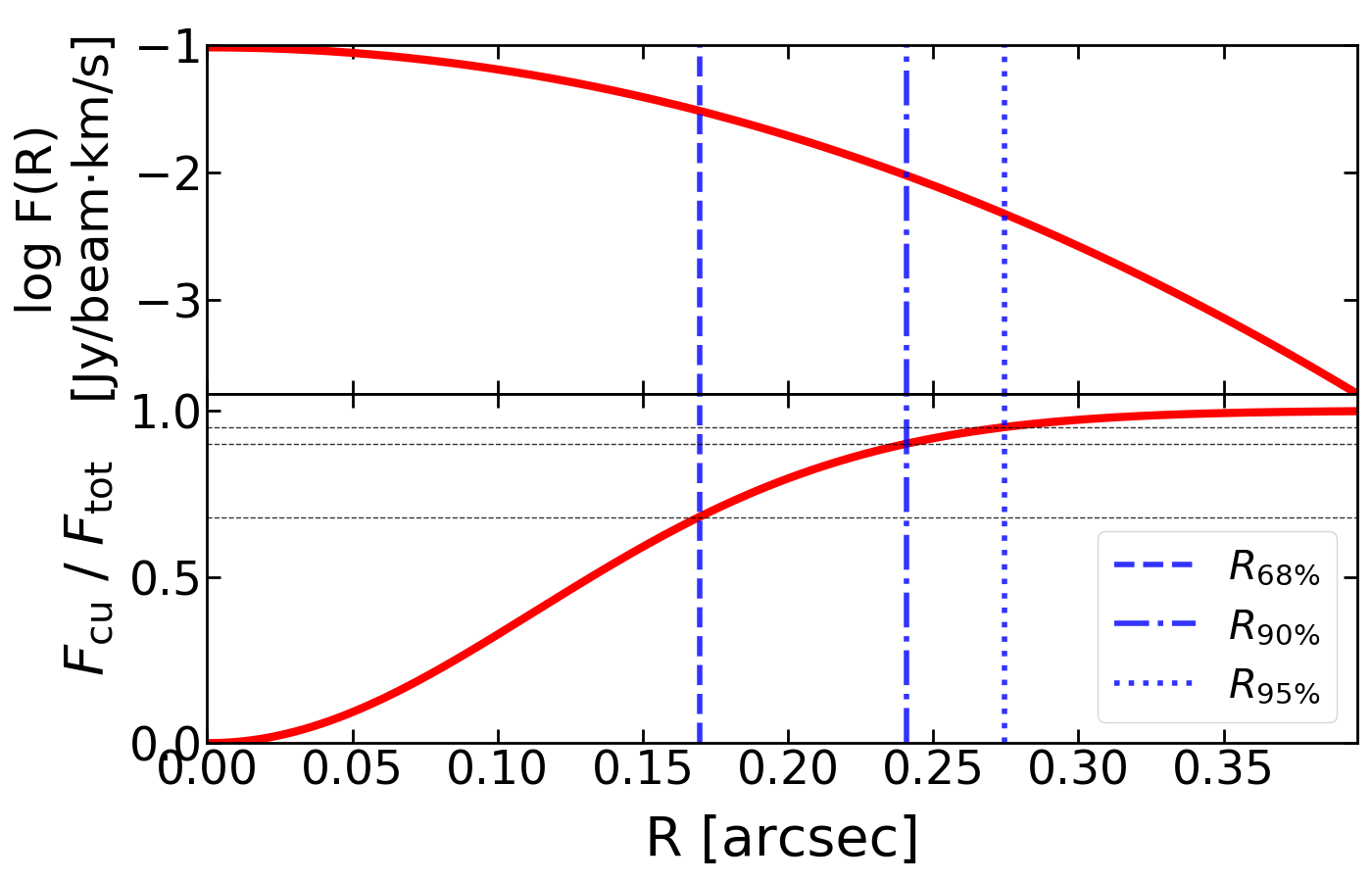}
    \end{center}
  \caption[]{
  Results of the CO modeling for every disk with measured CO size, following the methodology described in Section~\ref{sec:gasmodeling}. For each disk, the first three sub-panels show the observed, model and residual CO moment zero maps; solid (dashed) line contours are drawn at increasing (decreasing) $3\sigma$ intervals. The forth sub-panel represents the integrated spectrum enclosed by the $R_{68\%}^{\mathrm{CO}}$ of the source. Last sub-panel shows the radial brightness profile and the respective cumulative distribution of the CO model.
  }
  \label{fig:comodelresults_all_3}
\end{figure*}

\begin{figure*}
    \begin{center}
    \includegraphics[width=.490\textwidth]{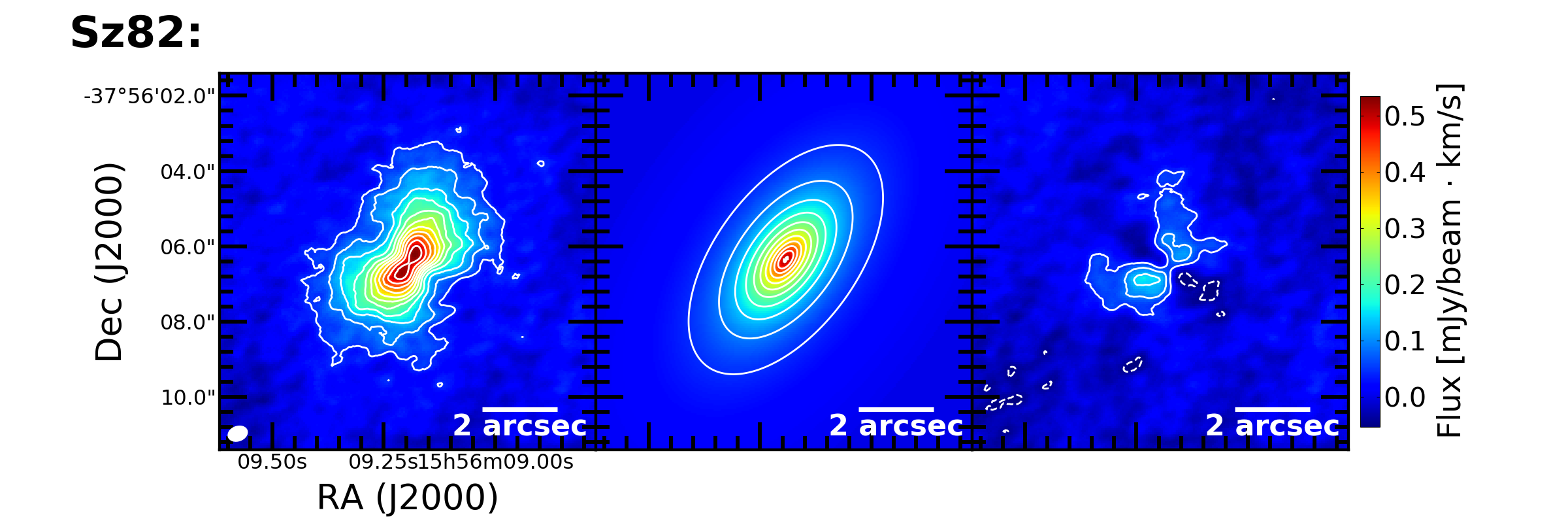}
    \includegraphics[width=.225\textwidth]{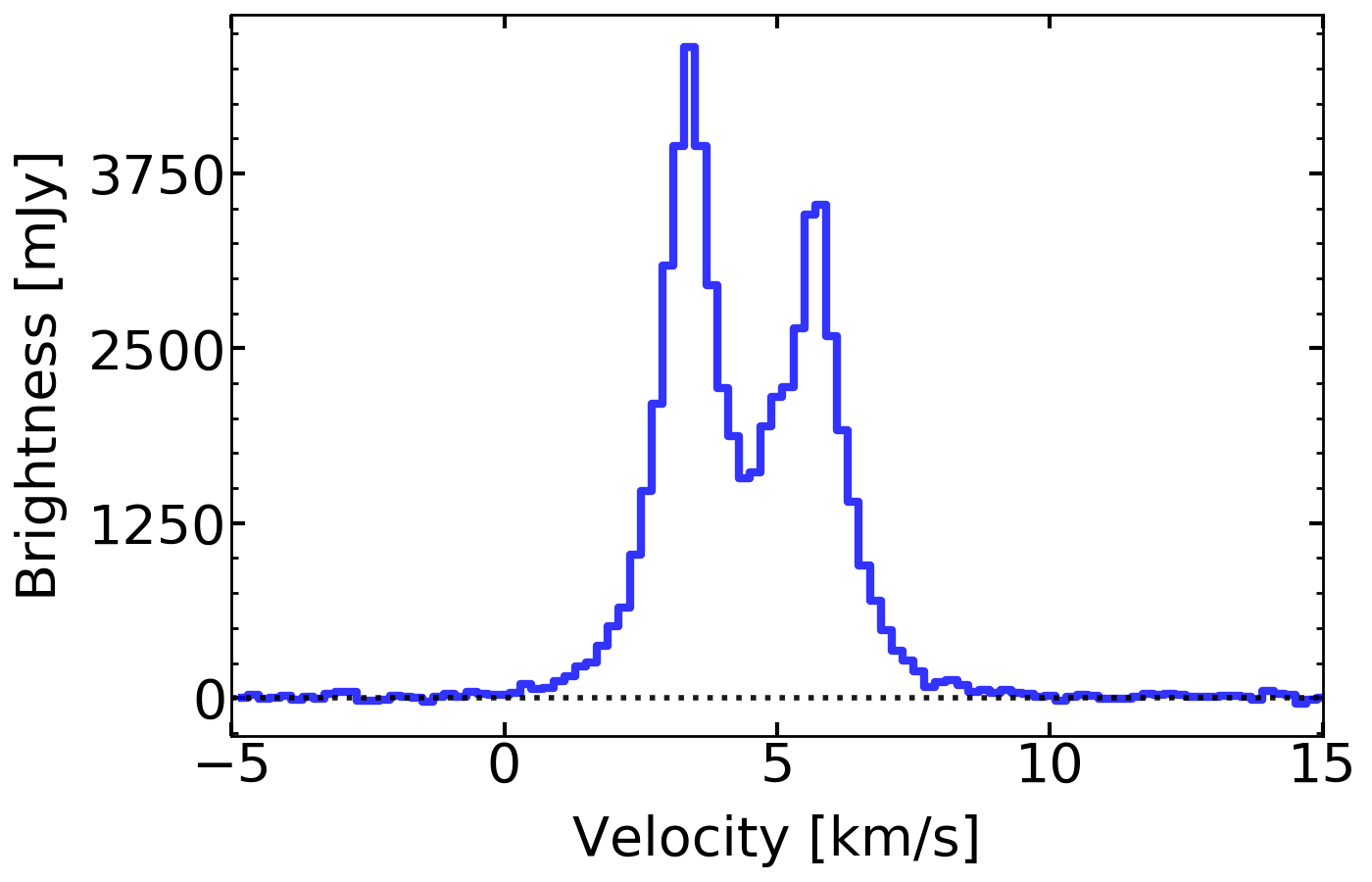}
    \includegraphics[width=.235\textwidth]{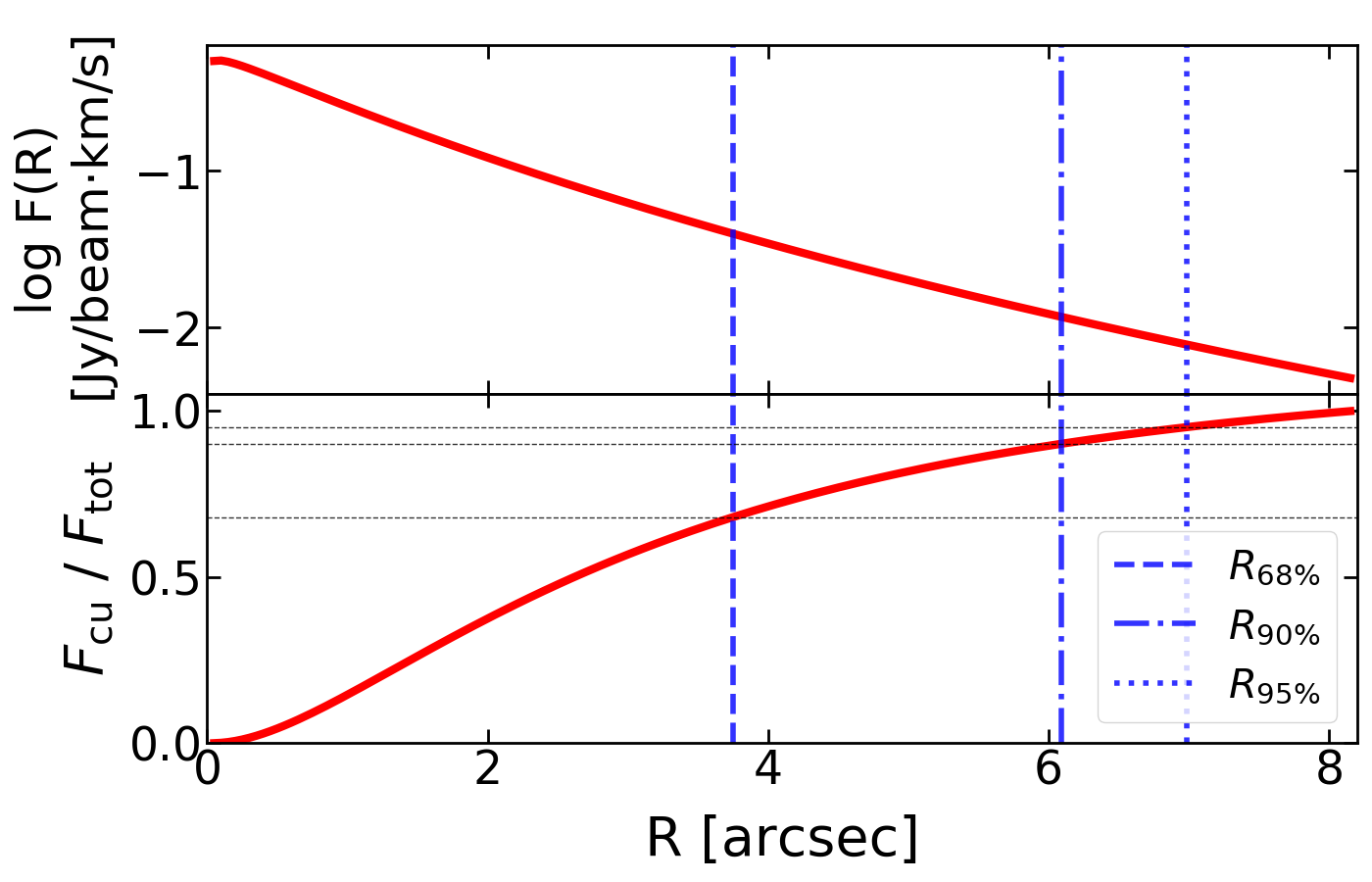}
    \end{center}
    \begin{center}
    \includegraphics[width=.490\textwidth]{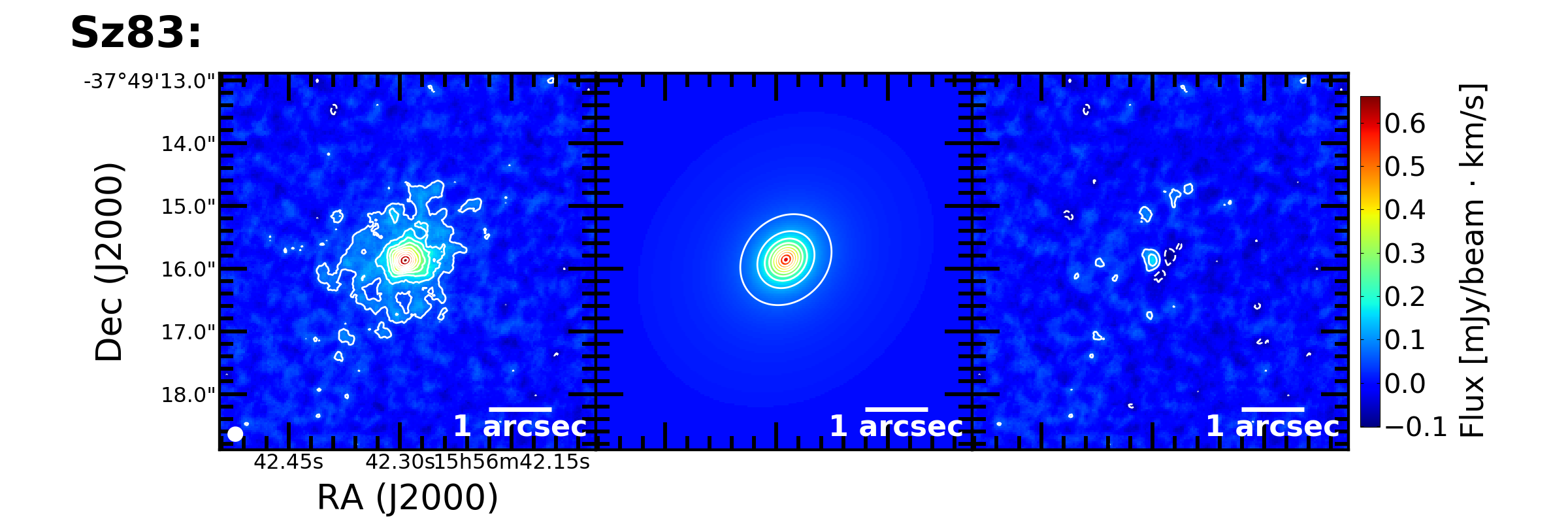}
    \includegraphics[width=.225\textwidth]{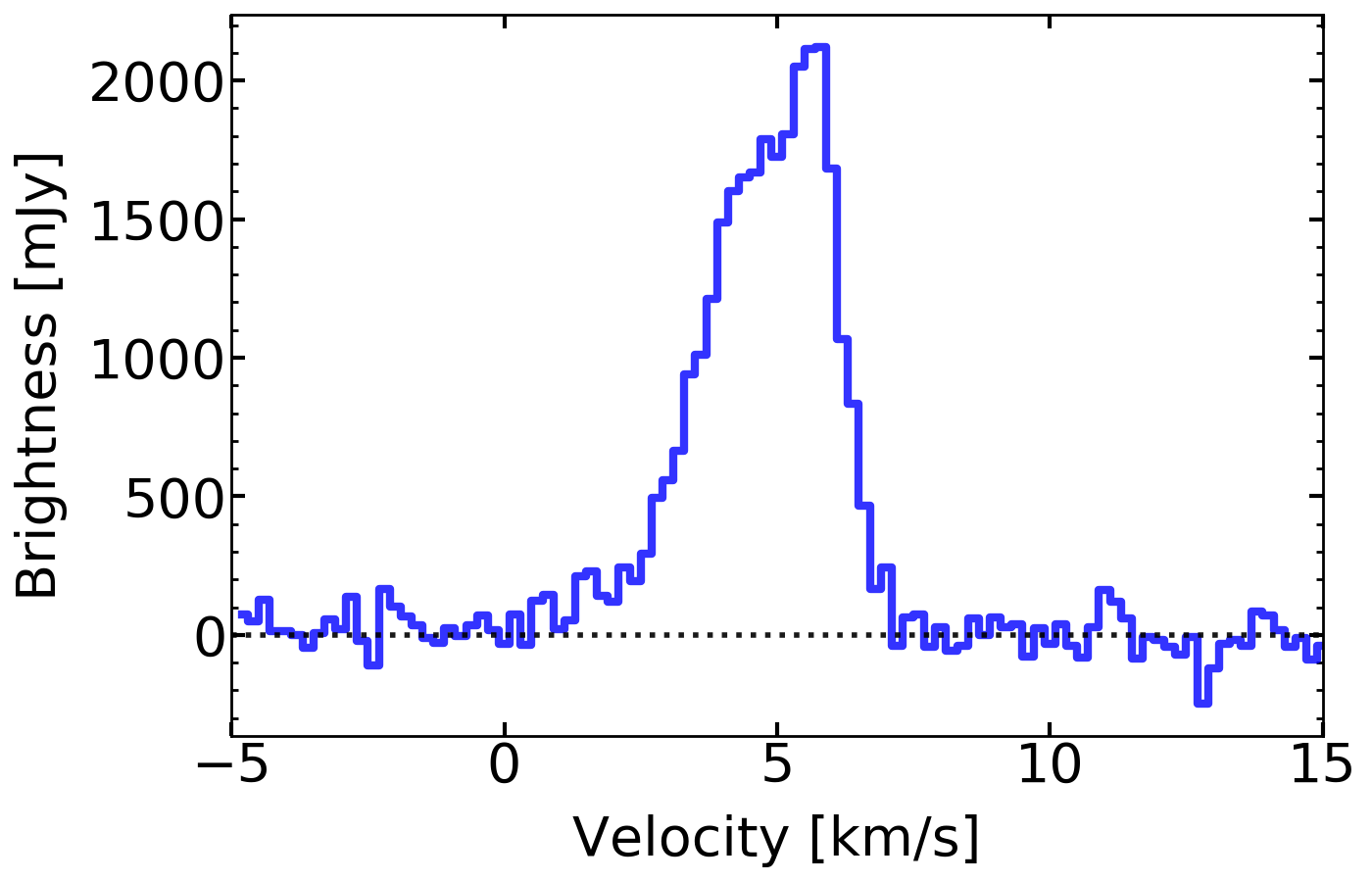}
    \includegraphics[width=.235\textwidth]{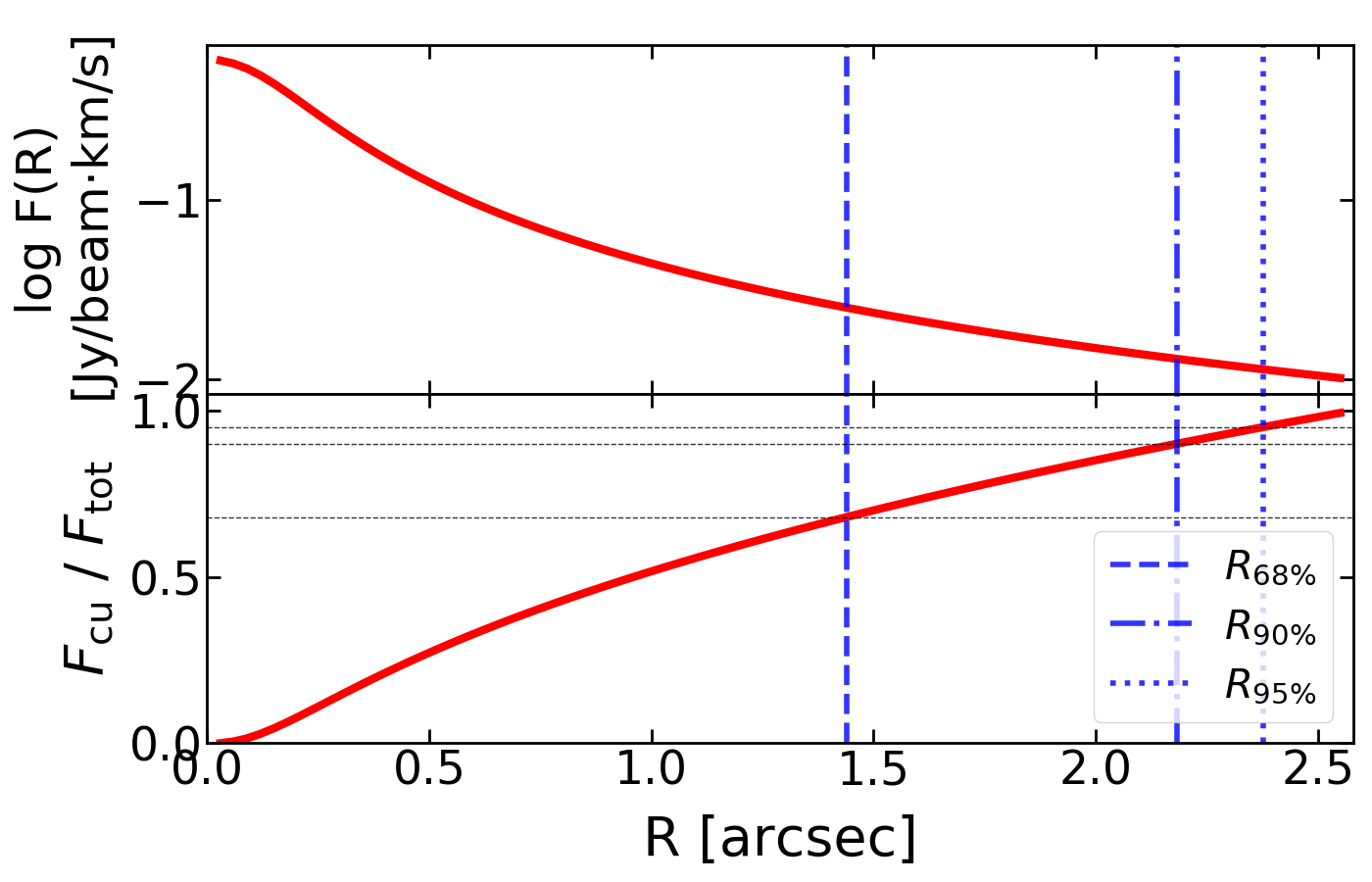}
    \end{center}
    \begin{center}
    \includegraphics[width=.490\textwidth]{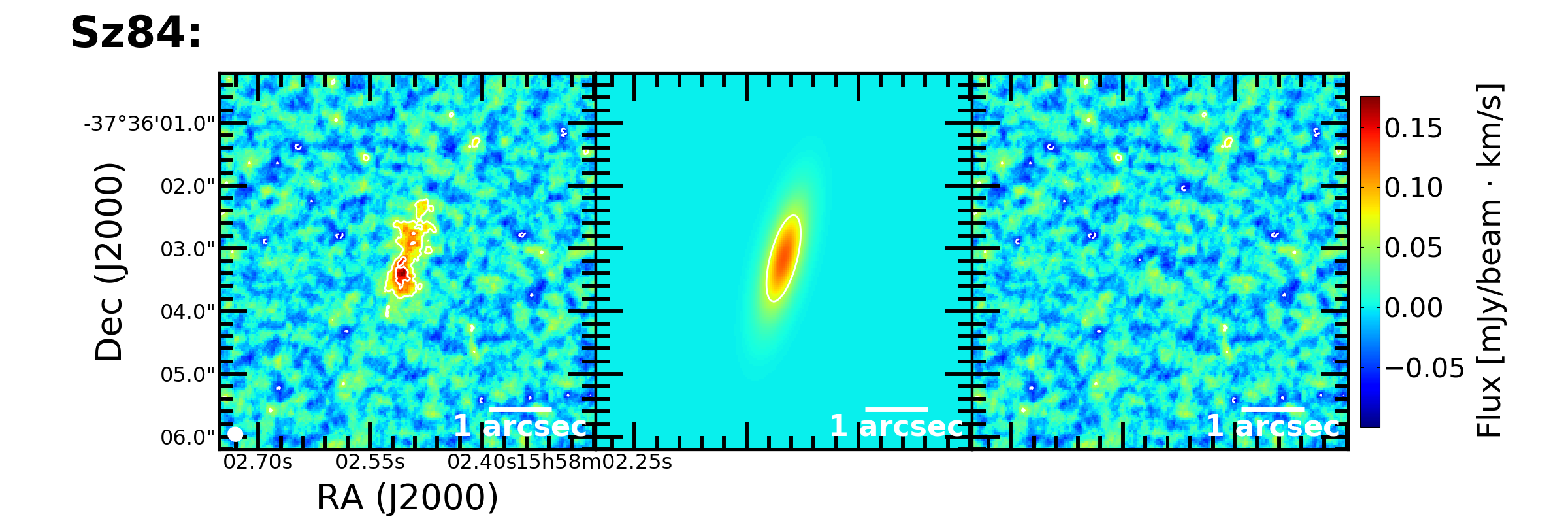}
    \includegraphics[width=.225\textwidth]{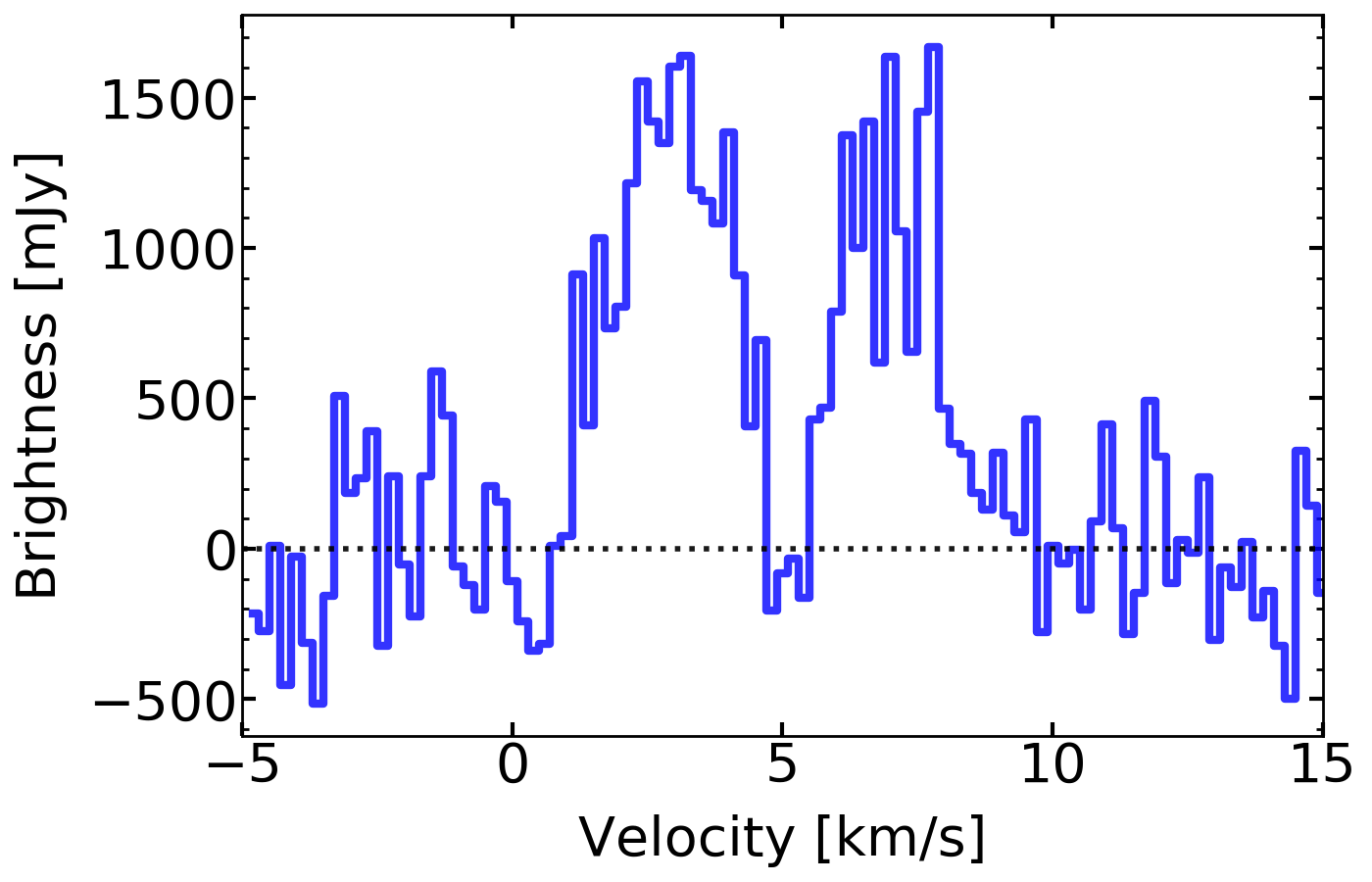}
    \includegraphics[width=.235\textwidth]{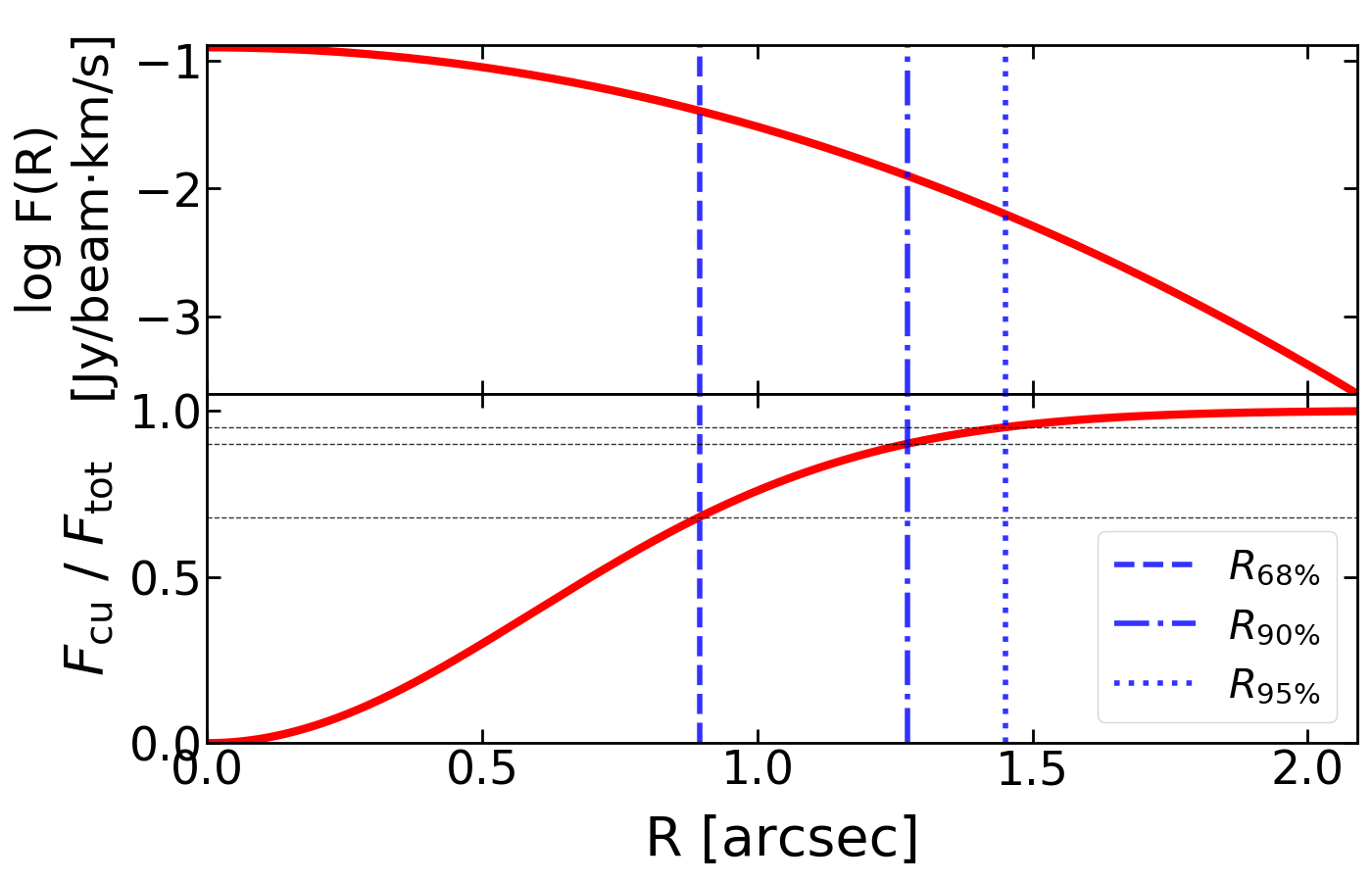}
    \end{center}
    \begin{center}
    \includegraphics[width=.490\textwidth]{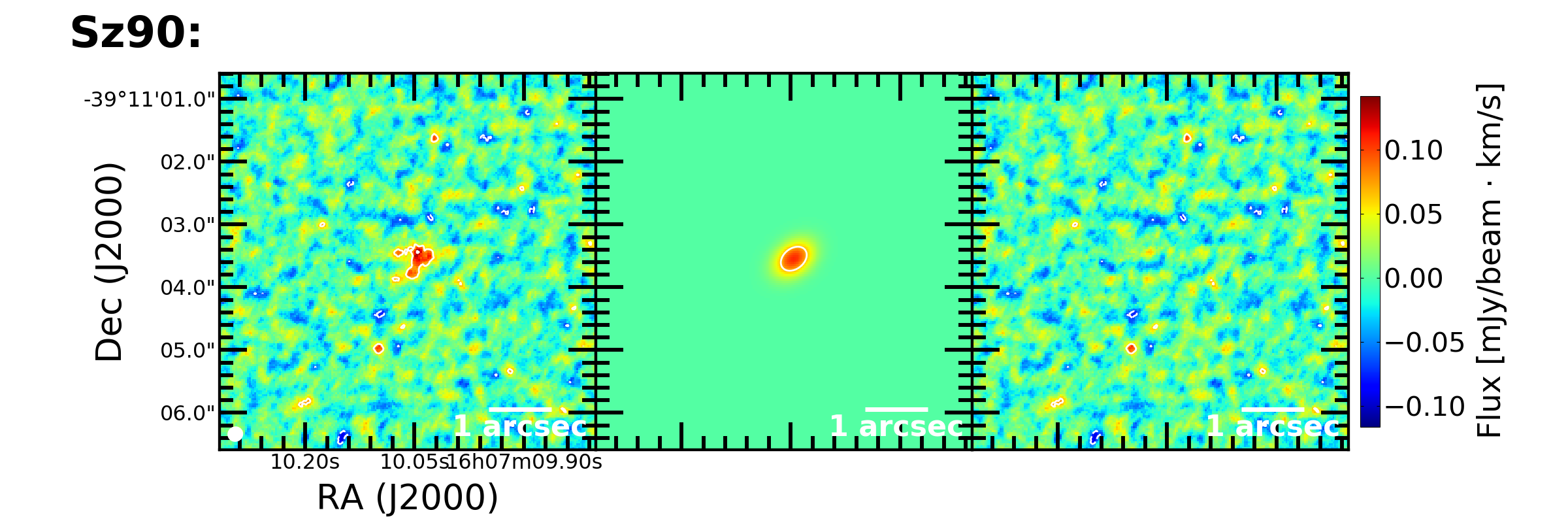}
    \includegraphics[width=.225\textwidth]{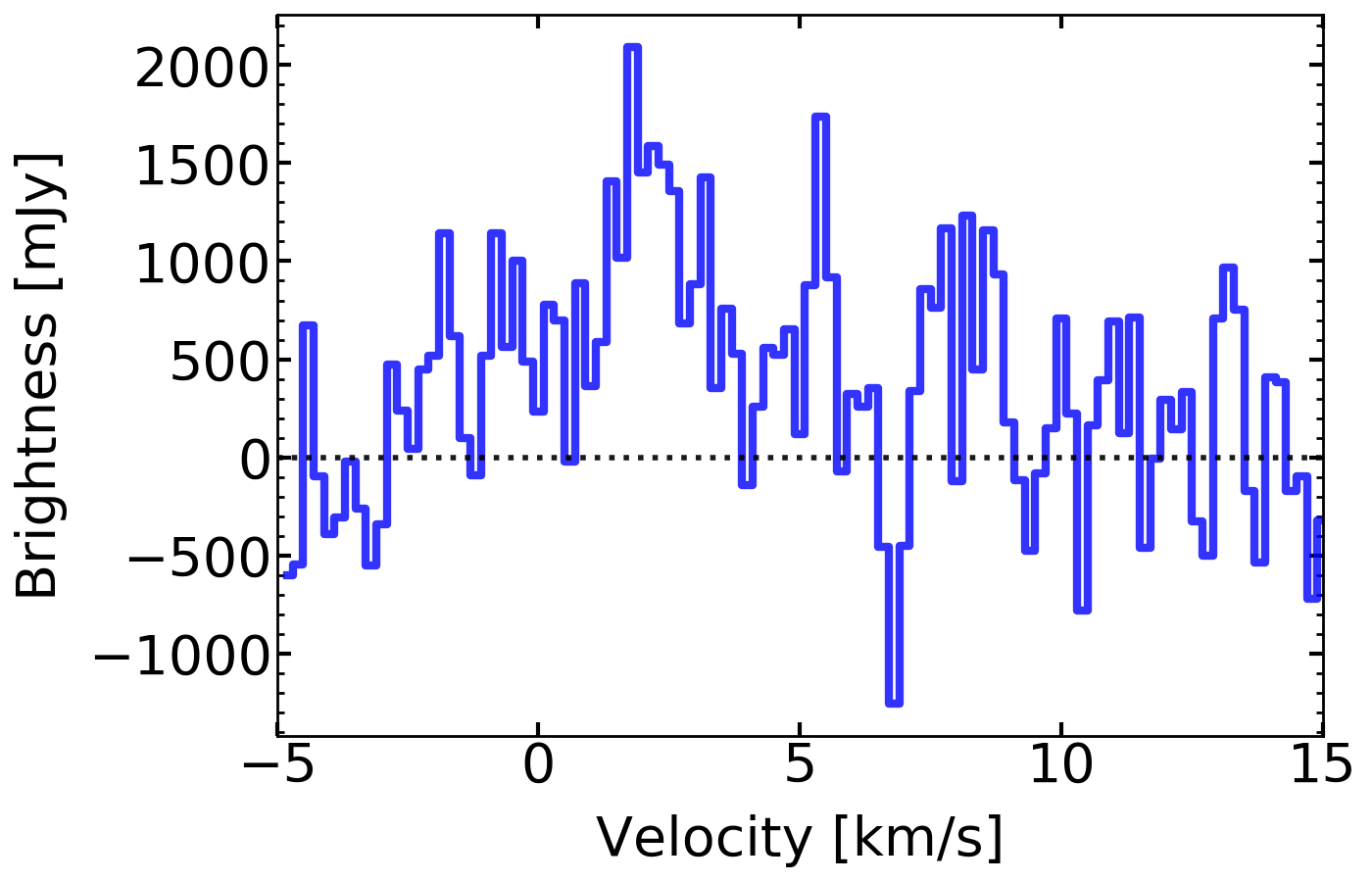}
    \includegraphics[width=.235\textwidth]{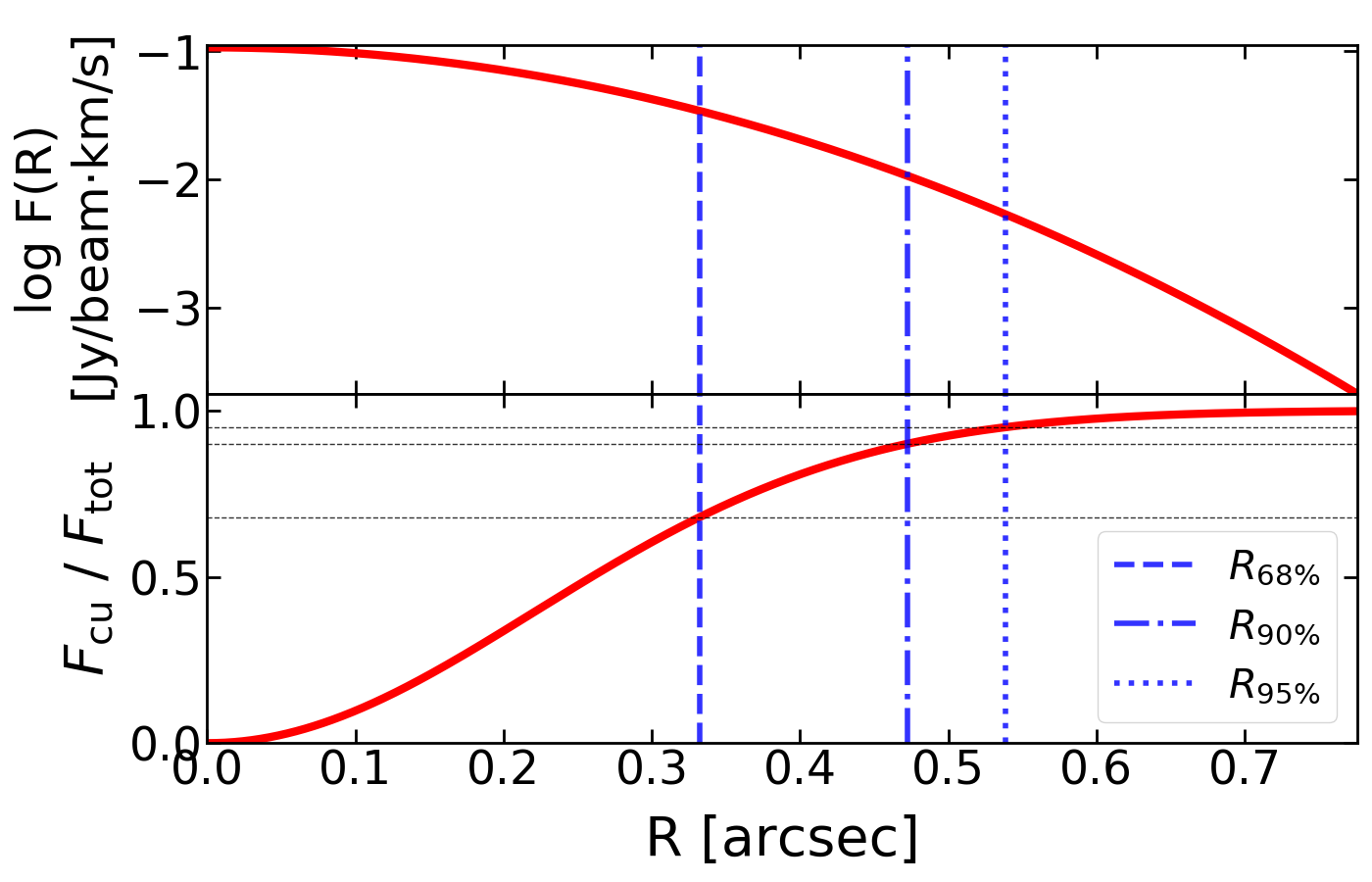}
    \end{center}
    \begin{center}
    \includegraphics[width=.490\textwidth]{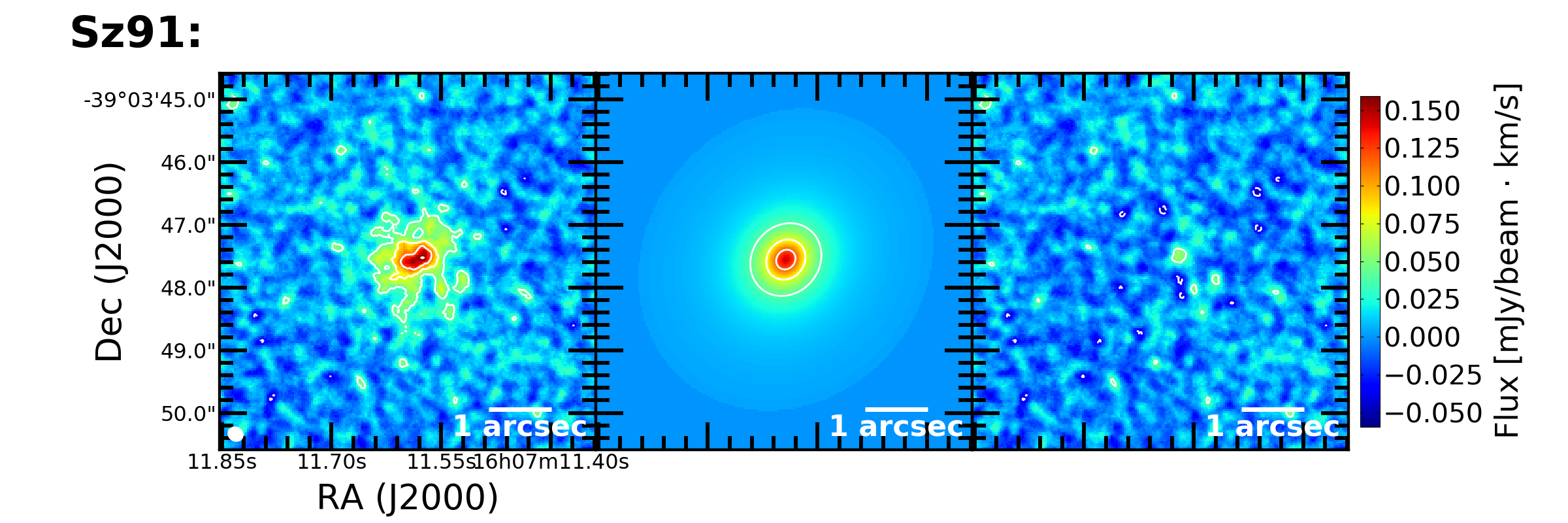}
    \includegraphics[width=.225\textwidth]{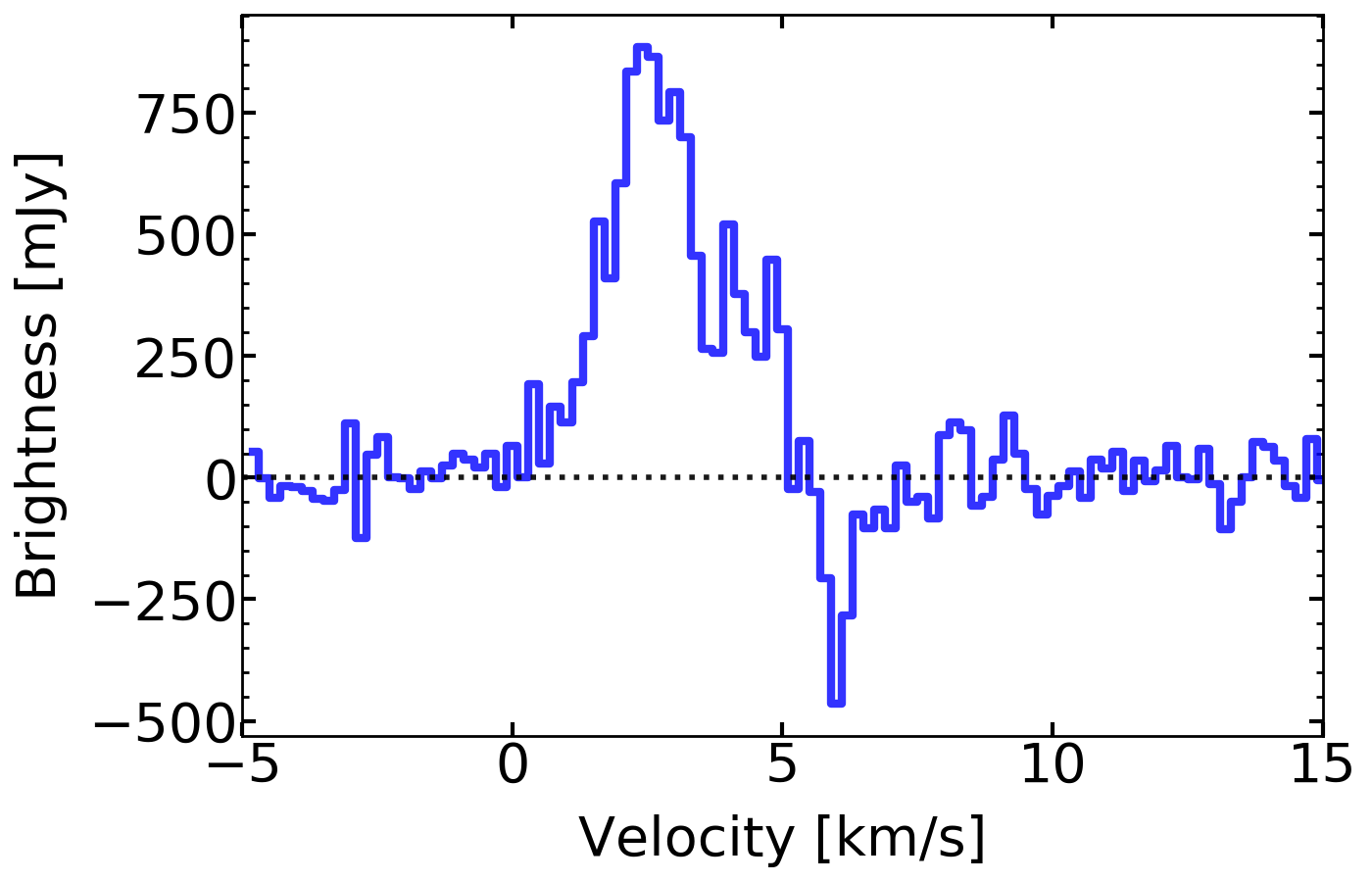}
    \includegraphics[width=.235\textwidth]{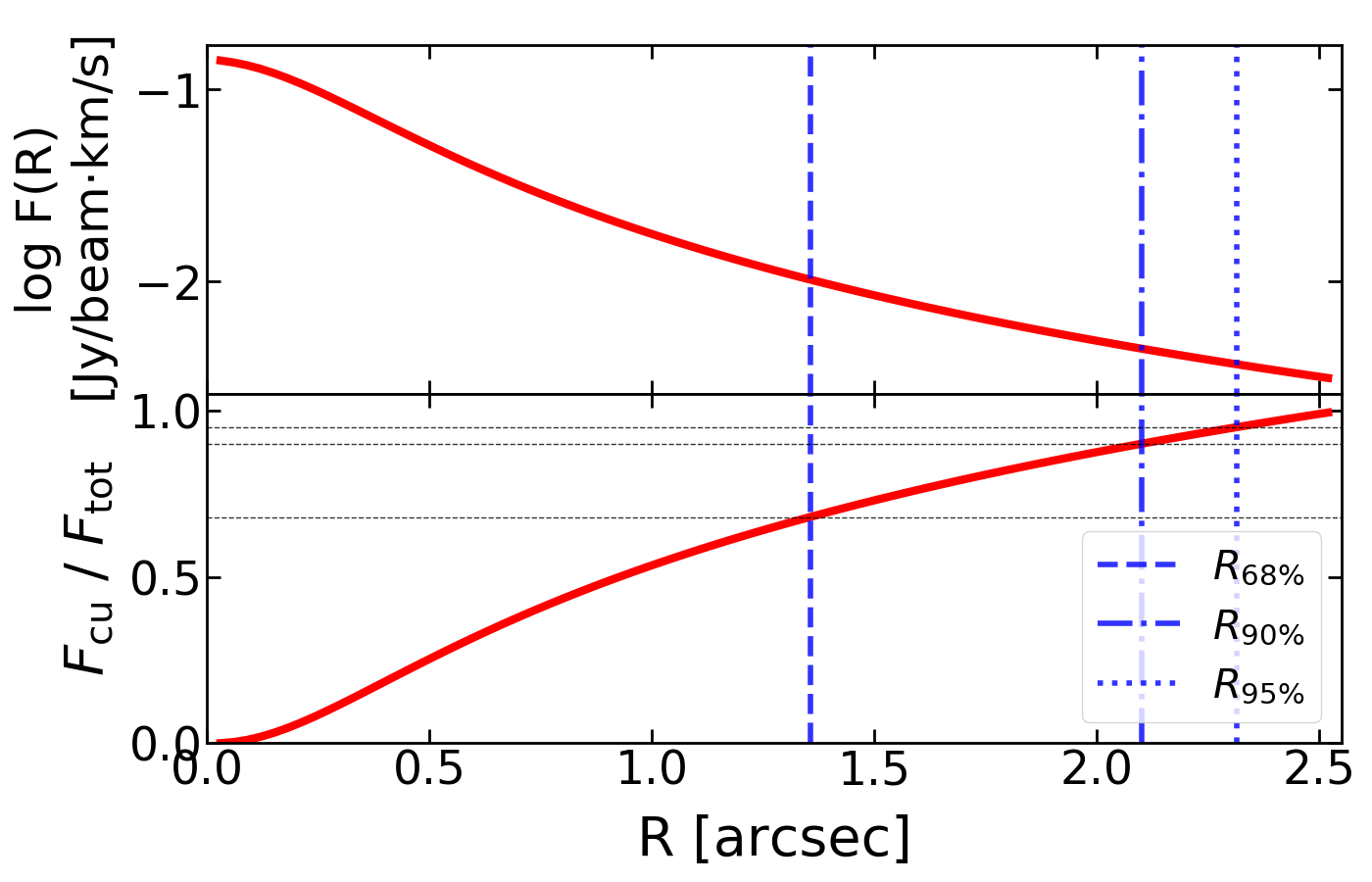}
    \end{center}
    \begin{center}
    \includegraphics[width=.490\textwidth]{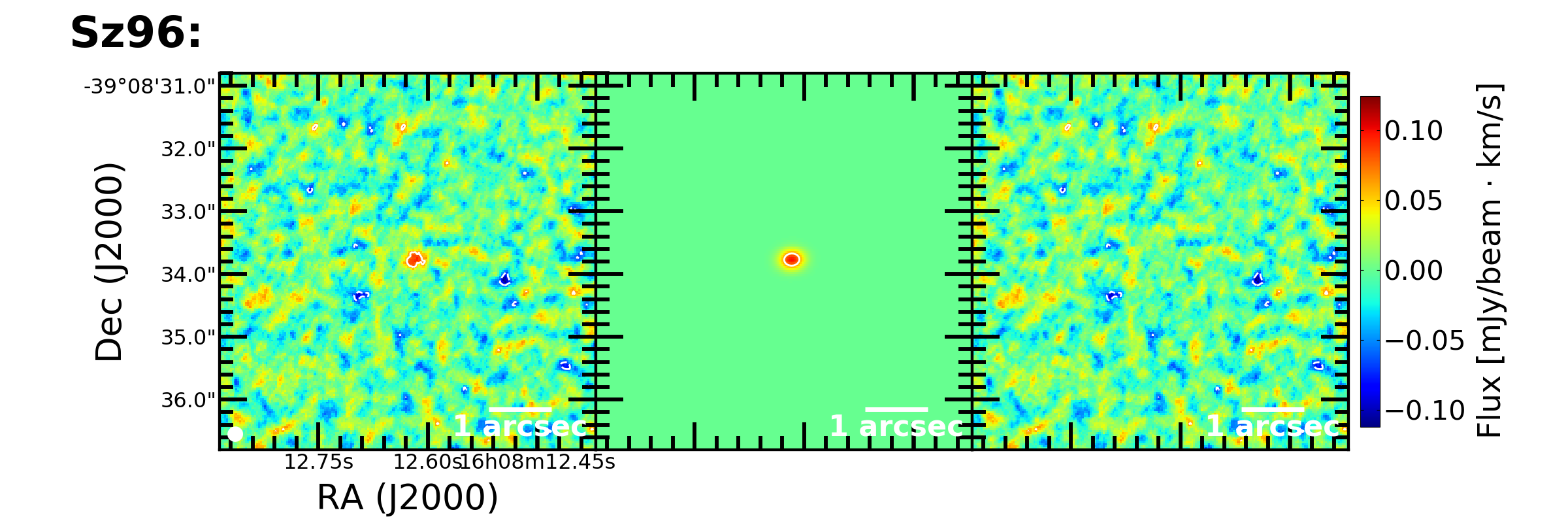}
    \includegraphics[width=.225\textwidth]{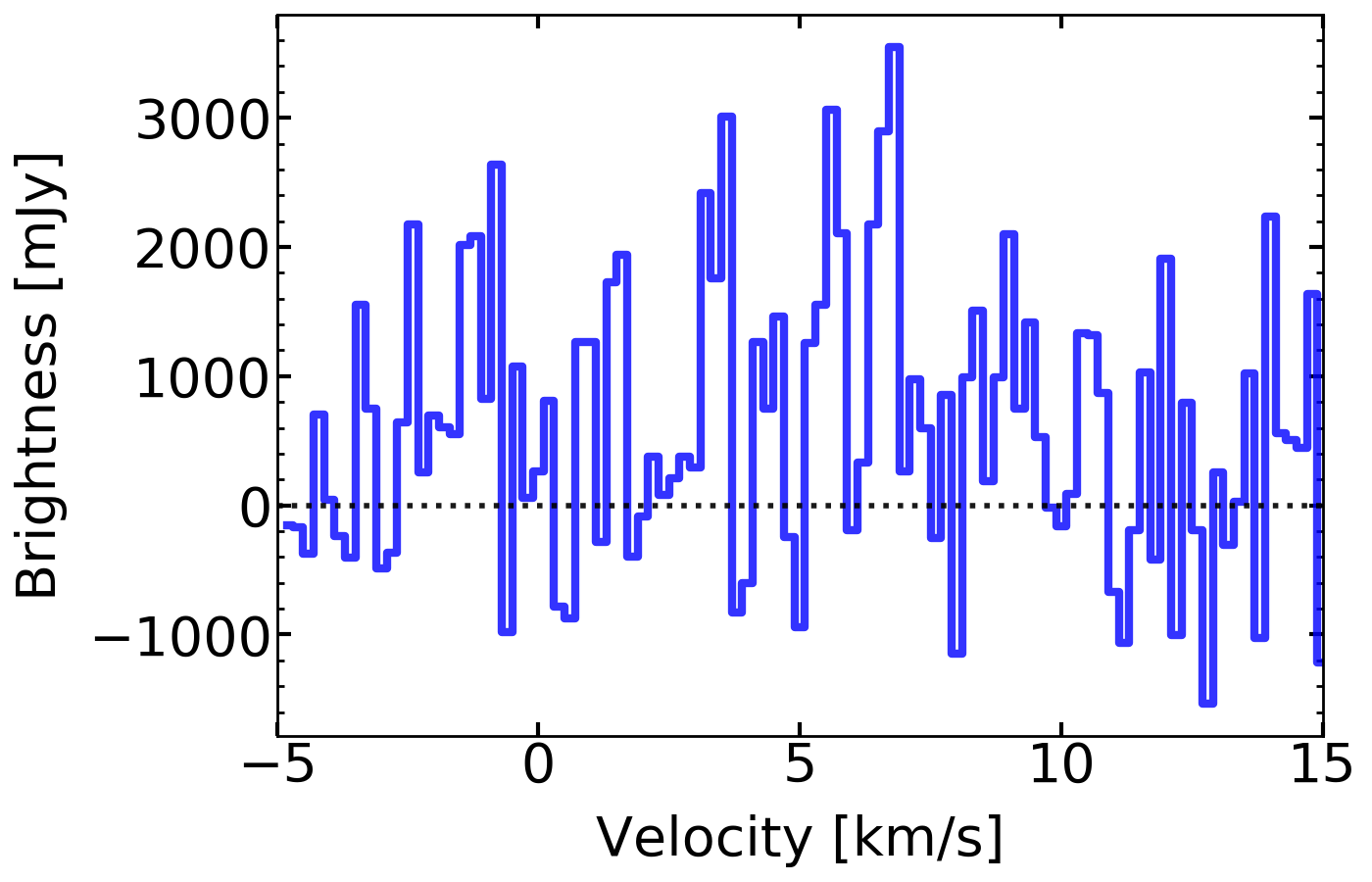}
    \includegraphics[width=.235\textwidth]{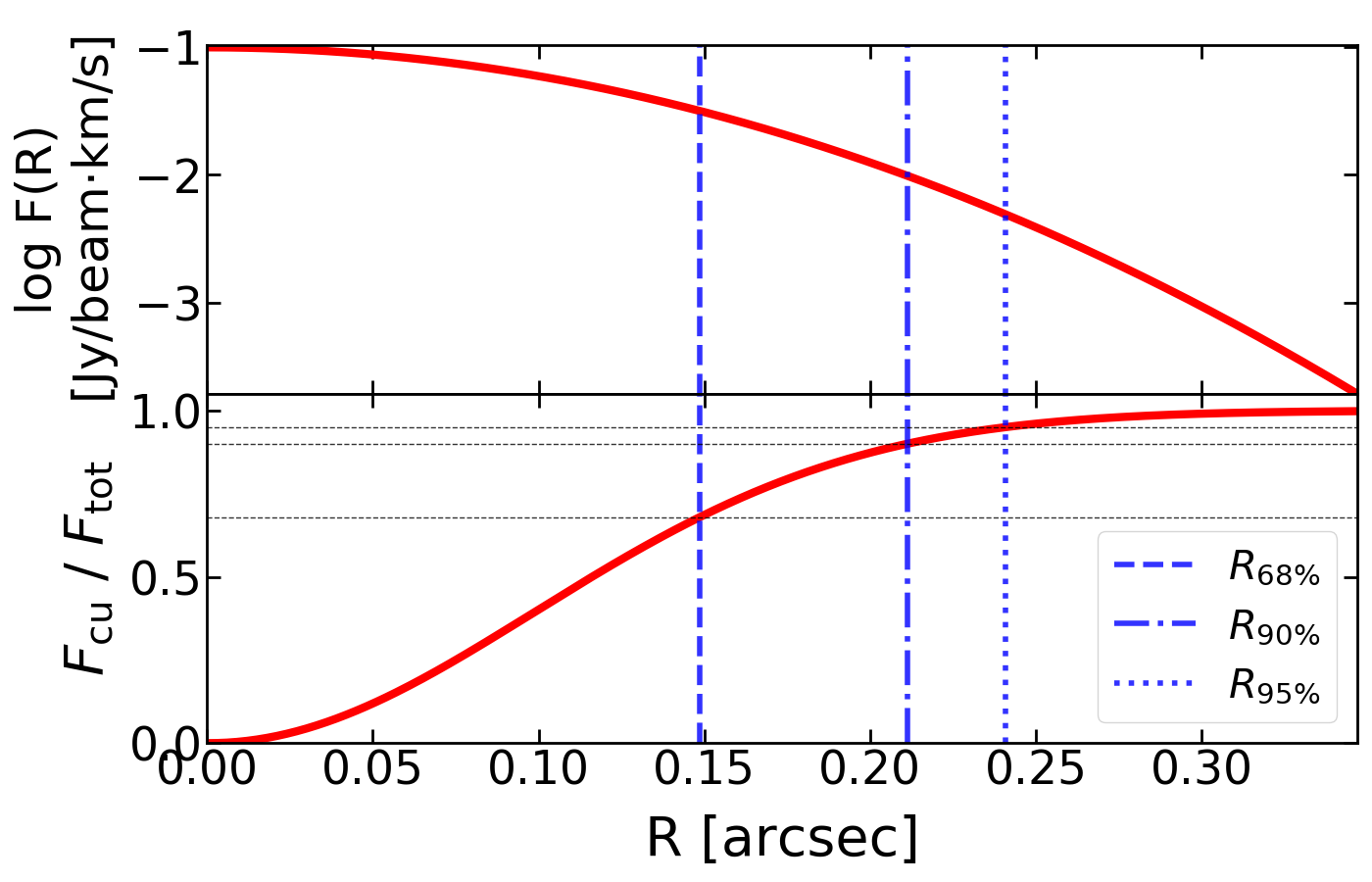}
    \end{center}
    \begin{center}
    \includegraphics[width=.490\textwidth]{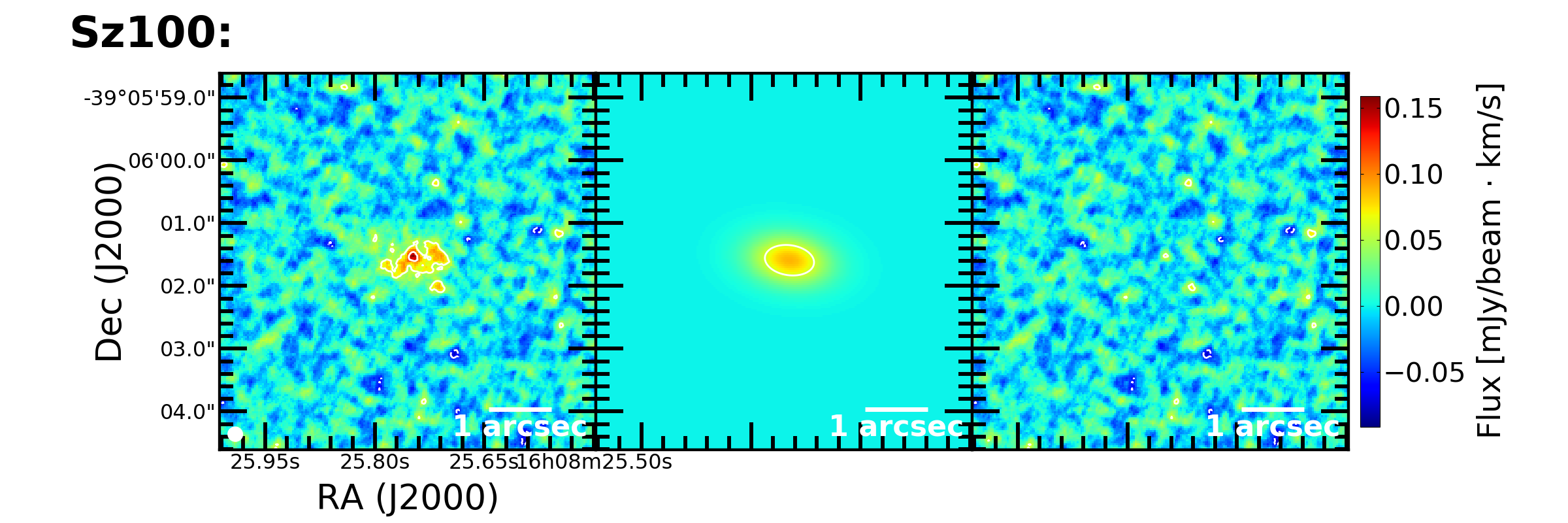}
    \includegraphics[width=.225\textwidth]{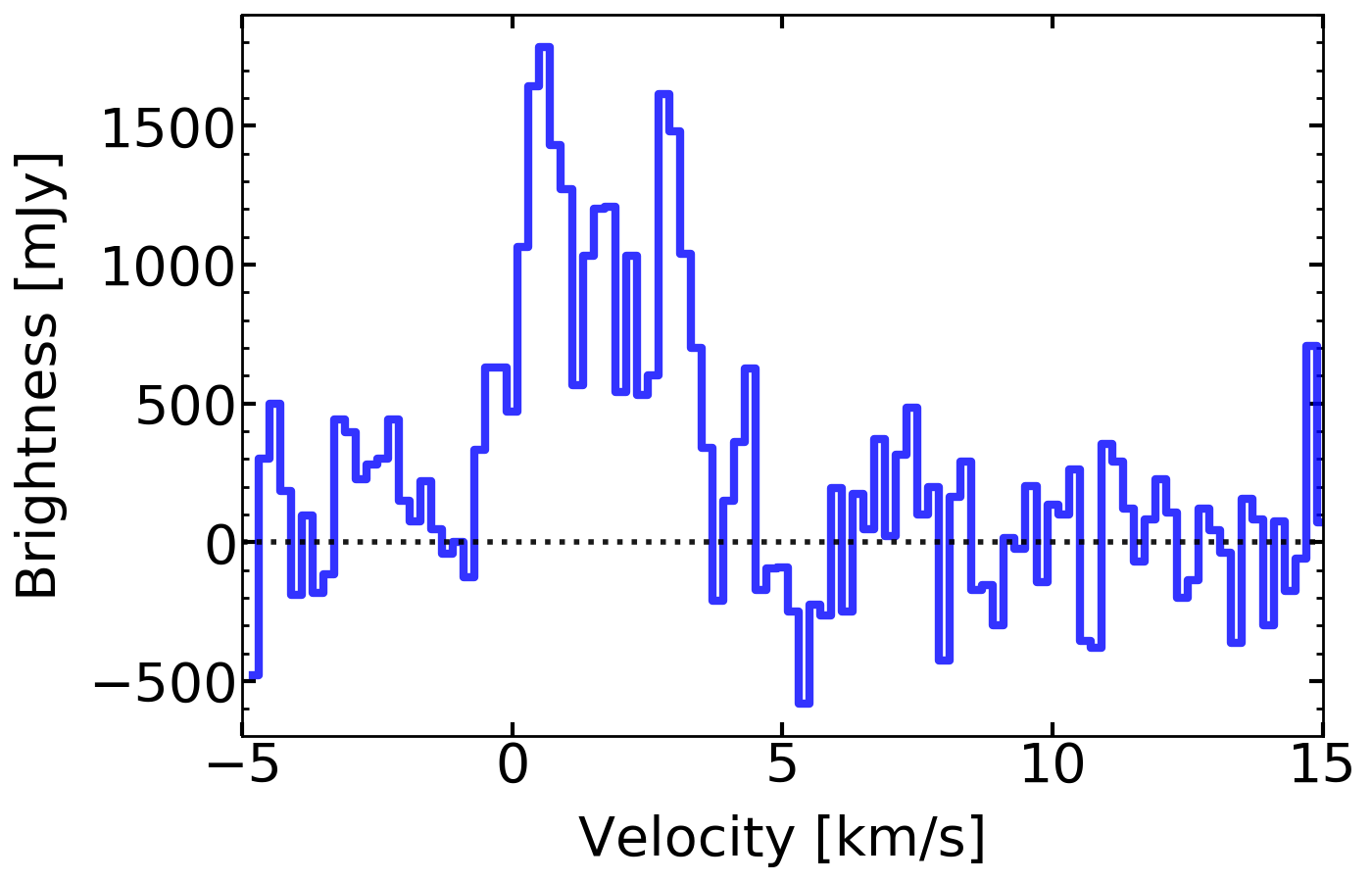}
    \includegraphics[width=.235\textwidth]{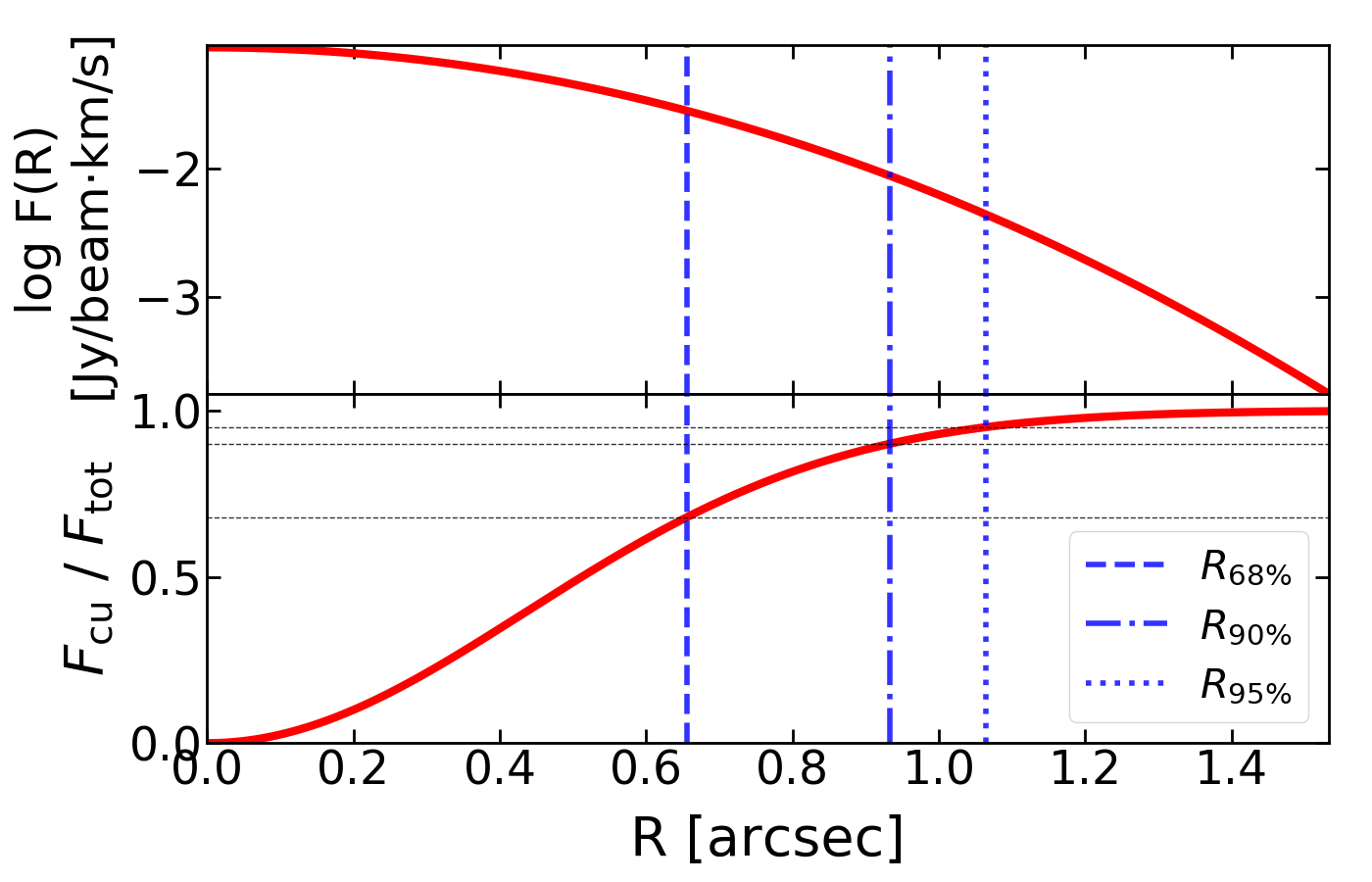}
    \end{center}
  \caption[]{
  Results of the CO modeling for every disk with measured CO size, following the methodology described in Section~\ref{sec:gasmodeling}. For each disk, the first three sub-panels show the observed, model and residual CO moment zero maps; solid (dashed) line contours are drawn at increasing (decreasing) $3\sigma$ intervals. The forth sub-panel represents the integrated spectrum enclosed by the $R_{68\%}^{\mathrm{CO}}$ of the source. Last sub-panel shows the radial brightness profile and the respective cumulative distribution of the CO model.
  }
  \label{fig:comodelresults_all_4}
\end{figure*}

\begin{figure*}
    \begin{center}
    \includegraphics[width=.490\textwidth]{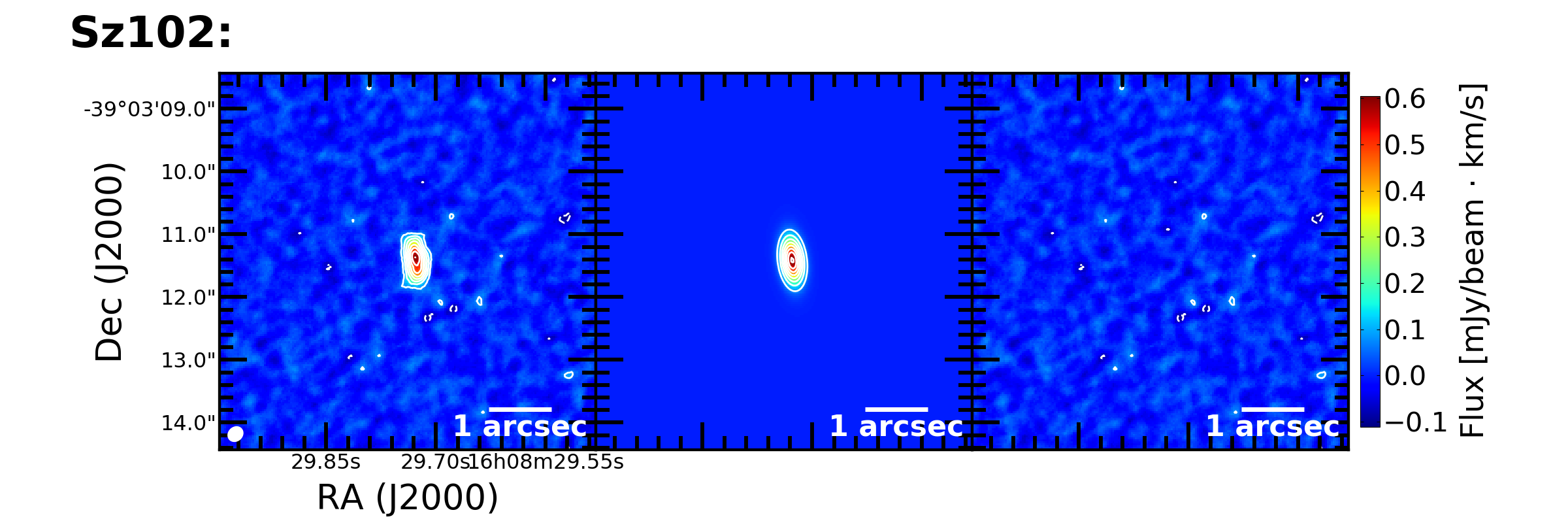}
    \includegraphics[width=.225\textwidth]{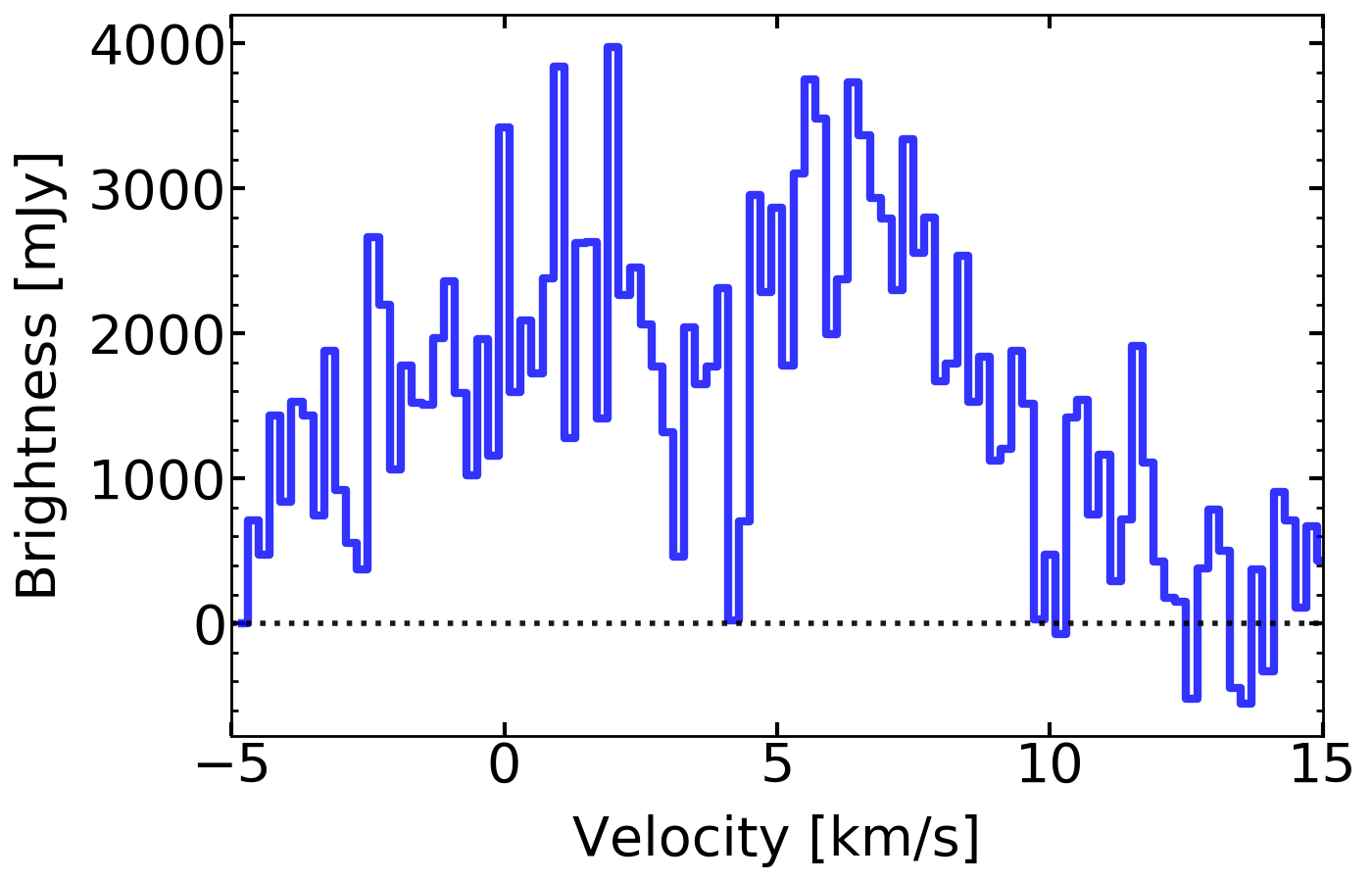}
    \includegraphics[width=.235\textwidth]{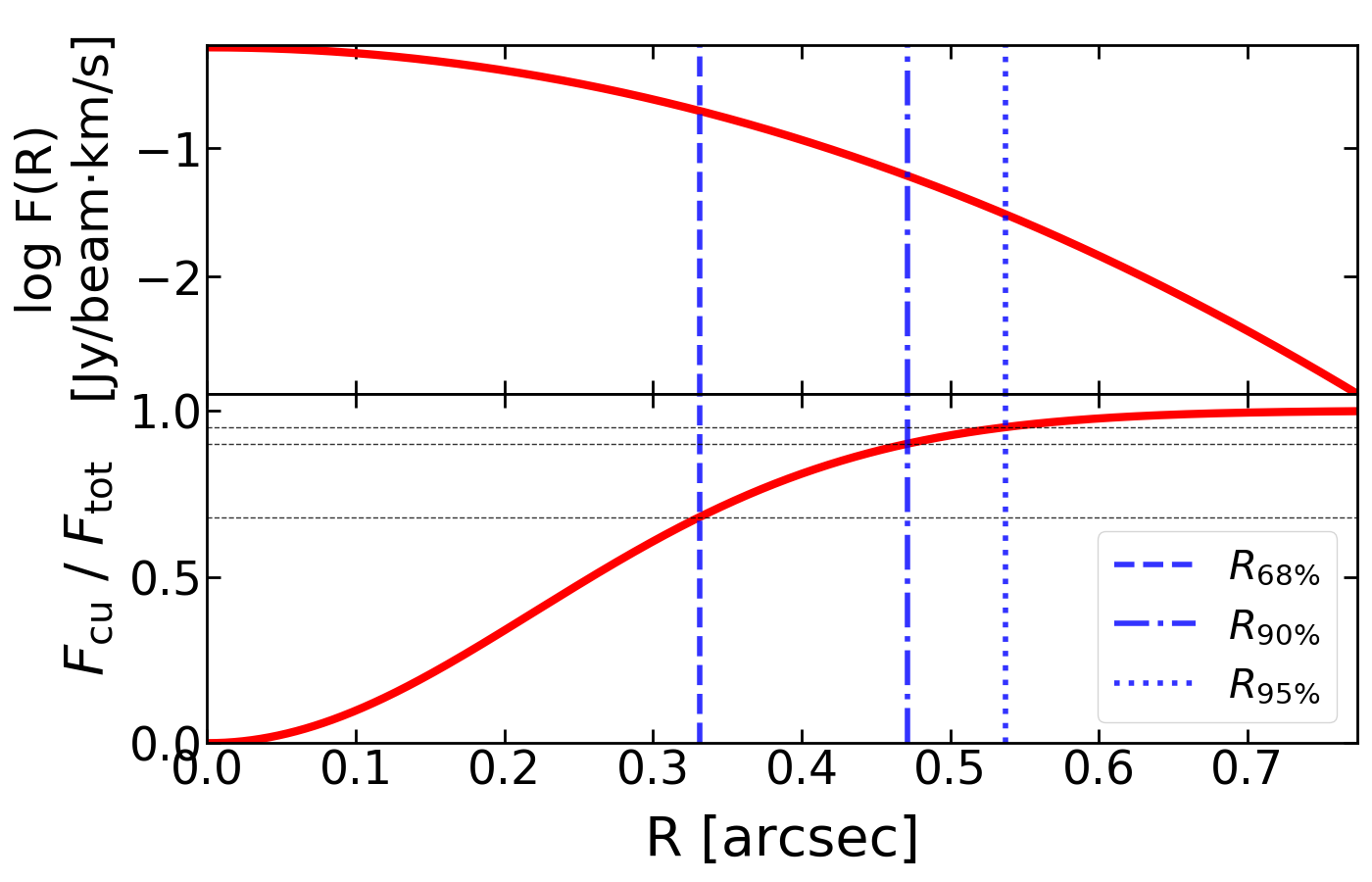}
    \end{center}
    \begin{center}
    \includegraphics[width=.490\textwidth]{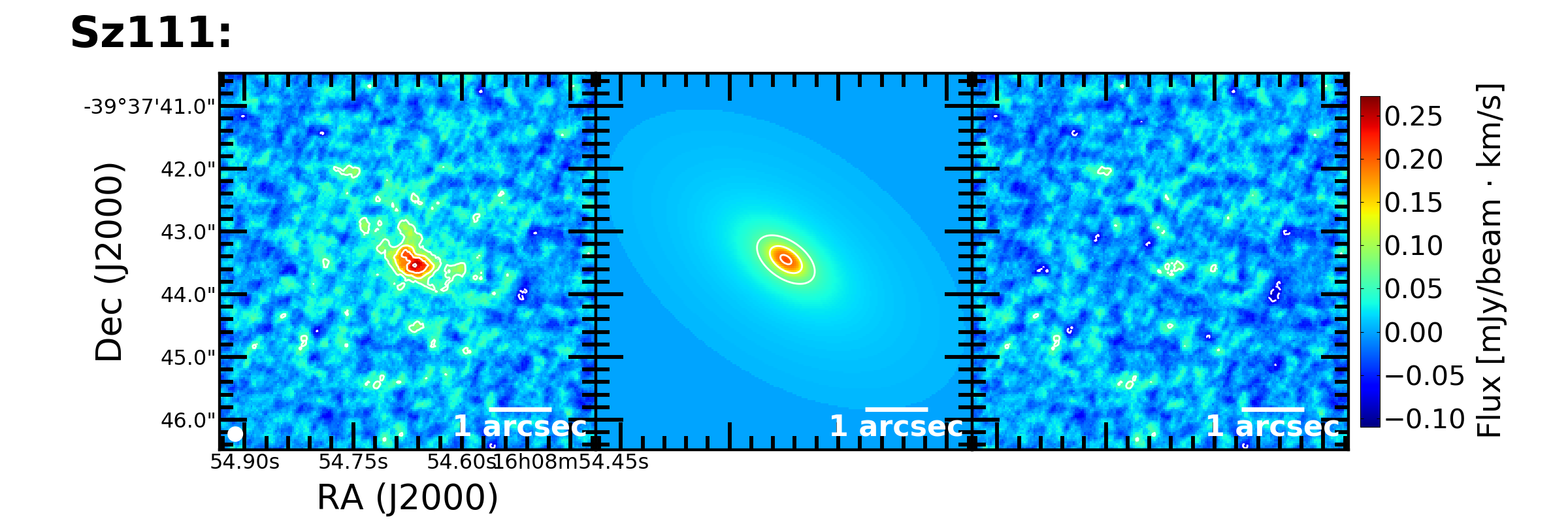}
    \includegraphics[width=.225\textwidth]{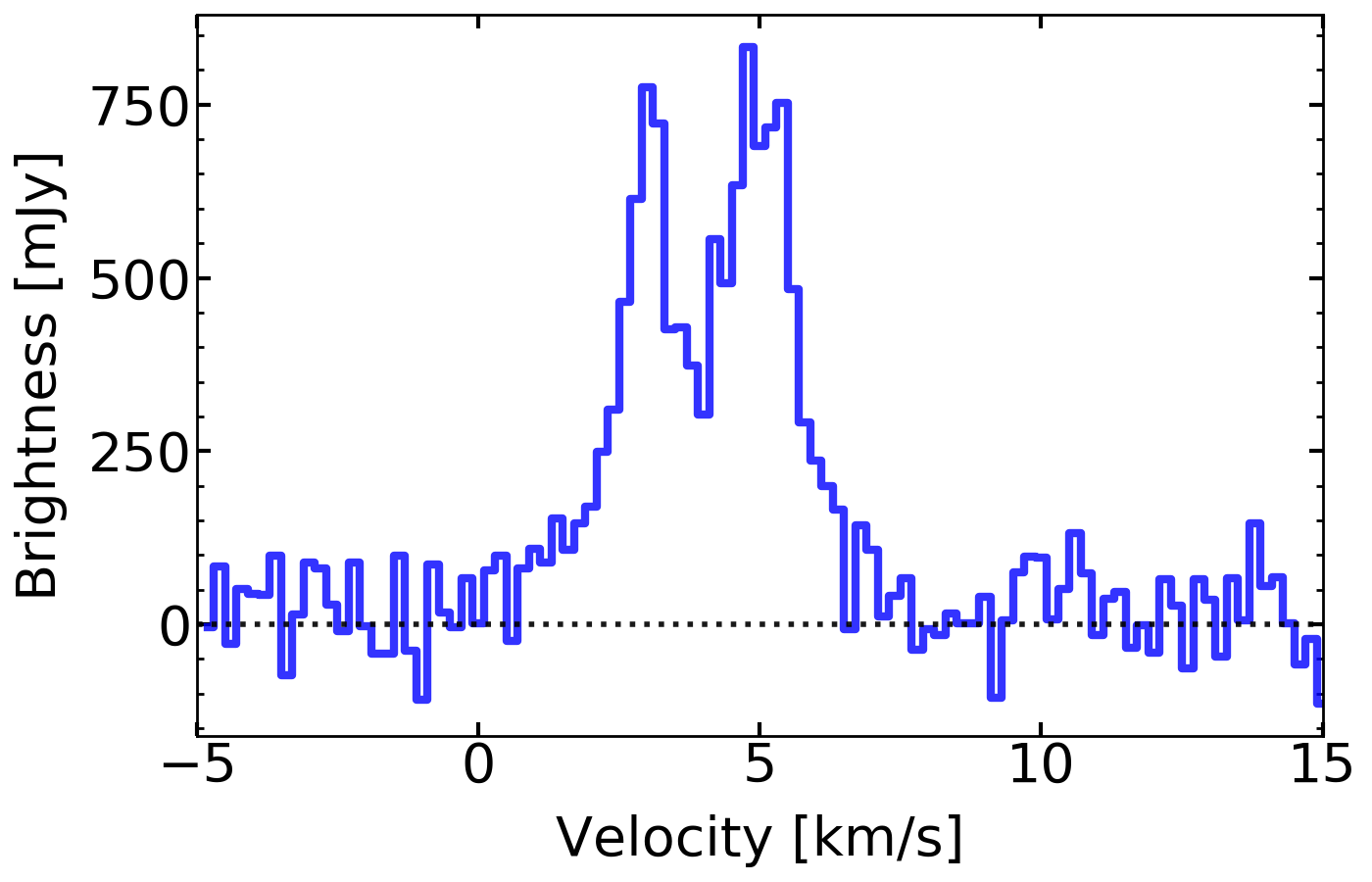}
    \includegraphics[width=.235\textwidth]{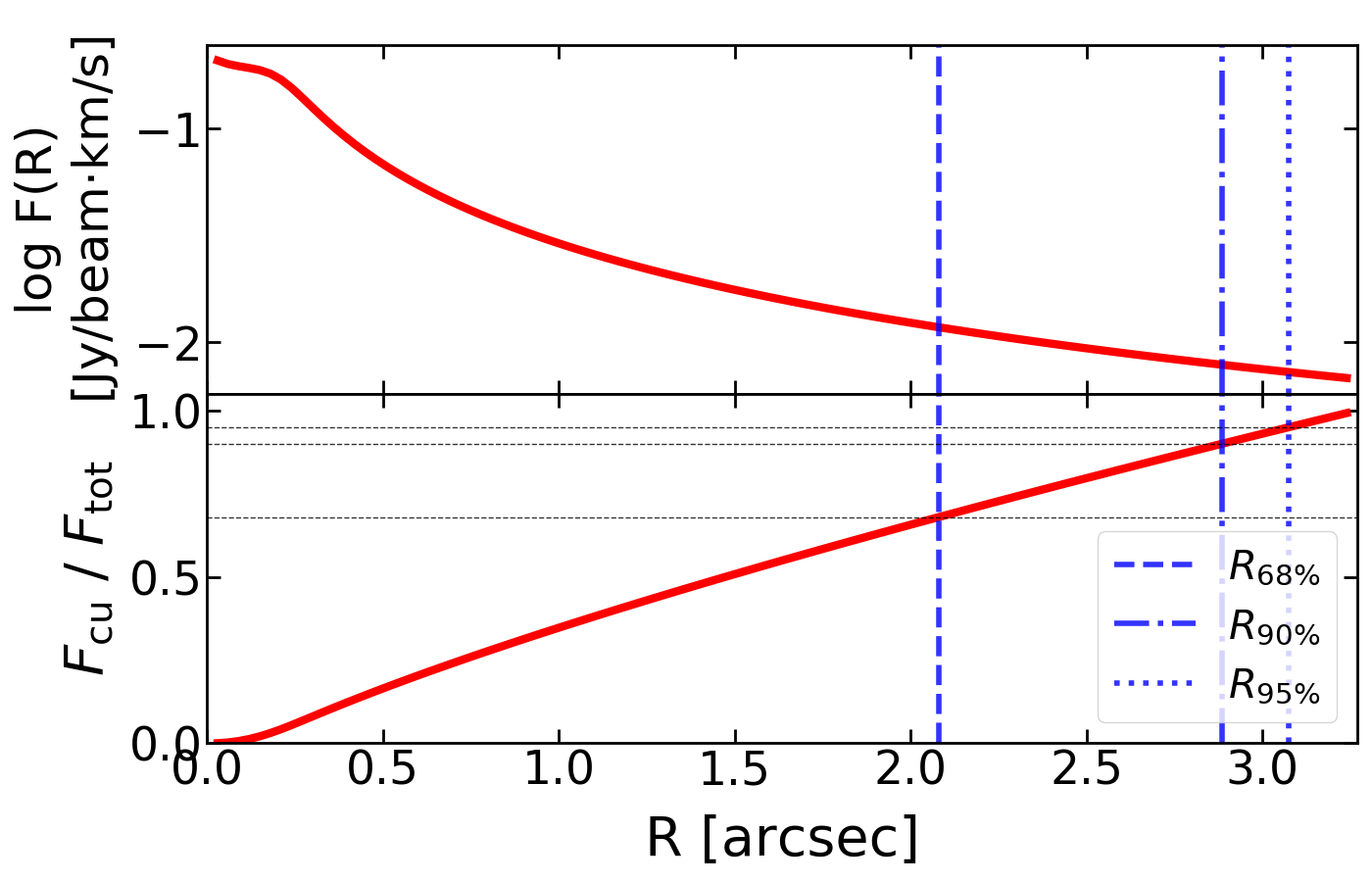}
    \end{center}
    \begin{center}
    \includegraphics[width=.490\textwidth]{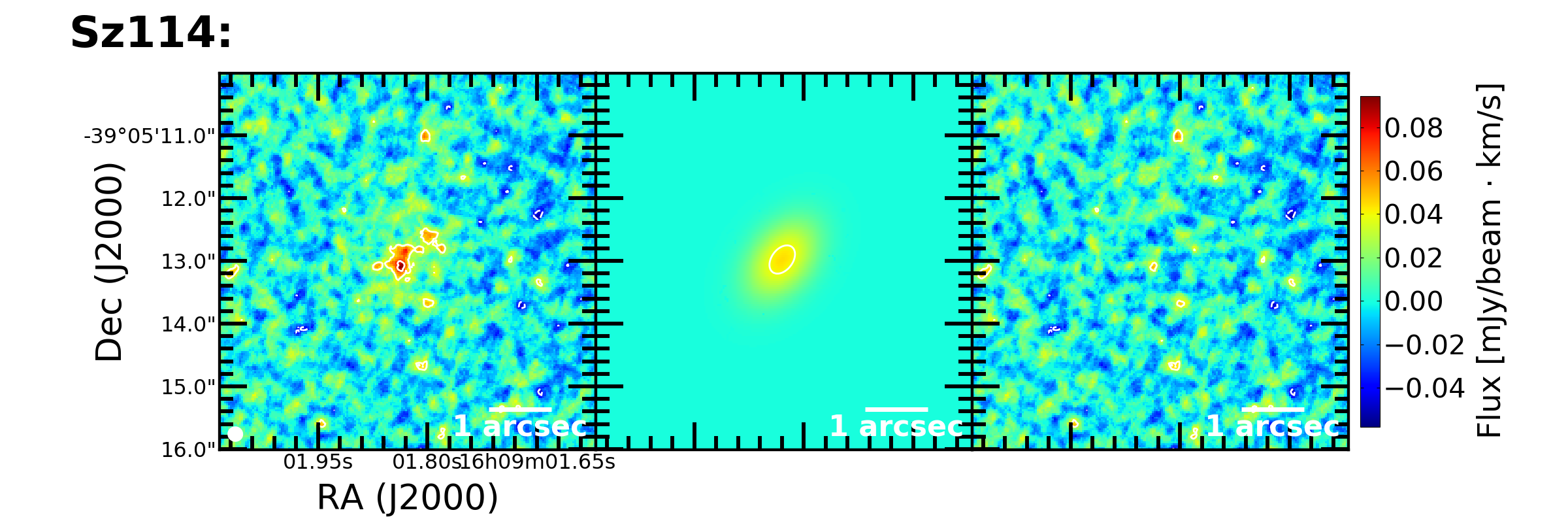}
    \includegraphics[width=.225\textwidth]{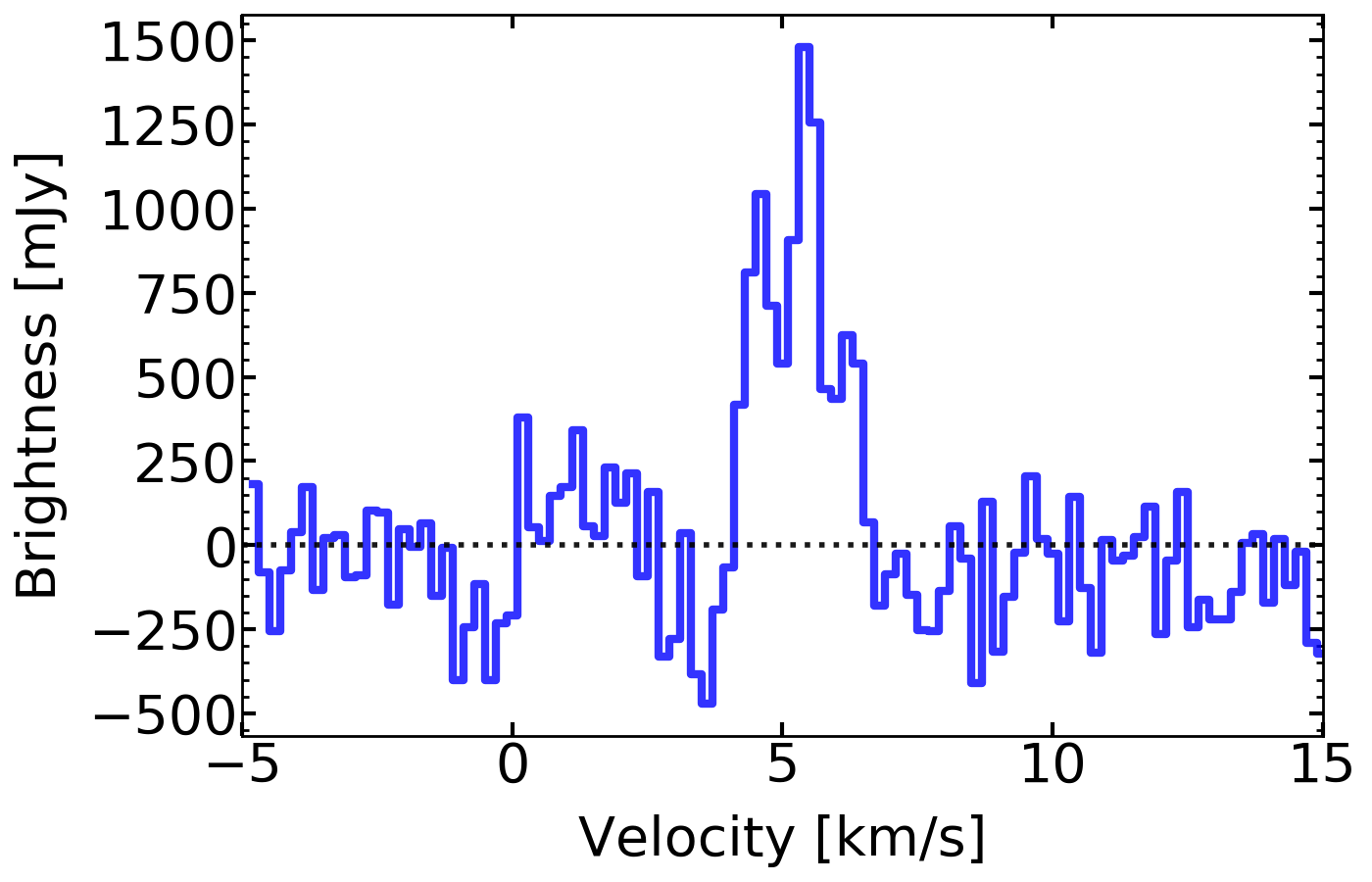}
    \includegraphics[width=.235\textwidth]{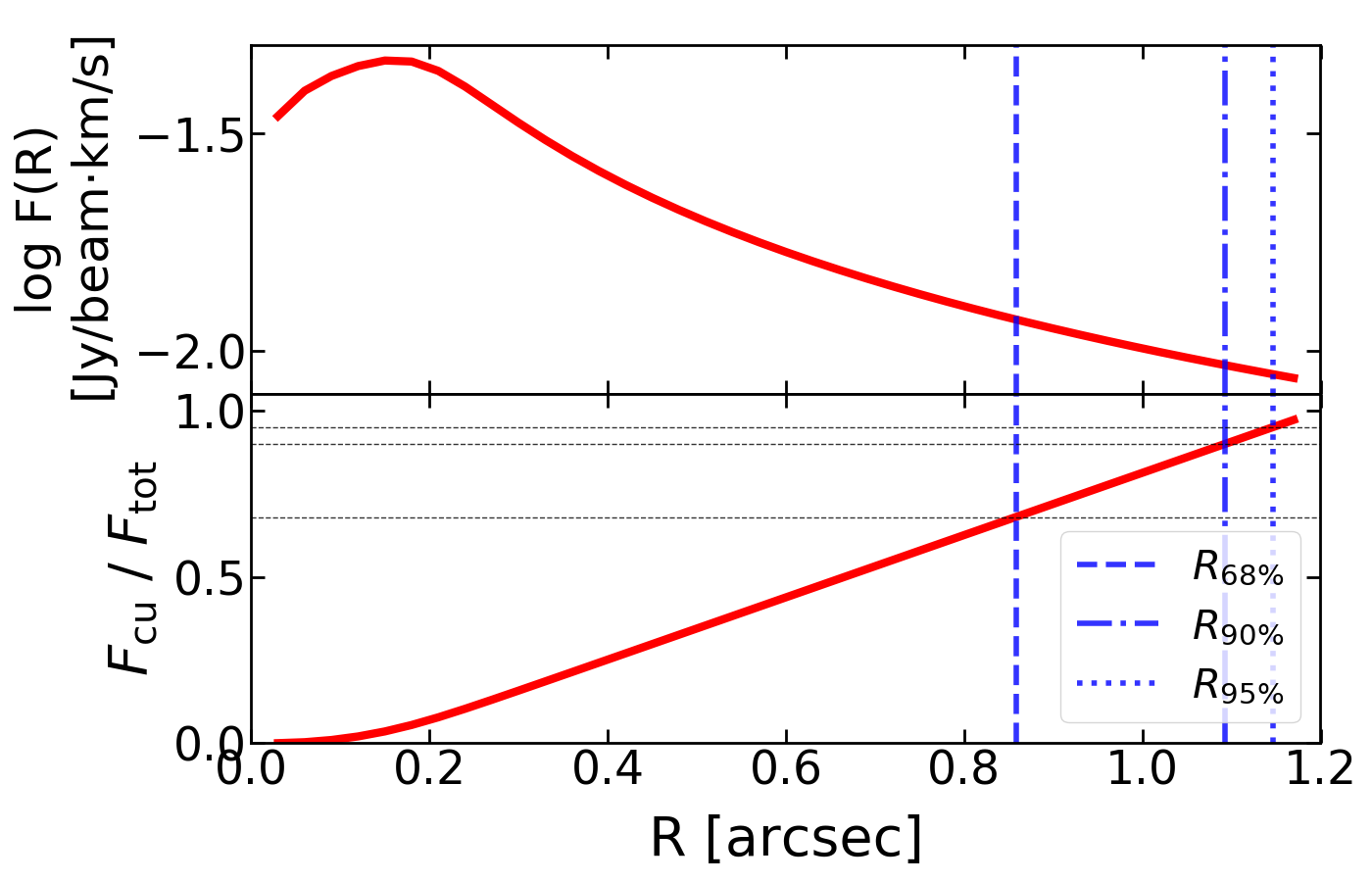}
    \end{center}
    \begin{center}
    \includegraphics[width=.490\textwidth]{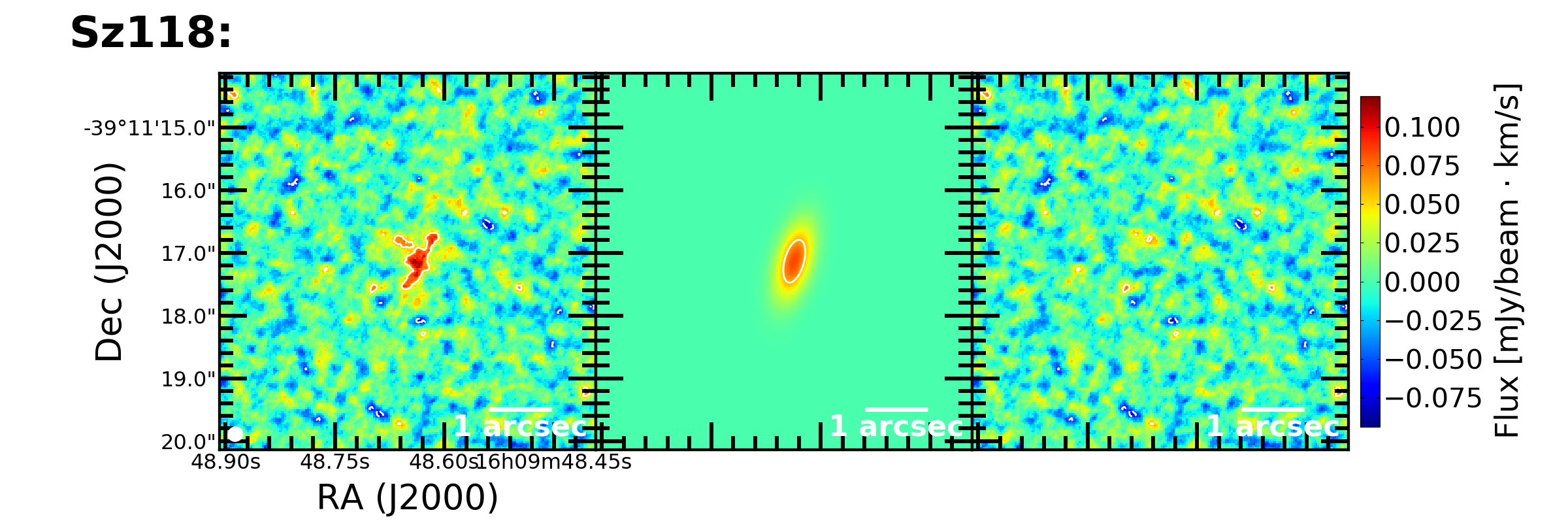}
    \includegraphics[width=.225\textwidth]{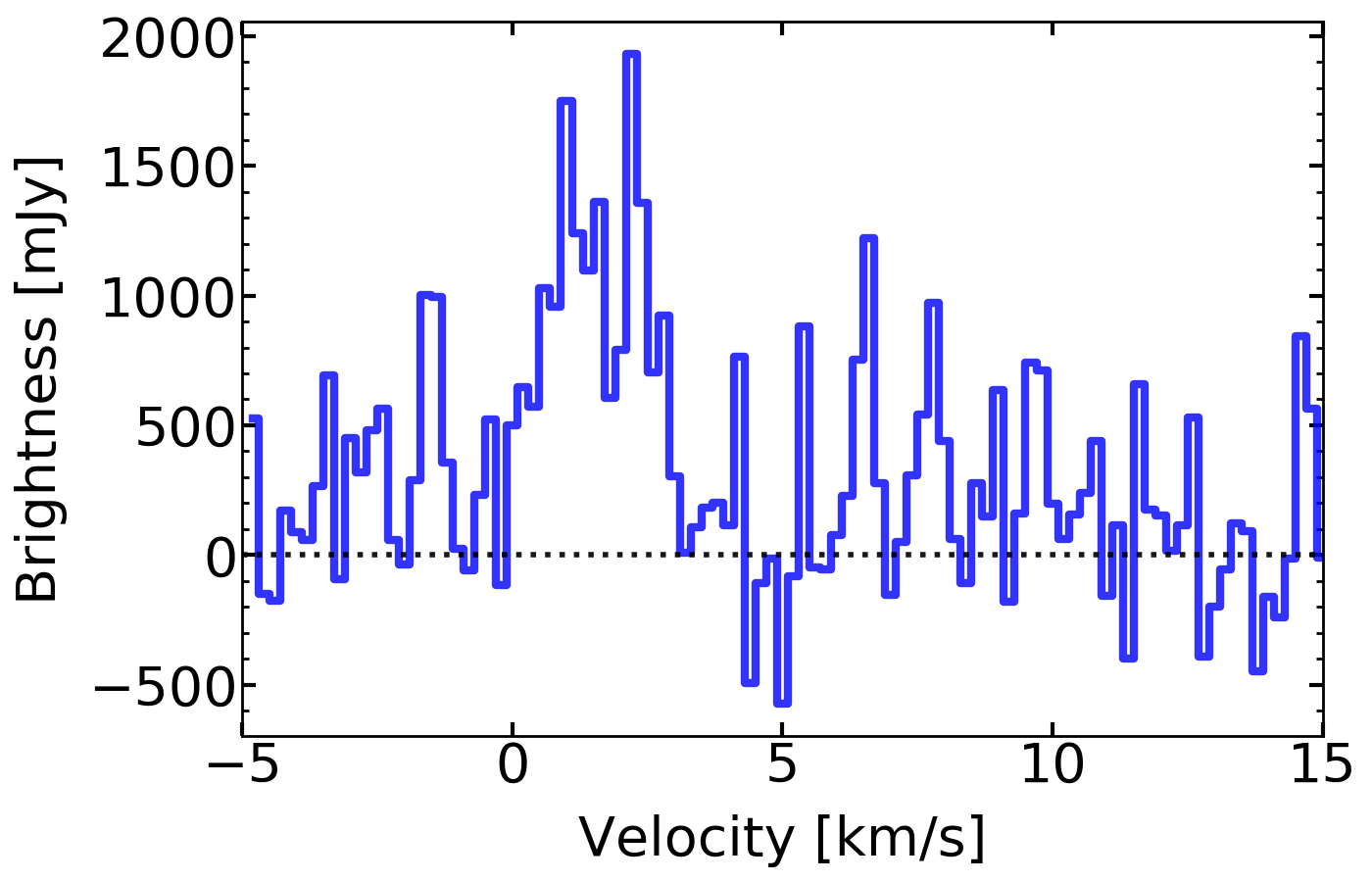}
    \includegraphics[width=.235\textwidth]{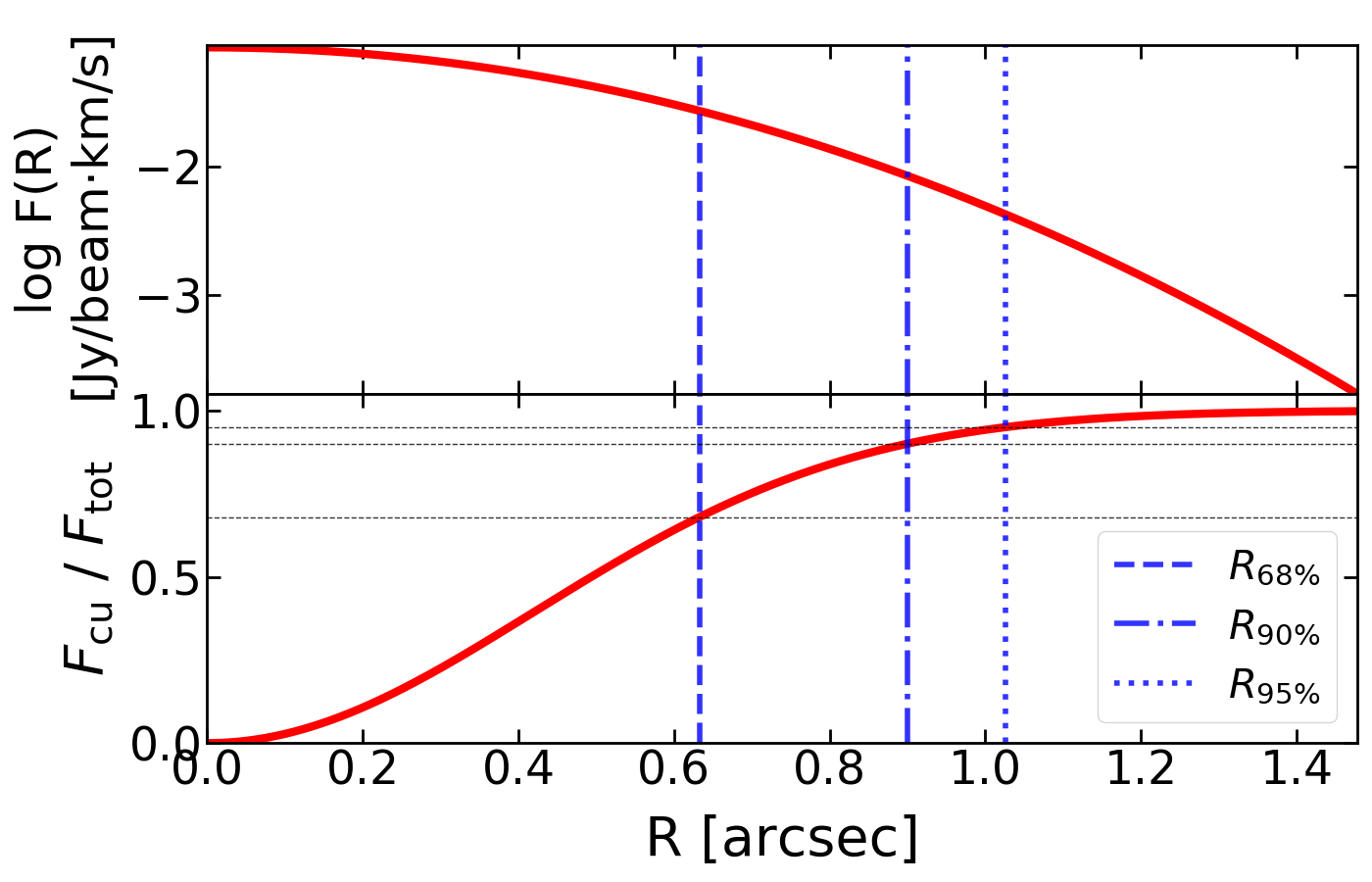}
    \end{center}
    \begin{center}
    \includegraphics[width=.490\textwidth]{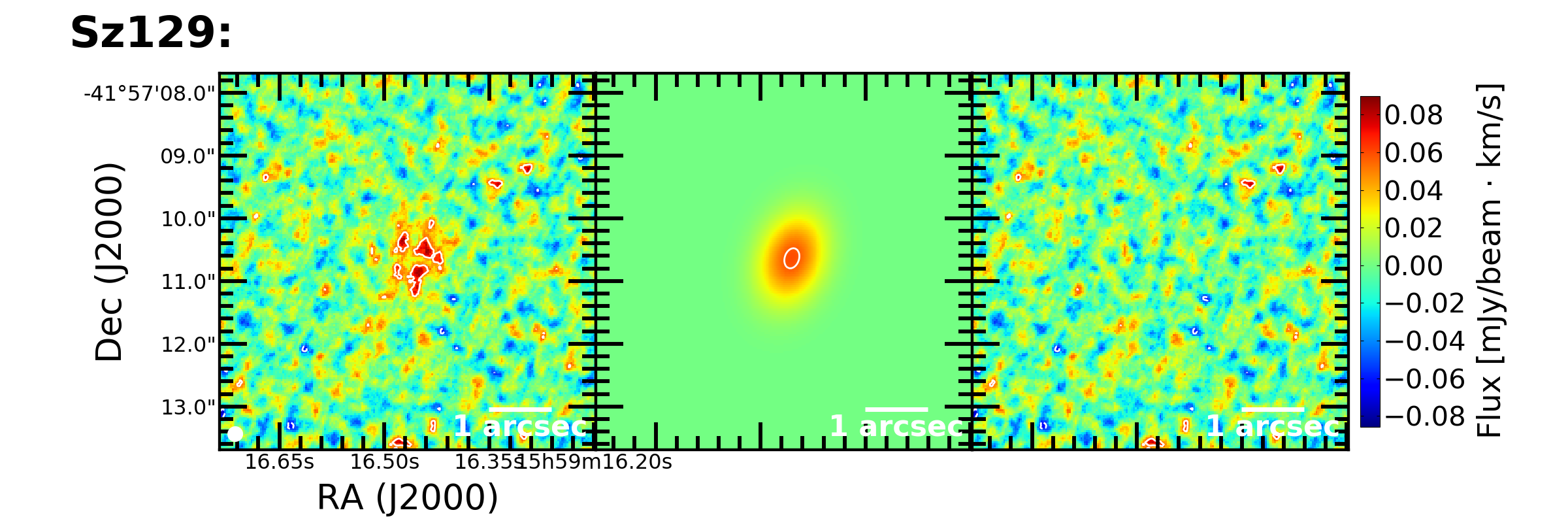}
    \includegraphics[width=.225\textwidth]{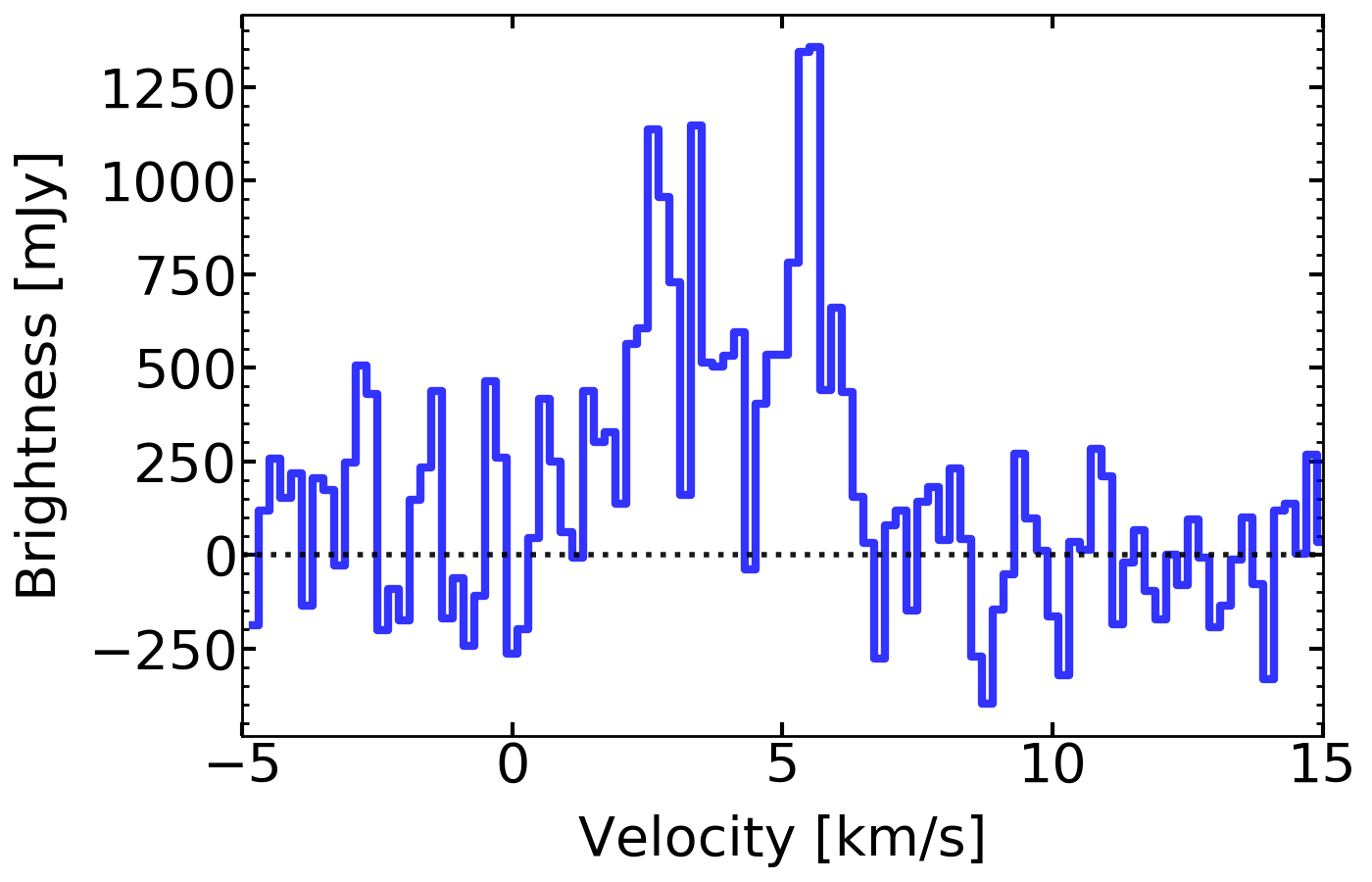}
    \includegraphics[width=.235\textwidth]{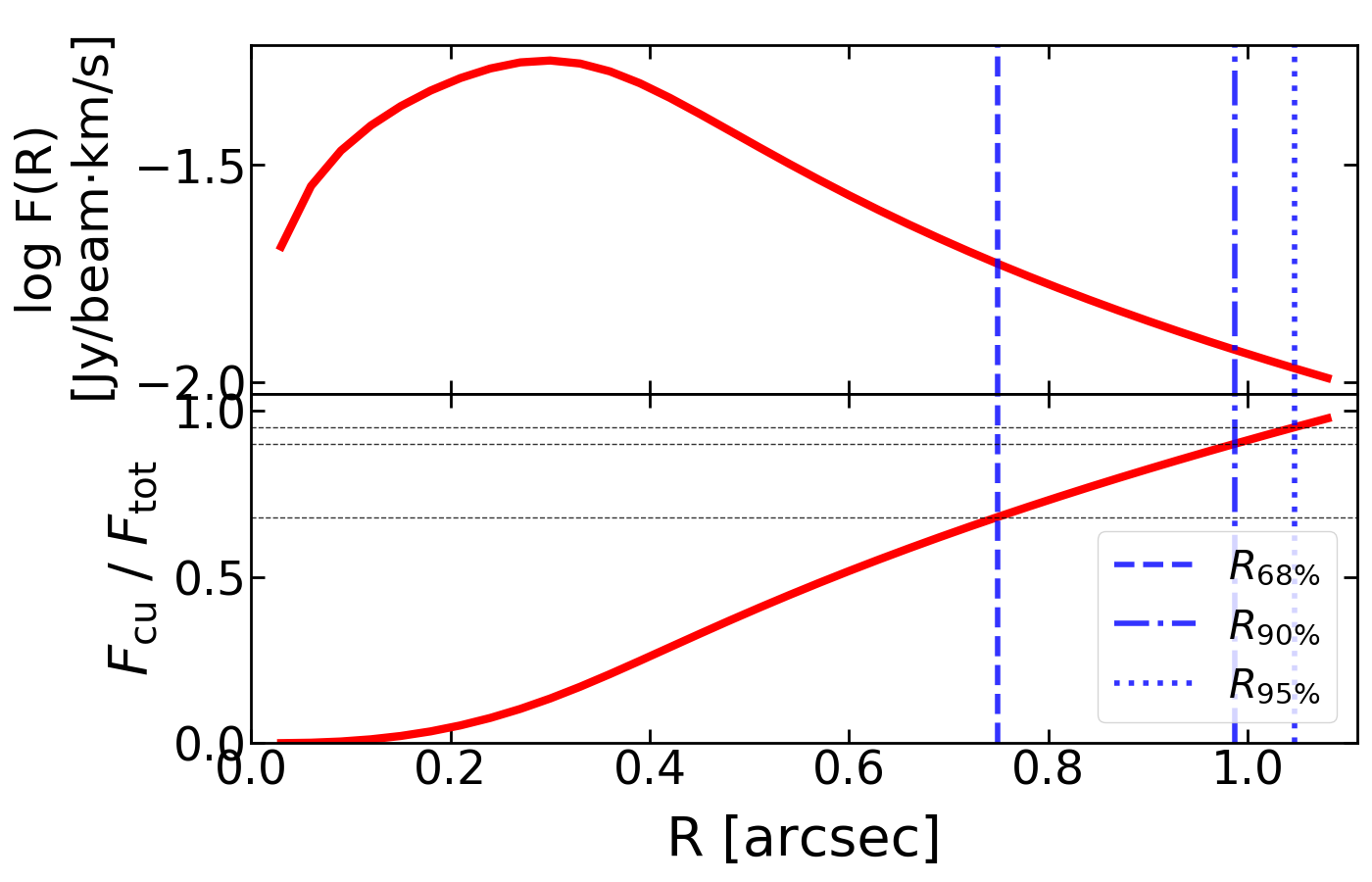}
    \end{center}
    \begin{center}
    \includegraphics[width=.490\textwidth]{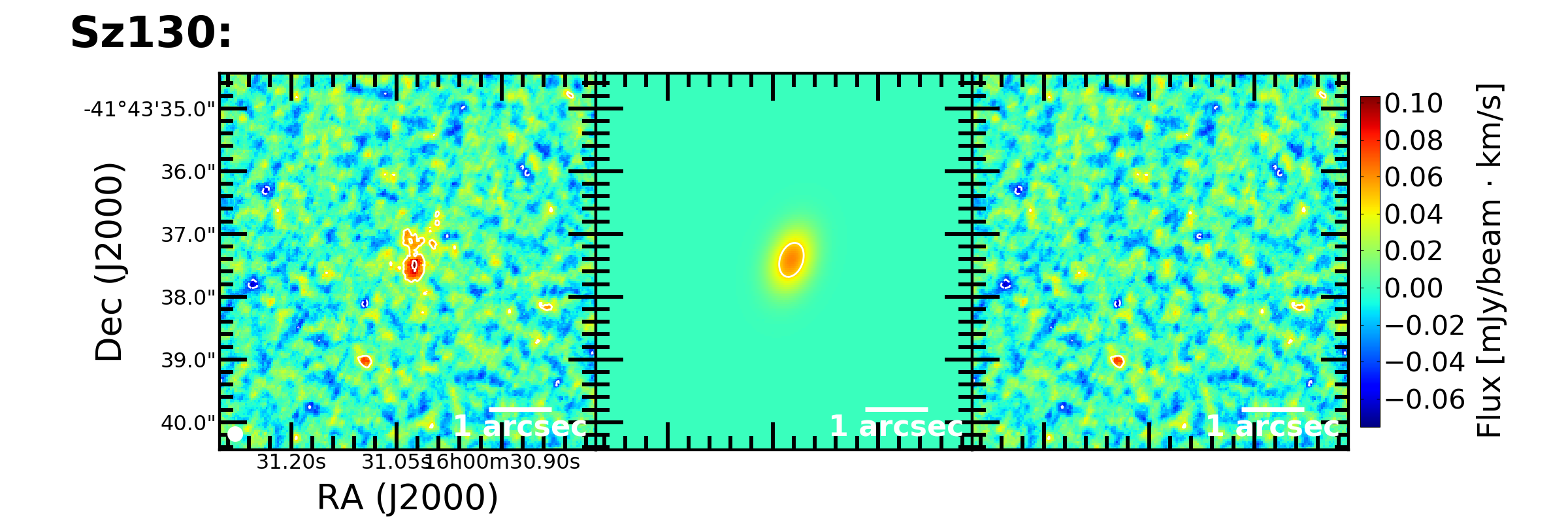}
    \includegraphics[width=.225\textwidth]{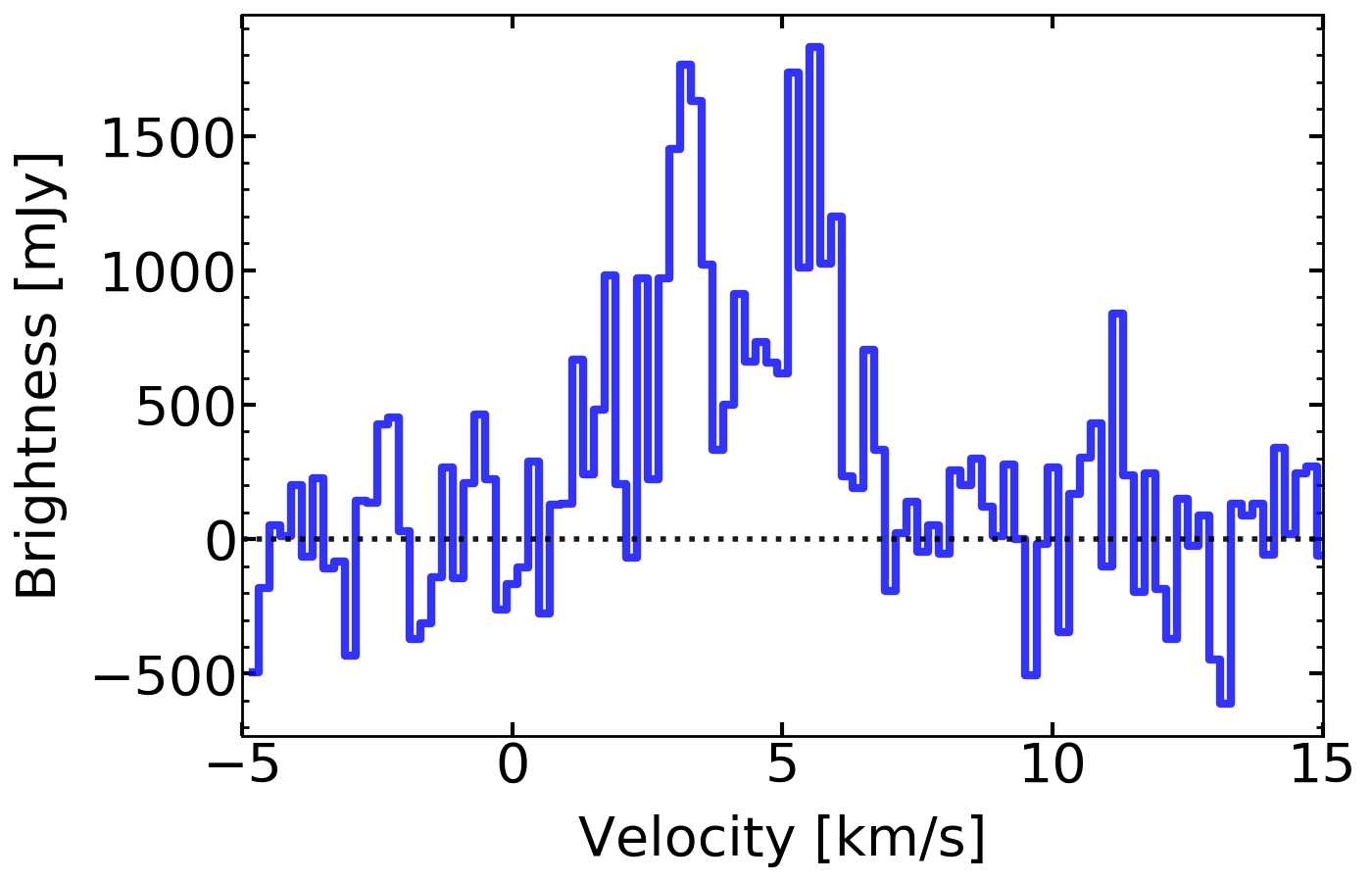}
    \includegraphics[width=.235\textwidth]{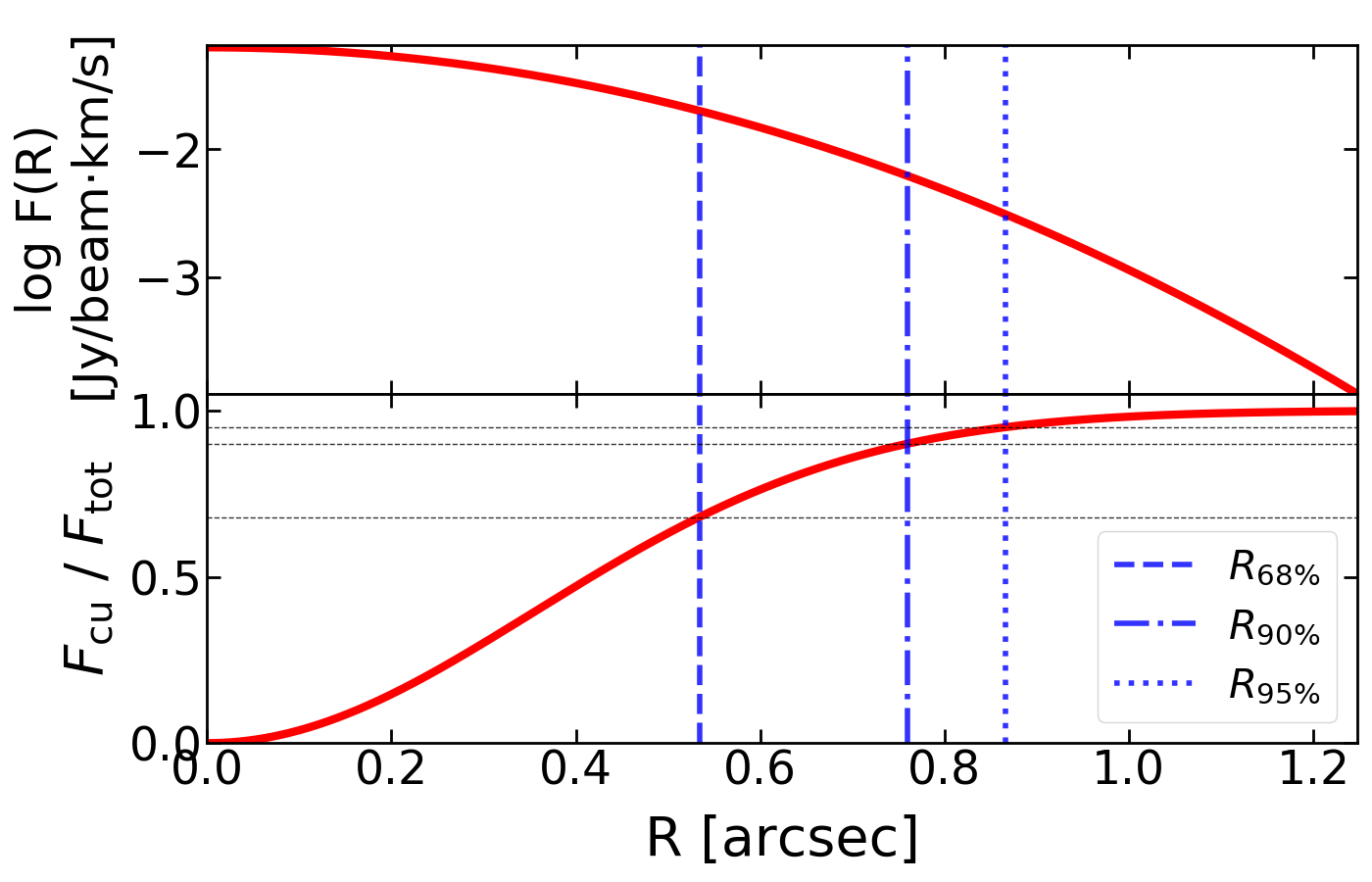}
    \end{center}
    \begin{center}
    \includegraphics[width=.490\textwidth]{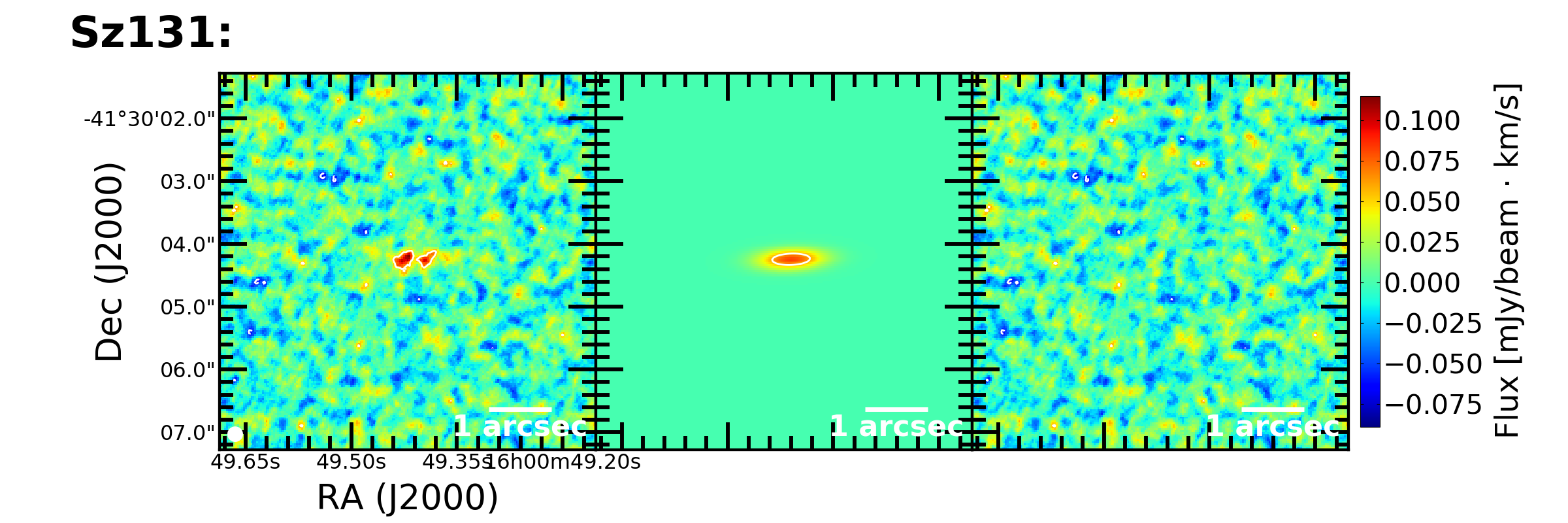}
    \includegraphics[width=.225\textwidth]{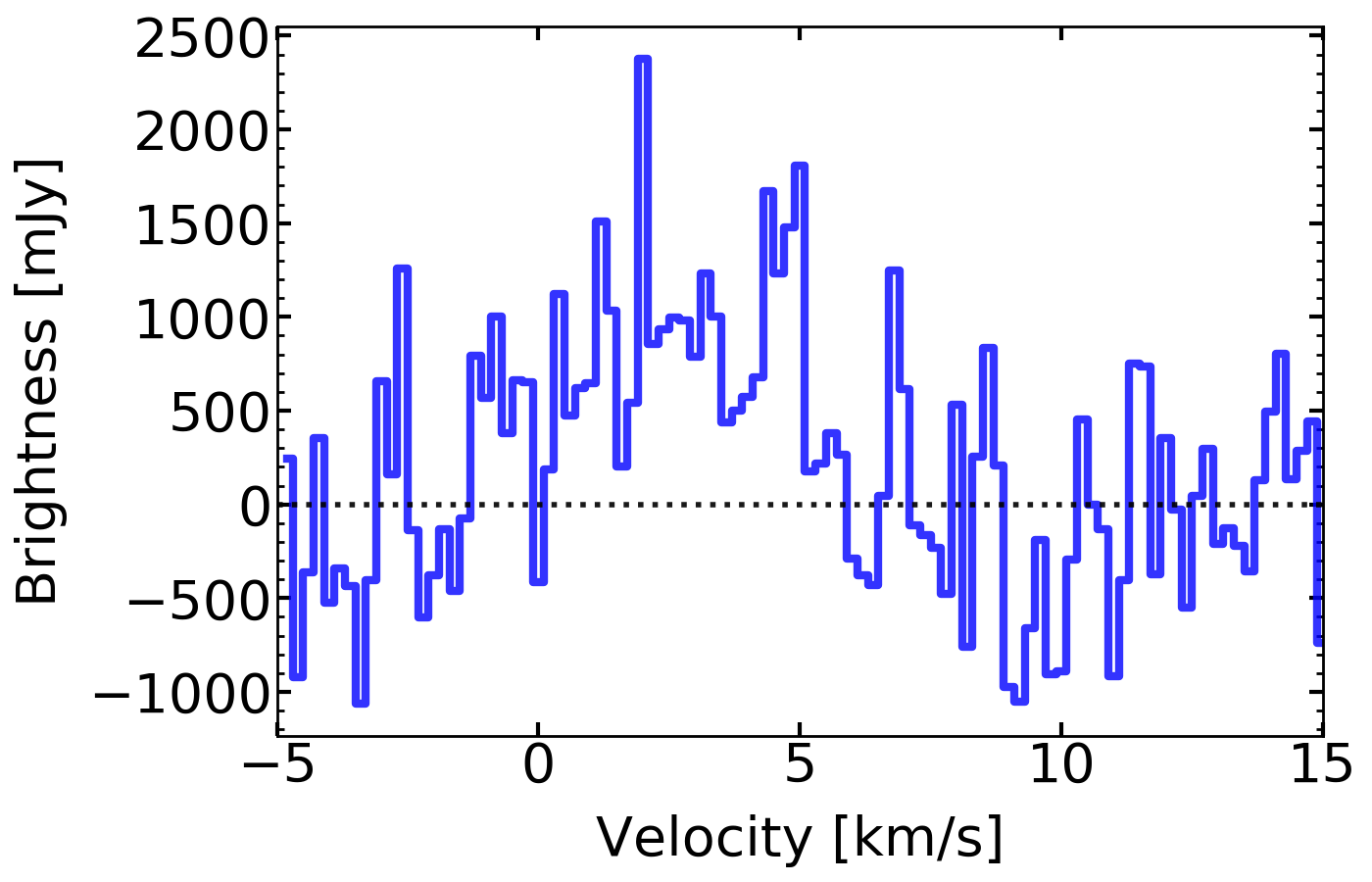}
    \includegraphics[width=.235\textwidth]{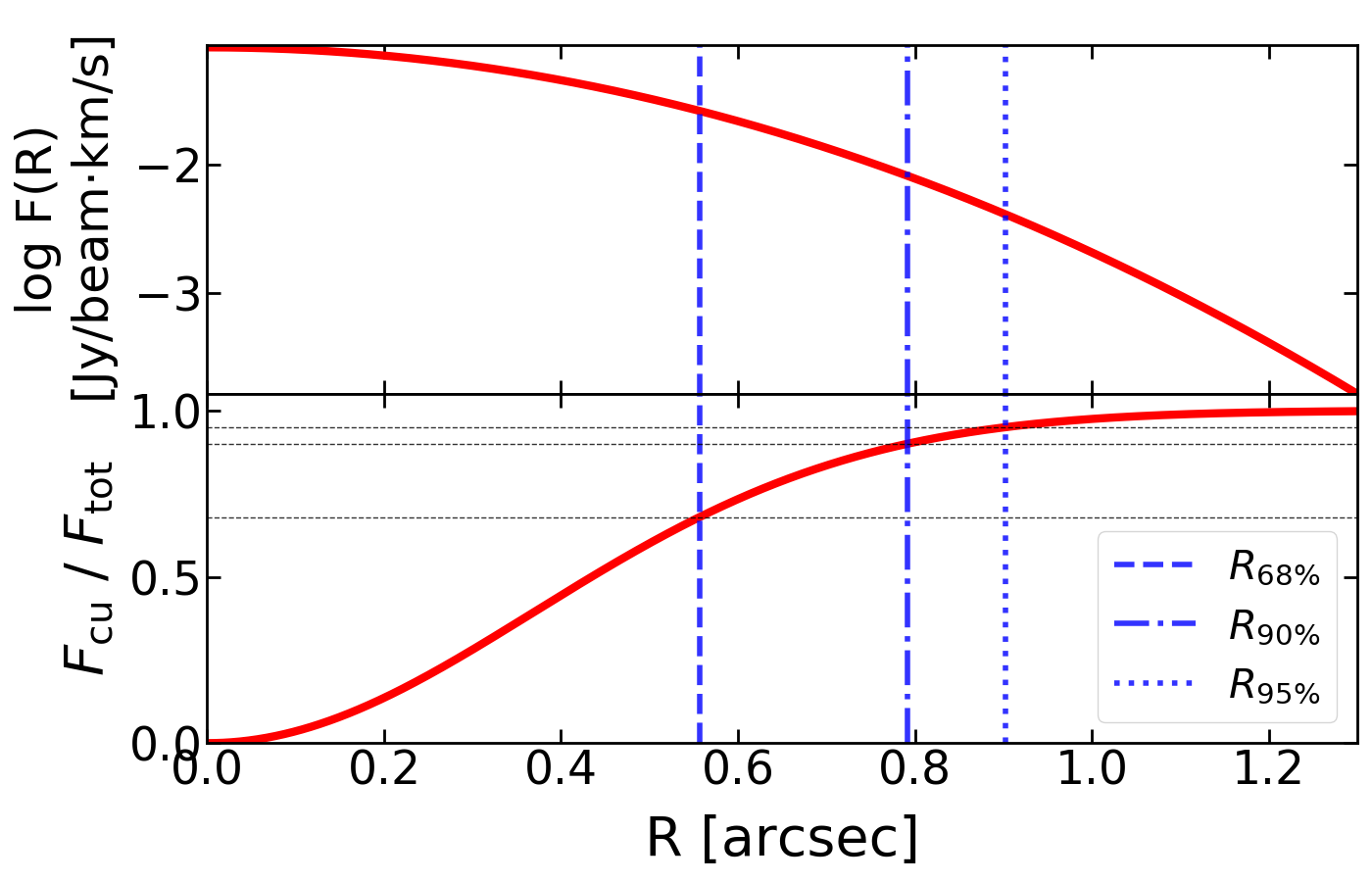}
    \end{center}
  \caption[]{
  Results of the CO modeling for every disk with measured CO size, following the methodology described in Section~\ref{sec:gasmodeling}. For each disk, the first three sub-panels show the observed, model and residual CO moment zero maps; solid (dashed) line contours are drawn at increasing (decreasing) $3\sigma$ intervals. The forth sub-panel represents the integrated spectrum enclosed by the $R_{68\%}^{\mathrm{CO}}$ of the source. Last sub-panel shows the radial brightness profile and the respective cumulative distribution of the CO model.
  }
  \label{fig:comodelresults_all_5}
\end{figure*}

\begin{figure*}
    \begin{center}
    \includegraphics[width=.490\textwidth]{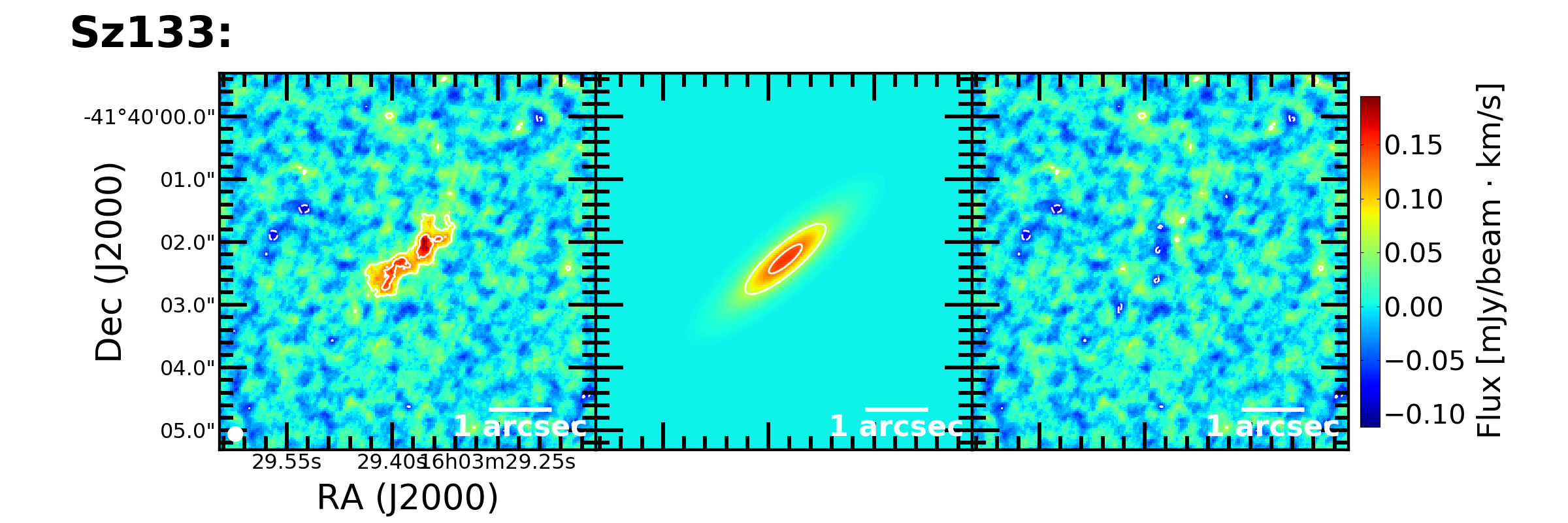}
    \includegraphics[width=.225\textwidth]{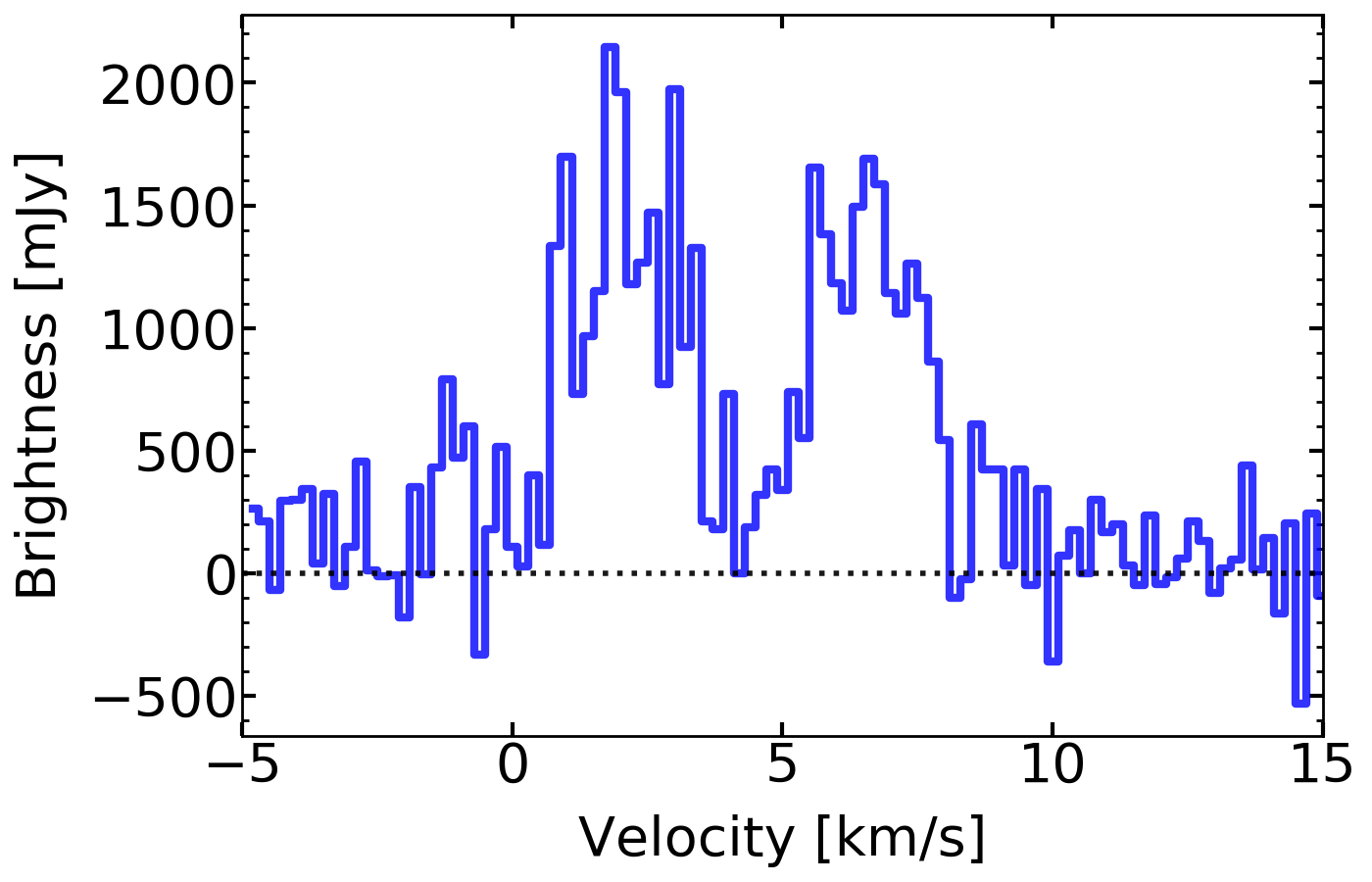}
    \includegraphics[width=.235\textwidth]{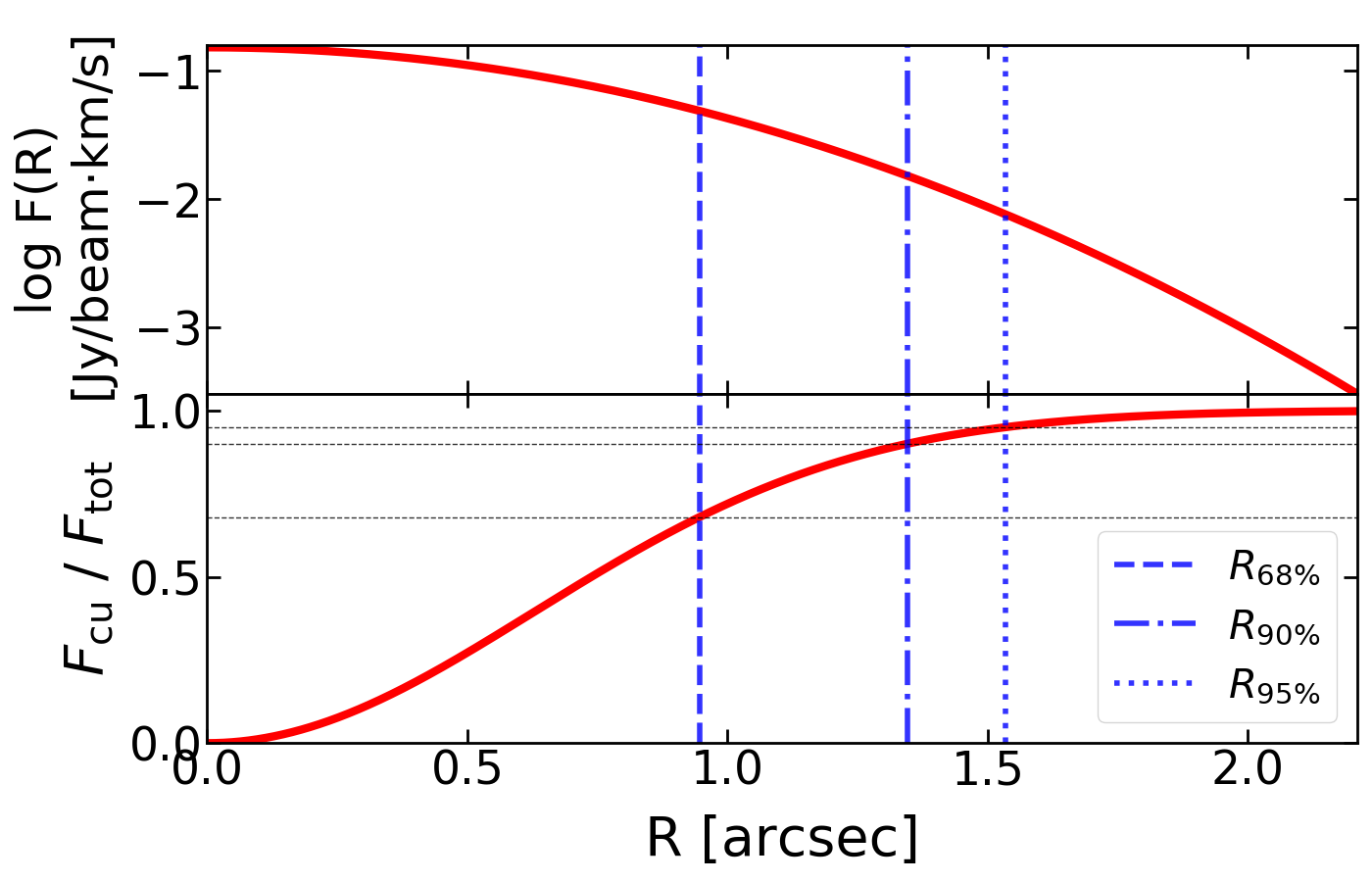}
    \end{center}
    \begin{center}
    \includegraphics[width=.490\textwidth]{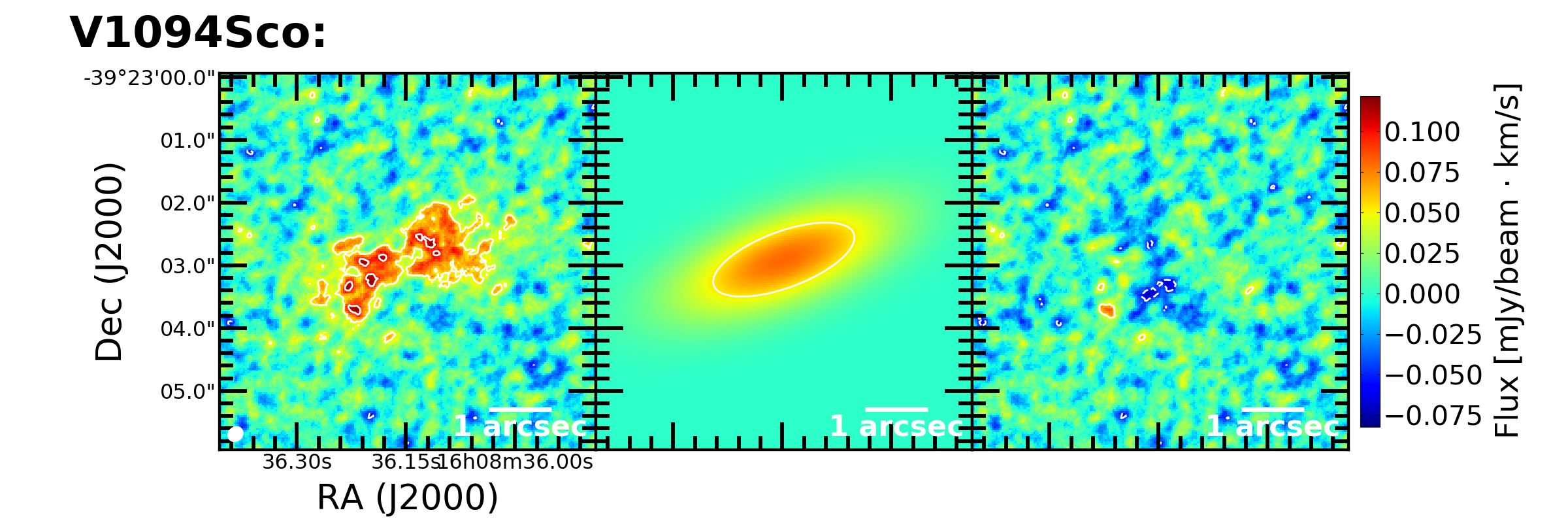}
    \includegraphics[width=.225\textwidth]{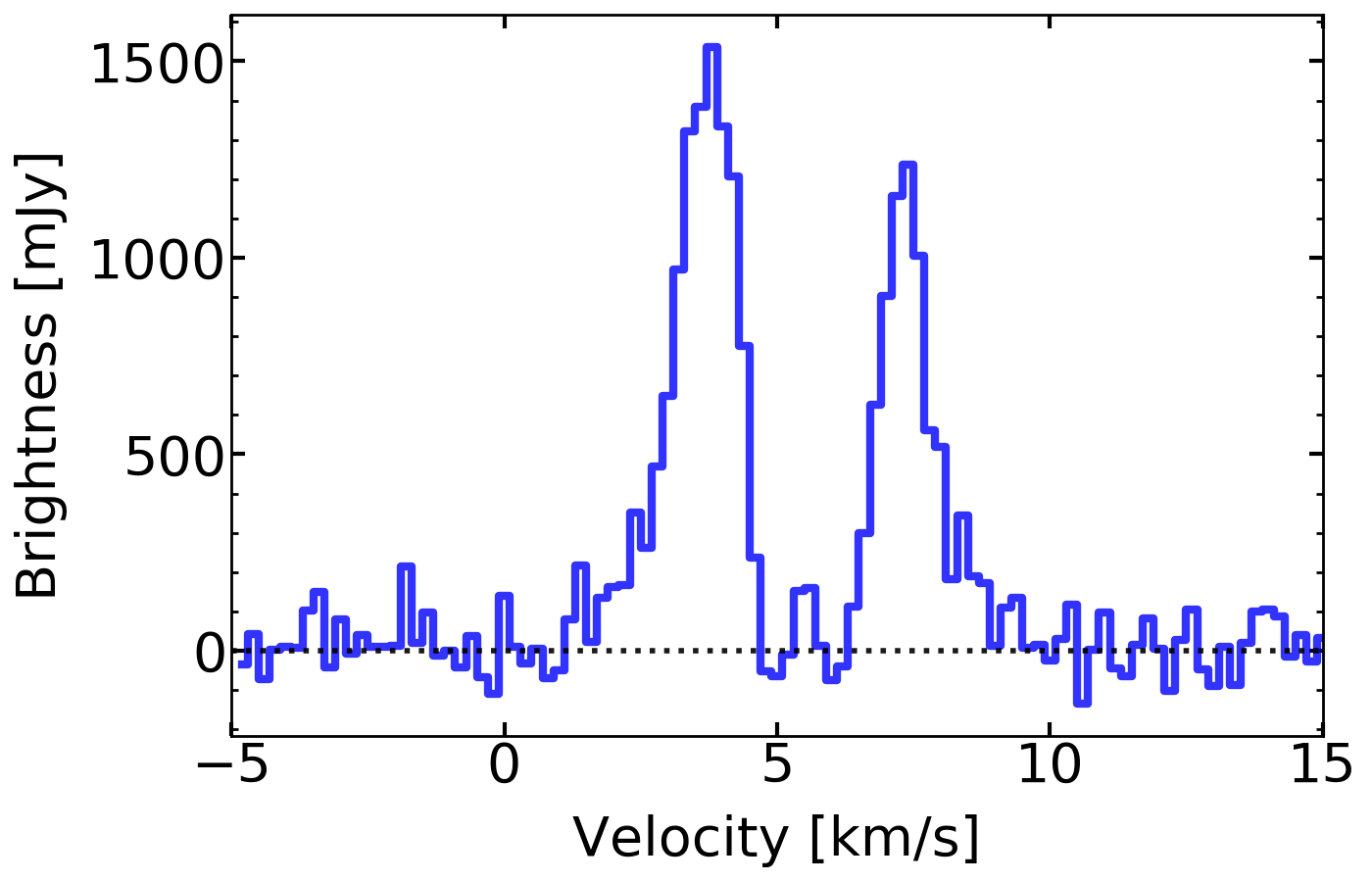}
    \includegraphics[width=.235\textwidth]{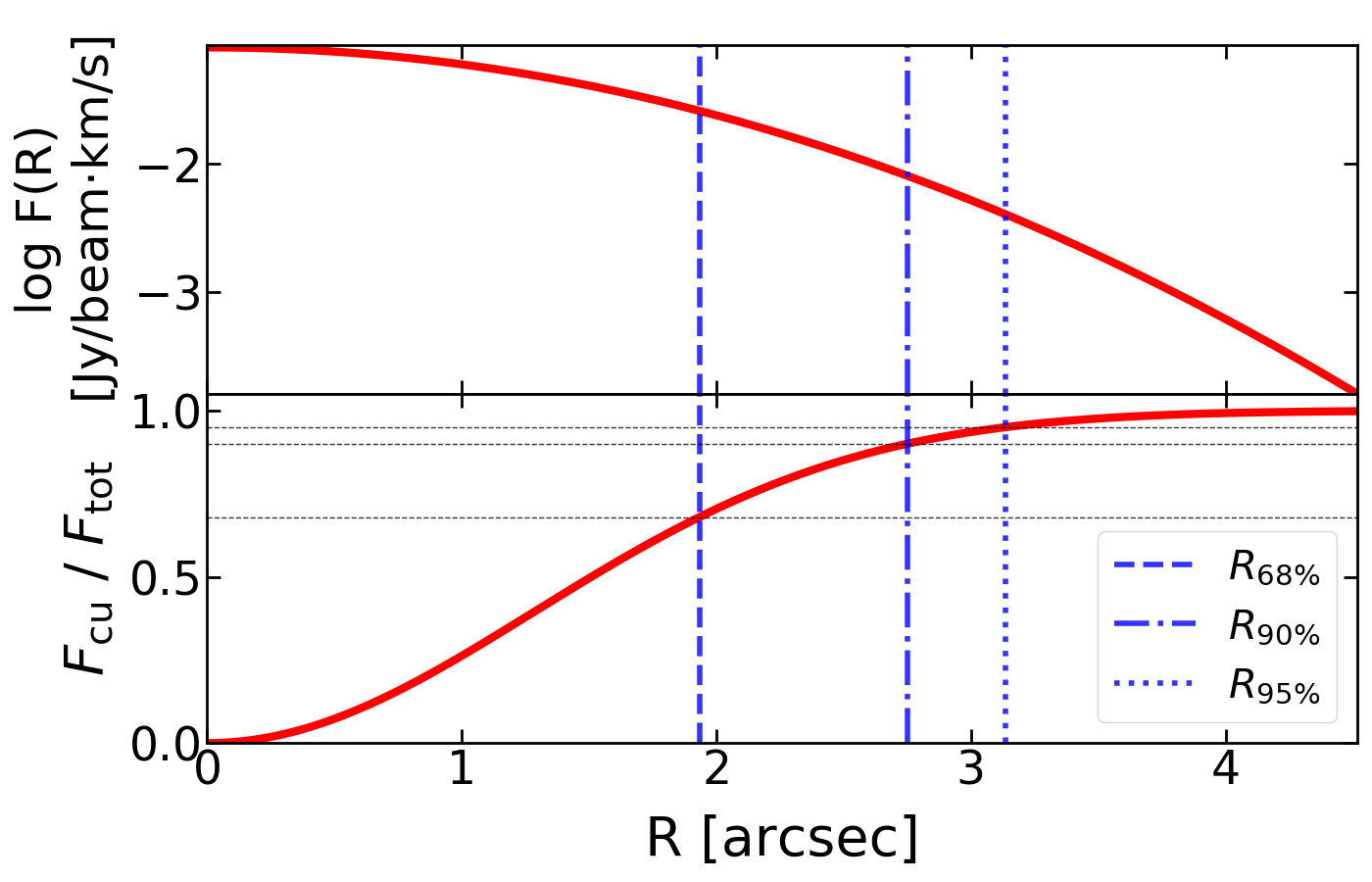}
    \end{center}
    \begin{center}
    \includegraphics[width=.490\textwidth]{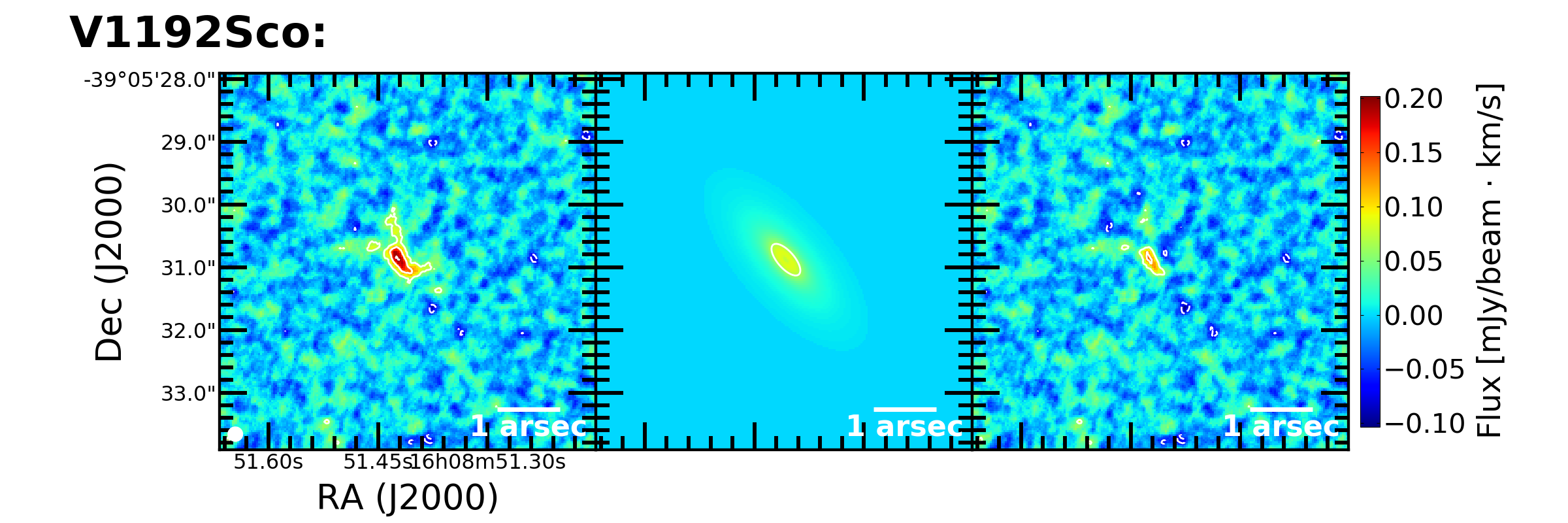}
    \includegraphics[width=.225\textwidth]{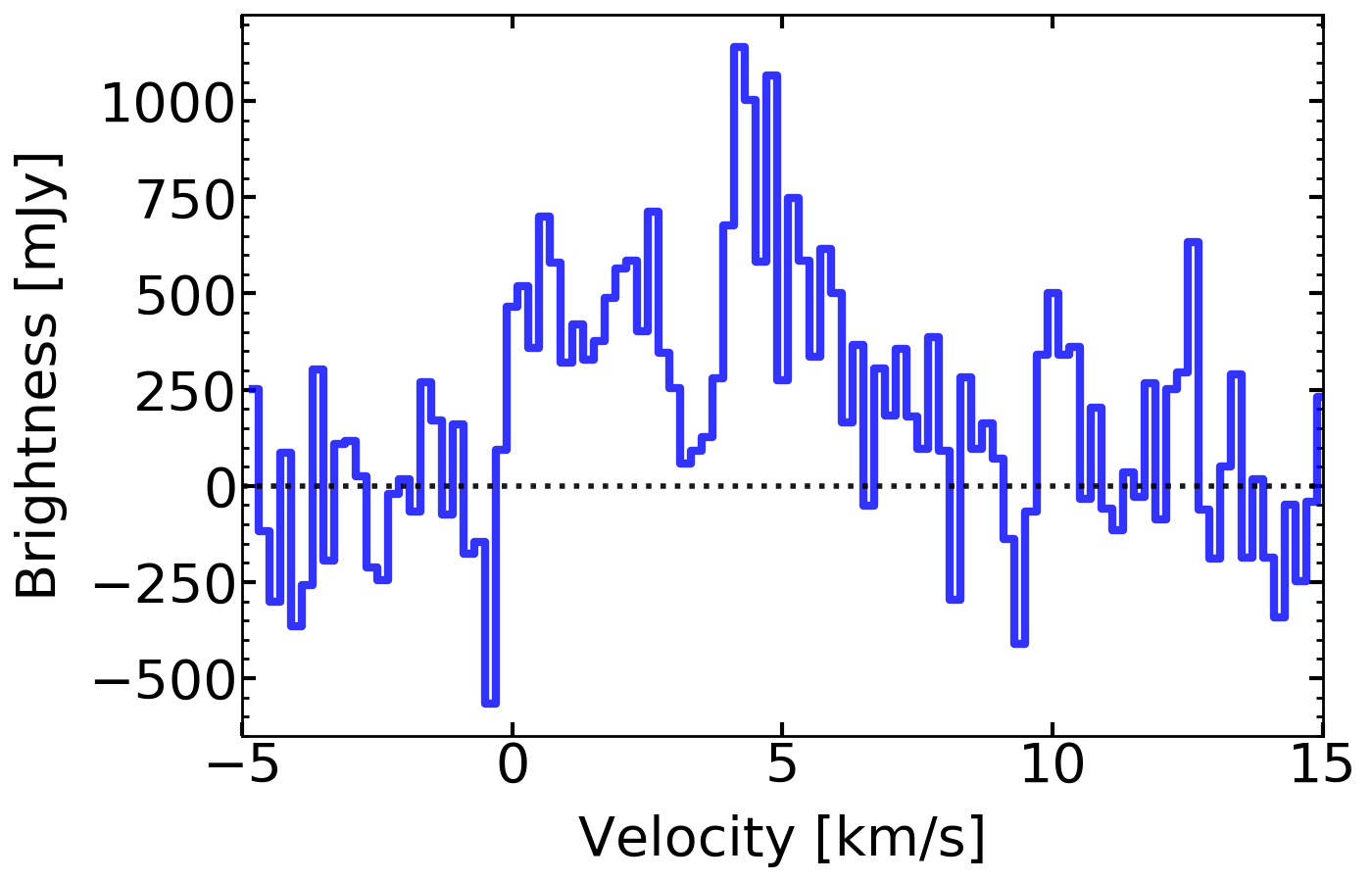}
    \includegraphics[width=.235\textwidth]{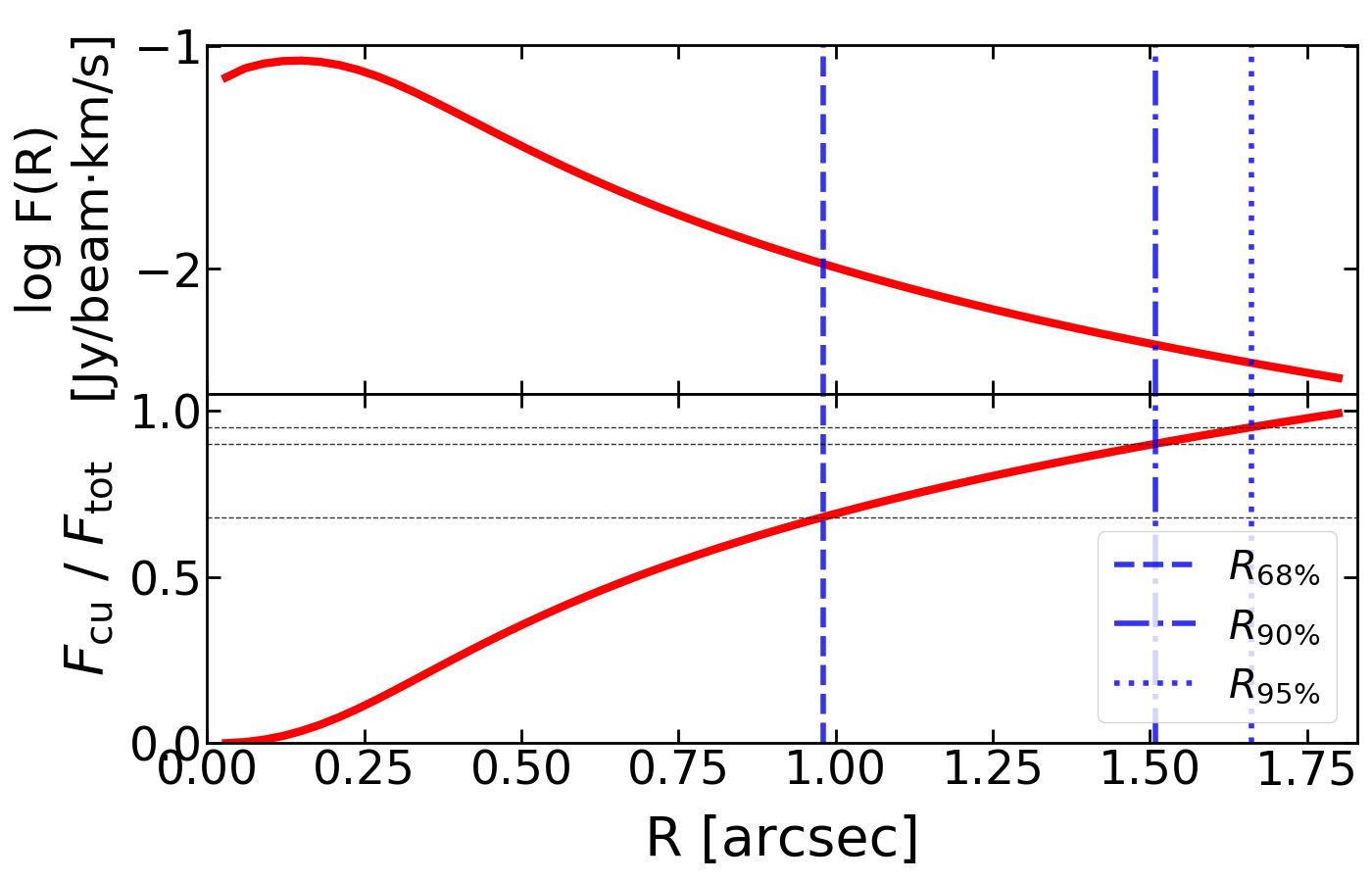}
    \end{center}
    \begin{center}
    \includegraphics[width=.490\textwidth]{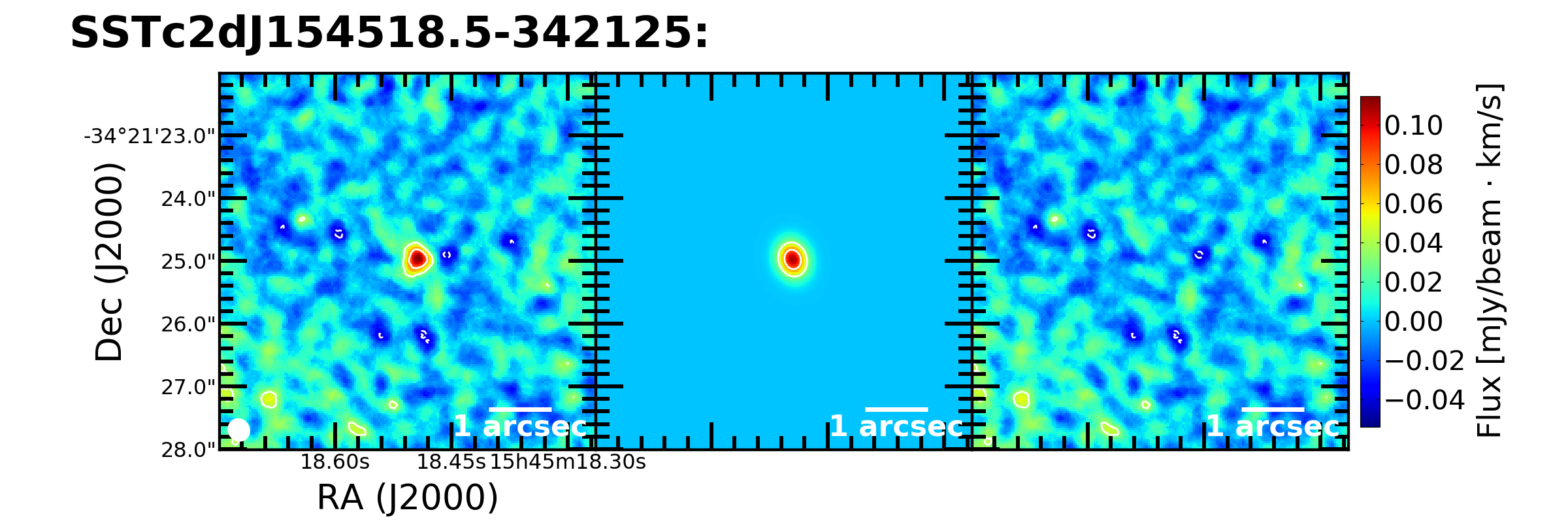}
    \includegraphics[width=.225\textwidth]{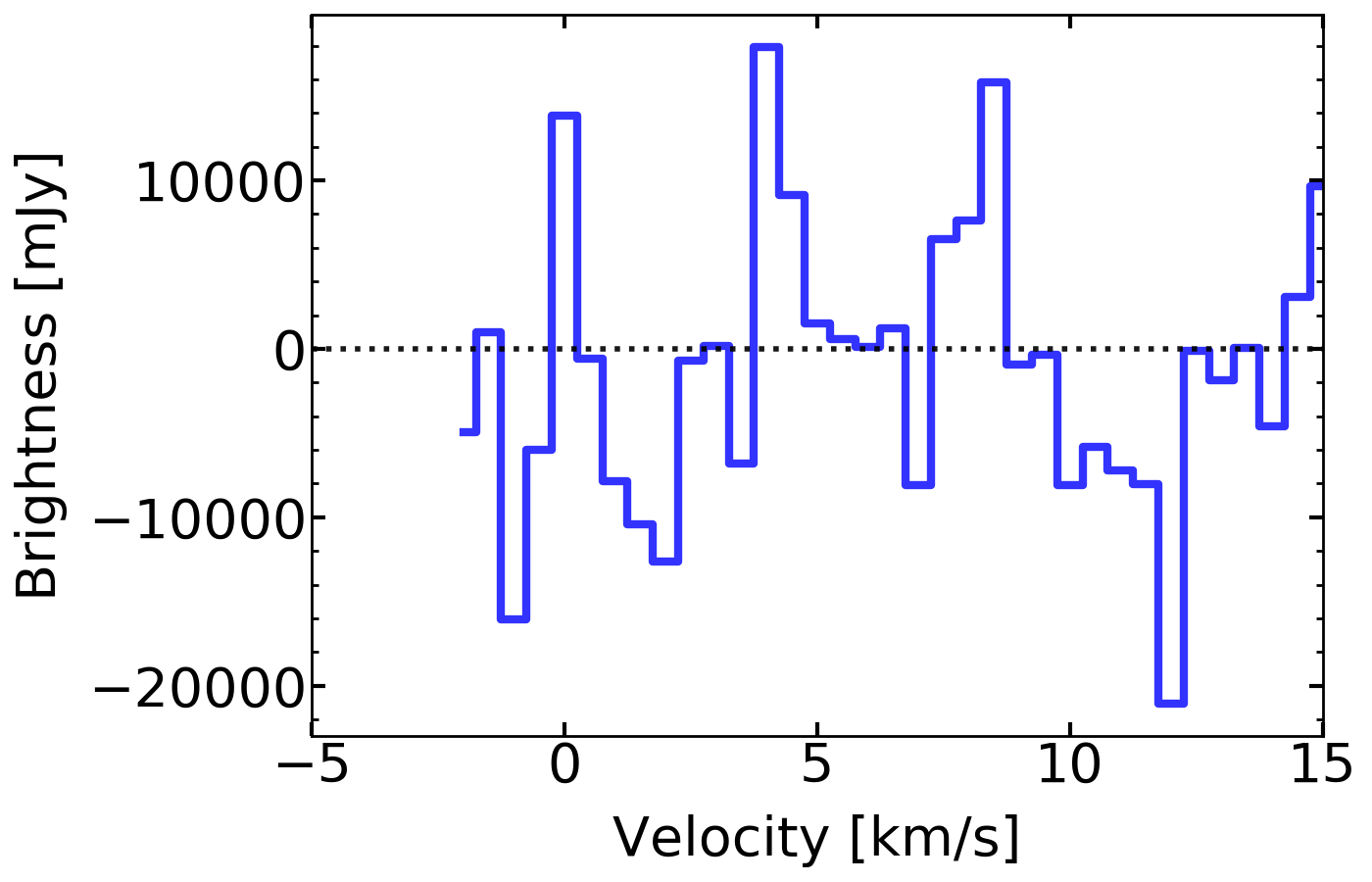}
    \includegraphics[width=.235\textwidth]{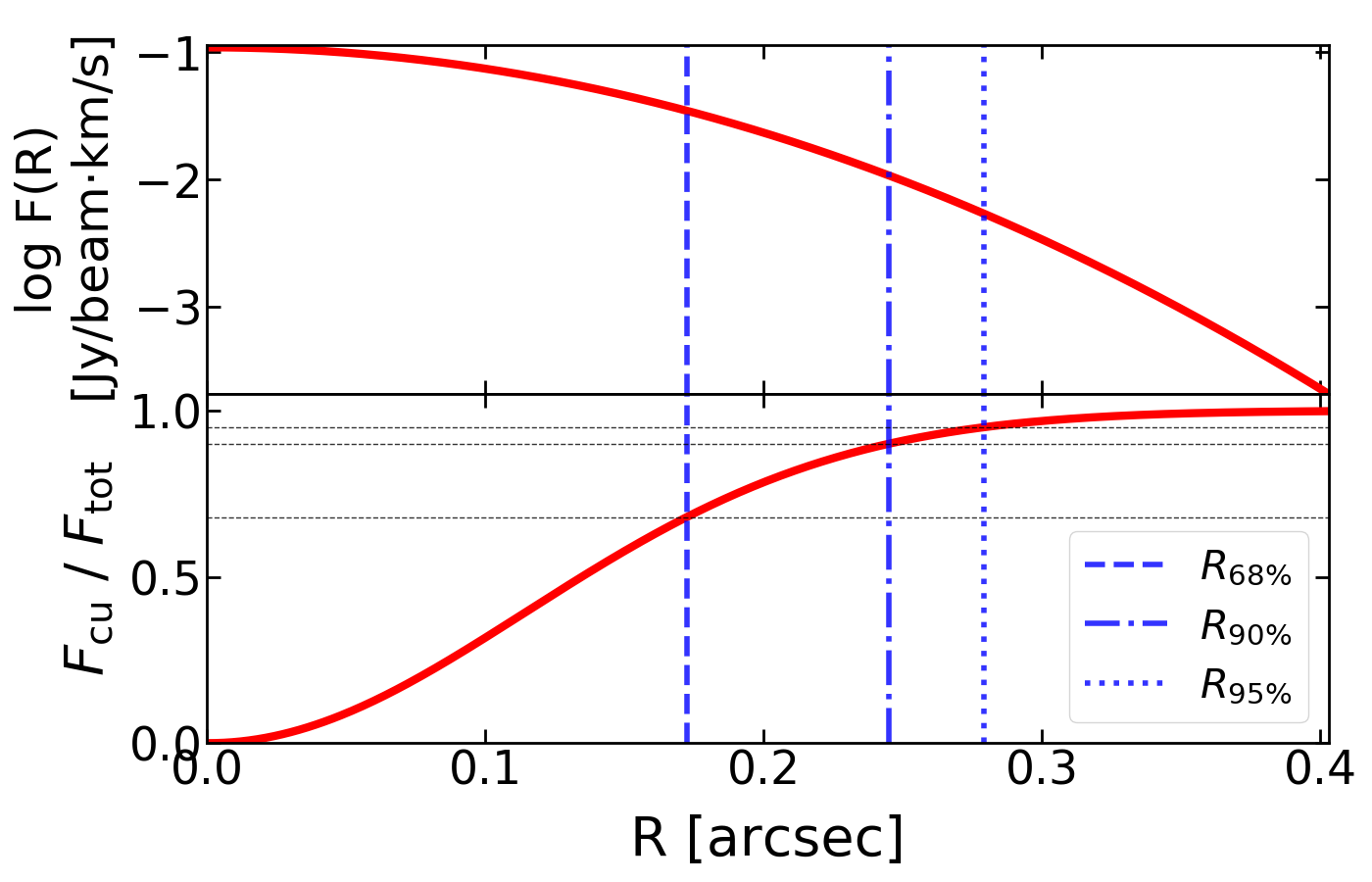}
    \end{center}
  \caption[]{
  Results of the CO modeling for every disk with measured CO size, following the methodology described in Section~\ref{sec:gasmodeling}. For each disk, the first three sub-panels show the observed, model and residual CO moment zero maps; solid (dashed) line contours are drawn at increasing (decreasing) $3\sigma$ intervals. The forth sub-panel represents the integrated spectrum enclosed by the $R_{68\%}^{\mathrm{CO}}$ of the source. Last sub-panel shows the radial brightness profile and the respective cumulative distribution of the CO model.
  }
  \label{fig:comodelresults_all_6}
\end{figure*}

\section{Singular objects}\label{sec:appendix_singularobjects}
In this Section we discuss unusual systems in the Lupus sample, namely, outliers of the size ratio distribution, or those with particular known features. 

\subsubsection{Sz~75}\label{sec:sz75}
The Sz~75 system (or GQ~Lup) has a central star of $0.8$ $M_{\odot}$, a sub-stellar companion at an angular separation of $\sim 0.7^{\prime\prime}$ \citep{neuhauser+2005} of uncertain mass \cite[likely in the BD regime,][]{seifahrt+2007,neuhauser+2008,lavigne+2009}, and a second companion candidate at $16^{\prime\prime}$ of $\sim 0.15$ $M_{\odot}$ that is likely gravitationally bound to the central object \citep{alcala+2020}. The first companion is within the pointing of the ALMA observations, but no $\ce{^{12}CO}$ or continuum emission is detected around it. The second companion candidate falls outside the ALMA pointing, however, its disk is detected in archival \textit{HST} and \textit{WISE} data \citep{lazzoni+2020}. 

The disk of the central star has the largest size-ratio of the entire Lupus population (size ratio of $\sim 8$). Previous ALMA observations \citep{macgregor+2017,long+2020}, yielded consistent results of the $\ce{^{12}CO}$ (3-2) and dust continuum extent. The compact continuum emission together with the large size ratio confirms that radial drift has been particularly efficient in the disk around the central star. Whether the presence of companions has boosted the radial drift process is unknown, follow-up studies on this system are needed in order to address this question. 

The CO disk around the central star (with $R_{95\%}^{\mathrm{CO}}$ = $1.7^{\prime\prime}$) extents beyond the deprojected distance between central star and the sub-stellar companion, considering the companion's orbital inclination to be $\sim{60}^{\circ}$ as suggested by \citet{schwarz+2016}. This is contrary to expected tidal truncation effects \citep[e.g.,][]{MartinLubow2011,bate2018}, possibly due to the sub-stellar nature of this companion. Additionally, the CO channel maps do not show any distortion due to the presence of the sub-stellar object. 

The formation mechanism of the system is unclear. A possible scenario, suggested by \cite{alcala+2020}, is that the central star and the second companion might be formed by fragmentation of a turbulent core.  Likewise, the formation of the primary companion is uncertain \citep{macgregor+2017}: if due to fragmentation of the circumprimary disk, it would result in a relatively massive disk around the sub-stellar companion; while formation close the the central star and posterior scattering outwards is a less favored explanation \citep{bryan+2016}.

\subsubsection{Sz~83}\label{sec:sz83}
This source (also known as RU~Lup) is one of the most active young stars of the Lupus clouds \citep{comeron2008}. The CO/dust size ratio measured is very large (of $\simeq5$). However, the result should be taken with caution, since the structure of this disk is highly complex, with outflows, jets and mild cloud contamination \citep{herczeg+2005,ansdell+2018,andrews+2018dsharp}. 

Recent high angular resolution observations of the $\ce{^{12}CO}$, $\ce{^{13}CO}$, $\ce{C^{18}CO}$ and $\ce{DCO^{+}}$ lines showed an intricate structure of the gas in this system, with a central keplerian disk, an extended diffuse emission, spiral arms, and various 'clumps' of emission \citep{huang+2020}. The CO size reported in that work distinguishes between a keplerian disk of $\sim{0.75}^{\prime\prime}$, non-keplerian CO emission of $\sim{1.6}^{\prime\prime}$, spiral structure up to $\sim{6}^{\prime\prime}$, and 'clumps' further out. Although they did not report values of $R_{68\%}$, our $R_{95\%}$ is slightly larger than the non-keplerian emission reported in \citep{huang+2020}. Since our measurement considers any detected emission (from the $\ce{^{12}CO}$ $J = 2-1$ line observed from the ALMA Band~6 Lupus disk survey), our size is likely accounting part of the spiral structure. 

However, the difference between the gas and the dust size is expected to be large, based on the recent studies of the system \cite[e.g.,][]{huang+2020},  and, in addition, our dust size coincides with previous works \citep[e.g.,][]{andrews2018A,hendler+2020}. For consistency, we use our inferred CO size along this manuscript, but we warn that, due to the complex structure of this system, the true extent of the CO disk might differ with respect our tabulated values.

\subsubsection{Sz~131}\label{sec:sz131}
This is a single star system ($0.3$ $M_{\odot}$ mass), and has as well a very high size ratio ($\sim 7.2$), although with high uncertainty. The large size ratio is driven by the very compact continuum emission, of only $0.08^{\prime \prime}$. The high uncertainty is caused by the very faint emission in both $\ce{^{12}CO}$ and continuum, resulting in large error bars of the $R_{\mathrm{CO}}$ and $R_{\mathrm{dust}}$ sizes. Even considering the upper and lower bounds of the dust and CO sizes respectively, the ratio lays above 4, setting a strong evidence for dust evolution and radial drift.

\subsubsection{Sz~111}\label{sec:sz111}
The Sz~111 system is another single star (of $0.5$ $M_{\odot}$ mass), with a bright disk in both continuum and $\ce{^{12}CO}$. Its size ratio is slightly above the 4 threshold, dominated by the CO disk extent, which is among the largest CO disks of the Lupus population ($R_{68\%}^{\mathrm{CO}}$ = $2.1\pm0.4^{\prime \prime}$). Due to its extended emission in both CO and dust continuum, this system is a good candidate for future ALMA observations at higher resolution and sensitivity to better constrain the radial drift, and to resolve possible substructures as the aftermath of dust evolution.

\subsubsection{Sz~69}\label{sec:sz69}
Sz~69 (HW~Lup) is part of a wide visual binary together with 2MASS J15451720-3418337, a source in Southwest direction at $\sim6.6^{\prime \prime}$ of angular separation \citep{merin+2008}. The second element of the binary is undetected in dust continuum nor in CO.

The dusty disk around Sz~69 is extremely compact and partially unresolved, with a $R_{68\%}^{\mathrm{dust}} < 0.09^{\prime \prime}$. Due to its compact continuum emission, the CO/dust size ratio is particularly high, with a lower bound value of 5.7. Using the dust disk size from \cite{andrews2018A}, the size ratio would be 9.3. Such a high value of the size ratio points towards extremely efficient radial drift. The angular separation between the binary components is larger than the distances at which dynamical interactions would typically alter the circumprimary disk  \citep[e.g.,][]{jensen+1996,bate2000,harris+2012}. 
A better characterization of the second source is necessary in order to understand this system in more detail.

\subsubsection{Sz~65}\label{sec:sz65}
The disk around Sz~65 ($0.7$ $M_{\odot}$) has a size ratio of 4.8, it is therefore another disk with ratio above the threshold value of 4. It forms a binary system together with Sz~66 ($0.3$ $M_{\odot}$), angular separation of $\sim 6.4^{\prime \prime}$. The disk around the second component is very faint in both $\ce{^{12}CO}$ and dust continuum (with a size ratio of about 2.5), the size of the CO disk could not be constrained due to its compactness.

This is another multiple system in which the primary element has a very large size ratio, this case is particularly appealing since CO and dust are detected in the two components. Observations at higher sensitivity and resolution of the two components will greatly help understanding the level of dust evolution of the two disks. Intriguingly, Sz~65 accretion is weak and considered to be only an upper limit \citep[its excess emission is close to the chromospheric levels,][]{alcala+2017}, while accretion in Sz~66 is slightly above the known correlation with the continuum flux \citep[see, e.g.,][]{manara+2016, alcala+2017,sanchis+2020a}.

\subsubsection{Sz~82}\label{sec:sz82}
Sz~82 (IM~Lup) is the brightest object of the entire Lupus disk population, both in $\ce{^{12}CO}$ and in dust continuum emission. It is therefore among the most studied protoplanetary disks. This disk is exceptionally large, its azimuthally averaged emission shows a plateau of emission that extents up to $\gtrsim1000$ AU. Extensive modeling of several CO lines at scales <450 was performed in \cite{pinte+2018a}, and suggested that UV photo-desorption from the interstellar radiation field could explain the further out CO emission. In \cite{cleeves+2016}, they discussed the possible origin of this diffuse emission: it could be the remnant of an envelope, a plausible explanation since the system is young \citep[$\leq 1$ Myr][]{mawet+2012}; gravitationally captured gas is another viable scenario, since the system has a Bondi radius of $\sim 3000$ AU, far beyond the diffuse emission extent; foreground emission is less likely since there is an offset between the cloud velocity and the object.

\subsubsection{Brown dwarfs/very low-mass stars}\label{sec:bdsvlms}
We would like to point out the results in the very low-mass range of the Lupus disk population. For that, we consider only disks orbiting around objects of masses below $0.2$ $M_{\odot}$. Several of these objects were targeted in a separate survey \citep[][]{sanchis+2020a}, and are of particular interest since their formation and evolution might differ from disks around more massive stars. Six disks in this mass range have inferred CO and dust sizes (Sz~84, Sz~100, Sz~114, SSTc2d J160703.9-391112, 2MASS J16081497-3857145, and SSTc2d J154518.5-342125). 
From theoretical modeling of the gas/dust sizes \citep{trapman+2019}, the measured sizes inferred from the emission extent can be affected by the resolution of the observations, and in objects whose emission extent is of the order of the beam size, the size ratios inferred could be lower than the true size ratios.

The median size ratio of disks around very low-mass stars is 3.1 (with a dispersion of 0.4), larger than the mean value of the entire population. If the low resolution/beam size effect is indeed affecting the inferred CO and dust sizes of these compact sources, the true size ratio may be even larger than the observed values. Due to the low number of sources in this range is not yet possible to address whether radial drift in disks around this mass range is more efficient than in disks around more massive stars, as predicted theoretically in \cite{pinilla+2013}. It is also worth noting that the only BD in the sample with inferred sizes has a ratio of 3.2, higher than the median of the entire disk population. Due to the faint and compact emission of this source, its size uncertainties are relatively large, thus it is difficult to address the true behavior of this disk. Follow-up observations at higher sensitivity and resolution for this source will allow us to better constrain its gas/dust size ratio.

\end{appendix}

\end{document}